\documentclass[12pt,a4paper,notitlepage,twoside,dvips,openright]{report}
\usepackage{epsf,rotate}
\usepackage{amsmath,amstext,amsbsy,amsopn,amsfonts,amssymb}
\usepackage[dvips]{graphicx}
\usepackage[latin1]{inputenc}
\usepackage[english]{babel}
\usepackage[T1]{fontenc}
\usepackage{feynmf}
\usepackage{geometry}
\usepackage{psfrag}
\newcommand{\eval}[2][\right]{\relax
  \ifx#1\right\relax \left.\fi#2#1\rvert}

\newcommand{\Log}{\mbox{Log}}

\newcommand{\RRe}{\mbox{Re}}
\newcommand{\IIm}{\mbox{Im}}

\newcommand{\eqdef}{\mbox{$ \stackrel{\rm def}{=}$}}
\newcommand{\as}{\ifmmode\alpha_{\rm s}\else{$\alpha_{\rm s}$}\fi}
\newcommand{\asbar}{\ifmmode\wbar {\alpha}_{\rm s}\else{$\wbar{\alpha}_{\rm s}$}\fi}

\psfrag{nu_h}[cc][bc]{$\mbox{\large$\nu_h$}$}
\psfrag{-E_3/4}[cc][cc]{$-E_3/4$}
\psfrag{-E_4/4}[cc][cc]{$-E_4/4$}
\psfrag{-E_N/4}[cc][cc]{$-E_N/4$}
\psfrag{q4/q2}[cc][cc]{$q_{_{\scriptstyle 4}}/q_{_{\scriptstyle 2}}$}
\psfrag{N}[cc][bc]{$N$}
\psfrag{qN/q2}[cc][cc]{$|q_{_{\scriptstyle N}}|/q_{_{\scriptstyle 2}}$}

\psfrag{Re(q4^(1/4))}[cc][bc]{$\Re[q_4^{1/4}]$}
\psfrag{Im(q4^(1/4))}[cc][cc]{$\Im[q_4^{1/4}]$}

\psfrag{Im(q3^(1/3))}[cc][cc]{$\Im[q_3^{1/3}]$}
\psfrag{Re(q3^(1/3))}[cc][bc]{$\Re[q_3^{1/3}]$}

\psfrag{Im[q3^(1/3)]}[cc][cc]{$\Im[q_3^{1/3}]$}
\psfrag{Re[q3^(1/3)]}[cc][bc]{$\Re[q_3^{1/3}]$}

\psfrag{a1}[cc][cc]{$\alpha_1$}
\psfrag{a2}[cc][bc]{$\alpha_{N-2}$}
\psfrag{b1}[cc][bc]{$\beta_1$}
\psfrag{b2}[cc][cc]{$\beta_{N-2}$}
\psfrag{g}[tc][br]{$\gamma_{P_0^-P_0^+}$}
\psfrag{d1}[cc][cc]{$\sigma_1$}
\psfrag{d2}[cc][cc]{$\sigma_2$}
\psfrag{d3}[cc][cc]{$\sigma_3$}
\psfrag{d4}[cc][cc]{$\sigma_{2N-4}$}
\psfrag{d5}[cc][cc]{$\sigma_{2N-3}$}
\psfrag{d6}[cc][cc]{$\sigma_{2N-2}$}
\psfrag{zero}[cc][cc]{$0$}

 \catcode`\@=11
\def \be  {\begin{equation}}
\def \ee  {\end{equation}}
\def \ba  {\begin{eqnarray}}
\def \ea  {\end{eqnarray}}
\def \baa {\begin{eqnarray*}}
\def \eaa {\end{eqnarray*}}
\def \bb  {\begin {thebibliography} }
\def \eb  {\end{thebibliography}}

\def \lab #1 {\label{#1}}
\newcommand \ci [1] {\cite{#1}}

\newcommand\re[1]{(\ref{#1})}
\def \qqquad {\qquad\quad}

\def \matrix #1 {\left(\begin{array}{cc} #1 \end{array}\right)}

\def \tr {\mathop{\rm tr}\nolimits}
\def \Im {\mathop{\rm Im}\nolimits}
\def \Re {\mathop{\rm Re}\nolimits}

\def \e  {\mathop{\rm e}\nolimits}
\def \B  {\mathop{\rm B}\nolimits}
\def \Mod  {\mathop{\rm mod}\nolimits}
\newcommand\lr[1]{{\left({#1}\right)}}

\newcommand \wbar [1] {\overline{#1}}

\newcommand \vev [1] {\langle{#1}\rangle}

\newcommand \ket [1] {|{#1}\rangle}

\font\cmss=cmss12 
\def\inbar{\,\vrule height1.5ex width.4pt depth0pt}
\def\IC{\relax\hbox{$\inbar\kern-.3em{\rm C}$}}
\def\IZ{\relax{\hbox{\cmss Z\kern-.4em Z}}}
\def\IR{{\hbox{{\rm I}\kern-.2em\hbox{\rm R}}}}

\def\IP{{\hbox{{\rm I}\kern-.2em\hbox{\rm P}}}}
\def\II{\hbox{{1}\kern-.25em\hbox{l}}}

\newcommand \Mybf[1] {\mbox{\boldmath$ {#1} $}}
\newcommand \mybf[1] {\mbox{\boldmath$ {#1} $}}

\newcommand \ot[1] {\Hat{t}_{#1}}
\newcommand \otb[1] {\Hat{\wbar{t}}_{#1}}
\newcommand \oq[1] {\Hat{q}_{#1}}
\newcommand \oqb[1] {\Hat{\wbar{q}}_{#1}}
\newcommand \sq[1] {\widetilde{q}_{#1}}

\newcommand \st[1] {\widetilde{t}_{#1}}

\def\numberbysection{\@addtoreset{equation}{section}
                     \def\theequation{\thesection\arabic{equation}}}

\begin{document}
\newpage
\thispagestyle{empty}
\begin{center}
\large{\textsc{Uniwersytet Jagiello\'nski w Krakowie}}
\end{center}
\begin{center}
\large{\textsc{Instytut Fizyki im.~Mariana Smoluchowskiego}}
\end{center}
\vspace{0.5cm}
\begin{center}
\includegraphics[scale=0.6]{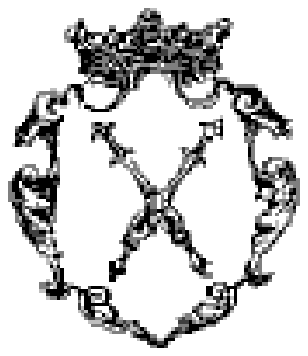}
\end{center}
\vspace{0.5cm}
\begin{center}
\LARGE{\textbf{Stany zreggeizowanych gluon\'ow w chromodynamice kwantowej}}
\end{center}
\vspace{0.5cm}
\begin{center}
\large{\textbf{JAN KOTA{\'N}SKI}}
\end{center}
\vspace{1.5cm}
\begin{center}
\large{PRACA DOKTORSKA WYKONANA POD KIERUNKIEM 
PROF.~DR~HAB.~MICHA{\L}A PRASZA{\L}OWICZA}
\end{center}
\vspace{3cm}
\vspace{3cm}
\begin{center}
KRAK\'OW, MAJ 2005
\end{center}

\newpage
\thispagestyle{empty}
$ $
\newpage
\thispagestyle{empty}
\begin{center}

\large{\textsc{Jagiellonian University in Krak{\'o}w}}
\end{center}
\begin{center}

\large{\textsc{Marian Smoluchowski  Institute of Physics}}
\end{center}
\vspace{0.5cm}
\begin{center}
\includegraphics[scale=0.6]{ujlogo.ps}
\end{center}
\vspace{0.5cm}
\begin{center}
\LARGE{\textbf{Reggeized gluon states in Quantum Chromodynamics}}
\end{center}
\vspace{0.5cm}
\begin{center}
\large{\textbf{JAN KOTA{\'N}SKI}}
\end{center}
\vspace{1.5cm}
\begin{center}
\large{DOCTORAL DISSERTATION PERFORMED UNDER GUIDANCE OF
PROF.~DR~HAB.~MICHA{\L} PRASZA{\L}OWICZ }
\end{center}
\vspace{3cm}
\vspace{3cm}
\begin{center}
KRAK{\'O}W, MAY 2005
\end{center}
\newpage
\thispagestyle{empty}
${}$



\tableofcontents

\listoffigures

\listoftables

\bibliographystyle{unsrt}

\newpage
\chapter{Preface}

The aim of this work is to present a role of the reggeized gluon states
in perturbative Quantum Chromodynamics.
The reggeized gluon states, also called Reggeons, appear in 
the Regge limit where the square of the total energy $s$ is large while
the transfer of four-momentum $t$ is low and fixed.
In this limit the leading contribution to the scattering amplitude
of hadrons
is dominated by the exchange of 
intermediate particles,
Reggeons, which are the compound states 
of gluons 
\ci{gell,gell2,Gribov:1968fc,Fadin:1975cb,Bartels:1977hz,Cheng:1977gt}.

Even for
the Regge limit 
in the generalized leading logarithm approximation  
\ci{Bartels:1980pe,Kwiecinski:1980wb,Jaroszewicz:1980mq}
this problem is 
technically very complicated
due to the
non-abelian structure of QCD.
Therefore, 
in order to simplify colour factors
the 't Hooft's multi-colour limit 
\ci{'tHooft:1973jz,Lipatov:1990zb,Lipatov:1993qn} 
is performed,
in which a number of colours $N_c \to \infty$.
This causes that the Reggeon wave-functions
become the eigenstates of a Hamiltonian which
is 
equivalent to the Hamiltonian for the non-compact Heisenberg
$SL(2,\mathbb{C})$ spin magnet.
Because of this symmetry the multi-Reggeon system with $N$ particles 
is completely solvable \ci{Lipatov:1993yb,Faddeev:1994zg}.
Thus, it possesses a complete set of integrals of motion,
conformal charges $(q_2,\wbar q_2,q_3, \wbar q_3,\ldots,q_N,\wbar q_N)$.
The eigenvalues of the $SL(2,\mathbb{C})$ Hamiltonian are
also called the energies of the Reggeons.
The Schr\"odinger equation for the lowest non-trivial case,
i.e. for $N=2$ Reggeons, was formulated and solved
by Balitsky, Fadin, Kureav and Lipatov 
\ci{Balitsky:1978ic,Kuraev:1977fs,Fadin:1975cb}.
They calculated the energy of the Pomeron state
with $N=2$ Reggeons.
An integral equation for three and more Reggeons
was formulated in Refs.  
\ci{Bartels:1980pe,Kwiecinski:1980wb,Jaroszewicz:1980rw}
in 1980. 
However, it took almost twenty years to obtain
the solution for $N=3$, which corresponds to the 
QCD odderon \ci{Janik:1998xj,Gauron:1987jt,Lukaszuk:1973nt}.
Finally, the solutions for higher $N=4,\ldots,8$
recently in series of papers 
\ci{Derkachov:2002pb,Korchemsky:2001nx,Derkachov:2002wz} 
written in collaboration with S.\'E. Derkachov,
G.P. Korchemsky and A.N. Manashov.

Description of scattering amplitudes in terms of 
the reggeized gluon states is most frequently used 
in two different kinematical
regions. The first region appears in calculation of the elastic 
scattering amplitude of two heavy hadrons whose masses are
comparable. Here the amplitude
is equal to a sum over the Regge poles. They give
behaviour in the Regge limit, 
$s \to \infty$ and $t=\mbox{const}$, like $s^{\alpha(t)}$ where 
$\alpha(0)$ is called the intercept and its value
is close to $1$. 
On the other hand, the intercept is related to the
minimum of the Reggeon energy defined by the $SL(2,\mathbb{C})$
Hamiltonian.
Thus, evaluating  the spectrum of the $SL(2,\mathbb{C})$ 
XXX Heisenberg model
we can calculate the behaviour of the hadron scattering amplitudes.

The second kinematical region is the region of deep inelastic
scattering (DIS) where a virtual photon
$\gamma^*(Q^2)$ is scattered off a (polarized) hadron with 
the mass $M^2=p_\mu^2$, such that
$\Lambda_{\rm QCD}^2\ll M^2\ll Q^2$
with $x=Q^2/2(p\cdot q)$ and $q_\mu^2=-Q^2$.
In this case the moments of the 
structure function $F_2(x,Q^2)$ can be expanded (OPE)
in a power series in $1/Q^2$ where the exponents of 
this expansion are related to the twist of Reggeons
and anomalous dimensions of QCD.
Analysing the spectrum of the $SL(2,\mathbb{C})$
Hamiltonian one can find a twist related to 
a given  Reggeon wave-function and calculate
the corresponding anomalous dimension.
This was done for the leading twist $n=2$ 
in Ref.~\ci{Jaroszewicz:1982gr}
and for other twists of two-Reggeon states in Ref. \ci{Lipatov:1985uk}.
The anomalous dimension and twists for more than two Reggeons were 
discussed
in Ref.~\ci{Korchemsky:2003rc}.

In this work 
we present the methods of constructing 
the Reggeon eigenstates. Moreover, we calculate
the reach spectrum of the the Reggeon energy
and the conformal charges $\{q_k,\wbar q_k\}$.
Finally, we calculate anomalous dimensions of QCD
and twists in operator product expansion (OPE) which 
are also provided by the Reggeon states.
In the first three Chapters we discuss the current state of knowledge.
The main results of the present work are collected in
Chapters $5$--$7$ and in Chapter $9$.
For completeness 
in Chapter $8$ we added description
of another method presented in Refs. \ci{DeVega:2001pu,deVega:2002im}.

Thus, in Chapter 2 we explain when the reggeization of the gluon appears 
\ci{Braun:1994ll}.
Next, we perform the multi-colour limit 
\ci{'tHooft:1973jz,Lipatov:1990zb,Lipatov:1993qn}, 
discuss the properties of the 
$SL(2,\mathbb{C})$ symmetry and construct invariants of this symmetry,
the conformal charges. In next Chapter  we introduce 
a Baxter
$Q-$operator method \ci{Baxter} with
Baxter equations which allows us to solve the Reggeon system, completely.
Chapter 4 contains the quasi-classical solution for the Baxter equations 
\ci{Derkachov:2002pb}.
It gives us the WKB approximation for the spectra of the Reggeon energy
and the conformal charges. It also explains the structure of these spectra.

In Chapter 5 we show construction of the Reggeon eigenfunctions
which consists in
solving the the eigenequations of the conformal charges.
We systematize the knowledge about the ansatzes for eigenstates
of $\{q_k,\wbar q_k\}$ 
with an arbitrary number of Reggeons, $N$,
extend the calculations to an arbitrary complex spins $s$ and
derive differential eigenequations for the conformal charges with $N=3,4$.
Moreover, we show solutions to the  differential eigenequations
with $N=3$ and $s=0$ and resum obtained series solutions for the $q_3=0$ case.
In next Chapter we present an exact solution to the Baxter equations.
It consists in rewriting the Baxter equation into the differential
equation, which may be solved by a series method, and finding
the quantization conditions for $\{q_k,\wbar q_k\}$ 
which comes from single-valuedness of the Reggeon wave-function
and analytical properties of the Baxter functions.
The numerical results of this method for $N=2,\ldots,8$ Reggeons are shown
in Chapter 7 \ci{Korchemsky:2001nx,Derkachov:2002wz}.
In the $N=3$ case we present quantized values of $q_3$
for different Lorentz spins $n_h=0,\ldots,3$ as well as corrections to 
the WKB approximation. 
For $N=4$ we show the resemblant and winding structure
of the $q_4$ and $q_3$ spectrum and 
also  corrections to 
the WKB approximation when $q_3=0$.
At the end we discuss the ground states
with $N=2,\ldots, 8$ Reggeons and describe their properties.
In Chapter $8$ we review another approach to the Baxter equation
which was presented in Refs. \ci{DeVega:2001pu,deVega:2002im}.
We discuss the advantages and disadvantages 
of this method and
compare it with the results
presented in previous Chapters.
Finally, in Chapter 9 we concentrate on deep inelastic scattering processes.
We calculate anomalous dimensions of QCD and twists 
coming from Reggeized gluon states \ci{Korchemsky:2003rc}. 
One can obtain them by performing an analytical
continuation of the Reggeon energy into the complex space
of the scaling dimension $\nu_h$ and  
analysing the pole structure of the energy.
At the end we make final conclusions.

\newpage
\chapter{Introduction}
\section{Hamiltonian for the $N$-Reggeon states}
\subsection{The gluon reggeization}

The Regge limit
can be conveniently illustrated by considering 
the elastic scattering amplitude of two hadrons
\unitlength=1mm
\begin{center}
\begin{fmffile}{elamp}
\begin{eqnarray}{
\begin{fmfgraph*}(30,15)
 \fmfstraight 
 \fmfleft{i1,i2}
 \fmfright{o1,o2}
 \fmf{plain_arrow}{i1,vp}
 \fmf{plain_arrow}{i2,vp}
 \fmf{plain_arrow}{vp,o1}
 \fmf{plain_arrow}{vp,o2}
 \fmffreeze
\fmfv{label=$\longrightarrow s$,label.dist=-1.w,
decor.shape=circle, decor.filled=shaded, decor.size=.4w}{vp}
\fmfv{label=$p_2$}{i1}
\fmfv{label=$p_1$}{i2}
\fmfv{label=$p_2'$}{o1}
\fmfv{label=$p_1'$}{o2}
\end{fmfgraph*}
} &{ \;\;\; \;\;\; \;\;\; \mbox{with} \;\; 
s=(p_1+p_2)^2, \;\; \mbox{and} \;\; t=(p_1-p_1')^2,
}
\lab{eq:scatpr}
\end{eqnarray}
\end{fmffile}
\end{center} 
in the kinematical region:
\begin{equation}
s\rightarrow \infty, \;\; t=\mbox{const}\,.
\lab{eq:rlim}
\end{equation} 

In this case the largest contribution to 
the scattering amplitude is provided by
a multi-gluon exchange. 
It was shown in the papers 
\ci{Gribov:1968fc,
Fadin:1975cb,Bartels:1977hz,Cheng:1977gt,
Kuraev:1977fs,Balitsky:1978ic,Kuraev:1976ge} 
that in the limit (\ref{eq:rlim})
so called gluon reggeization occurs.
It means that the contribution to the scattering amplitude 
may be written as a sum of ladder diagrams,
where compound states of the reggeized gluons, called Reggeons, 
are exchanged in the $t-$channel and interact with each other.

\begin{figure}[!h]
\begin{fmffile}{pczynnyg6}
\begin{displaymath}
\begin{array}{ccccc}
\begin{fmfgraph*}(20,15)
\fmfstraight
  \fmfleft{i3}
  \fmfbottom{o1,b1,b2,b3,i1}
  \fmftop{o2,t1,t2,t3,i2}
  \fmf{boson}{i1,o1}
  \fmf{boson}{i2,o2}
  \fmf{boson}{t1,b1}
  \fmf{boson}{t3,b3}
  \fmf{dots}{t2,b2}
  \fmfdot{b1}
  \fmfdot{b3}
  \fmfdot{t1}
  \fmfdot{t3}
\end{fmfgraph*}
 \;\;\; & \;\;\;   
\begin{fmfgraph*}(20,15)
  \fmfstraight
  \fmfleft{i3}
  \fmfbottom{o1,b1,b2,b3,b4,i1}
  \fmftop{o2,t1,t2,t3,t4,i2}
  \fmf{boson}{i1,o1}
  \fmf{boson}{i2,o2}
  \fmf{boson}{t1,b1}
  \fmf{boson}{t2,b2}
  \fmf{dots}{t3,b3}
  \fmf{boson}{t4,b4}
  \fmfdot{b1}
  \fmfdot{b2}
  \fmfdot{b4}
  \fmfdot{t1}
  \fmfdot{t2}
  \fmfdot{t4}
\end{fmfgraph*}
&
\begin{fmfgraph*}(20,15)
  \fmfstraight
  \fmfleft{i3}
  \fmfbottom{o1,b1,b2,b3,b4,i1}
  \fmftop{o2,t1,t2,t3,t4,i2}
  \fmf{boson}{i1,o1}
  \fmf{boson}{i2,o2}
  \fmf{boson}{t2,b1}
  \fmf{boson}{t1,b2}
  \fmf{dots}{t3,b3}
  \fmf{boson}{t4,b4}
  \fmfdot{b1}
  \fmfdot{b2}
  \fmfdot{b4}
  \fmfdot{t1}
  \fmfdot{t2}
  \fmfdot{t4}
\end{fmfgraph*}
\;\;\; &  \;\;\;
\begin{fmfgraph*}(20,15)
  \fmfstraight
  \fmfleft{i3}
  \fmfbottom{o1,b1,b2,b3,i1}
  \fmftop{o2,t1,t2,t3,i2}
  \fmf{boson}{i1,o1}
  \fmf{boson}{i2,o2}
  \fmf{boson}{t1,v2,b1}
  \fmf{boson}{t3,v1,b3}
  \fmf{boson,tension=0}{v1,v2}
  \fmf{dots}{t2,b2}
  \fmfdot{b1}
  \fmfdot{b3}
  \fmfdot{v1}
  \fmfdot{t1}
  \fmfdot{t3}
  \fmfdot{v2}
\end{fmfgraph*}
  &  
\begin{fmfgraph*}(20,15)
\fmfstraight
  \fmfleft{i3}
  \fmfbottom{o1,b1,b2,b3,b4,i1}
  \fmftop{o2,t1,t2,t3,t4,i2}
  \fmf{boson}{i1,o1}
  \fmf{boson}{i2,o2}
  \fmf{boson}{t1,v1,b1}
  \fmffreeze
  \fmf{boson,left=0.35,tension=1}{t3,v1}
  \fmf{dots}{t2,b2}
  \fmf{boson}{t4,b4}
  \fmfdot{b1}
  \fmfdot{t3}
  \fmfdot{b4}
  \fmfdot{t1}
  \fmfdot{v1}
  \fmfdot{t4}
\end{fmfgraph*}
\\
\mbox{a) Born approx.}& 
\multicolumn{2}{c}{\mbox{b) composite virtual and Born diagrams }}
&
\multicolumn{2}{c}{\mbox{c) gluon emission}} 
\\
\end{array}
\end{displaymath}
\end{fmffile}
\lab{rys:pczynny6}
\caption[Some diagrams 
which give contribution to the total cross section $\sigma_{2\rightarrow 2}$]{Some diagrams of the order $g^4$ and $g^6$
which give contribution to the total cross section $\sigma_{2\rightarrow 2}$.}
\end{figure}
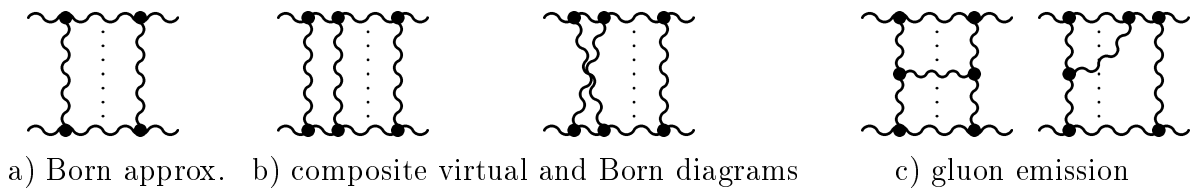

In order to illustrate what the gluon reggeization is,
we show in Figure~1
some diagrams of the order $g^4$ and $g^6$
which give contribution to the total cross section $\sigma_{2\rightarrow 2}$  
for two scattered gluons. 
The Born approximation diagram, Figure~1.a, appears
in the lowest order of the perturbative calculation. 
Next diagrams, Figure 1.b and 1.c, are of the order $g^6$.
They include virtual contributions and real 
gluon emissions.

The non-triviality of the virtual diagrams due to 
the colour factor causes modification
of the gluon propagator. This phenomenon is called gluon reggeization.
A leading contribution in $\ln s$
coming from the diagrams of the 1.b type, i.e.\ of the order $g^6$,
leads effectively to exchange of the gluon propagator
\begin{equation}
\frac{1}{t}\rightarrow\frac{\omega(t) \ln s}{t}\,,
\;\;\;\;\;\; \mbox{where} \;\;\;
\omega(t)=3 \alpha_s t \int \frac{d^2 l}{(2 \pi)^2}
\frac{1}{l^2 (\sqrt{-t}-l)^2}.
\lab{eq:omt}
\end{equation}
Note that $\omega(t)$ in (\ref{eq:omt}) is infrared divergent.
Here we perform regularization adding a small gluon mass 
\ci{Fadin:1975cb,Braun:1994ll}.
Summation of the leading terms obtained from the virtual diagrams
gives 
\begin{equation}
\frac{s^{\omega(t)}}{t}.
\end{equation}

Similarly the real gluon diagrams from Figure 1.c
lead to the modification of vertices responsible for the gluon emission.
The $j$-th vertex give contribution 
\vspace{-0.5cm}
\begin{center}
\begin{fmffile}{ver1}
\begin{eqnarray}{
g T_{i_{j+1} i_j}^{d_j} (e_j F(q_{j+1},q_j))\,,
\;\;\;\;\;\;\;\;\;\;\;\;
\;\;\;\;\;\;\;\;\;\;\;\;
} & {
\begin{fmfgraph*}(15,15)
 \fmfstraight 
 \fmfleft{i1,i2}
 \fmfright{o1}
 \fmf{dbl_wiggly}{i1,v1,i2}
 \fmffreeze
 \fmf{wiggly}{o1,v1}
 \fmfv{decor.shape=circle,decor.filled=hatched,decor.size=5thick}{v1}
 \fmfv{label=$q_{j+1} \;\; i_{j+1} $}{i1}
 \fmfv{label=$q_{j} \;\; i_{j}$}{i2}
 \fmfv{label=$k_j \;\; e_{j} \;\; d_j$}{o1}
\end{fmfgraph*}
}
\end{eqnarray}
\end{fmffile}
\end{center}
where  $F(q_{j+1},q_j)$ is a function defined  for example in Ref. 
\ci{Braun:1994ll},
where $T$ is the gluon colour matrix.  
These vertices with the emitted gluons serve to build the ladder rungs.

In the ladder diagrams there are also Reggeon couplings 
with two scattering particles. 
For gluons they look like

\vspace{-0.5cm}
\begin{center}
\begin{fmffile}{ver2}
\begin{eqnarray}{
2 p_1 g \delta_{\lambda_1^{} \lambda_1'} T_{a_1' a_1^{}}^{i_1} 
\;\;\;
\begin{fmfgraph*}(15,15)
 \fmfstraight 
 \fmftop{t1,t2}
 \fmfbottom{b1}
 \fmf{wiggly}{t1,v1,t2}
 \fmffreeze
 \fmf{dbl_wiggly}{b1,v1}
 \fmfv{decor.shape=circle,decor.filled=full,decor.size=5thick}{v1}
 \fmfv{label=$a_1'$}{t1}
 \fmfv{label=$a_1$}{t2}
 \fmfv{label=$i_1$}{b1}
\end{fmfgraph*}
} & 
{ \qquad \qquad
2 p_2 g \delta_{\lambda_2^{} \lambda_2'} T_{a_2' a_2^{}}^{i_{n+1}},
\;\;\;}  & {
\begin{fmfgraph*}(15,15)
 \fmfstraight 
 \fmfbottom{b1,b2}
 \fmftop{t1}
 \fmf{wiggly}{b1,v1,b2}
 \fmffreeze
 \fmf{dbl_wiggly}{t1,v1}
 \fmfv{decor.shape=circle,decor.filled=full,decor.size=5thick}{v1}
 \fmfv{label=$a_2'$}{b1}
 \fmfv{label=$a_2$}{b2}
 \fmfv{label=$i_{n+1}$}{t1}
\end{fmfgraph*}
}
\end{eqnarray}
\end{fmffile}
\end{center}
where $\delta_{\lambda \lambda'}$ means conservation of helicity.

One can see that the absorptive part of the gluon scattering amplitude 
for two gluons may be written in the form of the ladder diagrams
summed over the intermediate $n$-gluon states
\begin{equation}
\IIm A(s,t)=\frac{1}{2} \sum_{n}\int d\tau_{n+2} 
A^{\ast}_{2\rightarrow n+2}(p_1',p_2')
A_{2\rightarrow n+2}(p_1,p_2)\,,
\lab{eq:twopt}
\end{equation}
where $d\tau_{n+2}$ is an element of the phase space for $n+2$
intermediate particles, whereas $A_{2\rightarrow2+n}$
is an amplitude represented by an effective diagram in Figure~2.
\unitlength=1mm
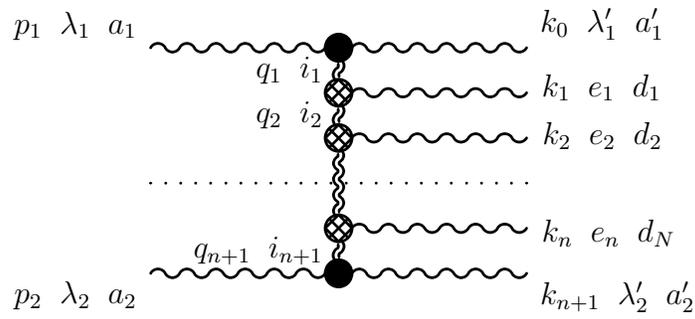
\begin{figure}[h]
\vspace{0.5cm}
\begin{center}
\begin{fmffile}{odd_diag2}
\begin{fmfgraph*}(50,30)
  \fmfstraight
  \fmfleft{i1,i3,i4,i5,i6,i2}
  \fmfright{on1,on,o3,o2,o1,o0}
  \fmfbottom{b2}
  \fmftop{t2}
  \fmf{boson}{i1,b2}
  \fmf{boson}{i2,t2}
  \fmf{dbl_wiggly,label=$q_1 \;\;  i_1$,label.side=right}{t2,vc1}
  \fmf{dbl_wiggly,label=$q_2 \;\;  i_2$,label.side=right}{vc1,vc2}
  \fmf{dbl_wiggly}{vc2,vc3}
  \fmf{dbl_wiggly}{vc3,vcn}
  \fmf{dbl_wiggly,label=$q_{n+1} \;\;  i_{n+1}$,label.side=right}{vcn,b2}
  \fmf{boson}{b2,on1}
  \fmf{boson}{t2,o0}
  \fmf{dots,tension=0}{i4,o3}
  \fmf{boson,tension=0}{vcn,on}
  \fmf{boson,tension=0}{vc2,o2}
  \fmf{boson,tension=0}{vc1,o1}
  \fmfv{label=$p_2 \;\; \lambda_2 \;\; a_2$}{i1}
  \fmfv{label=$p_1 \;\; \lambda_1\;\; a_1$}{i2}
  \fmfv{label=$k_0 \;\; \lambda_1' \;\; a_1'$}{o0}
  \fmfv{label=$k_1 \;\; e_1\;\; d_1$}{o1}
  \fmfv{label=$k_2 \;\; e_2 \;\; d_2$}{o2}
  \fmfv{label=$k_n \;\; e_n \;\; d_N$}{on}
  \fmfv{label=$k_{n+1} \;\; \lambda_2'\;\; a_2'$}{on1}
  \fmfv{decor.shape=circle,decor.filled=full,decor.size=5thick}{b2}
  \fmfv{decor.shape=circle,decor.filled=full,decor.size=5thick}{t2}
  \fmfv{decor.shape=circle,decor.filled=hatched,decor.size=5thick}{vc1}
  \fmfv{decor.shape=circle,decor.filled=hatched,decor.size=5thick}{vc2}
  \fmfv{decor.shape=circle,decor.filled=hatched,decor.size=5thick}{vcn}
\end{fmfgraph*}
\end{fmffile}
\end{center}
\lab{rys:A2n2}
\caption{Diagram symbolizing the scattering amplitude $A_{2\rightarrow n+2}$}
\end{figure}

In this diagram, momenta  $p_i$ and $k_0$, $k_{n+1}$ and also
helicities $\lambda_i$ and $\lambda_i'$, for $i=1,2$ characterize
two initial and two final states.
The remaining intermediate gluons with momenta $k_i$
numbered from  $1$ to $n$ are characterized by polarization vectors
$e_j$, where $j=1,\ldots,n$.
Similarly for the colour indices, the two initial gluons 
and the two final ones
have the colour indices $a_i$, $a_i'$ where $i=1,2$, whereas 
the indices 
of the produced gluons are denoted as $d_j$ where $j=1,\ldots,n$.
The Reggeons transfer squares of 4-momenta $t_j=q_j^2$ and possess 
colour indices $i_j$, where $j=1,\ldots,n+1$.

The total contribution of the diagram from the Figure 2 
to the scattering amplitude
has the form
\begin{multline}
A_{2\rightarrow n+2}=2 s g^{n+1} 
\delta_{\lambda_1^{ } \lambda_1'}
\delta_{\lambda_2^{ } \lambda_2'} \\
\times \sum_{i_1\ldots i_n}
 T_{a_1' a_1^{ }}^{i_1} 
\frac{s_{0,1}^{\omega(t_1)}}{t_1}
 T_{i_2 i_1}^{d_1} (e_1 F(q_2,q_1)) 
\frac{s_{1,2}^{\omega(t_2)}}{t_2}
 T_{i_3 i_2}^{d_2} (e_2 F(q_3,q_2)) \ldots
\frac{s_{n,n+1}^{\omega(t_{n+1})}}{t_{n+1}} 
  T_{a_2' a_2^{ }}^{i_{n+1}},
\lab{eq:Atree}
\end{multline}
where $s_{i,j}=(k_i+k_j)^2$.

One can decompose the momenta $k_i$ in a basis of the scattered particles
momenta 
\begin{equation}
k_i=\alpha_i p_1 + \beta_i p_2 + {k_i}_{\perp}\,.
\lab{eq:ki}
\end{equation}
In the limit (\ref{eq:rlim}) the largest contribution comes from 
the kinematical region where 
$\alpha_{j-1} \gg \alpha_j$, $\beta_{j-1} \ll \beta_j$
and small $q_j^2 \simeq {q_j^2}_{\perp}$.
Taking square of the diagram in Figure 2 one can perform integration over 
longitudinal 
degrees of freedom.
All non-trivial dynamics is then concealed in the $2$-dimensional
subspace of transverse variables.

In Eq. (\ref{eq:twopt}) one has to 
sum over polarizations and colours of intermediate particles.
The sum over polarizations gives 
\begin{equation}
\sum_{i} e^{\mu}_{\lambda_i} F_{\mu}(q_{i+1},q_i)
 e^{\nu}_{\lambda_i} F_{\nu}(q_{i+1}',q_i')=
-F_{\mu}(q_{i+1},q_i) g^{\mu \nu} F_{\nu}(q_{i+1}',q_i')
\equiv 2 K_q (q_i,q_{i+1}),
\end{equation}   
where
\begin{equation}
K_q(q_i,q_{i+1})=
q^2-\frac{q_{i \perp}^2 {q'}_{i+1 \perp}^2
+q_{i+1 \perp}^2 {q'}_{i \perp}^2}{k_{i \perp}^2}
\end{equation}
with $t = q^2$.
In order to perform the sum over colours
it is convenient
to introduce projectors $P^{(R)}$
onto irreducible representation $(R)$ composed of two colour octets
in the $t-$channel:
\begin{equation} 
(8)\otimes(8)=(1)\oplus(8)_S\oplus(8)_A
\oplus(10)\oplus(\overline{10})\oplus(27).
\lab{eq:8x8}
\end{equation}
The scattering interaction of two reggeized gluons 
may be then written as
\begin{equation}
A_{a_2^{ } a_2', a_1^{ } a_1'}(s,t)
=\sum_{(R)}P_{a_2^{ } a_2', a_1^{ } a_1'}^{(R)}A^{(R)}(s,t),
\end{equation} 
where $A^{(R)}(s,t)$ is the amplitude with specified colour state 
in the $t$ channel. For the $i$-th intermediate gluon we obtain a colour
part in the form 
\begin{equation}
\sum_{d_i} T_{b_{i+1},b_i}^{d_i}(T_{b_{i+1}',b_i'}^{d_i})^{\ast}=
-(T T')_{b^{}_{i+1} b^{}_i, b_{i+1}' b_i'}=
\sum_{R} \lambda_R 
P^{(R)}_{b^{}_{i+1} b^{}_i, b_{i+1}' b_i'}\,,
\end{equation}
where $\lambda_R$ is the colour factor.

Summarizing, the absorptive part of the scattering amplitude
for two gluons with conservation of helicities
is given as 
\ci{Bartels:1973pn,Fadin:1975cb,Kaidalov:1982xg}

\begin{multline}
\IIm A^{(R)}(s,t)=\pi \lambda_R^2 s \sum_{n=0}^{\infty}
g^{2(n+2)}\int \prod_{i=1}^{n+1}
\frac{d^2 q_i ds_{i-1,i}}{16 \pi^3}
\left(\frac{s_{i-1,i}}{s_0}\right)^{\omega(t_i)+\omega(t_i')}
\frac{1}{t^{}_i t_i'} \\
\times
\prod_{i=1}^{n} 2 K_q^{(R)}(q_i,q_{i+1})
\delta(\prod_{i=0}^{n} s_{i,i+1}-s \prod_{i=1}^{n} |k_{i\perp}^2| )\,.
\lab{eq:qcdopt}
\end{multline}

The reggeization may only occur in the non-abelian theories. In contrast to 
the quantum chromodynamics, the reggeization of particles does not appear
in such theories as quantum electrodynamics as well as scalar
field theories. These theories do possess neither 
a colour structure 
nor three-gauge-field vertices, which are responsible for gluon emission,
and as a result in these theories $\omega(t)=0$.

\subsection{Integral equation for the scattering amplitude}


Performing the Mellin transformation on (\ref{eq:qcdopt}),
the sum over $n$ and introducing new 2-dimensional variables
in the transverse subspace:
\begin{equation}
\begin{aligned}
l_1&=q_1, \;\;\quad l_2=-q_1'=q-q_1, \;\;\; \\
l_1'&=q_{n+1},\;\;\quad l_2'=-q_{n+1}'=q-q_{n+1}, \;\;\; \\
t_i&=-{l_i}^2<0, \;\;\quad t=-q^2<0  \;\;\; \mbox{for } i=1,2, 
\end{aligned}
\end{equation}
we obtain
\begin{equation}
a_j^{(R)}=
\int \frac{ds}{2 \pi i} s^{-j-1} \IIm A^{R}(s,t)=
2 g^2 \lambda_R^2 \int \frac{d^2 l_1}{16 \pi^3 t_1 t_2}
 \phi_{jq}^{(R)}(l_1),
\end{equation}
where the function $\phi_{jq}^{(R)}(l_1)$ describes the gluon ladder with two 
external reggeized gluons and with the fixed total angular momentum $j$ 
and the transferred momentum $q$.
This function satisfies the integral BFKL equation 
\ci{Balitsky:1978ic,Kuraev:1977fs,Fadin:1975cb}
\begin{equation}
(j-1-\omega(t_1)-\omega(t_2))\phi_{jq}^{(R)}(l_1)=
1+2 \alpha_s \lambda_R \int \frac{d^2 l_1'}{4 \pi^2} V_q(l_1,l_1')
\phi_{jq}^{(R)}(l_1'),
\lab{eq:lip}
\end{equation}
where  $\alpha_s=g^2/4 \pi$, and the interaction potential between
two Reggeons is
\begin{equation}
\hat{V}_{12}: \quad V_q(l_1,l_1')
=\frac{K_q (l_1, l_1')}{{l_1'}^2 {l_2'}^2}=
\left(\frac{{l_1}^2 {l_2'}^2+{l_2}^2 {l_1'}^2}{(l_1-l_1')^2}-q^2\right)
\frac{1}{{l_1'}^{2}}\frac{1}{{l_2'}^{2}}.
\lab{eq:Vik}
\end{equation}
Equation (\ref{eq:lip}) 
may be illustrated using the diagrammatic form
shown in Figure~3. 
Notice that Eq. (\ref{eq:lip}) 
does not contain the propagators of out-going gluons related to
the momenta $l_1$ and $l_2$.

\vspace{0.5cm}

\begin{figure}[h]
\begin{fmffile}{BFKL}
\begin{displaymath}
\begin{array}{ccccc}
\begin{fmfgraph*}(20,25)
  \fmfstraight
  \fmftop{t1,t2,t3,t4}
  \fmfbottom{b1,b2,b3,b4}
  \fmf{boson}{t1,t2}
  \fmf{boson}{t3,t4}
  \fmf{dbl_wiggly}{b2,v2a,v2b,v2c}
  \fmf{dbl_wiggly}{b3,v3a,v3b,v3c}
  \fmf{phantom}{t2,v2c}
  \fmf{phantom,label=$\phi$}{t3,v3c}
  \fmffreeze
  \fmf{plain,left=0.4}{t2,t3}
  \fmf{plain,left=0.4}{v3c,v2c}
  \fmf{plain,left=0.6}{t3,v3c}
  \fmf{plain,left=0.6}{v2c,t2}
  \fmfv{label=$(j-1)$}{t1}
  \fmfv{label=$l_1$}{v2b}
  \fmfv{label=$l_2$}{v3b}
\end{fmfgraph*}
  &  \;\;\;  
\begin{fmfgraph*}(20,25)
  \fmfstraight
  \fmftop{t1,t2,t3,t4}
  \fmfbottom{b1,b2,b3,b4}
  \fmf{boson}{t1,t2}
  \fmf{boson}{t3,t4}
  \fmf{dbl_wiggly}{b2,v2a,v2b,v2c}
  \fmf{dbl_wiggly}{b3,v3a,v3b,v3c}
  \fmf{phantom}{t2,v2c}
  \fmf{phantom}{t3,v3c}
  \fmffreeze
  \fmf{plain,left=0.4}{t2,t3}
  \fmf{plain,left=0.4}{v3c,v2c}
  \fmf{plain,left=0.6}{t3,v3c}
  \fmf{plain,left=0.6}{v2c,t2}
  \fmfv{label=$-$}{t1}
  \fmfv{decor.shape=circle, decor.filled=shaded, 
  decor.size=0.16w,label=$\omega(t_1)$}{v2b}
\end{fmfgraph*}
  &  \;\;\;  
\begin{fmfgraph*}(20,25)
  \fmfstraight
  \fmftop{t1,t2,t3,t4}
  \fmfbottom{b1,b2,b3,b4}
  \fmf{boson}{t1,t2}
  \fmf{boson}{t3,t4}
  \fmf{dbl_wiggly}{b2,v2a,v2b,v2c}
  \fmf{dbl_wiggly}{b3,v3a,v3b,v3c}
  \fmf{phantom}{t2,v2c}
  \fmf{phantom}{t3,v3c}
  \fmffreeze
  \fmf{plain,left=0.4}{t2,t3}
  \fmf{plain,left=0.4}{v3c,v2c}
  \fmf{plain,left=0.6}{t3,v3c}
  \fmf{plain,left=0.6}{v2c,t2}
  \fmfv{label=$-$}{t1}
  \fmfv{decor.shape=circle, decor.filled=shaded, 
  decor.size=0.16w,label=$\omega(t_2)$}{v3b}
\end{fmfgraph*}
  &  \;\;\;  
\begin{fmfgraph*}(20,25)
  \fmfstraight
  \fmftop{t1,t2,t3,t4}
  \fmfbottom{b1,b2,b3,b4}
  \fmf{boson}{t1,t2,t3,t4}
  \fmf{dbl_wiggly}{b2,v2a,v2b,v2c,t2}
  \fmf{dbl_wiggly}{b3,v3a,v3b,v3c,t3}
  \fmfv{label=$=$}{t1}
\end{fmfgraph*}
  &  \;\;\;  
\begin{fmfgraph*}(20,25)
  \fmfstraight
  \fmftop{t1,t2,t3,t4}
  \fmfbottom{b1,b2,b3,b4}
  \fmf{boson}{t1,t2}
  \fmf{boson}{t3,t4}
  \fmf{dbl_wiggly}{b2,v2a}
  \fmf{plain}{v2a,v2b}
  \fmf{dbl_wiggly,label.side=left,label=$l_1'$}{v2b,v2c}
  \fmf{dbl_wiggly}{b3,v3a}
  \fmf{plain}{v3a,v3b}
  \fmf{dbl_wiggly,label=$l_2'$}{v3b,v3c}
  \fmf{phantom}{t2,v2c}
  \fmf{phantom}{t3,v3c}
  \fmffreeze
  \fmf{plain,label=$V$}{v3a,v2a}
  \fmf{plain}{v3b,v2b}
  \fmf{plain,left=0.4}{t2,t3}
  \fmf{plain,left=0.4}{v3c,v2c}
  \fmf{plain,left=0.6}{t3,v3c}
  \fmf{plain,left=0.6}{v2c,t2}
  \fmfv{label=$+$}{t1}
\end{fmfgraph*}
\\
\end{array}
\end{displaymath}
\end{fmffile}
\lab{rys:pczynny6b}
\caption{Illustration of the BFKL equation by means of Feynman diagrams.}
\end{figure}
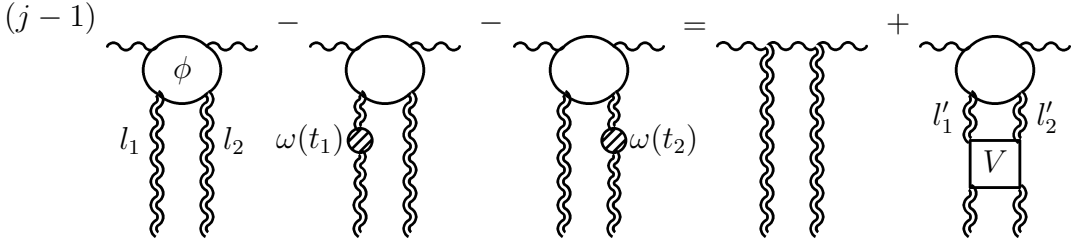

Equation  (\ref{eq:lip}) contains infrared singularities.
The Reggeons in the $t-$channel may form different colour representations
(\ref{eq:8x8}). 
For the singlet $\lambda_{R=1}=3$. In this case, for $q\ne0$ all infrared 
singularities cancel. For $q=0$,
the integrand kernel in Eq. (\ref{eq:lip}) contains singularities at
$l_1'=l_2'=0$. 
Equation (\ref{eq:lip}) describes the process with 
a gluon of momentum $p_2$ being a target.
In reality we are not interested in the scattering on a gluon, but  on 
a physical colourless particle. In this case thanks to inhomogeneity 
in Equation (\ref{eq:lip}),
which describes the Reggeon interaction with a target,
all infrared singularities disappear.  

Equation (\ref{eq:lip}) has the form of the inhomogeneous 
integral Schr\"odinger equation 
\begin{equation}
(\Hat{\cal H}_2-{\cal E})\otimes \phi=f, \;\;\; \mbox{ where }   \;\;\;
\Hat{\cal H}_2= \omega(t_1)+\omega(t_2)+2 \alpha_s \lambda_1 \Hat{V}_{12}, 
\;\;\; {\cal E}=j-1,
\lab{eq:inSch}
\end{equation}
whereas $\otimes$ 
means convolution of the Hamiltonian with the wave-function $\phi$.

Equation (\ref{eq:inSch}) describes exchange of two reggeized gluons.
It is possible to generalize it to more gluons.
The kinetic energy of the $i$-th Reggeon is equal to $\omega(t_i)$.
The potential of the interaction between two Reggeons has the form
$6 \alpha_s \Hat{V}_{12}$, where $\Hat{V}_{12}$  (\ref{eq:Vik}) 
is an integral operator. 
For $N$ Reggeons we have
\begin{equation}
(\Hat{\cal H}_N-{\cal E})\otimes \phi=f,
\lab{eq:Sch}
\end{equation}   
where the Hamiltonian, which was firstly derived in the works 
 \ci{Bartels:1980pe,Kwiecinski:1980wb,Jaroszewicz:1980rw} 
takes a form
\begin{equation}
\begin{aligned}
\Hat{\cal H}_N=&\sum_{i=1}^{N}\omega(t_i)
-2\alpha_s \sum_{i<k=2}^{N} T_i T_k \Hat{V}_{ik}\\
=-& \frac{1}{3} \sum_{i<k=2}^{N} T_i T_k
(\omega(t_i)+\omega(t_k)+6 \alpha_s \Hat{V}_{ik})
=- \frac{1}{3}\sum_{i<k=2}^{N} T_i T_k \Hat{H}_{ik}.
\lab{eq:NHam}
\end{aligned} 
\end{equation}
The interaction potential  $\Hat{V}_{ik}$
between $i$-th and $k$-th reggeized gluons
is defined by Eq. (\ref{eq:Vik}).


\subsection{Impact parameter representation}

The Hamiltonian (\ref{eq:NHam}) was defined in two-dimensional
space of transverse Reggeon momenta. It appears that it is 
more convenient to perform calculations using variables in
the impact parameter space using the transverse spatial 
coordinates of Reggeons $(x_j,y_j)$. 
We pass to this representation performing Fourier transform of 
Eq. (\ref{eq:NHam}).
After introducing holomorphic and 
anti-holomorphic complex coordinates for the $j$-th reggeized gluon 
\begin{equation}
z_j=x_j+i y_j\mbox {,}\qquad \wbar{z}_j=x_j-i y_j
\lab{eq:zik}
\end{equation}
 the Hamiltonian
for a pair of Reggeons $\Hat{H}_{ik}$
defined in Eq. (\ref{eq:NHam})
separates into two independent parts:
\begin{equation}
\Hat{H}_{ik}=-\frac{\alpha_s}{2 \pi}
(\Hat{H}(z_i,z_k)+\Hat{H}(\wbar{z}_i,\wbar{z}_k)).
\lab{eq:Hik}
\end{equation}
Now, one can see that (\ref{eq:Sch}) was reduced
to two one-dimensional Schr\"odinger equations.

The holomorphic part and analogously the anti-holomorphic one has the form
\begin{equation}
\Hat{H}(z_j,z_k)=-P_j^{-1} \ln(z_{jk}) P_j-P_k^{-1} \ln(z_{jk}) P_k
-\ln(P_i P_k)+2 \psi(1),
\lab{eq:homH} 
\end{equation}
where $P_j=-i \partial_{z_j}$, and $z_{jk}=z_j-z_k$. 

The Hamiltonian (\ref{eq:homH}) is $SL(2,\mathbb{C})$ conformal invariant.
It means that is invariant under translation, scaling and inversion
operation \ci{Lipatov:1985uk}.
An eigenstate of the full Hamiltonian (\ref{eq:NHam})
may be written as the bilinear form 
$\varPhi=\overline{\varPsi } \times \varPsi$.
In this work we demand that this function is
single-valued and normalized.

The colour factor $T_i T_k$ in the Hamiltonian (\ref{eq:NHam})
may be calculated exactly for $N=2$ (i.e.\ the leading order of 
Pomeron exchange)
and for $N=3$ (i.e.\ the leading order of odderon 
and the next to leading order of Pomeron exchange).
For $N>3$ we use an approximation $N_c\rightarrow \infty$.
Finally we obtain the Hamiltonian of the nearest neighbour interaction
\begin{equation}
\Hat{\cal H}_N=\Hat{H}_N+\Hat{\overline{H}}_N=
\frac{\wbar{\alpha}_s}{4} \sum_{k=1}^{N}
\left[
\Hat{H}(z_k,z_{k+1})+\Hat{H}(\wbar{z}_k,\wbar{z}_{k+1})
\right]
\lab{eq:HHam}
\end{equation}
with $\wbar{\alpha}_s=\alpha_s N_c/\pi$
where $N_c$ is the number of colours,
and $z_1$ and $z_{N+1}$ correspond to the same Reggeon.

The total cross section for the exchange of two reggeized gluons
in the singlet representation (Pomeron)
is proportional to  $s^{\omega(t)}$.
This behaviour is in contradiction with unitarity
of the $S$ matrix. 
There is a conjecture \ci{Bartels:1980pe,Kovchegov:2002qu} 
that in order to conserve unitarity 
one has to sum up diagrams for all $N$-Reggeon
exchanges. 
Additionally, as discussed in the papers 
\ci{Bartels:1993ke,Bartels:1994jj,Kovchegov:2002qu} 
the multiple Pomeron
exchanges, so called the fan diagrams, 
are more likely to unitarize the total hadronic cross section 
\ci{Mueller:1996hm}. In the latter diagrams the number of the Reggeons 
varies in the $t-$channel. Those diagrams are beyond the scope of this work.


\section{System with $SL(2,\mathbb{C})$ symmetry}
\subsection{Symmetry $SL(2,\mathbb{C})$}

The Hamiltonian (\ref{eq:HHam}) is invariant under 
the coordinate transformation of the $SL(2,\mathbb{C})$ group
\begin{equation}
z_{k}^{\prime }=\frac{az_{k}+b}{cz_{k}+d}\mbox {,}\qquad 
\wbar{z}_{k}^{\prime }=
\frac{\wbar{a}\wbar{z}_{k}+\wbar{b}}{\wbar{c}\wbar{z}_{k}+\wbar{d}}
\lab{eq:trcoords}
\end{equation}
with $k=1,\ldots ,N$ and  $ad-bc=\wbar{a}\wbar{d}-\wbar{b}\wbar{c}=1$.

Now one may associate with each particle 
generators of this transformation \ci{Derkachov:2001yn}.
This generators are a pair of mutually commuting holomorphic
and anti-holomorphic spin operators, 
$S_{\alpha }^{(k)}$ and $\wbar{S}_{\alpha }^{(k)}$.
They satisfy the standard commutation relations 
$\left[S_{\alpha }^{(k)},S_{\beta }^{(n)}\right]
=i\epsilon _{\alpha \beta \gamma }\delta ^{kn}S_{\gamma }^{(k)}$
and similarly for $\wbar{S}_{\alpha }^{(k)}$.
The generators act on the quantum
space of the $k$-th particle, $V^{(s_{k},\wbar{s}_{k})}$ as 
differential operators 
\begin{equation}
\begin{array}{ccc}
 S_{0}^{k}=z_{k}\partial _{z_{k}}+s_{k}\,, \quad  & 
S_{-}^{(k)}=-\partial _{z_{k}}\,, \quad  & 
S_{+}^{(k)}=z_{k}^{2}\partial _{z_{k}}+2s_{k}z_{k}\,,\\
 \wbar{S}_{0}^{k}=
\wbar{z}_{k}\partial _{\wbar{z}_{k}}+\wbar{s}_{k}\,, \quad  & 
\wbar{S}_{-}^{(k)}=-\partial _{\wbar{z}_{k}} \,, \quad  & 
\wbar{S}_{+}^{(k)}=
\wbar{z}_{k}^{2}\partial _{\wbar{z}_{k}}+2\wbar{s}_{k}\wbar{z}_{k}\,,\\
\end{array}
\lab{eq:spins}
\end{equation}
 where $ S_{\pm}^{(k)}=S_{1}^{(k)}\pm i S_{2}^{(k)}$
while the complex parameters, $s_{k}$ and $\wbar{s}_{k}$,
are called the complex spins. Thus, the Casimir operator reads 
\begin{equation}
\sum_{j=0}^2(S_j^{(k)})^{2}=(S_{0}^{(k)})^{2}+(S_{+}^{(k)}S_{-}^{(k)}
+S_{-}^{(k)}S_{+}^{(k)})/2=s_{k}(s_{k}-1)
\lab{eq:casimir}
\end{equation}
and similarly for the anti-holomorphic operator $(\wbar{S}^{(k)})^{2}$.
The eigenstates of $SL(2,\mathbb{C})$ invariant system transform
as \ci{CFT,Zuber:1995rj}
\begin{equation}
\Psi (z_{k},\wbar{z}_{k})\rightarrow \Psi ^{\prime }(z_{k},\wbar{z}_{k})
=(cz_{k}+d)^{-2s_{k}}(\wbar{c}\wbar{z}_{k}
+\wbar{d})^{-2\wbar{s}_{k}}\Psi 
(z_{k}^{\prime },
\wbar{z}_{k}^{\prime })\mbox {.}
\lab{eq:trpsi}
\end{equation}

Due to the invariance (\ref{eq:trcoords}) of the system
we can rewrite the Hamiltonian (\ref{eq:HHam})
\begin{equation}
\mathcal{H}_{N}=H_{N}+\wbar{H}_{N}\mbox {,}\qquad 
[H_{N},\wbar{H}_{N}]=0\mbox { }\lab{eq:sepH}
\end{equation}
in terms of the conformal spins (\ref{eq:spins})
\begin{equation}
H_{N}=\sum _{k=1}^{N}H(J_{k,k+1})\mbox {,}\quad 
\wbar{H}_{N}=\sum _{k=1}^{N}H(\wbar{J}_{k,k+1})\mbox {,}\quad 
H(J)=\psi (1-J)+\psi (J)-2\psi (1)
\lab{eq:Ham}
\end{equation}
 with $\psi (x)=d\ln \Gamma (x)/dx$ being the Euler function and
$J_{N,N+1}=J_{N,1}$. Here operators, $J_{k,k+1}$ and $\wbar{J}_{k,k+1}$,
are defined through the Casimir operators for the sum of the spins
of
the neighbouring Reggeons
\begin{equation}
J_{k,k+1}(J_{k,k+1}-1)=(S^{(k)}+S^{(k+1)})^{2}
\lab{eq:jvss}
\end{equation}
with $S_{\alpha }^{(N+1)}=S_{\alpha }^{(1)}$, and $\wbar{J}_{k,k+1}$ is
defined similarly. 

In statistical physics  (\ref{eq:Ham}) is called the Hamiltonian of 
the non-compact $SL(2,\mathbb{C})$
XXX Heisenberg spin magnets.
It describes the  nearest neighbour interaction
between $N$ non-compact $SL(2,\mathbb{C})$ spins attached to the particles
with periodic boundary conditions.

For the homogeneous spin chain we have to take $s_{k}=s$ and 
$\wbar{s}_{k}=\wbar{s}$. In QCD values of $(s,\wbar{s})$ depend on 
a chosen scalar product in the space of the wave-functions (\ref{eq:trpsi}) 
and they are usually equal to $(0,1)$ or $(0,0)$ 
\ci{Derkachov:2001yn,DeVega:2001pu}.

\subsection{Scalar product}

In order to find the high energy behaviour of the scattering amplitude 
we have to solve the Schr\"{o}dinger equation 
\begin{equation}
\mathcal{H}^{(s=0,\wbar s=1)}_{N}
\Psi (\vec{z}_{1},\vec{z}_{2},\ldots ,\vec{z}_{N})
=E_{N}\Psi (\vec{z}_{1},\vec{z}_{2},\ldots ,\vec{z}_{N})
\lab{eq:Schr}
\end{equation}
with the eigenstate $\Psi (\vec{z}_{1},\vec{z}_{2},\ldots ,\vec{z}_{N})$
being single-valued function on the plane $\vec{z}=(z,\wbar{z})$,
normalizable with respect to the $SL(2,\mathbb{C})$ invariant scalar
product\begin{equation}
||\Psi ||^{2}=\vev{\Psi|\Psi}=
\int d^{2}z_{1}d^{2}z_{2}\ldots d^{2}z_{N}
|\Psi (\vec{z}_{1},\vec{z}_{2},\ldots ,\vec{z}_{N})|^{2}\,,
\lab{eq:norm}
\end{equation}
where $d^{2}z_i=dx_idy_i=dz_id\wbar{z}_i/2$ with $\wbar z_i={z_i}^{\ast}$.

It is possible to use other scalar products \ci{Derkachov:2002wz}.
Let us consider the amplitude for the scattering of two colourless objects $A$
and $B$. In the Regge limit, the contribution
to the scattering amplitude from $N-$gluon exchange in the $t-$channel takes
the form 
\begin{equation}
\mathcal{A}(s,t)=is \sum_N (i {\alpha}_s)^N \mathcal{A}_N(s,t)\,.
\lab{eq:a-st}
\end{equation}
Using the $SL(2,\mathbb{C})$ scalar product (\ref{eq:norm}) we have
\be
\mathcal{A}_N(s,t) = s \int d^2 z_0\,
\e^{i\vec z_0 \cdot \vec p}
\vev{\tilde{\Phi}_A(\vec z_0)|
\e^{- \wbar{\alpha}_s Y \tilde{\cal H}_N/4 }
\lr{\vec{\partial}_1^2 \ldots {{\vec{\partial}}_N}^2 }^{-1}
 | \tilde{\Phi}_B(0)}\,,
\lab{A}
\ee
where $\partial_k=\partial/\partial z_k$, the rapidity $Y=\ln s$
and $\lr{{\vec{\partial}_1}^2 \ldots {\vec{\partial}_N}^2}^{-1}$ 
are gluon propagators omitted in (\ref{eq:lip}). 
Here the Hamiltonian $\widetilde{\cal H}_N$ is related to (\ref{eq:HHam}) as
$\Hat{\cal H}_N=-\wbar{\alpha}_s \widetilde{\cal H}_N/4$, 
so it is given by the sum of $N$ BFKL kernels
corresponding to nearest neighbour interaction between $N$ reggeized gluons.
The wave-functions 
$\ket{\Phi_{A(B)}(\vec z_0)}\equiv \Phi_{A(B)}(\vec z_i-\vec z_0)$
describe the coupling of $N-$ gluons to the scattered particles.
The $\vec z_0 -$ integration fixes the momentum transfer $t=-\vec p^2$ 
whereas the operators $1/\vec \partial_k^2
$  
stand for two-dimensional 
transverse propagators.

Defining the functions
 $\Phi(\vec z)$ as
\be
\widetilde\Phi(\vec z)=(-i)^N\partial_{z_1}
\partial_{z_2}\ldots \partial_{z_N}\Phi(\vec z)
\ee
the scalar product in the amplitude (\ref{A}) can be rewritten as
\be
\vev{\tilde{\Phi}_A(\vec z_0)|
\e^{- \wbar{\alpha}_s Y  \tilde{\cal H}_N/4 }
\lr{{\vec \partial_1}^2\ldots \vec \partial_N^2}^{-1}
 | \tilde{\Phi}_B(0)}=
\vev{\Phi_A(\vec z_0)|
\e^{- \wbar{\alpha}_s Y {\cal H}^{(s=0,\wbar s=1)}_N/4 }
 | \Phi_B(0)}\,.
\lab{vev}
\ee
The Hamiltonians, ${\cal H}_N$ and $\widetilde{\cal H}_N$ are
related to each other as
\be
{\cal H}^{(s=0,\wbar s=1)}_N=
\lr{\wbar \partial_1\ldots \wbar \partial_N}\,\widetilde{\cal H}_N\,\lr{\wbar
\partial_1\ldots \wbar \partial_N}^{-1}\,.
\lab{H-rel}
\ee
From the
point of view of the $SL(2,\mathbb{C})$ spin chain
\be
\widetilde{\cal H}_N=\mathcal{H}_N^{(s=0,\wbar s=0)}\,.
\lab{HH}
\ee
Indeed, the transformation $\wbar S_\alpha
\to\lr{\wbar\partial_1\ldots \wbar\partial_N}\,\wbar S_\alpha\,
\lr{\wbar \partial_1\ldots \wbar \partial_N}^{-1}$
maps the $SL(2,\mathbb{C})$ generators of the spin $\wbar s=0$ 
into those with the spin
$\wbar s=1$. One concludes from \re{HH} that the Hamiltonian 
${\cal H}^{(s=0,\wbar s=1)}_N$ is
advantageous with respect to $\widetilde{\cal H}_N$ as it has the quantum numbers
of the principal series representation of the $SL(2,\mathbb{C})$ group.

However, one can also use $\mathcal{H}_N^{(s=0,\wbar s=0)}$ 
\ci{DeVega:2001pu} 
or even 
$\mathcal{H}_N^{(s=1,\wbar s=1)}$ \ci{deVega:2002im}. 
Then the factor $\lr{\partial_1\ldots \partial_N}^{\mp 1}$
or $\lr{\wbar \partial_1\ldots \wbar \partial_N}^{\mp 1}$ may 
be included into the scalar product, i.e. for $(s=1,\wbar s=1)$ we have 
\begin{equation}
||\Psi ||^{2}=
\int d^{2}z_{1}d^{2}z_{2}\ldots d^{2}z_{N}
|\lr{\wbar \partial_1\ldots \wbar \partial_N}^{- 1}
\Psi (\vec{z}_{1},\vec{z}_{2},\ldots ,\vec{z}_{N})|^{2}\,.
\lab{eq:norml}
\end{equation}
Here
the scalar product is no longer in 
the principal series representation of the $SL(2,\mathbb{C})$ group.
All these Hamiltonians with the corresponding scalar products are equivalent
up to the zero modes of the  
$\lr{\partial_1\ldots \partial_N}^{\mp 1}$ and
$\lr{\wbar \partial_1\ldots \wbar \partial_N}^{\mp 1}$
operators.

\subsection{Conformal charges $q_k$ and the conformal spins}
The Hamiltonian (\ref{eq:sepH}) possesses 
a complete set of the integrals of motion 
$\{\vec p, \vec q_k\}$ where 
$\vec{q}_k=\{q_k,\wbar{q}_k \}$ with  $k=2,\ldots N$ are called 
conformal charges while $\vec p=\{p,\wbar p\}$ is 
the total momentum of the system. 
In order to construct them we introduce the Lax operators 
\ci{Sklyanin:1991ss,Faddeev:1994nk,Faddeev:1996iy,Faddeev:1979gh} 
in holomorphic and anti-holomorphic sectors: 
\begin{eqnarray}
 L_{k}(u)&=&u+i(\sigma \cdot S^{(k)})=\left(\begin{array}{cc}
 u+iS_{0}^{(k)} & iS_{-}^{(k)}\\
 iS_{+}^{(k)} & u-iS_{0}^{(k)}\end{array}\right)\,,
\nonumber\\
 \wbar{L}_{k}(\wbar{u})&=&
\wbar{u}+i(\sigma \cdot \wbar{S}^{(k)})=\left(\begin{array}{cc}
 \wbar{u}+i\wbar{S}_{0}^{(k)} & i\wbar{S}_{-}^{(k)}\\
 i\wbar{S}_{+}^{(k)} & \wbar{u}-i\wbar{S}_{0}^{(k)}
\end{array}\right)
\lab{eq:Lax}
\end{eqnarray}
 with $u$ and $\wbar{u}$ being arbitrary complex parameters  called 
the spectral parameters and 
$\sigma _{\alpha }$ being
Pauli matrices.

To identify the total set of the integrals of motion of the model,
one constructs the auxiliary holomorphic monodromy matrix
\begin{equation}
T_{N}(u)=L_{1}(u)L_{2}(u)\ldots L_{N}(u)
\lab{eq:LLL}
\end{equation}
and similarly for the anti-holomorphic monodromy 
operator $\wbar{T}_{N}(\wbar{u})$.
Taking the trace of the monodromy matrix we define the auxiliary transfer
matrix (spectral invariants)
\begin{equation}
\ot{N}(u)
=\mbox {tr}T_{N}(u)=2u^{N}+\oq{2} u^{N-2}+ \ldots +\oq{N}
\lab{eq:tnu}
\end{equation}
and similarly for $\otb{N}(\wbar{u})$. 
We see from (\ref{eq:tnu})
the advantage of using the transfer matrix $\ot{N}(u)$:
that is 
a polynomial in $u$ with coefficients given in terms of conformal charges
$\oq{k}$ and $\oqb{k}$, 
which are expressed as linear
combinations of the product of $k$ spin operators:
\begin{eqnarray}
\displaystyle
\oq{2} & = & 
-2\sum _{i_{2}>i_{1}=1}^{N}\sum _{j_{1}=0}^{2}
\left(S_{j_{1}}^{(i_{1})}S_{j_{1}}^{(i_{2})}\right)\\
\displaystyle
\oq{4} & = & 
-\sum _{i_{2}>i_{1}=1}^{N} \sum _{i_{4}>i_{3}=1}^{N}\sum _{j_{1},j_{2}=0}^{2}
\varepsilon _{i_{1}i_{2}i_{3}i_{4}}
\left(S_{j_{1}}^{(i_{1})}S_{j_{1}}^{(i_{2})}\right)
\left(S_{j_{2}}^{(i_{3})}S_{j_{2}}^{(i_{4})}\right)\\
\displaystyle
\nonumber
\oq{6} & = & 
-\frac{1}{3}\sum _{i_{2}>i_{1}=1}^{N}\sum _{i_{4}>i_{3}=1}^{N}
\sum _{i_{6}>i_{5}=1}^{N}\sum _{j_{1},j_{2},j_{3}=0}^{2}
\varepsilon _{i_{1}i_{2}i_{3}i_{4}i_{5}i_{6}}
\left(S_{j_{1}}^{(i_{1})}S_{j_{1}}^{(i_{2})}\right)\\
& & \times
\left(S_{j_{2}}^{(i_{3})}S_{j_{2}}^{(i_{4})}\right)
\left(S_{j_{3}}^{(i_{5})}S_{j_{3}}^{(i_{6})}\right)\,,
\lab{eq:q2q4q6}
\end{eqnarray}
 where $\varepsilon _{i_{1}i_{2}\ldots i_{k}}$ 
is completely anti-symmetric
tensor and $\varepsilon _{i_{1}i_{2}\ldots i_{k}}=1$ 
for $i_{1}<i_{2}<\ldots <i_{k}$.
So for even $k$ we have a formula for conformal charges 
\begin{align}
\oq{k} & =-\frac{2}{(k/2)!}
\sum_{
\begin{array}{c}
 i_{2}>i_{1}=1\\
 i_{4}>i_{3}=1\\
 \ldots \\
 i_{n}>i_{n-1}=1
\end{array}}^{N}
\sum _{j_{1},j_{2},\ldots ,j_{k/2}=0}^{2}
\varepsilon _{i_{1}i_{2}\ldots i_{k}}
\prod_{n=1}^{k/2}
\left(S_{j_{n}}^{(i_{2 n-1})}S_{j_{n}}^{(i_{2 n})} \right)\,.
\lab{eq:sqen}
\end{align}

For odd $k$'s we have
\begin{eqnarray}
\nonumber
\oq{3}&=&2\sum _{i_{1},i_{2},i_{3}=1}^{N}
\varepsilon _{i_{1}i_{2}i_{3}}
\left(S_{0}^{(i_{1})}S_{1}^{(i_{2})}S_{2}^{(i_{3})}\right)\\
& = &
2\sum _{i_{3}>i_{2}>i_{1}=1}^{N}
\sum _{j_{1},j_{2},j_{3}=0}^{2}
\varepsilon _{i_{1}i_{2}i_{3}}
\varepsilon _{j_{1}j_{2}j_{3}}
\left(S_{j_{1}}^{(i_{1})}S_{j_{2}}^{(i_{2})}S_{j_{3}}^{(i_{3})}\right)\,,\\
\oq{5} & = & -2\sum _{i_{3},i_{2},i_{1}=1}^{N}
\sum _{i_{5}>i_{4}=1}^{N}\sum _{j_{4}=0}^{2}
\varepsilon _{i_{1}i_{2}i_{3}i_{4}i_{5}}
\left(S_{0}^{(i_{1})}S_{1}^{(i_{2})}S_{2}^{(i_{3})}\right)
\left(S_{j_{4}}^{(i_{4})}S_{j_{4}}^{(i_{5})}\right)\,,\\
\nonumber
\oq{7} & = & \sum _{i_{3},i_{2},i_{1}=1}^{N}
\sum _{i_{5}>i_{4}=1}^{N}\sum _{i_{7}>i_{6}=1}^{N}
\sum _{j_{4},j_{5}=0}^{2}
\varepsilon _{i_{1}i_{2}i_{3}i_{4}i_{5}i_{6}i_{7}}
\left(S_{0}^{(i_{1})}S_{1}^{(i_{2})}S_{2}^{(i_{3})}\right) \\
& & \times
\left(S_{j_{4}}^{(i_{4})}S_{j_{4}}^{(i_{5})}\right)
\left(S_{j_{5}}^{(i_{6})}S_{j_{5}}^{(i_{7})}\right)
\lab{eq:q3q5q7}
\end{eqnarray}
and the general expression for an odd number of the conformal spins is 
\begin{eqnarray}
\nonumber
\oq{k} & = & \frac{2 (-1)^{(k+1)/2}}{\left(\frac{k-3}{2}\right)!}
\sum _{i_{3},i_{2},i_{1}=1}^{N}\sum _{\begin{array}{c}
 i_{5}>i_{4}=1\\
 i_{7}>i_{6}=1\\
 \ldots \\
 i_{k}>i_{k-1}=1\end{array}}^{N}\sum _{j_{1},j_{2},
\ldots ,j_{(k-3)/2}=0}^{2}
\varepsilon _{i_{1}i_{2}\ldots i_{k}}
\left(S_{0}^{(i_{1})} S_{1}^{(i_{2})} S_{2}^{(i_{3})}\right)\\
 &  & \times \, 
\prod_{n=1}^{(k-3)/2}
\left(S_{j_{n}}^{(i_{2n+2})}S_{j_{n}}^{(i_{2n+3})}\right)\,.
\lab{eq:sqok}
\end{eqnarray}

In the above formulae we have two basic blocks 
$\left(S_{j_{1}}^{(i_{1})}S_{j_{1}}^{(i_{2})}\right)$
and $\left(S_{0}^{(i_{1})}S_{1}^{(i_{2})}S_{2}^{(i_{3})}\right)$ 
whose products are summed with antisymmetric tensor 
$\varepsilon _{i_{1}i_{2}\ldots i_{k}}$.

\subsection{Two-dimensional Lorentz spin and the scaling dimension}

The Hamiltonian (\ref{eq:Ham})
is a function of two-particle Casimir operators \ci{Derkachov:2001yn}, and,
therefore, it commutes with the operators of the total spin 
$S_{\alpha }=\sum _{k}S_{\alpha }^{(k)}$
and $\wbar{S}_{\alpha }=\sum _{k}\wbar{S}_{\alpha }^{(k)}$, acting
on the quantum space of the system 
$V_{N}\equiv V^{(s_{1},\wbar{s}_{1})}\otimes V^{(s_{2},\wbar{s}_{2})}\otimes 
\ldots \otimes V^{(s_{N},\wbar{s}_{N})}$.
This implies that the eigenstates can be classified according to the
irreducible representations of the $SL(2,\mathbb{C})$ group, 
$V^{(h,\wbar{h})}$,
parameterized by spins $(h,\wbar{h})$ \ci{Derkachov:2001yn}. 

The Hamiltonian depends on differences of particle coordinates 
so eigenfunctions can be written as 
\begin{equation}
\Psi _{\vec{p}}(\vec{z}_{1},\vec{z}_{2},\ldots ,\vec{z}_{N})=
\int d^{2}z_{0}e^{i\vec{z}_{0}\cdot \vec{p}}
\Psi (\vec{z}_{1}-\vec{z}_{0},\vec{z}_{2}-\vec{z}_{0},\ldots ,
\vec{z}_{N}-\vec{z}_{0})\,.
\lab{eq:Psip}
\end{equation}
The eigenstates $\Psi (\vec{z}_{1},\vec{z}_{2},\ldots ,\vec{z}_{N})$
belonging to $V^{(h,\wbar{h})}$ are labelled by the centre-of-mass coordinate
$\vec{z}_{0}$ and can be chosen to have the following  the $SL(2,\mathbb{C})$
transformation properties
\begin{equation}
\Psi (\{\vec{z}_{k}^{\ \prime }-\vec{z}_{0}^{\ \prime }\})=
(cz_{0}+d)^{2h}(\wbar{c}\wbar{z}_{0}+\wbar{d})^{2\wbar{h}}
\left(\prod _{k=1}^{N}(cz_{k}+d)^{2s_{k}}(\wbar{c}\wbar{z}_{k}
+\wbar{d})^{2\wbar{s}_{k}}
\right)
\Psi (\{\vec{z}_{k}-\vec{z}_{0}\})
\lab{eq:trpsiz0}
\end{equation}
 with $z_{0}$ and $\wbar{z}_{0}$ transforming in the same way as $z_{k}$
and $\wbar{z}_{k}$, (\ref{eq:trcoords}).
As a consequence, they diagonalize the Casimir operators:
\begin{equation}
(S^2-h(h-1)) 
\Psi(\vec{z}_{1},\vec{z}_{2},\ldots ,\vec{z}_{N})=0
\lab{eq:SCas}
\end{equation}
corresponding to the total spin of the system, 
\begin{equation}
S^2= 
\sum _{i_{2},i_{1}=1}^{N}
\sum _{j=0}^{2}S_{j}^{(i_{1})}S_{j}^{(i_{2})}=
-\oq{2}-\sum _{k=1}^{N}s_{k}(s_{k}-1)\,.
\lab{eq:S2}
\end{equation}

The complex parameters $(s_{k},\wbar{s}_{k})$ and $(h,\wbar{h})$ parameterize
the irreducible representations of the $SL(2,\mathbb{C})$ group. 
For principal series representation
they satisfy the conditions 
\begin{equation}
s_{k}-\wbar{s}_{k}=n_{s_{k}}\mbox {,}\qquad s_{k}+(\wbar{s}_{k})^{*}=1
\lab{eq:ssbar}
\end{equation}
 and have the following form
\begin{equation}
s_{k}=\frac{1+n_{s_{k}}}{2}+i\nu _{s_{k}}\mbox {,}\qquad \wbar{s}_{k}=
\frac{1-n_{s_{k}}}{2}+i\nu _{s_{k}}
\lab{eq:spar}
\end{equation}
 with $\nu _{s_{k}}$being real and $n_{s_{k}}$ being integer or
half-integer. The spins $(h,\wbar{h})$ are given by similar expressions
with $n_{s_{k}}$and $\nu _{s_{k}}$ replaced by $n_{h}$ and $\nu _{h}$,
respectively
\begin{equation}
h=\frac{1+n_{h}}{2}+i\nu _{h}\mbox {,}\qquad \wbar h=
\frac{1-n_{h}}{2}+i\nu _{h}.
\lab{eq:hpar}
\end{equation}
 The parameter $n_{s_{k}}$ has the meaning of the two-dimensional
Lorentz spin of the particle, whereas $\nu _{s_{k}}$ defines its
scaling dimension. To see this one can perform a $2\pi $-rotation of
the particle on the plane, 
and find from eigenstates transformations 
(\ref{eq:trpsiz0}) that the wave-function acquires a phase. Indeed
\begin{equation}
z_{k}\rightarrow z_{k}\exp (2\pi i) 
\quad 
\mbox{and} 
\quad
\wbar{z}_{k}\rightarrow \wbar{z}_{k}\exp (-2\pi i)
\qquad
\mbox{gives}
\qquad
\Psi (z_{k},\wbar{z}_{k})\rightarrow 
(-1)^{2 n_{s_{k}}}\Psi (z_{k},\wbar{z}_{k}).
\end{equation}
For half-integer $n_{s_{k}}$ it changes the sign and the corresponding
representation is spinorial. Similarly, to define scaling dimension,
$s+\wbar{s}=1+2i\nu _{s_{k}}$
one performs the transformation 
\begin{equation}
z\rightarrow \lambda z
\quad
\mbox{and} 
\quad
\wbar{z}\rightarrow \lambda \wbar{z}
\qquad
\mbox{giving} 
\qquad
\Psi (z_{k},\wbar{z}_{k})\rightarrow \lambda ^{1+2i\nu _{s_{k}}}
\Psi (z_{k},\wbar{z}_{k})\,.
\end{equation}
Because the scale product for the wave-functions is 
invariant under $SL(2,\mathbb{C})$ transformations, 
(\ref{eq:trcoords}), 
the parameter $\nu _{s_{k}}$ is real.

We notice that the holomorphic and anti-holomorphic spin generators
as well as Casimir operators (\ref{eq:jvss}) are conjugated to each
other with respect to the scalar product (\ref{eq:norm}):
\begin{equation}
[S_{\alpha }^{(k)}]^{\dagger }
=-\wbar{S}_{\alpha }^{(k)}\mbox {,}\qquad [J_{k}]^{\dagger }
=1-\wbar{J}_{k}\,.
\lab{eq:sjdag}
\end{equation}
Moreover, because of the transformation law (\ref{eq:sjdag}),
$h^{\ast}=1-\wbar{h}$
\footnote{$ ^\ast$ -- denotes complex conjugation}.
This implies that $H_{N}^{\dagger }=\wbar{H}_{N}$ and, as a consequence,
the Hamiltonian is hermitian on the space of the functions endowed
with the $SL(2,\mathbb{C})$ scalar product, 
$\mathcal{H}_{N}^{\dagger }=\mathcal{H}_{N}$.

\section{Conformal charges $\oq{k}$ as a differential operators}

We noticed in the previous paragraphs 
that the conformal charge operators $\oq{k}$
are given by invariant sum of linear combinations of the product of
$k$ spin operators. They can be rewritten as $k$-th order differential
operators acting on (anti)holomorphic coordinates $(z,\wbar{z})$.

Two particle spin square can be written as
\begin{eqnarray}
\sum _{j_{1}=0}^{2}S_{j_{1}}^{(i_{1})}S_{j_{1}}^{(i_{2})} 
 & = & 
-\frac{1}{2}(z_{i_{1}}-z_{i_{2}})^{2}
\partial _{z_{i_{1}}}\partial _{z_{i_{2}}}
+(z_{i_{1}}-z_{i_{2}})
(s_{i_{2}}\partial _{z_{i_{1}}}+s_{i_{1}}\partial _{z_{i_{2}}})
+s_{1}s_{2}\,.
\end{eqnarray}
For homogeneous spins $s=s_{1}=s_{2}=\ldots =s_{N}$,
what is also the QCD case,
we have
\begin{eqnarray}
\displaystyle
\nonumber
\oq{2}&=&-2\sum _{i_{2}>i_{1}=1}^{N}
\left(\sum _{j_{1}=0}^{2}S_{j_{1}}^{(i_{1})}S_{j_{1}}^{(i_{2})}\right)
=\sum _{i_{2}>i_{1}=1}^{N}\left((z_{i_{2}i_{1}})^{2(1-s)}
\partial _{z_{i_{2}}}\partial _{z_{i_{1}}}
(z_{i_{2}i_{1}})^{2s}+2s(s-1)\right)\,,
\\
\nonumber
\displaystyle
\oq{3} & = & 2\sum _{i_{1},i_{2},i_{3}=1}^{N}
\varepsilon _{i_{1}i_{2}i_{3}}
S_{0}^{(i_{1})}S_{1}^{(i_{2})}S_{2}^{(i_{3})}
=i^3 \sum _{i_{3}>i_{2}>i_{1}=1}^{N}
\left(z_{i_{1}i_{2}}z_{i_{2}i_{3}}z_{i_{3}i_{1}}
\partial _{z_{i_{3}}}\partial _{z_{i_{2}}}\partial _{z_{i_{1}}}+\right.\\
\nonumber
 &  &+  \left.sz_{i_{1}i_{2}}(z_{i_{2}i_{3}}-z_{i_{3}i_{1}})
\partial _{z_{i_{2}}}\partial _{z_{i_{1}}}
+sz_{i_{2}i_{3}}(z_{i_{3}i_{1}}-z_{i_{1}i_{2}})
\partial _{z_{i_{3}}}\partial _{z_{i_{2}}} \right.\\
&& \left. +sz_{i_{3}i_{1}}(z_{i_{3}i_{1}}-z_{i_{1}i_{2}})
\partial _{z_{i_{3}}}\partial _{z_{i_{2}}}
 -  2s^{2}z_{i_{1}i_{2}}\partial _{z_{i_{3}}}
-2s^{2}z_{i_{2}i_{3}}\partial _{z_{i_{1}}}
-2s^{2}z_{i_{3}i_{1}}\partial _{z_{i_{2}}}\right)\,,
\lab{eq:q2q3}
\end{eqnarray}
where $z_{ij}=z_{i}-z_{j}$. Similar relations hold for 
the anti-holomorphic sector.

In that way one can also construct operators for the higher conformal charges.
They have a particularly simple form for the $SL(2,\mathbb{C})$ spins $s=0$
\begin{equation}
\oq{k}=i^k\sum_{1\le j_1 < j_2 < \ldots  < j_k\le N}
z_{j_1j_2}\ldots z_{j_{k-1},j_k}z_{j_k,j_1}\partial_{z_{j_1}}\ldots 
\partial_{z_{j_{k-1}}}\partial_{z_{j_k}}
\lab{eq:qks0}
\end{equation}
as well as for $s=1$
\begin{equation}
\oq{k}=i^k\sum_{1\le j_1 < j_2 < \ldots  < j_k\le N}
\partial_{z_{j_1}}\ldots \partial_{z_{j_{k-1}}}\partial_{z_{j_k}}
z_{j_1j_2}\ldots z_{j_{k-1},j_k}z_{j_k,j_1}\,.
\lab{eq:qks1}
\end{equation}

\section{Other symmetries}

The states (\ref{eq:Psip}) have additional symmetries \ci{Derkachov:2001yn}:
\begin{equation}
\begin{array}{rcl}
 \mathbb{P}\Psi _{q,\wbar q}(\vec{z}_{1},\vec{z}_{2},\ldots,\vec{z}_{N}) & 
\eqdef & \Psi _{q,\wbar q}(\vec{z}_{2},\vec{z}_{3},\ldots,\vec{z}_{1})
= e^{i\theta_N (q,\wbar q)}\Psi _{q,\wbar q}(\vec{z}_{1},\vec{z}_{2},
\dots,\vec{z}_{N})\,,
\\
 \mathbb{M}\Psi ^{\pm }(\vec{z}_{1},\vec{z}_{2},\ldots,\vec{z}_{N}) & 
\eqdef & \Psi ^{\pm }(\vec{z}_{N},\vec{z}_{N-1},\ldots,\vec{z}_{1}) 
= \pm \Psi ^{\pm }(\vec{z}_{1},\vec{z}_{2},\ldots,\vec{z}_{N})
\end{array}
\lab{eq:PMsym}
\end{equation}
so called cyclic and mirror permutation
where the conformal charges are denoted by $q\equiv(q_2,q_3,\ldots,q_n)$ 
and $\wbar q\equiv(\wbar q_2,\wbar q_3,\ldots,\wbar q_n)$.
They generators  $\mathbb{P}$ and  $\mathbb{M}$, respectively, commute 
with the Hamiltonian $\cal H$ but they 
do not commute with each other. They satisfy  relations
\begin{equation}
 \mathbb{P}^N = \mathbb{M}^2=1, \quad
 \mathbb{P}^{\dagger} = \mathbb{P}^{-1}= \mathbb{P}^{N-1},  \quad
 \mathbb{M}^{\dagger} = \mathbb{M},  \quad
 \mathbb{P M} = \mathbb{M \, P}^{-1}= \mathbb{M \, P}^{N-1}\,.
\end{equation}

The phase $\theta_N (q)$ which is connected with eigenvalues of $\mathbb{P}$ 
is called 
the quasimomentum. It takes the following values 
\begin{equation}
\theta_N(q,\wbar{q})=2 \pi \frac{k}{N}, \qquad \mbox{for } k=0,1,\ldots,N-1\,.
\lab{eq:quask}
\end{equation}
The eigenstates of the conformal
charges $\oq{k}$ diagonalize ${\cal H}$ and $\mathbb{P}$. 

The transfer matrices (\ref{eq:tnu}) are invariant under 
the cyclic permutations 
$\mathbb{P}^{\dagger}\, \ot{N}(u)\, \mathbb{P} = \ot{N}(u)$ whereas
they transform under the mirror transformation as
\begin{equation}
\mathbb{M}\, \ot{N}(u)\, \mathbb{M} = (-1)^N \ot{N}(-u).
\lab{eq:MtM}
\end{equation}
Substituting (\ref{eq:tnu}) into (\ref{eq:MtM}) one 
derives a transformation law of the conformal charges $q_k$
\begin{equation}
\mathbb{P}^{\dagger}\, \oq{k}\, \mathbb{P} = \oq{k}
\quad
\mathbb{M}\, \oq{k}\, \mathbb{M} = (-1)^k \oq{k}
\lab{eq:PMq}
\end{equation}
and similarly for the anti-holomorphic charges.
Since the Hamiltonian (\ref{eq:Ham}) is invariant under 
the mirror permutation, it has to satisfy 
\begin{equation}
{\cal H} (\oq{k},\oqb{k})=
\mathbb{M} {\cal H} (\oq{k},\oqb{k}) \mathbb{M}=
{\cal H} (\mathbb{M}\, \oq{k} \,\mathbb{M}, \mathbb{M}\, \oqb{k}\, \mathbb{M})=
{\cal H} ((-1)^k \oq{k},(-1)^k \oqb{k})\,.
\end{equation}
This implies that the eigenstates of the Hamiltonian (\ref{eq:Ham})
corresponding to two
different sets of the quantum number $\{q_k,\wbar{q}_k \}$ and
 $\{(-1)^k q_k,(-1)^k \wbar{q}_k \}$ have the same energy
\begin{equation}
E_N (q_k,\wbar{q}_k) = E_N ((-1)^k q_k,(-1)^k \wbar{q}_k)\,.
\end{equation}
Similarly one can derive relation for quasimomentum
\begin{equation}
\theta_N (q_k,\wbar{q}_k) = -\theta_N ((-1)^k q_k,(-1)^k \wbar{q}_k)\,.
\end{equation}

The cyclic and mirror permutation symmetries come
from the Bose symmetry and they appear
after performing the multi-colour limit \ci{'tHooft:1973jz}.
Physical states should possess both symmetries. 

\section{Pomeron and odderon}

In the previous Sections we introduced the wave-functions
of the Reggeized gluon states. Now, 
we answer how one can distinguish
Pomeron states from odderon states \ci{Ewerz:2003xi}. 
Since the Pomeron states
have conjugation charge parity $C=1$ and the odderon states have
$C=-1$, we have to check the $C-$parity of studied states.
Let us consider for simplicity purely perturbative gluon fields 
${\bf A}_{\mu}(x)=A_{\mu}^a(x) t^a$, where $t^a$ are the generators 
of the gauge group $SU(N_c)$ and $x$ are coordinates of the gluon.
Thus, under a charge conjugation transformation this field
transforms as
\begin{equation}
{\bf A}_{\mu}(x) \to - {\bf A}^T_{\mu}(x)\,,
\lab{eq:Acrg}
\end{equation}
where $T$ denotes matrix transposition.

For $N=2$ we have only one possibility to form a colour singlet state
from two gluon operators
\begin{equation}
{\cal P}_{\mu \nu}(x,y)
=\tr({\bf A}_{\mu}(x),{\bf A}_{\nu}(y))
=\frac1{2} \delta_{a b} A^a_{\mu}(x) A^b_{\nu}(y)\,.
\lab{eq:N2P}
\end{equation}
This state is invariant under charge conjugation (\ref{eq:Acrg})
so that it has $C=+1$. Therefore, for two gluons 
we have only the Pomeron states.

In the $N=3$ case they are two ways of constructing 
a colour singlet state.
Firstly, one can build
\begin{equation}
{\cal P}_{\mu \nu \rho}(x,y,z)
=-i \tr(\left[{\bf A}_{\mu}(x),{\bf A}_{\nu}(y)\right]{\bf A}_{\nu}(y))
=\frac1{2} f_{abc} A^a_{\mu}(x) A^b_{\nu}(y) A^c_{\rho}(z)
\lab{eq:N3P}
\end{equation}   
with the total antisymmetric structure constants
$f_{abc}$ defined via the Lie algebra of $SU(N_c)$:
\begin{equation}
\left[ t^a, t^b\right]= i f_{abc} t^c
\,.
\lab{eq:fabc}
\end{equation}
Using (\ref{eq:Acrg}), we find that 
${\cal P}_{\mu \nu \rho}(x,y,z)$ have also 
$C=+1$ so for $N=3$ the state (\ref{eq:N3P}) gives a subleading correction
to the $N=2$ Pomeron state.

The other possibility to form colour singlet states is to use
the totally symmetric constant
\begin{equation}
d_{abc}=2\left[\tr(t^a t^b t^c) + \tr(t^c t^b t^a) \right].
\lab{eq:dabc}
\end{equation}
In this way we obtain a state
\begin{equation}
{\cal O}_{\mu \nu \rho}(x,y,z)
= \tr(\left\{{\bf A}_{\mu}(x),{\bf A}_{\nu}(y)\right\}{\bf A}_{\nu}(y))
=\frac1{2} d_{abc} A^a_{\mu}(x) A^b_{\nu}(y) A^c_{\rho}(z)
\,.
\lab{eq:N3O}
\end{equation}   
Similarly, applying (\ref{eq:Acrg}) we find that (\ref{eq:N3O})
is the odd under $C-$parity. Thus, it is the leading contribution to
the odderon state.  
This state only appears for the gauge groups $SU(N_c)$ with $N_c>2$.

In order to find the eigenstates of the Hamiltonian 
(\ref{eq:NHam})
we perform the multi-colour limit (\ref{eq:HHam}).
In this limit the colour factor of the Hamiltonian 
is reduced to a trivial delta-function. 
Thus, a solution to the Schr\"odinger equation (\ref{eq:Schr})
does not have a colour factor. 
On the other hand, the Reggeon wave-function should be 
invariant under Bose symmetry. 
Adding a proper colour factor into
the Reggeon wave-functions we are able to restore Bose symmetry.
In the multi-colour limit the Bose symmetry reduce into
the cycle and mirror permutation symmetries (\ref{eq:PMsym}).
Thus, we have states which under the mirror permutations are
either odd or even. 
For example in the $N=3$ case
it turns out that adding colour factor:
corresponding to the antisymmetric constant $f_{abc}$
for the odd-mirror states and the symmetric one $d_{abc}$
for the even-mirror states, we are able to restore Bose symmetry.
As we have said in (\ref{eq:N3P}) and (\ref{eq:N3O})
$f_{abc}$ corresponds to  the Pomeron states while
$d_{abc}$ is related to the odderon states. 
One may expect similar relationship for $N>3$ states.
Therefore, in order to check
$C-$parity of a given state we need to study its parity under 
the mirror permutation $\mathbb{M}$. 
The states $\Psi$ satisfying $\mathbb{M}\, \Psi=-\Psi$ are the Pomeron states
whereas states for which $\mathbb{M}\, \Psi=+\Psi$ are the odderon states.

\chapter{Baxter $Q$-operator}
\section{Definition of the Baxter $Q$-operator}
The Schr\"odinger equation (\ref{eq:Schr}) may be solved applying the powerful method 
of the Baxter $Q$-operator \ci{Baxter}.
This operator depends on two complex spectral parameters $u$, $\wbar u$ and  
in the following will be denoted as
$\mathbb{Q}(u,\wbar{u})$.  
This operator has to satisfy the following relations
\begin{itemize}
\item Commutativity:
\begin{equation}
\left[ \mathbb{Q} (u,\wbar{u}) ,
 \mathbb{Q}(v,\wbar{v}) \right]=0\,,
\lab{eq:comQQ}
\end{equation}
\item $Q-t$ relation:
\begin{equation}
\left[\ot{N}(u), \mathbb{Q}(u,\wbar{u}) \right]=
\left[\otb{N}(\wbar{u}), \mathbb{Q}(u,\wbar{u}) \right]=0\,,
\lab{eq:comQt}
\end{equation}
\item Baxter equation:
\begin{equation}
\ot{N}(u) \mathbb{Q}(u,\wbar{u})  =
(u + i s)^N \mathbb{Q}(u+i,\wbar{u})  +
(u - i s)^N \mathbb{Q}(u-i,\wbar{u})  \,,
\lab{eq:Baxeq}
\end{equation}
\begin{equation}
\otb{N}(\wbar{u}) \mathbb{Q}(u,\wbar{u})  =
(\wbar{u} + i \wbar{s})^N \mathbb{Q}(u,\wbar{u}+i)  +
(\wbar{u} - i \wbar{s})^N \mathbb{Q}(u,\wbar{u}-i)  \,,
\lab{eq:Baxbeq}
\end{equation}
\end{itemize}
where $\ot{N}(u)$ and $\otb{N}(\wbar{u})$ are the auxiliary transfer matrices 
(\ref{eq:tnu}). 
According to (\ref{eq:comQt}) the Baxter $\mathbb{Q}(u,\wbar{u})$-operator
and the auxiliary transfer matrices as well as the Hamiltonian 
(\ref{eq:HHam}) share the common set of the eigenfunctions
\begin{equation}
 \mathbb{Q}(u,\wbar{u}) 
\Psi _{q,\wbar q}(\vec{z}_{1},\vec{z}_{2},\ldots,\vec{z}_{N}) 
=  Q_{q,\wbar q}(u,\wbar{u}) 
\Psi _{q,\wbar q}(\vec{z}_{1},\vec{z}_{2},\dots,\vec{z}_{N})\,.
\lab{eq:eeqQ}
\end{equation} 
The eigenvalues of the $Q$-operator satisfy the same Baxter equation
(\ref{eq:Baxeq}) and (\ref{eq:Baxbeq})
with the auxiliary transfer matrices replaced by their corresponding 
eigenvalues.

In the paper \ci{Derkachov:2001yn} the $Q$-operator was constructed 
as an $N-$fold integral operator
\begin{multline}
\mathbb{Q}(u,\wbar{u}) \Psi(\vec{z}_{1},\vec{z}_{2},\ldots,\vec{z}_{N}) =
\int d^2 w_1 \int d^2 w_2 \ldots \\
\ldots \int d^2 w_N
Q_{u,\wbar{u}} 
(\vec{z}_{1},\vec{z}_{2},\ldots,\vec{z}_{N} |
\vec{w}_{1},\vec{w}_{2},\ldots,\vec{w}_{N})
\Psi(\vec{w}_{1},\vec{w}_{2},\ldots,\vec{w}_{N}) \,,
\lab{eq:convQ}
\end{multline}
where the integrations are performed over two-dimensional
$\vec{w}_i-$planes.
The integral kernel in (\ref{eq:convQ}) takes two different
forms:
\begin{equation}
Q^{(+)}_{u,\wbar{u}}(\vec{z}|\vec{w})=
a(2-2s,s+iu,\wbar{s}-i\wbar{u})^N {\pi}^N
\prod_{k=1}^{N}
\frac{\left[z_k-z_{k+1} \right]^{1-2s}}
{\left[w_k-z_k \right]^{1-s-iu}
\left[w_k-z_{k+1} \right]^{1-s+iu}}
\lab{eq:Qpker}
\end{equation}
and
\begin{equation}
Q^{(-)}_{u,\wbar{u}}(\vec{z}|\vec{w})=
\prod_{k=1}^{N}
\frac{\left[w_k-w_{k+1} \right]^{2s-2}}
{\left[z_k-w_k \right]^{s+iu}
\left[z_k-w_{k+1} \right]^{s-iu}}\,,
\lab{eq:Qmker}
\end{equation}
which appear to be equivalent \ci{Derkachov:2001yn}.
In Eqs. (\ref{eq:Qpker}) and (\ref{eq:Qmker}) $a(\ldots)$ factorizes as 
\begin{equation}
a(\alpha,\beta,\ldots)=a(\alpha)a(\beta)\ldots 
\quad
\mbox{and} 
\quad
a(\alpha)=\frac{\Gamma(1-\wbar{\alpha})}{\Gamma(\alpha)}
\lab{eq:aalph}
\end{equation}
and
$\wbar{\alpha}$ is an anti-holomorphic partner of $\alpha$ satisfying 
$\alpha- \wbar{\alpha} \in \mathbb{Z}$.
Moreover, the two-dimensional propagators are defined as 
\begin{equation}
\left[z_k-w_k \right]^{-\alpha}=
\left(z_k-w_k \right)^{-\alpha}
\left(\wbar{z}_k-\wbar{w}_k \right)^{-\wbar{\alpha}}\,.
\lab{eq:2dprop}
\end{equation}
In order for the Baxter $Q-$operators to be well defined, 
(\ref{eq:Qpker}) and (\ref{eq:Qmker}), should be  single-valued functions.
In this way we can find that the spectral parameters $u$ and $\wbar{u}$
have to satisfy the condition
\begin{equation}
i(u-\wbar{u})=n
\lab{eq:uubarn}
\end{equation}
with $n$ being an integer.

The Baxter $Q$-operator has a defined pole structure. For  
$\mathbb{Q}^{(+)}(u,\wbar{u})$ we have an infinite set of poles
of the order not higher than $N$ situated at
\begin{equation}
\left\{\ u_m^+=i(s-m), \wbar{u}_{\wbar{m}}^+=i(\wbar{s}-\wbar{m})\right\};
\qquad
\left\{\ u_m^-=-i(s-m), \wbar{u}_{\wbar{m}}^-=-i(\wbar{s}-\wbar{m})\right\}
\lab{eq:upoles}
\end{equation}
with  $m,\wbar m=1,2,\ldots \,$.
The behaviour of $Q_{q,\wbar q}(u,\wbar u)$,
i.e. an eigenvalue of $\mathbb{Q}^{(+)}(u,\wbar{u})$,  
in the vicinity of the pole at $m=\wbar
m=1$ can be parameterized as
\begin{equation}
Q_{q,\wbar q}(u_{1}^{\pm}+\epsilon,{\wbar u}_{1}^{\pm}+\epsilon)=R^\pm(q,\wbar
q)\left[\frac1{\epsilon^N} +\frac{i\,E^\pm(q,\wbar q)}{\epsilon^{N-1}}+ \ldots 
\,\right]\!.
\lab{eq:Q-R,E}
\end{equation}
\\[-3mm]
The functions $R^\pm(q,\wbar q)$ fix an overall normalization of the Baxter
operator, while the residue functions $E^\pm(q,\wbar q)$ define the energy of the
system (see Eqs.~(\ref{eq:RR}) and (\ref{eq:energy}) below).
It has also specified asymptotic behaviour. For $|\IIm \lambda| < 1/2$
and $\RRe \lambda \rightarrow \infty$
\begin{equation}
Q_{q,\wbar q}(\lambda-i n/2,\lambda+i n/2) \sim
\e^{i \Theta_h(q,\wbar q)} \lambda^{h +\wbar h-N(s-\wbar s)}+
\e^{-i \Theta_h(q,\wbar q)} \lambda^{1-h +1-\wbar h-N(s-\wbar s)} \,,
\lab{eq:Qanalb}
\end{equation}
where $\Theta_h$ is a phase which should not be confused with quasimomentum 
$\theta_N(q,\wbar q)$.

\section{Observables}

The Hamiltonian (\ref{eq:sepH}) may be written in terms of 
the Baxter $Q$-operator \ci{Derkachov:2001yn}:
\begin{equation}
{\cal H}_N=\epsilon_N + \left. i \frac{d}{du} 
\ln \mathbb{Q}^{(+)}(u+is,\wbar{u}+i \wbar{s})\right|_{u=0}  
-\left( \left. i \frac{d}{du} 
\ln \mathbb{Q}^{(+)}(u-is,\wbar{u}-i \wbar{s})\right|_{u=0}\right)^{\dagger}\,,
\lab{eq:HNQQ}
\end{equation}
where the additive normalization constant is given as
\begin{equation}
\epsilon_N = 2 N \, \RRe [\psi(2 s)+ \psi(2-2s)-2 \psi(1)]\,.
\lab{eq:epsnor}
\end{equation}

Applying to (\ref{eq:HNQQ}) the eigenstate $\Psi_q$ we obtain the
energy 
\begin{equation}
E_N(q,\bar q)=\varepsilon_N + i\frac{d}{du} \ln\left[ Q_{q,\bar
q}^{}(u+is,u+i\bar s)\, \left(Q_{q,\bar q}^{}(u-is,u-i\bar
s)\right)^*\right]\bigg|_{u=0},
\lab{eq:enQ}
\end{equation}
or equivalently
\begin{eqnarray}
E_N(q,\bar q) &=&- \Im\frac{d}{du}\ln \bigg[u^{2N}Q_{q,\bar
q}(u+i(1-s),u+i(1-\bar s))\,
\lab{eq:enQ2}
\\[-1mm]
&&\hspace*{38mm}
\times Q_{-q,-\bar q}(u+i(1-s),u+i(1-\bar s))\bigg]\bigg|_{u=0} \,,
\nonumber
\end{eqnarray}
where
$Q_{q_,\wbar{q}}(u,\wbar{u})\equiv Q^{(+)}_{q_,\wbar{q}}(u,\wbar{u})$ 
is eigenvalue
of the $ \mathbb{Q}_+(u,\wbar{u})$ operator,
while 
\begin{equation}
\pm q=(q_2,\pm q_3,\ldots,(\pm)^N q_N) 
\lab{eq:pmq}
\end{equation}
are 
the conformal charges.

It is also possible to rewrite the quasimomentum operator 
in terms of  $ \mathbb{Q}_+(u,\wbar{u})$:
\begin{equation}
  \Hat{{\theta}}_N=-i \ln \mathbb{P} 
= i \ln \frac{ \mathbb{Q}_{+}(is,i\wbar{s})}{
\mathbb{Q}_{+}(-is,-i\wbar{s})}\,.
\lab{eq:quasQ}
\end{equation}
Moreover, using the mirror permutation (\ref{eq:PMq}) one finds 
the following parity relations for the residue functions $R^+(q,\wbar q)$
defined in (\ref{eq:Q-R,E}):
\begin{equation}
R^+(q,\wbar q)/R^+(-q,-\wbar q)=\e^{2i\theta_N(q,\wbar q)}
\lab{eq:RR}
\end{equation}
and for the eigenvalues of the Baxter operator:
\begin{equation}
Q_{q,\wbar q}(-u,-\wbar u) = \e^{i\theta_N(q,\wbar q)} 
Q_{-q,-\wbar q}(u,\wbar u)\,,
\lab{eq:Q-symmetry}
\end{equation}
where $-q\equiv(q_2,-q_3,\ldots ,(-1)^n q_n)$ and similarly for $\wbar q$. 
Examining the behaviour of (\ref{eq:Q-symmetry}) 
around the pole
at $u=u_{1}^{\pm}$ and $\wbar u={\wbar u}_{1}^{\pm}$ and making use of
Eq.~(\ref{eq:Q-R,E}) one gets
\begin{equation}
R^\pm(q,\wbar q)=(-1)^N\e^{i\theta_N(q,\wbar q)}R^\mp(-q,-\wbar q)\,,\qquad
E^\pm(q,\wbar q)=-E^\mp(-q,-\wbar q)\,.
\lab{eq:sym}
\end{equation}

To obtain the expression for the energy $E_N(q,\wbar q)$, we apply 
(\ref{eq:enQ})
and
replace the function $Q_{q,\wbar q}(u\pm i(1-s),u\pm i(1-\wbar
s))$ by its pole expansion (\ref{eq:Q-R,E}). 
Then, applying the second relation in
(\ref{eq:sym}), one finds
\begin{equation}
E_N(q,\wbar q)= E^+(-q,-\wbar q)+ (E^+(q,\wbar q))^*=\Re\left[E^+(-q,-\wbar q)+
E^+(q,\wbar q)\right]\,,
\lab{eq:energy}
\end{equation}
where the last relation follows from hermiticity of the Hamiltonian 
(\ref{eq:Ham}). We
conclude from Eqs.~(\ref{eq:energy}) and (\ref{eq:Q-R,E}), 
that in order to find the energy
$E_N(q,\wbar q)$, one has to calculate the residue of 
$Q_{q,\wbar q}(u,\wbar u)$ at
the $(N-1)$-th order pole at $u=i(s-1)$ and $\wbar u=i(\wbar s-1)$.

\section{Construction of the eigenfunction}

The Hamiltonian eigenstate $\Psi_{\vec{p},\{q,\wbar{q}\}}(\vec{z})$
is a common state of the total set of the integrals of motion,
$\vec{p}$ and $\{q,\wbar{q}\}$ as well as the Baxter $Q$-operator.
Thanks to the method of the Separation of Variables (SoV) developed by
Sklyanin \ci{Sklyanin:1991ss,Derkachov:2001yn} we can write the eigenstate  
using separated coordinates
$\vec{\mybf x}=(\vec{x}_1,\ldots,\vec{x}_{N-1})$ as
\begin{equation}
\Psi_{\vec{p},\{q,\wbar{q}\}}(\vec{z})=
\int d^{N-1} \vec{\mybf x} \,
\mu(\vec{x}_1,\ldots,\vec{x}_{N-1})
\, U_{\vec{p},\vec{x}_1,\ldots,\vec{x}_{N-1}}(\vec{z}_1,\ldots,\vec{z}_{N})
\left(\Phi_{q,\wbar{q}}(\vec{x}_1,\ldots,\vec{x}_{N-1})\right)^{\ast}\,,
\lab{eq:PsiU}
\end{equation}
where $U_{\vec{p},\vec{\mybf x}}$ is the kernel of the unitary operator while
\begin{equation}
\left(\Phi_{q,\wbar{q}}(\vec{x}_1,\ldots,\vec{x}_{N-1})\right)^{\ast}
=e^{i \theta_N(q,\wbar{q})/2}
\prod_{k=1}^{N-1}
\left(
\frac{\Gamma(s+ix_k)\Gamma(\wbar{s}-i\wbar{x}_k)}{
\Gamma(1-s+ix_k)\Gamma(1-\wbar{s}-i\wbar{x}_k)} 
\right)^N
Q_{q,\wbar{q}}(x_k,\wbar{x}_k)\,.
\lab{eq:PhiQ}
\end{equation}
The functions
$Q_{q,\wbar{q}}(x_k,\wbar{x}_k)$ are eigenstates of the Baxter $Q$-operator.
In contrast to the $\vec{z}_i=(z_i,\wbar{z}_i)$ - coordinates, 
the allowed values
of separated coordinates are
\begin{equation}
x_k=\nu_k-\frac{i n_k}{2}\,, \qqquad
\wbar{x}_k=\nu_k+\frac{i n_k}{2}
\lab{eq:xcoords}
\end{equation} 
with $n_k$ integer and $\nu_k$ real. Integration over the space of separated
variables implies summation over integer $n_k$ and integration over 
continuous $\nu_k$
\begin{equation}
\int d^{N-1} \vec{\mybf x}=
\prod_{k=1}^{N-1} 
\left(\sum_{n_k=-\infty}^{\infty} \int_{-\infty}^{\infty} d \nu_k \right),
\qqquad
\mu(\vec{\mybf x})=\frac{2 \pi^{-N^2}}{(N-1)!}
\prod_{\shortstack{$\scriptstyle j,k=1$\\$\scriptstyle j>k$}}^{N-1}
\left| \vec{x}_k - \vec{x}_j \right|^2 \,,
\lab{eq:xmes}
\end{equation}
where $\left| \vec{x}_k - \vec{x}_j \right|^2=(\nu_k-\nu_j)^2+
(n_h-n_j)^2/4$.

The integral kernel  $U_{\vec{p},\vec{\mybf x}}$ can be written as
\begin{equation}
U_{\vec{p},\vec{\mybf x}}(\vec{z}_1,\ldots,\vec{z}_N)
=c_N(\vec{\mybf x})(\vec{p}^{\, \,2})^{(N-1)/2}
\int d^2 w_N e^{2 i \vec{p} \cdot \vec{w}_N}
U_{\vec{\mybf x}}(\vec{z}_1,\ldots,\vec{z}_N;\vec{w}_N)\,,
\lab{eq:Upker}
\end{equation}
where $2 \vec{p} \cdot \vec{w}_N=p \, w_N+ \wbar{p} \, \wbar{w}_N$,
\begin{equation}
U_{\vec{\mybf x}}(\vec{z}_1,\ldots,\vec{z}_N;\vec{w}_N)=
\left[
\Lambda_{N-1,(\vec{x}_1)}^{(s,\wbar{s})}
\Lambda_{N-2,(\vec{x}_2)}^{(1-s,1-\wbar{s})}
\ldots
\Lambda_{1,(\vec{x}_{N-1})}^{(s,\wbar{s})}
\right](\vec{z}_1,\ldots,\vec{z}_N|\vec{w}_N)
\lab{eq:Uweker}
\end{equation}
for even $N$, and
\begin{equation}
U_{\vec{\mybf x}}(\vec{z}_1,\ldots,\vec{z}_N;\vec{w}_N)=
\left[
\Lambda_{N-1,(\vec{x}_1)}^{(s,\wbar{s})}
\Lambda_{N-2,(\vec{x}_2)}^{(1-s,1-\wbar{s})}
\ldots
\Lambda_{1,(\vec{x}_{N-1})}^{(1-s,1-\wbar{s})}
\right](\vec{z}_1,\ldots,\vec{z}_N|\vec{w}_N)
\lab{eq:Uwoker}
\end{equation}
for odd $N$. Here the convolution involves the product of $(N-1)$
functions $\Lambda_{N-k,(\vec{x}_k)}$ with alternating spins $(s,\wbar{s})$
and $(1-s,1-\wbar{s})$. They are defined as
\begin{multline}
\Lambda_{N-n,(\vec{x})}^{(s,\wbar s)}
(\vec{z}_n,\ldots,\vec{z}_N|\vec{y}_{n+1},\ldots,\vec{y}_N)=
\left[z_1-y_2\right]^{-x+iu} \\
\times \left(
\prod_{k=n+1}^{N-1}
\left[z_k-y_k\right]^{-x-iu}
\left[z_k-y_{k+1}\right]^{-x+iu}
\right)
\left[z_N-y_N\right]^{-x-iu}\,,
\lab{eq:Lambker}
\end{multline}
where the convolution  $[\Lambda_{N-k,(\vec{x}_{k})}
\Lambda_{N-k+1,(\vec{x}_{k-1})}]$
contains $(N-k)$ two-dimensional integrals.
The coefficient $c_N(\vec{\mybf x})$  is given for $N \ge 3$
\begin{equation}
 c_N(\vec{\mybf x})=
\prod_{k=1}^{[(N-1)/2]}
\left( a(s+ix_{2k},\wbar{s}-i\wbar{x}_{2k})\right)^{N-k}
\prod_{k=1}^{[N/2-1]}
\left( a(s+ix_{2k+1},\wbar{s}-i\wbar{x}_{2k+1})\right)^{k}
\lab{eq:cnx}
\end{equation}
while the sums goes over integer numbers lower than upper limit.
For $N=2$ we have $c_2(\vec{x}_1)=1$.

\chapter{WKB approximation}

\section{WKB approximation for Reggeons}

The WKB approach was presented in a series of papers 
\ci{Korchemsky:1995be,Korchemsky:1997ve,Derkachov:2002pb}. 
We will use this method
to find an approximation set of conformal charges
as well as to understand the structure of the $q_k$'s values.
In the representation of the separated coordinates
(\ref{eq:PsiU}) -- (\ref{eq:xcoords}) 
the Baxter equations 
(\ref{eq:Baxeq}), (\ref{eq:Baxbeq}) 
play the role of one-dimensional Schr\"odinger equations
where the eigenfunctions are the eigenvalues of 
the Baxter operator (\ref{eq:PhiQ}).
These equations may be solved 
in the quasi-classical limit \ci{Derkachov:2002pb} 
which corresponds to large values 
of the conformal charges, $\oq{k}$ and $\oqb{k}$, (\ref{eq:tnu}).

In order to apply the  WKB limit we introduce an auxiliary parameter
$\eta$ and rescale simultaneously the coordinates, $x\rightarrow x/\eta$, and
the charges. Thus we define
\begin{equation} 
\st{N}(x)=\eta^N t_N(x/\eta)=2x^N+\sq{2}x^{N-2}+\ldots+\sq{N} x^N
\lab{eq:stn}
\end{equation}
with $\sq{k} \equiv q_k \eta^k = {\cal O}(\eta^0)$ as $\eta\rightarrow0$.
The small parameter $\eta$ appears in the Baxter equation 
\begin{equation}
\left(x -i \,\eta\,s \right)^N \, Q((x-i \eta)/\eta) +
\left(x +i \,\eta\,s \right)^N \, Q((x+i \eta)/\eta)  = Q(x/\eta)\,\st{N}(x)
\lab{eq:sBax}
\end{equation}
which allows us to perform 
the WKB expansion
\begin{equation}
Q(x/\eta)=\exp\left(\frac{i}{\eta}\int_{x_0}^x dx \, S'(x)\right)
\quad
\mbox{with}
\quad
S(x)=S_0(x)+\eta S_1(x)+\eta^2 S_2(x)+{\cal O}(\eta^3),
\lab{eq:WKB}
\end{equation}
where $S'(x)=dS(x)/dx$ and $x_0$ is an arbitrary reference point.
Substituting (\ref{eq:WKB}) into the Baxter equation (\ref{eq:sBax})
we obtain relations
\begin{equation}
2 \cosh S_0'(x)=\frac{\st{N}(x)}{x^N},
\qquad
S_1'(x)=\frac{i}{2}(\ln \sinh S_0'(x))'+\frac{i s N}{x}.
\lab{eq:S0S1}
\end{equation}
In this way one can obtain further relations for higher $S_k(x)$.

The first relation in (\ref{eq:S0S1}) can be rewritten as
\begin{equation}
y^2(x)=\st{N}\negthickspace{}^2(x)-4x^{2 N}\,,
\lab{eq:yx}
\end{equation}
where
\begin{equation}
y(x)=2 x^N \sinh p_x
\quad
\mbox{with}
\quad
p_x=S_0\negthickspace '(x)
\lab{eq:yS}
\end{equation}
is a hyperelliptic curve.
Solving this relation (\ref{eq:S0S1}) we obtain 
the leading term of (\ref{eq:WKB}) 
\begin{equation}
S_0(x)=\int_{x_0}^x dx \, p_x=
\int_{x_0}^x \frac{dx}{y(x)}(N \st{N}(x)-x\st{N}\negthickspace '(x))
+x p_x \bigg|_{x_0}^{x}.
\lab{eq:S0x}
\end{equation}
The hyperelliptic curve, $y(x)$, as well as  $S'(x)$, are double-valued 
functions on the complex $x$-plane. Two different branches of $S_{0}'(x)$ 
will be denoted by 
$(\pm)$, so we have $S_{0,\pm}'(x)$.
To specify them  
one makes cuts on the $x$-plane in an arbitrary way between  the $2(N-1)$ 
branching points $\sigma_j$ where $y(\sigma_j)=0$.
Taking (\ref{eq:S0S1}) and (\ref{eq:S0x}) we have 
the first non-leading correction
\begin{equation}
S_1'(x)=\frac{i}{2}\left(\ln \frac{y(x)}{2 x^N}\right)^{\prime}+\frac{iNs}{x}
=\frac{i}{4}\sum_{j=0}^{2 N -2}\frac{1}{x-\sigma_j}
+\frac{i N}{x}\left(s-\frac12\right).
\lab{eq:S1x}
\end{equation}
which is a single-valued function on the complex $x-$plane.

Combining together (\ref{eq:S0x}) and (\ref{eq:S1x}) we obtain
two different solutions to the holomorphic Baxter equation (\ref{eq:sBax})
\begin{equation}
Q_{\pm}(x/\eta)=\exp\left(\frac{i}{\eta}\int_{x_0}^{x} dx S_{\pm}'(x)\right)\,,
\quad
\mbox{where}
\quad
S_{\pm}'=S_{0,\pm}+\eta S_1'(x)+{\cal O}(\eta^2)
\lab{eq:Qpm}
\end{equation} 
with asymptotics at $x\rightarrow \infty$:
\begin{equation}
Q_+(x/\eta)\sim x^{1-h-N s},
\qquad
Q_-(x/\eta)\sim x^{h-N s}.
\lab{eq:Qasm}
\end{equation}
Similarly, one gets solutions in the anti-holomorphic sector
with 
\begin{equation}
\wbar Q_+(\wbar{x}/\eta)\sim \wbar x^{1-\wbar{h}-N \wbar{s}},
\qquad
\wbar Q_-(\wbar x/\eta)\sim \wbar x^{\wbar h-N \wbar s},
\lab{eq:Qbasm}
\end{equation}
as  $\wbar x\rightarrow \infty$.

\section{Quantization conditions with WKB}
The quasi-classical solution of the Baxter equation (\ref{eq:sBax})
is given as a bilinear combination of the chiral solutions $Q_{\pm}(x/\eta)$
and $\wbar Q_{\pm}(\wbar x/\eta)$:
\begin{equation}
Q(x/\eta,\wbar x/\eta)=
c_+ Q_+(x/\eta)\wbar Q_+(\wbar x/\eta)+
c_- Q_-(x/\eta)\wbar Q_-(\wbar x/\eta).
\lab{eq:sQxx}
\end{equation}
The cross-terms $Q_{\pm} \wbar Q_{\mp}$ do not enter (\ref{eq:sQxx})
since they do not satisfy (\ref{eq:Qanalb}).

The function (\ref{eq:sQxx}) should not depend on reference point $x_0$.
This gives condition
\begin{equation}
c_{\pm}(x_0')=c_{\pm} \exp
\left(
\frac{i}{\eta}
\int_{x_0}^{x_0'} dx S_{\pm}'(x)+
\frac{i}{\eta}
\int_{\wbar x_0}^{\wbar x_0'} d\wbar x \wbar S_{\pm}'(\wbar x)
\right)
\lab{eq:cx0}
\end{equation}
with $\wbar x_0=x_0^{\ast}$. 
Additionally, taking into account that
\begin{equation} 
c_+(\sigma_j)=c_-(\sigma_j)
\quad
\mbox{with}
\quad
j=1,2,\ldots,2(N-1)
\lab{eq:cpcm}
\end{equation}
one can derive \ci{Derkachov:2002pb} the quantization conditions
\begin{equation}
\RRe \oint_{\alpha_k} dx S_0'(x)/\eta=\pi \ell_{2k-1}+{\cal O(\eta)},
\qquad
\RRe \oint_{\beta_k} dx S_0'(x)/\eta=\pi \ell_{2k}+{\cal O(\eta)},
\lab{eq:sqcond}
\end{equation}
where $k=1,\ldots,N-2$ and $\mybf{\ell}=(l_1,\ldots,l_{2N-4})$ being integer 
and the integrals are performed over cycles around branching points $\sigma_j$,
Fig.\ \ref{fg:cuts}.The relations (\ref{eq:sqcond}) give the WKB quantization conditions for 
the integrals of motion, $q_k$ and $\wbar q_k$.
\begin{figure}[ht]
\vspace*{3mm}
\centerline{{\epsfysize8cm \epsfbox{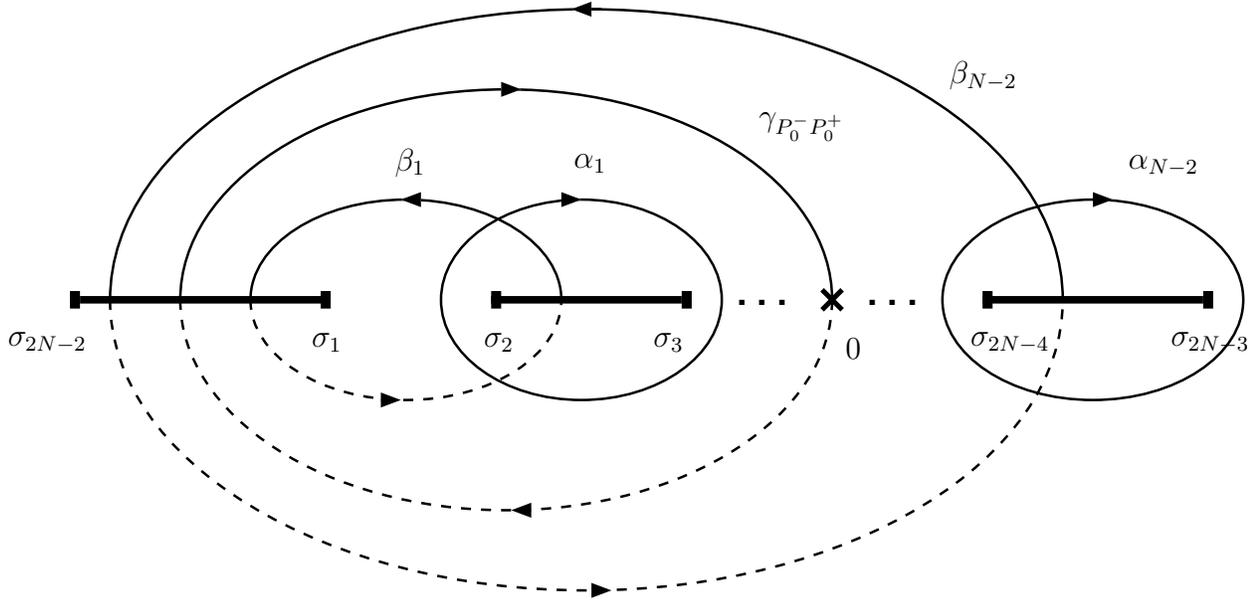}}}
\caption[The Riemann surface $\Gamma_N$ with integration cycles,
 $\alpha_j$ and $\beta_j$.]{The Riemann 
surface $\Gamma_N$ with integration cycles,
 $\alpha_j$ and $\beta_j$. The dashed lines represent the contours
on the lower sheet \ci{Derkachov:2002pb}. The contour $\gamma_{P_0^- P_0^+}$
used in (\ref{eq:quasas}), goes from the point $x=0$ on the lower sheet
to $x=0$ on the upper sheet.}
\lab{fg:cuts}
\end{figure}

\section{Lattice structure of the conformal charges spectrum}
Additionally \ci{Derkachov:2002pb}, 
one can obtain the quasimomentum, 
(\ref{eq:quasQ}) as 
\begin{equation}
\theta_N= -2 Re \oint_{\gamma_{P_0^- P_0^+}} dx S_0'(x)/\eta
=\frac{2 \pi}{N}\ell
\quad (\Mod 2 \pi)
\lab{eq:quasas}
\end{equation}
and the energy (\ref{eq:enQ}) in the WKB approximation
\begin{multline}
E_N^{(as)}=4 \ln 2 +
2 \RRe \sum_{\IIm \lambda_k \ge 0}
\left(
\psi(1-s-i\lambda_k)+\psi(s-i\lambda_k)-2 \psi(1)
\right)\\
+2 \RRe \sum_{\IIm \lambda_k < 0}
\left(
\psi(1-s+i\lambda_k)+\psi(s+i\lambda_k)-2 \psi(1)
\right)\,,
\lab{eq:enas}
\end{multline}
where $t_N(\lambda_k)=0$ for $k=1,\ldots,N$.

The further analysis of (\ref{eq:sqcond}) and (\ref{eq:quasas}) 
where we fix $\eta = (q_N/4)^{-1/N}$,
which is equivalent to expansion for large conformal charges,
gives the system of $(N-2)$ equations for the integrals of motion
\begin{equation}
\begin{split}
& u_N \cdot N \B\left(\frac{1}{2},\frac{1}{N}\right)
-(u_2 u_N)^{\ast} \cdot \B\left(\frac{1}{2},\frac{N-1}{N}\right)=
\pi\sum_{k=1}^{N} \e^{-i \pi (2k -1)/N} n_k, \\
& (u_{N+1-m} u_N) \cdot  \B\left(\frac{1}{2},\frac{1}{N}\right)
-(u_{N+1-m} u_N)^{\ast} \cdot \B\left(\frac{1}{2},\frac{N-m}{N}\right)=
\pi\sum_{k=1}^{N} \e^{-i \pi (2k -1)m/N} n_k, 
\end{split}
\lab{eq:ukcond}
\end{equation}
with $m=2,\ldots,N-2$ and
\begin{equation}
u_n=\frac{q_n}{4} \left(\frac{q_N}{4}\right)^{-n/N}
\quad 
\mbox{for} 
\; n=2,\ldots,N-2,
\quad
u_N=\left(\frac{q_N}{4}\right)^{1/N}.
\lab{eq:unuN}
\end{equation}
The Euler beta-function $\B(x,y)=\Gamma(x)\Gamma(y)/\Gamma(x+y)$ 
and $n_k$ are integer numbers related to the quasimomentum (\ref{eq:quasas}) by
\begin{equation}
\ell=-\sum_{k=1}^{N} n_k 
\qquad 
(\Mod N).
\end{equation}
The equations (\ref{eq:ukcond}) give only $(N-1)$ conditions 
so the system 
is underdetermined. 

Solving (\ref{eq:ukcond}) one can obtain the quantized values
of the highest conformal charge $q_N$ as
\begin{equation}
q_N^{1/N}=\pi \frac{\Gamma(1+2/N)}{\Gamma^2(1/N)}
{\cal Q}(\mybf n)
\left(
1+\frac{q_2^{\ast}}{2 \pi}\frac{N^2}{(N-2)} 
\cot(\pi/N)|{\cal Q}(\mybf n)|^{-2}
+{\cal O}({\cal Q}(\mybf n)|^{-4})
\right)\,,
\lab{eq:qNqnt}
\end{equation}
where
\begin{equation}
{\cal Q}(\mybf n)=\sum_{k=1}^{N}n_k \e^{-i \pi (2k-1)/N}.
\lab{eq:calq}
\end{equation}
with ${\mybf n}=\{n_1,\ldots,n_N\}$ integer 
forming a lattice structure on the spectrum
of $q_N$ (\ref{eq:qNqnt}).

\chapter{Eigenfunctions for the conformal charges {$\oq{k}$} }

In this Chapter we present the general method
of constructing the ansatzes for eigenfunction
of the conformal charges $\oq{k}$.
These ansatzes have been commonly used before  
\ci{Wosiek:1996bf,Praszalowicz:1998pz,Kotanski:2001iq,Lipatov:1998as}.
Here we systematize the knowledge about 
this ansatzes and extend our calculations 
to an arbitrary complex spin $s$.
Moreover, we derive  
differential eigenequations for the conformal 
charges. 
Next,
we show solutions to the differential 
equation for $N=3$ and $s=0$ using 
the series method
and present derivation of the quantization conditions 
for the integrals of motion $q_k$.
Finally, we resum obtained series solutions for 
the $q_3=0$ case.

\section{Solution of the $\oq{2}$ eigenproblem}

In this Section we solve the $\oq{2}$ eigenproblem.
This gives us not only the solution for
the $N=2$ Pomeron problem but also  provides us with 
a structure of the ansatz
for the eigenfunctions of 
the conformal charges, $\oq{k}$ for $N>2$. 

As one can see from Eq. (\ref{eq:Psip}),
the eigenfunctions (\ref{eq:trpsiz0}) of the Hamiltonian (\ref{eq:sepH})
depend on differences of coordinates, $z_i-z_0$.
They may contain two types of factors.
The first ones are arbitrary functions of $(N-2)$ 
variables, 
$x_{i}=\frac{(z_{i-1}-z_{r})(z_{i+1}-z_{0})}{(z_{i-1}-z_{0})(z_{i+1}-z_{r})}$, 
which are $SL(2,\mathbb{C})-$invariant. 
They are defined in Appendix A.
The other ones are
some products of the coordinate differences
that satisfy the transformation law (\ref{eq:trpsiz0}).
Let us consider the general product 
\begin{equation}
\prod _{i_{2}>i_{1}=0}^{N}(z_{i_{2}i_{1}})^{k_{i_{1}i_{2}}}\,,
\lab{eq:zprod}
\end{equation} 
where $z_{i_{2}i_{1}}=z_{i_2}-z_{i_1}$ and $k_{i_{1}i_{2}}$ is arbitrary.

It is convenient to parameterize the eigenvalues of the lowest conformal
charge as 
\begin{equation}
q_{2}=-h(h-1)+\sum _{j=1}^{N}s_{j}(s_{j}-1)\,,
\lab{eq:q2}
\end{equation} 
where $h$
and $s_{j}$ are the conformal spins (\ref{eq:SCas}). 
Now one can act 
with the $\oq{2}$ operator (\ref{eq:q2q3}) 
on the product of holomorphic coordinates (\ref{eq:zprod}):
\begin{eqnarray}
\hat{q}_{2}\prod _{i_{2}>i_{1}=0}^{N}(z_{i_{2}i_{1}})^{k_{i_{1}i_{2}}} & = & 
q_2
\prod _{i_{2}>i_{1}=0}^{N} 
(z_{i_{2}i_{1}})^{k_{i_{2}i_{1}}}\,.
\lab{eq:egq2}
\end{eqnarray}
In this way we obtain the equation for $k_{i_{2}i_{1}}$
which can be easily solved. 

For each $N$ we obtain two sets of solutions: the first ones, with 
the $z_0$ coordinate,
and the second ones, without the $z_0$ coordinate. 
The latter solutions may be calculated by integrating the first ones  
over $z_0$ with $\vec{p}=0$ with the help of (\ref{eq:Psip}). 
They don't transform
as ({\ref{eq:trpsiz0}}) so they will not be further considered. 
The integration with an arbitrary $\vec{p}$ (\ref{eq:Psip}) for two Reggeons
was performed in \ci{Navelet:1997xn}.
In the first set we obtain two groups of solutions related 
by the symmetry $h\rightarrow 1-h$, which comes from that the form of the 
$q_2$ eigenvalue (\ref{eq:q2}). More detailed
studies of this problem are included in Appendix B.

As a result we obtain an ansatz for the eigenfunction.
For an arbitrary $N$ it has a form 
\begin{equation}
\Psi(z_{10},z_{20},\ldots,z_{N0})  
=\frac{1}{(z_{10})^{2s_{1}}(z_{20})^{2s_{2}}\ldots (z_{N0})^{2s_{N}}}
\left(\frac{z_{31}}{z_{10}z_{30}}\right)^{h-s_{1}-s_{2}-\ldots -s_{N}}
F(x_{1},x_{2},\ldots ,x_{N-2})\,.
\end{equation}
For $N$ particles we have $N-2$ invariant independent variables,
$x_{i}$.
For the homogeneous spin, i.e. with $s_1=s_2=\ldots=s_N=s$, chains we have
\begin{equation}
\Psi (z_{10},z_{20},\ldots,z_{N0}) 
=\frac{1}{(z_{10}z_{20}\ldots z_{N0})^{2s}}\left(\frac{z_{31}}{z_{10}
z_{30}}\right)^{h-Ns}F(x_{1},x_{2},\ldots ,x_{N-2})\,.
\end{equation}

\section{Various ansatzes}

Using the above ansatzes
we can construct other ansatzes which are equivalent to the original
ones. For example, for $N=3$ \ci{Lipatov:1998as},
\begin{equation}
 \Psi (z_{10},z_{20},z_{30})   =  
\frac{1}{(z_{10})^{2s_{1}}(z_{20})^{2s_{2}}
(z_{30})^{2s_{3}}}
\left(\frac{z_{31}}{z_{10}z_{30}}\right)^{h-s_{1}-s_{2}-s_{3}}F(x)\,,
\lab{eq:orgAn}
\end{equation}
where  $x=x_{2}$ we substitute 
$F(x)=G(x) \left(\frac{(x-1)^2}{-x}\right)^{h/3-s_1/3-s_2/3-s_3/3}$ 
what gives an ansatz
\begin{equation}
 \Psi (z_{10},z_{20},z_{30})   = 
 \frac{1}{(z_{10})^{2s_{1}}(z_{20})^{2s_{2}}(z_{30})^{2s_{3}}}
\left(\frac{z_{31}z_{12}z_{23}}{(z_{10})^{2}
(z_{20})^{2}(z_{30})^{2}}\right)^{\frac{h}{3}-\frac{s_{1}}{3}
-\frac{s_{2}}{3}-\frac{s_{3}}{3}}G(x)
\end{equation}
or equivalently
\begin{multline}
\Psi (z_{10},z_{20},z_{30})   = 
\left(\frac{(z_{20})^{2}(z_{30})^{2}}{(z_{10})^{4}z_{12}z_{23}z_{31}}
\right)^{\frac{s_{1}}{3}}
\left(\frac{(z_{10})^{2}
(z_{30})^{2}}{(z_{20})^{4}z_{12}z_{23}z_{31}}\right)^{\frac{s_{2}}{3}}
\left(\frac{(z_{20})^{2}(z_{10})^{2}}{(z_{30})^{4}z_{12}z_{23}z_{31}}
\right)^{\frac{s_{3}}{3}}\\
\times
\left(\frac{z_{31}z_{12}z_{23}}{(z_{10})^{2}
(z_{20})^{2}(z_{30})^{2}}\right)^{\frac{h}{3}}G(x)\,.
\lab{eq:symAn}
\end{multline}
As we can see we obtained a totally symmetric
ansatz which is equivalent to original one. For the homogeneous spin chains
we have
\begin{equation}
\Psi (z_{10},z_{20},z_{30})
=\frac{1}{(z_{12}z_{23}z_{31})^{s}}
\left(\frac{z_{31}z_{12}z_{23}}{(z_{10})^{2}
(z_{20})^{2}(z_{30})^{2}}\right)^{\frac{h}{3}}G(x)\,.
\end{equation}

Both ansatzes have advantages and disadvantages. The symmetric one 
(\ref{eq:symAn})
is appropriate if we want to deal with the particle symmetries.
The original one (\ref{eq:orgAn})
has a simpler structure, it contains powers of
$h$ (not $h/3$), so we can use it when we want to construct proper
single-valuedness conditions. 
One can easily notice that $z^{h}\wbar{z}^{\wbar{h}}$
is single-valued because $h-\wbar{h}=n_{h}\in \mathbb{Z}$.

Similarly, we can go from the original ansatz 
(\ref{eq:orgAn})
to an ansatz with different
permutation of particles
\begin{equation}
\begin{array}{ccc}
 \Psi(z_{10},z_{20},z_{30}) & = & 
\frac{1}{(z_{10})^{2s_{1}}(z_{20})^{2s_{2}}(z_{30})^{2s_{3}}}
\left(\frac{z_{31}}{z_{10}z_{30}}\right)^{h-s_{1}-s_{2}-s_{3}}F(x)=\\
  & = & \frac{1}{(z_{10})^{2s_{1}}(z_{20})^{2s_{2}}(z_{30})^{2s_{3}}}
\left(\frac{z_{12}}{z_{10}z_{20}}\right)^{h-s_{1}-s_{2}-s_{3}}H(x)\\
  & = & \frac{1}{(z_{10})^{2s_{1}}(z_{20})^{2s_{2}}(z_{30})^{2s_{3}}}
\left(\frac{z_{32}}{z_{20}z_{30}}\right)^{h-s_{1}-s_{2}-s_{3}}J(x)\,.
\end{array}
\lab{eq:n3psiz0b}
\end{equation}

Now we can generalize our ansatz for different number of particles, $N$.
In this case we have
\begin{equation}
\begin{array}{ccl}
 \Psi (z_{10},z_{20},\ldots,z_{N0}) & = & 
\left(\frac{(z_{10})^{2}(z_{20})^{2}\ldots (z_{N0})^{2}}{(z_{10})^{2N}
z_{12}z_{23}\ldots z_{N1}}\right)^{\frac{s_{1}}{N}}
\left(\frac{(z_{10})^{2}(z_{20})^{2}\ldots (z_{N0})^{2}}{(z_{20})^{2N}
z_{12}z_{23}\ldots z_{N1}}\right)^{\frac{s_{2}}{N}}\ldots \\
  &  & \ldots \left(\frac{(z_{10})^{2}(z_{20})^{2}\ldots 
(z_{N0})^{2}}{(z_{N0})^{2N}z_{12}z_{23}
\ldots z_{N1}}\right)^{\frac{s_{N}}{N}}
\left(\frac{z_{12}z_{23}\ldots z_{N1}}{(z_{10})^{2}(z_{20})^{2}
\ldots (z_{N0})^{2}}\right)^{\frac{h}{N}}G(x_{1},x_{2},\ldots ,x_{N-2})\,.
\end{array}
\lab{eq:nNpsiz0}
\end{equation}
And for the homogeneous spin chains, $s_i=s$, it has a form
\begin{equation}
\Psi (z_{10},z_{20},\ldots,z_{N0})=
\frac{1}{(z_{12}z_{23}\ldots z_{N1})^{s}}
\left(\frac{z_{12}z_{23}\ldots z_{N1}}{(z_{10})^{2}(z_{20})^{2}
\ldots (z_{N0})^{2}}\right)^{\frac{h}{N}}G(x_{1},x_{2},\ldots ,x_{N-2})\,.
\end{equation}

To sum up, we have many equivalent ansatzes. 
Here we presented two of them,
(\ref{eq:orgAn}) and (\ref{eq:symAn}). They are
most frequently used due to their simplicity 
and symmetry properties.

\section{Solutions of the $\hat{q}_{3}$ eigenproblem for $N=3$}
In this Section we solve
the $\hat{q}_{3}$ eigenproblem for three Reggeons.
We derive differential equations for
the integral of motion $q_3$ for an arbitrary
complex spin $s$
and solve them for $s=\wbar s=0$ making use of the series method.
We also construct the quantization conditions for $q_3$
and 
in the end we resum the series for zero-modes of $\oq{3}$.
We obtain the known solution (\ref{eq:Psib}) as well as some others
with $\Log(x)-$terms (\ref{eq:Psiq0}).

\subsection{Derivation of differential equations for $N=3$}

In the previous Section we solved the eigenproblem for $\hat{q}_{2}$.
However, for $N$ particles there are $N-1$ conformal charges which 
commute with the Hamiltonian. 
Thus, for $N=3$ we have $\hat{q}_{2}$ and $\hat{q}_{3}$.
In order to to solve the eigenproblem for $\hat{q}_{3}$ 
we can act with $\oq{3}$ on our
ansatz 
\begin{equation}
\Psi(z_{10},z_{20},z_{30})
=\frac{1}{(z_{10})^{2s_{1}}(z_{20})^{2s_{2}}(z_{30})^{2s_{3}}}
\left(\frac{z_{32}}{z_{20}z_{30}}\right)^{h-s_{1}-s_{2}-s_{3}}F(x)
\lab{eq:Anz32}
\end{equation}
and derive a differential equation for $F(x)$ with
$x=x_{2}=\frac{(z_{1}-z_{2})(z_{3}-z_{0})}{(z_{1}-z_{0})(z_{3}-z_{2})}$
. Thus we obtain
\begin{equation}
\begin{array}{c}
i q_{3}F(x)=-s_{1}(s_{1}+s_{2}+s_{3}-h)(h-1-s_{1}-s_{2}+s_{3}-2hx+2(1+s_{1})x)F(x)+\\
+((s_{1}+s_{2})(1-h+s_{1}+s_{2}-s_{3})+\\
-(2+h^{2}+7s_{1}+3s_{2}+(s_{1}+s_{2})(5s_{1}+s_{2})-h(3+6s_{1}+2s_{2})+s_{3}-s_{3}^{2})x+\\
+(h^{2}+2(1+s_{2}+s_{3})-h(3+6s_{1}+s_{2}+s_{3})+s_{1}(7+5s_{1}+3s_{2}+3s_{3}))x^{2})F^{\prime}(x)+\\
-(x-1)x(2+2s_{1}+2s_{2}-s_{3}-(4+4s_{1}+s_{2}+s_{3})x+h(2x-1))
F^{\prime\prime}(x)+\\
+(x-1)^{2}x^{2}F^{(3)}(x)\,.
\end{array}
\end{equation} 
Using various ansatzes we obtain equivalent differential equations.

For the homogeneous chain our equation looks like
\begin{equation}
\begin{array}{c}
i q_{3}F(x)=
(3s-h)(h-1-s)(1-2x)F(x)
+(((h-2)(h-1)(x-1)x+s^{2}(2+11(x-1)x)+\\
+s(2-2h(1-2x)^{2}+11(x-1)x))F^{\prime}(x)
+(2+h-3s)(1-x)x(2x-1)F^{\prime\prime}(x)+\\
+(x-1)^{2}x^{2}F^{(3)}(x)\,.
\end{array}
\lab{eq:qF}
\end{equation}
From the QCD point of view the most interesting cases are for $s=0$:
\begin{equation}
\begin{array}{c}
i q_{3}F(x)=
(h-1)(h-2)x(x-1)F^{\prime}(x)
+(h-2)(x-1)x(1-2x)F^{\prime\prime}(x)
+x^{2}(x-1)^{2}F^{(3)}(x)
\end{array}\lab{eq:ees0z32}
\end{equation}
and for $s=1$
\begin{equation}
\begin{array}{c}
i q_{3}F(x)=
(h-3)(h-2)(2x-1)F(x)+\\
+((4-2h-(h-8)(h-3)x+(h-8)(h-3)x^{2})F^{\prime}(x)+\\
+(5-h)(x-1)x(2x-1)F^{\prime\prime}(x)
+(x-1)^{2}x^{2}F^{(3)}(x)\,.
\end{array}
\lab{eq:qFs1}
\end{equation}
The first such solution was derived and found numerically in 
\ci{Janik:1998xj}.

\subsection{Solutions for $s=0$}

Now, we will solve the above equations by the series method. 
All these equations
have three regular singular points at $x=0$, $x=1$ and $x=\infty$.
Let us take an ansatz (\ref{eq:Anz32}) which for $s=0$ can be 
rewritten as
\begin{equation}
\Psi(z_{10},z_{20},z_{30})=w^h F(x)\,,
\end{equation} 
where $w$ is defined similarly to (\ref{eq:wdef}) with $z_3$, $z_2$ 
and $z_0$. 
Thus, we obtained (\ref{eq:ees0z32})
\begin{equation}
\begin{array}{c}
i q_{3}F(x)=
(h-1)(h-2)x(x-1)F^{\prime}(x)
+(x-1)x(h-2)(1-2x)F^{\prime\prime}(x)
+x^{2}(x-1)^{2}F^{(3)}(x)\,.
\end{array}\lab{eq:ees0z32b}
\end{equation}
This equation can be used to obtain solution around $x=0^{+}.$ To
generate solutions around other singular points we exchange variables. For
the case $x=1^{-}$ we can use a substitution $x=1-y$:
\begin{equation}
\begin{array}{c}
i q_{3}G(y)=
(h-1)(h-2)y(1-y)G^{\prime}(y)
+(h-2)(1-y)y(1-2y)G^{\prime\prime}(y)
-y^{2}(y-1)^{2}G^{(3)}(y).
\end{array}
\lab{eq:ees0z32x1m}
\end{equation}
Moreover, for $x=1^{+}$ we can use $x=y+1$:
\begin{equation}
\begin{array}{c}
q_{3}G(y)=i\left(y(y+1)(h-1)(h-2)G^{\prime}(y)-(y+1)y(h-2)(1+2y)
G^{\prime\prime}(y)+y^{2}(y+1)^{2}G^{(3)}(y)\right)
\end{array}\lab{eq:ees0z32x1p}
\end{equation}
and for $x=\infty^{-}$ we have $x=1/y$:
\begin{multline}
q_{3}G(y)=i\left((y-1)(h+1)(h-2y)G^{\prime}(y)+(y-1)y(2(h+1)-(h+4)y)
G^{\prime\prime}(y) 
\right. \\ \left.
+y(y-1)^{2}G^{(3)}(y)\right)\,.
\lab{eq:ees0z32xi}
\end{multline}
The upper-scripts \textit{plus} and \textit{minus} correspond to case
where $\RRe[x]$ is above and below the singular point, respectively.

\subsection{Wave-function for $s=\wbar{s}=0$ around $x=\wbar{x}=0^+$}

In order to obtain the full-complex solution containing
the holomorphic and anti-holomorphic parts we have to glue together
solutions from these parts:
\begin{equation}
\Phi_{q,\wbar{q}}(\{ z_{i}\},\{\wbar{z}_{i}\})=
\wbar{u}_{\wbar{q}}(\{\wbar{z}_{i}\})^{T}\cdot 
A^{(0)}(h,\wbar{h},q_{3},\wbar{q}_{3})\cdot u_{q}(\{ z_{i}\})\,,
\lab{eq:Psizz}
\end{equation}
where we 
use a ($N\times N$) 
mixing-matrix, 
$A_{q,\wbar{q}}^{(0)}$, \ci{Janik:1998xj}
which does not depend on particle coordinates but only
on $q\equiv\{ q_{2},q_{3}\}$.
From the QCD point of view we have two possibilities of gluing solutions:
$(s=0,\wbar{s}=0)$ \ci{DeVega:2001pu} 
and $(s=0,\wbar{s}=1)$ \ci{Derkachov:2001yn}.
These two cases are equivalent except zero modes of the highest conformal
charge $\hat{q}_{N}$. Let us consider the first case, $s=\wbar{s}=0$.

The conformal charges are related by conditions $\wbar{h}=1-h^{*}$
and $\wbar{q}_{k}=q_{k}^{*}$. The wave-function has to be single-valued.
This condition defines the structure of the mixing-matrix. 

For $h\not\in\mathbb{Z}$ and $\hat{q}_{3}\ne0$ we have solutions
of the following type
\begin{equation}
\begin{array}{ccc}
u_{1}(x) & = & x^{h}\sum_{n=0}^{\infty}a_{n,r_{1}}x^{n}\,,\\
u_{2}(x) & = & x^{1}\sum_{n=0}^{\infty}a_{n,r_{2}}x^{n}\,,\\
u_{3}(x) & = & x^{0}\sum_{n=0}^{\infty}b_{n,r_{3}}x^{n}+x^{1}
\sum_{n=0}^{\infty}a_{n,r_{2}}x^{n}\mbox{Log}(x)
\end{array}
\lab{eq:u-zero}
\end{equation}
and similarly in the anti-holomorphic sector. 
The coefficient recurrence relations for the $a_{n,r_i}$ 
are given in Appendix C.
One can notice that $x^{a}\wbar{x}^{b}$
is single-valued only if $a-b\in\mathbb{Z}$. Moreover we have also
terms with $\mbox{Log}(x)$ which have to give in a sum
 $\mbox{Log}(x\wbar{x})$.
So in this case we have a mixing matrix of the form
\begin{equation}
A^{(0)}(h,\wbar{h},q_{3},\wbar{q}_{3})=
\left[
\begin{array}{ccc}
\alpha & 0 & 0\\
0 & \beta & \gamma\\
0 & \gamma & 0
\end{array}
\right]\,,
\end{equation}
where $\alpha$, $\beta$, $\gamma$ are arbitrary.
In the above matrix we have $A_{12}=A_{13}=A_{21}=A_{31}=0$ in order
to eliminate multi-valuedness coming from the power-terms, $A_{23}=A_{32}$
to obtain single-valuedness in $\Log(x)-$terms and $A_{33}=0$ because the term
$\mbox{Log}(x)\mbox{Log}(\wbar{x})$ is not single-valued on
the $\vec{x}-$plane.

In the case of $q_{3}=0$ and $h\not\in\{0,1\}$ we don't have any 
$\Log(x)-$terms
so the mixing matrix looks like 
\footnote{{\it Greek} variables in each  $A-$matrix have different 
numerical values}
\begin{equation}
A^{(0)}(h,\wbar{h},q_{3}=0,\wbar{q}_{3}=0)=
\left[
\begin{array}{ccc}
\beta & 0 & 0\\
0 & \alpha & \delta\\
0 & \varepsilon & \gamma
\end{array}
\right]\,.
\end{equation}

For $q_{3}\ne0$ and $h\in\mathbb{Z}$ we have a solution with all power
in $x$ integer and solutions without Log, with one-Log and with double-Log.
The structure of the matrix is
\begin{equation}
A^{(0)}(h\in\mathbb{Z},
\wbar{h}\in\mathbb{Z}
,q_{3},\wbar{q}_{3})
=\left[\begin{array}{ccc}
\alpha & \beta & \gamma\\
\beta & 2\gamma & 0\\
\gamma & 0 & 0
\end{array}\right]\,.
\end{equation}

In the last case for $q_{3}=0$ and $h\in\{0,1\}$ we have all powers
of $x$ integer and the third solution with one-Log term. The matrix 
has a form
\begin{equation}
A^{(0)}(h\in\{0,1\},
\wbar{h}\in\{0,1\}
,q_{3}=0,\wbar{q}_{3}=0)=
\left[
\begin{array}{ccc}
\alpha & \gamma & 0\\
\beta & \delta & \varepsilon\\
0 & \varepsilon & 0
\end{array}
\right]\,.
\end{equation}

\subsection{Wave-function for $s=\wbar{s}=0$ around $x=\wbar{x}=1^{-}$
and $x=\wbar{x}=\infty^-$}

Around the other singular point we construct the wave-function exactly
in the same way obtaining matrices 
$A^{(1^{-})}(h,\wbar{h},q_{3},\wbar{q}_{3})$,
$A^{(1^{+})}(h,\wbar{h},q_{3},\wbar{q}_{3})$ (around $1$) and 
$A^{(\infty^{-})}(h,\wbar{h},q_{3},\wbar{q}_{3})$ (around $\infty$).

Thus, we have the wave-function similar to
(\ref{eq:Psizz}).
For $h\not\in\mathbb{Z}$ and $\hat{q}_{3}\ne0$ we take solutions
\begin{equation}
\begin{array}{ccc}
u_{1}(x) & = & (1-x)^{h}\sum_{n=0}^{\infty}a_{n,r_{1}}(1-x)^{n}\,,\\
u_{2}(x) & = & (1-x)^{1}\sum_{n=0}^{\infty}a_{n,r_{2}}(1-x)^{n}\,,\\
u_{3}(x) & = & (1-x)^{0}\sum_{n=0}^{\infty}b_{n,r_{3}}(1-x)^{n}+(1-x)^{1}
\sum_{n=0}^{\infty}a_{n,r_{2}}(1-x)^{n}\mbox{Log}(1-x)
\end{array}
\lab{eq:u-one}
\end{equation}
and similarly in the anti-holomorphic sector. 
Our the wave-functions have to be single-valued.
Thus, the mixing matrices take the following forms

\begin{equation}
A^{(1)}(h,\wbar{h},q_{3},\wbar{q}_{3})=\left[\begin{array}{ccc}
\alpha & 0 & 0\\
0 & \beta & \gamma\\
0 & \gamma & 0\end{array}\right]\,,
\quad 
A^{(1)}(h,\wbar{h},q_{3}=0,\wbar{q}_{3}=0)=
\left[
\begin{array}{ccc}
\beta & 0 & 0\\
0 & \alpha & \delta\\
0 & \varepsilon & \gamma
\end{array}
\right]\,,
\end{equation}
%

\begin{equation}
A^{(1)}(h\in\mathbb{Z},
\wbar{h}\in\mathbb{Z}
,q_{3},\wbar{q}_{3})
=\left[
\begin{array}{ccc}
\alpha & \beta & \gamma\\
\beta & 2\gamma & 0\\
\gamma & 0 & 0
\end{array}
\right]\,,
\end{equation}

\begin{equation}
A^{(1)}(h\in\{0,1\},
\wbar{h}\in\{0,1\}
,q_{3}=0,\wbar{q}_{3}=0)
=\left[\begin{array}{ccc}
\alpha & \gamma & 0\\
\beta & \delta & \varepsilon\\
0 & \varepsilon & 0\end{array}\right]\,.
\end{equation}


Similarly, we proceed around $x=\infty^-$.
For $h\not\in\mathbb{Z}$ and $\hat{q}_{3}\ne0$ we have solutions
of type
\begin{equation}
\begin{array}{ccc}
u_{1}(x) & = & (1/x)^{0}\sum_{n=0}^{\infty}a_{n,r_{1}}(1/x)^{n}\,,\\
u_{2}(x) & = & (1/x)^{1-h}\sum_{n=0}^{\infty}a_{n,r_{2}}(1/x)^{n}\,,\\
u_{3}(x) & = & (1/x)^{-h}\sum_{n=0}^{\infty}b_{n,r_{3}}x^{n}
+(1/x)^{1-h}\sum_{n=0}^{\infty}a_{n,r_{2}}x^{n}\mbox{Log}(x)
\end{array}
\lab{eq:u-inf}
\end{equation}
and similarly in the anti-holomorphic region.
In this case we have the matrix 
\begin{equation}
A^{(\infty)}(h,\wbar{h},q_{3},\wbar{q}_{3})=\left[\begin{array}{ccc}
\alpha & 0 & 0\\
0 & \beta & \gamma\\
0 & \gamma & 0\end{array}\right]\,,
\quad
A^{(\infty)}(h,\wbar{h},q_{3}=0,\wbar{q}_{3}=0)=\left[\begin{array}{ccc}
\beta & 0 & 0\\
0 & \alpha & \delta\\
0 & \varepsilon & \gamma\end{array}\right]\,,
\end{equation}

\begin{equation}
A^{(\infty)}(h\in\mathbb{Z},
\wbar{h}\in\mathbb{Z}
,q_{3},\wbar{q}_{3})=
\left[
\begin{array}{ccc}
\alpha & \beta & \gamma\\
\beta & 2\gamma & 0\\
\gamma & 0 & 0
\end{array}
\right]\,,
\end{equation}

\begin{equation}
A^{(\infty)}(h\in\{0,1\},
\wbar{h}\in\{0,1\}
,q_{3}=0,\wbar{q}_{3}=0)=\left[\begin{array}{ccc}
\alpha & \gamma & 0\\
\beta & \delta & \varepsilon\\
0 & \varepsilon & 0\end{array}\right]\,.
\end{equation}

\subsection{Transition matrices between solutions around different poles}

The above solutions  around $x=0,1,\infty$ 
have a convergence radius equal to the 
difference between the two singular points:
the points around which the solution is defined and the nearest
of the remaining two.
In order to define a global solution which is convergent
in the entire complex plane we have to glue
the solutions defined around different singular points. 
This can be done by expanding one set of solutions 
in terms of the other solutions in the overlap region of
the two considered solutions.
Thus, in the overlap region we can define the 
transition matrices $\varDelta$, $\varGamma$, where
\begin{equation}
\begin{array}{ccc}
\vec{u}^{(0)}(x,q) &=&\varDelta(q) \vec{u}^{(1)}(x,q)\,, \\
\vec{u}^{(1)}(x,q) &=&\varGamma(q) \vec{u}^{(\infty)}(x,q).
\end{array}
\lab{eq:u01}
\end{equation}

Matrices, $\varDelta$ and  $\varGamma$, are constructed in terms of the ratios
of certain determinants \ci{Wosiek:1996bf}.
For example, to calculate the matrix $\varDelta$ we construct Wro\'nskian
\begin{equation}
W=
\begin{vmatrix}
u_1^{(1)}(x;q)& u_2^{(1)}(x;q)& u_3^{(1)}(x;q) \\
{u'}_1^{(1)}(x;q)& {u'}_2^{(1)}(x;q)& {u'}_3^{(1)}(x;q) \\ 
{u''}_1^{(1)}(x;q)& {u''}_2^{(1)}(x;q)& {u''}_3^{(1)}(x;q) 
\end{vmatrix}.
\lab{eq:wron}
\end{equation}
Next we construct determinants $W_{ij}$ which are obtained from $W$ 
by replacing 
$j$-th column by the $i$-th solution around $x=0$, i.e. for 
$i=1$ and $j=2$ we have
\begin{equation}
W_{12}=
\begin{vmatrix}
u_1^{(1)}(x;q_3)& u_1^{(0)}(x;q_3)& u_3^{(1)}(x;q_3) \\
{u'}_1^{(1)}(x;q_3)& {u'}_1^{(0)}(x;q_3)& {u'}_3^{(1)}(x;q_3) \\ 
{u''}_1^{(1)}(x;q_3)& {u''}_1^{(0)}(x;q_3)& {u''}_3^{(1)}(x;q_3) 
\end{vmatrix}.
\end{equation}
The matrix elements $\varDelta_{ij}$ 
are given by
\begin{equation}
\varDelta_{ij}=\frac{W_{ij}}{W}.
\lab{eq:Dij}
\end{equation}
Matrix 
$\varDelta$ 
does not depend on $x$, but only on $q_k$.
In the similar way we can get the matrices
$\varGamma$ 
and their anti-holomorphic equivalents:
$\overline{\varDelta}$, $\overline{\varGamma}$.

Substituting  equation (\ref{eq:u01}) into the wave-function
(\ref{eq:Psizz}), one finds the following conditions for
continuity of the matrix $A(\overline{q},q)$:
\begin{eqnarray}
\overline{\varDelta}^T(\overline{q}_3)A^{(0)}(\overline{q}_3,q_3)
\varDelta(q_3)&=
A^{(1)}(\overline{q}_3,q_3)\,, 
\lab{eq:azao}
\\
\overline{\varGamma}^T(\overline{q}_3)A^{(1)}(\overline{q}_3,q_3)
\varGamma(q_3)&=
A^{(\infty)}(\overline{q}_3,q_3)\,.
\lab{eq:aoai}
\end{eqnarray}
Each Equation, (\ref{eq:azao},\ref{eq:aoai}), 
consists of nine equations.   
Solving them numerically, we obtain values of parameters 
$\alpha,\beta,\gamma,\ldots$
as well as quantized values of the conformal charges, $q_k$ and 
$\wbar{q}_k$. Numerical calculations tell us that the spectrum
of $q_k$ obtained using this method is equivalent 
to the spectrum obtained using the 
Baxter $Q-$operator method which is presented in next Chapter.

\subsection{Additional conditions coming from the particle permutation 
invariance}

Our states have additional symmetries: the cyclic and mirror permutation
(\ref{eq:PMsym}). The conformal charges commute only with $\mathbb{P}$.
Thus, our eigenstates are hardly ever eigenstates of $\mathbb{M}$,
so they usually have mixed $C$-parity. 

For $q_3=0$ we can easily resum the series solutions, see Appendix C.
Let us take a case for $h\not\in\{0,1\}$.
The eigenequation
for the cyclic permutation with a quasimomentum $\theta_3(q)$ gives 
a following condition
\begin{equation}
\begin{array}{ccl}
& & w^{h}\wbar{w}^{\wbar{h}}\left(\beta
+\gamma(-x)^{h}(-\wbar{x})^{\wbar{h}}+\alpha(x-1)^{h}(\wbar{x}-1)^{\wbar{h}}
+\delta(-x)^{h}(\wbar{x}-1)^{\wbar{x}}
+\varepsilon(x-1)^{h}(-\wbar{x})^{\wbar{h}}\right)=\\
& = & \e^{i\theta_3(q)}w^{h}\wbar{w}^{\wbar{h}}
\left(\alpha+\beta(-x)^{h}(-\wbar{x})^{\wbar{h}}+\gamma(x-1)^{h}
(\wbar{x}-1)^{\wbar{h}}+\delta(x-1)^{h}
+\varepsilon(\wbar{x}-1)^{\wbar{h}}\right)\,.
\end{array}
\lab{eq:PPsiz}
\end{equation} 
Here we have used cyclic transformation laws  (\ref{eq:wxcyc}).

Comparing these two lines we obtain conditions: 
$\alpha=e^{-i\theta_3(q)}\beta$,
$\beta=e^{-i\theta_3(q)}\gamma$, $\gamma=e^{-i\theta_3(q)}\alpha$ and
$\delta=\varepsilon=0$. One can derive that $\exp(3i\theta_3(q))=1$
so $\theta_3(q)=\frac{2k\pi}{3}$ where $k=0,1,2$ ($k=0$ for physical
states). Thus we have an eigenstate of $\mathbb{P}$ \ci{Bartels:1999yt}
\begin{equation}
\Psi(\vec{z}_{10},\vec{z}_{20},\vec{z}_{30})=w^{h}\wbar{w}^{\wbar{h}}
\left(1+\exp(i\frac{2\pi k}{3})(-x)^{h}(-\wbar{x})^{\wbar{h}}
+\exp(i\frac{4\pi k}{3})(x-1)^{h}(\wbar{x}-1)^{\wbar{h}}\right)\,,
\lab{eq:Psib}
\end{equation}
where we have omitted the normalization constant.

Now we can act with a mirror permutation operator on (\ref{eq:Psib})
and test its eigenequation (\ref{eq:PMsym}).
Using the mirror transformations (\ref{eq:wxmir}), similarly to 
(\ref{eq:PPsiz}),
we obtain the following relation
\begin{equation}
\begin{array}{ccc}
&&w^{h}\wbar{w}^{\wbar{h}}(-1)^{n_{h}}
\left(\exp(i\frac{2\pi k}{3})+(-x)^{h}(-\wbar{x})^{\wbar{h}}
+\exp(i\frac{4\pi k}{3})(x-1)^{h}(\wbar{x}-1)^{\wbar{h}}\right)=\\
& &= \pm w^{h}\wbar{w}^{\wbar{h}}\left(1
+\exp(i\frac{2\pi k}{3})(-x)^{h}(-\wbar{x})^{\wbar{h}}
+\exp(i\frac{4\pi k}{3})(x-1)^{h}(\wbar{x}-1)^{\wbar{h}}\right)\,.
\end{array}
\lab{eq:MPsi}
\end{equation}
Comparing both sides
of (\ref{eq:MPsi})
gives  
$(-1)^{n_{h}}\exp(i\frac{2\pi k}{3})=\pm1$,
$(-1)^{n_{h}}=\pm\exp(i\frac{2\pi k}{3})$ and $(-1)^{n_{h}}=\pm1$
where the $SL(2,\mathbb{C})$ Lorentz spins $n_h=h-\wbar h$.
These conditions are consistent with $k=0,\frac{3}{2}$. 
Only the first case
agrees with the cyclic permutation condition. As we can see for odd $n_{h}$
we have \emph{minus} sign, so taking into account colour factors $(-1)^{N}$,
solution (\ref{eq:MPsi}) is C-even. For even $n_{h}$ we have 
\emph{plus}
sign thus solution is C-odd. The last case is unnormalizable
because when $x\rightarrow0$ or $x\rightarrow1$ it does not vanish so
the norm, with $(s=0,\wbar s=0)$,  is divergent \ci{Bartels:1999yt}.

Using the duality symmetry \ci{Lipatov:1998as,Bartels:1999yt,Bartels:2001hw,
Vacca:2000bk,Kovchegov:2003dm},
which corresponds to  $h\rightarrow1-h$,
Bartels, Lipatov and Vacca constructed 
an eigenstate with $q_{3}=0$ and $C=-1$
\begin{equation}
\Psi(\vec{z}_{10},\vec{z}_{20},\vec{z}_{30}) =
w^{h}\wbar{w}^{\wbar{h}}
x(1-x) \wbar{x}(1-\wbar{x})
\left(
\delta^{(2)} (x)-\delta^{(2)} (1-x)
+\frac{x^h \wbar{x}^h}{x^3 \wbar{x}^3}\delta^{(2)} \left(\frac{1}{x} \right)
\right)\,.
\lab{eq:BLV}
\end{equation} 
This wave-function cannot 
be constructed using the method described here because
it contains non-analytical functions, the Dirac delta $\delta(x)$. 


Now, let us take the second wave-function with five parameters i.e.
for $q_{3}=0$ and $h\in\{0,1\}$. Choosing $h=1$ we have 
\begin{equation}
u(x)=[1,(-x),(-x)\mbox{Log}(-x)+(x-1)\mbox{Log}(x-1)]^{T}
\end{equation}
 and for $\wbar{h}=0$ it is 
\begin{equation}
\wbar{u}(\wbar{x})=[\mbox{Log}(\wbar{x}-1),1,\mbox{Log}(-\wbar{x})]^{T}\,.
\end{equation}
Combining them we obtain
\begin{equation}
\begin{array}{ccc}
\Psi(\vec{z}_{10},\vec{z}_{20},\vec{z}_{30}) & 
= & w^{h}\wbar{w}^{\wbar{h}}\left(\alpha\mbox{Log}(\wbar{x}-1)+\beta
+\gamma(-x)\mbox{Log}(\wbar{x}-1)+\delta(-x)+\right.\\
 &  & \left.+\varepsilon((-x)\mbox{Log}(-\wbar{x})+(-x)\mbox{Log}(-x)
+(x-1)\mbox{Log}(x-1))\right)\,.
\end{array}
\lab{eq:ePsiq0}
\end{equation}
Like in the previous case we write the eigenequation for the 
cyclic permutation
\begin{equation}
\begin{array}{ccc}
& 
& w^{h(=1)}\wbar{w}^{\wbar{h}(=0)}
\left((\alpha-\gamma-\varepsilon)\mbox{Log}(x-1)+(\delta-\beta)
+\alpha(-x)\mbox{Log}(\wbar{x}-1)+(-\beta)(-x)\right.+\\
 &  & \left.+\varepsilon(-x)\mbox{Log}(-x)
+\alpha(x-1)\mbox{Log}(-\wbar{x})+\varepsilon(-x)\mbox{Log}(-x)+\right.\\
 &  & \left.+\gamma\mbox{Log}(-\wbar{x})
+\varepsilon(x-1)\mbox{Log}(x-1)\right)=\\
 & &=  e^{i\theta_3(q)}w\left(\alpha\mbox{Log}(\wbar{x}-1)+\beta
+\gamma(-x)\mbox{Log}(\wbar{x}-1)+\delta(-x)+\right.\\
 &  & \left.+\varepsilon((-x)\mbox{Log}(-\wbar{x})+(-x)\mbox{Log}(-x)
+(x-1)\mbox{Log}(x-1))\right)\,.
\end{array}
\end{equation}
Thus we get conditions:
$\alpha=e^{-i\theta_3(q)}(\alpha-\gamma-\varepsilon)$,
$\beta=e^{-i\theta_3(q)}(\delta-\beta)$, $\gamma=e^{-i\theta_3(q)}\alpha$,
$\delta=e^{-i\theta_3(q)}(-\beta)$, $0=\e^{-i\theta_3(q)}(\gamma-\alpha)$
and $\varepsilon=e^{-i\theta_3(q)}\varepsilon$. 
We have two types of solutions.

The first one with $\theta_3(q)=0$ when  $\alpha=\gamma=-\varepsilon$
and $\beta=\delta=0$. It has a form
\begin{equation}
\begin{array}{ccc}
\Psi(\vec{z}_{10},\vec{z}_{20},\vec{z}_{30}) 
& = & 
w\left((-x)\mbox{Log}((-\wbar{x})(-x))
+(x-1)\mbox{Log}((\wbar{x}-1)(x-1))\right)\,.
\end{array}
\lab{eq:Psiq0}
\end{equation}
We obtained in this way solutions with $\Log(x)-$terms which have
not been presented before.The similar expressions 
were shown in \ci{DeVega:2001pu} 
as asymptotics of the $\oq{3}$ eigenfunction.
Acting with the mirror permutation operator on (\ref{eq:Psiq0}) we get
\begin{equation}
\begin{array}{ccc}
& & \mathbb{M}w\left((-x)\mbox{Log}((-\wbar{x})(-x))
+(x-1)\mbox{Log}((\wbar{x}-1)(x-1))\right)=\\
 & &=  -w\left((-x)\mbox{Log}((-\wbar{x})(-x))
+(x-1)\mbox{Log}((\wbar{x}-1)(x-1))\right)\,.
\end{array}
\lab{eq:MPsilog}
\end{equation}
We obtained \emph{minus} sign so this state is also symmetric under
C parity.

Other solutions have $\theta_3(q)=2 \pi/3, 4\pi/3$.
Thus, $\alpha=\gamma=\varepsilon=0$ and 
$\delta=-\e^{-i\theta_3(q)}\beta$. The wave-function has a form
\begin{equation}
\Psi(\vec{z}_{10},\vec{z}_{20},\vec{z}_{30}) =
w (1+x\e^{i \theta_3(q)})\,.
\lab{eq:Psiq0b}
\end{equation}
These solutions are not eigenstates of the $\mathbb{M}$ operator.

\section{Set of differential equations for $N=4$}

Similarly to the case for three particles we can derive a set of differential
equations for more particles. For $N=4$ with the following ansatz 
\begin{multline}
\Psi(z_{10},z_{20},z_{30},z_{40})=
\frac{1}{{z_{10}}^{2s_{1}}{z_{20}}^{2s_{2}}{z_{30}}^{2s_{3}}
{z_{40}}^{2s_{4}}}
\left(\frac{z_{31}}{z_{10}z_{30}}\right)^{h-s_{1}-s_{2}-s_{3}-s_{4}} \\
\times
F \left(x_{1}=\frac{z_{20}z_{41}}{z_{21}z_{40}},x_{2}=
\frac{z_{12}z_{30}}{z_{10}z_{32}}\right)
\end{multline}
we have two equations for $\hat{q}_{3}$ and for $\hat{q}_{4}$. 
Acting with $\hat{q}_{3}$ we obtain 
\begin{equation}
\begin{array}{c}
t_{0,0}F^{(0,0)}(x_{1},x_{2})+t_{1,0}F^{(1,0)}(x_{1},x_{2})
+t_{0,1}F^{(0,1)}(x_{1},x_{2})+t_{2,0}F^{(2,0)}(x_{1},x_{2}) +\\
+t_{1,1}F^{(1,1)}(x_{1},x_{2})+t_{0,2}F^{(0,2)}(x_{1},x_{2})
+t_{3,0}F^{(3,0)}(x_{1},x_{2})+t_{2,1}F^{(2,1)}(x_{1},x_{2})+ \\
+t_{1,2}F^{(1,2)}(x_{1},x_{2})+t_{0,3}F^{(0,3)}(x_{1},x_{2})=0\,,
\end{array}
\lab{eq:N4eqt}
\end{equation}
where coefficients are defined in Appendix D.

Moreover, acting with $\oq{4}$ on our ansatz we have
\begin{equation}
\begin{array}{c}
f_{0,0}F^{(0,0)}(x_{1},x_{2})+f_{1,0}F^{(1,0)}(x_{1},x_{2})
+f_{0,1}F^{(0,1)}(x_{1},x_{2})+f_{2,0}F^{(2,0)}(x_{1},x_{2}) +\\
+f_{1,1}F^{(1,1)}(x_{1},x_{2})+f_{0,2}F^{(0,2)}(x_{1},x_{2})
+f_{3,0}F^{(3,0)}(x_{1},x_{2})+f_{2,1}F^{(2,1)}(x_{1},x_{2}) +\\
+f_{1,2}F^{(1,2)}(x_{1},x_{2})+f_{0,3}F^{(0,3)}(x_{1},x_{2})
+f_{4,0}F^{(0,2)}(x_{1},x_{2})+f_{3,1}F^{(3,0)}(x_{1},x_{2}) +\\
+f_{2,2}F^{(2,1)}(x_{1},x_{2})+f_{1,3}F^{(1,2)}(x_{1},x_{2})
+f_{0,4}F^{(3,0)}(x_{1},x_{2})=0\,.
\end{array}
\lab{eq:N4eqf}
\end{equation}
These equations are not symmetric in $x_{1}$, $x_{2}$ because
the ansatz is also not symmetric. 

Equations, (\ref{eq:N4eqt}) and (\ref{eq:N4eqf}), are very hard to solve even
numerically, thus for the states with $N\ge4$ we use the $Q$-Baxter
method.

\chapter{Quantization conditions in the $Q$-Baxter method}

In Ref. \ci{Derkachov:2002wz}
the authors describe a construction of 
the solution to the Baxter equations, 
(\ref{eq:Baxeq}) and (\ref{eq:Baxbeq}),
which satisfies additionally the conditions, (\ref{eq:upoles}) 
and (\ref{eq:Qanalb}).
This can be done by means of
the following integral representation for $Q_{q,\wbar q}(u,\wbar u)$
\begin{equation}
Q_{q,\wbar q}(u,\wbar u)= \int\frac{d^2 z}{z\wbar z}\, 
z^{-i u} {\wbar z}^{-i\wbar u}\, Q(z,\wbar z)\,,
\lab{eq:Q-R}
\end{equation}
where we integrate over the two-dimensional 
$\vec z-$plane with $\wbar z=z^*$
and $Q(z,\wbar z)$ depends on $\{q,\wbar q\}$.
The advantages of this ansatz are:
\begin{itemize}
\item the functional Baxter equation on $Q_{q,\wbar q}(u,\wbar u)$ is
transformed into the $N-$th order differential equation for the function  
$Q(z,\wbar z)$
\begin{equation}
\left[z^s\lr{z\partial_z}^{N}z^{1-s}+z^{-s}\lr{z\partial_z}^{N}z^{s-1}
-2\lr{z\partial_z}^{N}-\sum_{k=2}^N i^{k}q_k\lr{z\partial_z}^{N-k}
\right]Q(z,\wbar z)=0\,.
\lab{eq:Eq-1}
\end{equation}
A similar equation holds in the anti-holomorphic sector 
with $s$ and $q_k$ replaced by $\wbar s=1-s^*$ and
$\wbar q_k=q_k^*$, respectively. 
\item 
the condition (\ref{eq:uubarn})
is automatically satisfied since 
the $z-$integral in the r.h.s.\ of (\ref{eq:Q-R}) 
is well-defined 
only for 
$i(u-\wbar u)=n$.
\item the remaining two
conditions for the analytical properties and asymptotic behaviour of 
$Q_{q,\wbar q}(u,\wbar u)$, Eqs.~(\ref{eq:upoles}) and (\ref{eq:Qanalb}), 
become equivalent to a
requirement for $Q(z,\wbar z=z^*)$ to be a single-valued function 
on the complex
$z-$plane.
\end{itemize}

Analogically to the eigenequations 
for $\oq{3}$ (\ref{eq:qF}), 
the differential equation (\ref{eq:Eq-1})
is of Fuchsian type. It possesses three regular singular points located 
at $z=0$, $z=1$ and $z=\infty$.
Moreover, it has
$N$ linearly independent solutions,
$Q_a(z)$. 
The anti-holomorphic equation has also
$N$ independent solutions,
$\wbar Q_b(\wbar z)$.

Now, similarly to (\ref{eq:Psizz}), 
we construct the general expression for the
function $Q(z,\wbar z)$ as
\begin{equation}
Q(z,\wbar z) = \sum_{a,b=1}^N Q_a(z)\, C_{ab}\, \wbar Q_b(\wbar z)\,,
\lab{eq:general-sol}
\end{equation}
where $C_{ab}$ is an arbitrary mixing matrix. The functions $Q_a(z)$ and
$\wbar Q_b(\wbar z)$ 
have a nontrivial monodromy\footnote{The monodromy matrix around 
$z=0$ is defined as $Q_n^{(0)}(z\e^{2\pi i})=M_{nk}Q_k^{(0)}(z)$
and 
similarly for the other singular points.} 
around three singular points, 
$z,\,\wbar z=0$, $1$
and $\infty$. 
In order to be well-defined on the whole plane, functions $Q(z,\wbar z=z^*)$ 
should be single-valued and
their
monodromy should cancel in the r.h.s.\ of (\ref{eq:general-sol}). 
This condition allows us to determine the values of the mixing coefficients,
$C_{ab}$, and also to calculate the quantized values of the 
conformal charges $q_k$.

The differential equation (\ref{eq:Eq-1}) is also 
symmetric under the  transformation
$z\to 1/z$ and $q_k\to (-1)^k q_k$. 
This property is related to 
Eq.~(\ref{eq:Q-symmetry}) and leads to
\\[1mm]
\begin{equation}
Q_{q,\wbar q}(z,\wbar z)=\e^{i\theta_N(q,\wbar q)}
Q_{-q,-\wbar q}(1/z,1/\wbar z)\,,
\lab{eq:Qz-symmetry}
\end{equation}
\\[0mm]
where $\pm q=(q_2,\pm q_3,\ldots,(\pm)^N q_N)$ denotes 
the integrals of motion corresponding to the function $Q(z,\wbar{z})$.
The above formula allows us to define the solution 
$Q(z,\wbar{z})$ around 
$z=\infty$ from the solution at $z=0$. Thus, applying (\ref{eq:Qz-symmetry})
we are able to find $Q(z,\wbar{z})$ and analytically continue it to the whole 
$z-$plane.

\section{Solution around $z=0$}

We find  a solution $Q(z)\sim z^a$ by the series method.
The indicial equation for the solution of Eq. (\ref{eq:Eq-1})
around $z=0$ reads as follows
\begin{equation}
(a-1+s)^N=0\,.
\lab{eq:indicial-0}
\end{equation}
and the solution, $a=1-s$, is $N-$times degenerate. This leads to
terms $\sim \Log^k(z)$ with $k\le N-1$.
We define the
fundamental set of linearly independent solutions to (\ref{eq:Eq-1}) 
around $z=0$ as
\begin{eqnarray}
&&Q_1^{(0)}(z) = z^{1-s} u_1(z)\,,
\nonumber\\
&&Q_m^{(0)}(z)=z^{1-s}\left[u_1(z) \Log^{m-1}(z)+\sum_{k=1}^{m-1}c_{m-1}^k
u_{k+1}(z) \Log^{m-k-1}(z)\right],
\lab{eq:Q-0-h}
\end{eqnarray}
with $2\le m\le N$ and where for the later
convenience 
$c_{m-1}^k=(m-1)!/(k!(m-k-1)!)$.
The functions $u_m(z)$ are defined
inside the region $|z|<1$
and have a form
\begin{equation}
u_m(z) = 1+\sum_{n=1}^\infty z^n\,u^{(m)}_{n}(q)\,.
\lab{eq:power-series-0}
\end{equation}
Inserting (\ref{eq:Q-0-h}) and
(\ref{eq:power-series-0}) into (\ref{eq:Eq-1}), 
one derives recurrence relations for $u^{(m)}_n(q)$.
However, in order to save space, we do not show here their explicit form.

In the anti-holomorphic sector the 
fundamental set of solutions 
can be obtained from (\ref{eq:Q-0-h}) 
by substituting $s$
and $q_k$ by $\wbar s=1-s^*$ and $\wbar q_k=q_k^*$, respectively. 
Sewing the two sectors we obtain
 the general
solution for $Q(z,\wbar z)$ around $z=0$ as
\begin{equation}
Q(z,\wbar z) \stackrel{|z|\to 0}{=} \sum_{m,\wbar m=1}^N Q^{(0)}_m(z)\,
C^{(0)}_{m\wbar m}\,\wbar{Q}^{(0)}_{\wbar m}(\wbar z)\,.
\lab{eq:Q-0}
\end{equation}
The above solution (\ref{eq:Q-0}) should be single-valued on the $z-$plane.
Thus, similarly to (\ref{eq:Psizz}) 
we find a structure of the mixing matrix $C^{(0)}_{m\wbar m}$
which for $n+m\le N+1$
\begin{equation}
C^{(0)}_{nm}=\frac{\sigma}{(n-1)!(m-1)!}
\sum _{k=0}^{N-n-m+1}{\frac {(-2)^{k}}{k!}\,\alpha_{k+n+m-1}}
\lab{eq:C0}
\end{equation}
with $\sigma, \alpha_1,\ldots ,\alpha_{N-1}$ being arbitrary 
complex parameters and $\alpha_N=1$. Below the main anti-diagonal,
that is for $n+m> N+1$, $C^{(0)}_{nm}$ vanish.

The mixing matrix $C^{(0)}_{m\wbar m}$ depends on $N$ arbitrary complex parameters
$\sigma$ and $\alpha_k$. 
However, two parity relations, Eqs.~(\ref{eq:RR}) and (\ref{eq:Qz-symmetry}), 
fix $\sigma=\exp(i\theta_N(q,\wbar q))$, 
with $\theta_N(q,\wbar q)$ being
the quasimomentum,
and lead to
the quantization of the quasimomentum. 
Later, we will use 
(\ref{eq:Qz-symmetry}) to
calculate the eigenvalues of $\theta_N(q,\wbar q)$ 
(see Eq.~(\ref{eq:parity-qc})).

The leading
asymptotic behaviour of $Q(z,\wbar{z})$ 
for $z\to 0$ can be obtained 
by substituting (\ref{eq:C0}) and (\ref{eq:Q-0-h}) into (\ref{eq:Q-0}). 
It has a form
\begin{multline}
Q_{q,\wbar q}(z,\wbar z)=z^{1-s}\wbar z^{1-\wbar s}\e^{i\theta_N(q,\wbar
q)}\left[\frac{\Log^{N-1} (z\wbar z)}{(N-1)!} +
\frac{\Log^{N-2} (z\wbar z)}{(N-2)!}\, \alpha_{N-1}+ \right.\\
\left. \ldots  +
\frac{\Log (z\wbar z)}{1!}\, \alpha_{2}+\alpha_1\right]
\lr{1+{\cal O}(z,\wbar z)}\,.
\lab{eq:Q-small}
\end{multline}
Making use of the integral identity 
\begin{equation}
\int_{|z|<\rho}\frac{d^2 z}{z\bar z} z^{-iu}\bar z^{-i\bar u}
\ln^n(z\bar z) z^{m-s} \bar z^{\bar m-\bar s}=\pi\delta_{m-s-iu,\bar m-\bar s-i\bar u}\left[
\frac{(-1)^n\,n!}{(m-s-iu)^{n+1}}+{\cal O}((m-s-iu)^0)\right],
\lab{eq:intident}
\end{equation}
with $m$ and $\wbar m$ positive integer,
we can calculate the contribution of the small$-z$ region to the eigenvalue 
of the Baxter equation (\ref{eq:Q-R}). The
function $Q_{q,\wbar q}(u,\wbar u)$ has poles of the order $N$ in the points
$u=i(s-m)$ and $\wbar u=i(\wbar s-\wbar m)$ what agrees with (\ref{eq:upoles}). 
For $m=\wbar m=1$ one finds from
(\ref{eq:Q-small})
\begin{equation}
Q_{q,\wbar q}(u_{1}^{+}+\epsilon,{\wbar u}_{1}^{+}+\epsilon)=-\frac{\pi
\e^{i\theta_N(q,\wbar
q)}}{(i\epsilon)^N}
\left[1+i\epsilon\,\alpha_{N-1}+\ldots +(i\epsilon)^{N-2}
\,\alpha_2+(i\epsilon)^{N-1}\,\alpha_1+{\cal O}(\epsilon^N)
\right]\,,
\lab{eq:Q-pole-0}
\end{equation}
where $u_{1}^{+}$ and ${\wbar u}_{1}^{+}$ are defined in (\ref{eq:upoles}). 
One can see that
the integration in (\ref{eq:Q-R}) over the region of large $z$
with (\ref{eq:Qz-symmetry}) and (\ref{eq:Q-small}) 
gives the second set of poles for  $Q_{q,\wbar q}(u,\wbar u)$ 
located at $u=-i(s-m)$ and $\wbar u=-i(\wbar s-\wbar m)$.

Comparing (\ref{eq:Q-pole-0}) with (\ref{eq:Q-R,E}) one obtains
\begin{equation}
R^+(q,\wbar q)=-\frac{\pi}{i^N}\e^{i\theta_N(q,\wbar q)}\,,\qquad E^+(q,\wbar
q)=\alpha_{N-1}(q,\wbar q)\,.
\lab{eq:E=alpha}
\end{equation}
Now, we may derive expression for the energy 
\begin{equation}
E_N(q,\wbar q)=\Re\left[\alpha_{N-1}(-q,-\wbar q)+\alpha_{N-1}(q,\wbar q)\right]\,.
\lab{eq:E-fin}
\end{equation}
The arbitrary complex parameters $\alpha_n$, defined in (\ref{eq:C0}),
will be fixed by 
the quantization conditions below.

In this Section we have obtained following Ref. \ci{Derkachov:2002wz} the 
expression for the energy spectrum 
$E_N(q,\wbar q)$, as a function of the matrix
elements of the mixing matrix (\ref{eq:C0}) in the fundamental basis 
(\ref{eq:Q-0-h}). Moreover, we have defined the solution to the 
Baxter equation
$Q(u,\wbar{u})$ and reproduced the analytical properties
of the eigenvalues of the Baxter operator, 
Eq.~(\ref{eq:upoles}).

\section{Solution around $z=1$}

Looking for a solution of (\ref{eq:Eq-1}) 
around $z=1$ in a form $Q(z)\sim(z-1)^b$
we obtain the following indicial equation
\begin{equation}
(b+1+h-Ns)(b+2-h-Ns)\prod_{k=0}^{N-3}(b-k)=0\,,
\lab{eq:b-exponents}
\end{equation}
where $h$ is the total $SL(2,\mathbb{C})$ spin  defined in (\ref{eq:q2}).
Although the solutions $b=k$ with $k=0,\ldots ,N-3$ differ from each other 
by an integer, for $h\neq (1+n_h)/2$, no logarithmic terms appear.
The $\Log(z)-$terms are only needed for $\IIm h = 0$ where the additional 
degeneration occurs. 

Thus, we define 
the fundamental set of solutions to Eq.~(\ref{eq:Eq-1}) around $z=1$. 
For $\IIm h\neq 0$ it has the form
\begin{eqnarray}
&&Q_1^{(1)}(z) = z^{1-s} (1-z)^{Ns-h-1}v_1(z)\,,
\nonumber
\\[2mm]
&&Q_2^{(1)}(z) = z^{1-s} (1-z)^{Ns+h-2}v_2(z)\,,
\nonumber
\\[2mm]
&&Q_m^{(1)}(z) = z^{1-s} (1-z)^{m-3} v_m(z)\,,
\lab{eq:set-1}
\end{eqnarray}
with $m=3,\ldots ,N$. The functions $v_{i}(z)$ $(i=1,2)$ and $v_m(z)$ given by the
power series
\begin{equation}
v_i(z)=1+\sum_{n=1}^\infty (1-z)^n \,v^{(i)}_{n}(q)\,,
\qquad
v_m(z)=1+\sum_{n=N-m+1}^\infty (1-z)^n\, v^{(m)}_{n}(q)\,,
\lab{eq:v-series}
\end{equation}
which converge inside the region $|1-z|<1$
and where 
the expansion coefficients
 $v^{(i)}_{n}$ and $v^{(m)}_{n}$ satisfy the $N-$term 
recurrence relations\footnote{
The factor $z^{1-s}$ was included in the r.h.s.\ of 
(\ref{eq:set-1}) and (\ref{eq:Q-deg})
to simplify the form of the recurrence relations.}
with respect to the index $n$.
For $h=(1+n_h)/2 \in 2 \mathbb{Z}+1$, one $\Log(z)-$terms appears
so for
 $n_h\ge 0$, 
\begin{equation}
Q_1^{(1)}(z)\bigg|_{h=(1+n_h)/2} = z^{1-s} (1-z)^{Ns-(n_h+3)/2}\left[(1-z)^{n_h}
\Log (1-z)\, v_2(z)+ \widetilde v_1(z)\right]\,,
\lab{eq:Q-deg}
\end{equation}
where the function $v_2(z)$ is the same as before, $\widetilde
v_1(z)=\sum_{k=0}^\infty\tilde v_k z^k$ and the coefficients $\tilde v_k$ 
satisfy
the $N-$term recurrence relations with the boundary condition 
$\tilde v_{n_h}=1$.
For $h \in \mathbb{Z}$ we have two additional terms:
$\Log(z)$ and $\Log^2(z)$.

Similar calculations have to be performed in the anti-holomorphic sector
with $s$ and $h$ replaced by
$\wbar s=1-s^*$ and $\wbar h=1-h^*$, respectively. 
A general solution for $Q(z,\wbar{z})$ for $\IIm(h) \ne 0$
with respect to the single-valuedness 
can be constructed as 
\begin{equation}
Q(z,\wbar z)\stackrel{|z|\to 1}{=}\beta_h Q_1^{(1)}(z)\wbar Q_1^{(1)}(\wbar
z)+\beta_{1-h} Q_2^{(1)}(z)\wbar Q_2^{(1)}(\wbar z) +\sum_{m,\wbar m=3}^N
Q_m^{(1)}(z)\,\gamma_{m\wbar m}\,\wbar Q_{\wbar m}^{(1)}(\wbar z)\,.
\lab{eq:Q-1}
\end{equation}
Here
the parameters $\beta_h$ and $\gamma_{m \wbar m}$ build the $C^{(1)}$ matrix
where $Q(z,\wbar{z})=Q^{(1)}_m C^{(1)}_{m\wbar{m}} \wbar{Q}^{(1)}_{\wbar{m}}$.
The $\beta-$coefficients depend, in general, on the total spin $h$ (and
$\wbar h=1-h^*$). They are chosen in (\ref{eq:Q-1}) 
in such a way that the symmetry of
the eigenvalues of the Baxter operator under $h\to 1-h$ becomes manifest. 
Thus, the mixing matrix $C^{(1)}$ defined in (\ref{eq:Q-1}) 
depends on $2+(N-2)^2$ complex parameters $\beta_h$, $\beta_{1-h}$
and $\gamma_{m\wbar m}$ which are some functions 
of the integrals of
motion $(q,\wbar q)$, so, they can be fixed by the quantization conditions. 

For $h=(1+n_h)/2$ the first two terms in the
r.h.s.\ of (\ref{eq:Q-1}) look differently in virtue of (\ref{eq:Q-deg})
\begin{equation}
Q(z,\wbar z)\bigg|_{h=(1+n_h)/2}=\beta_1\! \left[Q_1^{(1)}(z)\wbar
Q_2^{(1)}(\wbar z)+Q_2^{(1)}(z)\wbar Q_1^{(1)}(\wbar z)\right]+\beta_2\,
Q_2^{(1)}(z)\wbar Q_2^{(1)}(\wbar z) + \ldots \, ,
\end{equation}
where ellipses denote the remaining terms. Substituting 
(\ref{eq:Q-1}) into (\ref{eq:Q-R})
and performing integration over the region of $|1-z|\ll 1$, one can find the
asymptotic behaviour of $Q(u,\wbar u)$ at large $u$. 


Let us consider the duality relation
(\ref{eq:Qz-symmetry}). 
Using the function $Q(z,\wbar{z})$  
we evaluate (\ref{eq:Q-1}) in the limit $|z|\rightarrow1$. 
In this way, we obtain set of relations for
the functions
$\beta_i(q,\wbar q)$ and $\gamma_{m\wbar m}(q,\wbar q)$. 
The derivation is based
on the following property 
\begin{equation}
Q^{(1)}_a(1/z;-q)= \sum_{b=1}^N S_{ab} \,Q^{(1)}_b(z;q)\,,
\lab{eq:S-def}
\end{equation}
with $\Im(1/z)>0$
and where the dependence on the integrals of motion
was explicitly indicated.
Here taking limit $z\rightarrow 1$ in (\ref{eq:set-1}) and 
(\ref{eq:v-series}) and substituting them to (\ref{eq:S-def}) 
we are able to
evaluate the $S-$matrix
\begin{equation}
S_{11}=\e^{-i\pi(Ns-h-1)}\,,\qquad S_{22}=\e^{-i\pi(Ns+h-2)}
\,,\qquad
S_{k,k+m}=(-1)^{k-3}\frac{(k-2s-1)_m}{m!}
\lab{eq:S-matrix}
\end{equation}
with $(x)_m\equiv\Gamma(x+m)/\Gamma(x)$, $3\le k \le N$ and $0\le m \le N-k$.
Similar relations hold in the anti-holomorphic sector, 
\begin{equation}
\wbar S_{11}=\e^{i\pi(N\wbar s-\wbar h-1)}\,,
\qquad \wbar S_{22}=\e^{i\pi(N\wbar s+\wbar
h-2)}\,,\qquad
\wbar S_{k,k+m}=(-1)^{k-3}\frac{(k-2\wbar s-1)_m}{m!}\,.
\end{equation}
The $S-$matrix does not depend on $z$
because
the $Q-$functions on the both sides of  relation  
(\ref{eq:S-def}) satisfy the
same differential equation (\ref{eq:Eq-1}).

Now, substituting (\ref{eq:Q-1}) and 
(\ref{eq:S-def}) into (\ref{eq:Qz-symmetry}), we find
\\[-1mm]
\begin{eqnarray}
\beta_h(q,\wbar q)&=& \e^{i\theta_N(q,\wbar q)} (-1)^{Nn_s+n_h}\beta_h(-q,-\wbar q)\,,\qquad
\nonumber
\\[1mm]
\gamma_{m\wbar m}(q,\wbar q) &=& \e^{i\theta_N(q,\wbar q)}
\sum_{n,\wbar n\ge 3}^N S_{nm} \gamma_{n\wbar n}(-q,-\wbar q)\,\wbar S_{\wbar n\wbar
m}\,.
\lab{eq:parity-qc}
\end{eqnarray}
In this way, similarly to the energy, Eq.~(\ref{eq:E-fin}),
which was calculated from the mixing matrix at $z=0$, the
eigenvalues of the quasimomentum, $\theta_N(q,\wbar q)$, maybe calculated from
the mixing matrix at $z=1$, 
from the first relation in
(\ref{eq:parity-qc}). 
In the special case when
$q_{2k+1}=\wbar q_{2k+1}=0$
$(k=1,2\ldots )$, this means $\beta_h(q,\wbar q)=\beta_h(-q,-\wbar q)$, the 
quasimomentum
is equal to
\begin{equation}
\e^{i\theta_N(q,\wbar q)}=(-1)^{Nn_s+n_h}\,.
\lab{eq:quasi-0}
\end{equation}

\section{Transition matrices}

In the previous Sections we constructed
the solutions $Q(z,\wbar z)$ to (\ref{eq:Eq-1}) in
the vicinity of $z=0$ and $z=1$.
Now, we sew these solutions inside the region
$|1-z|<1,\,|z| < 1$ and, then analytically continue the
resulting expression for $Q(z,\wbar z)$ into the whole complex $z-$plane 
by making use of the duality relation (\ref{eq:Qz-symmetry}). 

The sewing procedure is similar to that 
described by (\ref{eq:u01})--(\ref{eq:azao}).
Firstly, we define
the transition matrices $\Omega(q)$ and $\wbar \Omega(\wbar q)$:
\begin{equation}
Q_n^{(0)}(z)=\sum_{m=1}^N \Omega_{nm}(q)\, Q_m^{(1)}(z)\,,\qquad
\wbar Q_{n}^{(0)}(\wbar z)=\sum_{m=1}^N\wbar \Omega_{nm}(\wbar q)
 \,\wbar Q_{m}^{(1)}(\wbar z)\,.
\lab{eq:Omega-def}
\end{equation}
which are uniquely fixed (\ref{eq:Dij}).
The resulting expressions for the matrices
$\Omega(q)$ and $\wbar\Omega(\wbar q)$ take the form of infinite series in $q$
and $\wbar q$, respectively. 
Substituting 
(\ref{eq:Omega-def}) into
(\ref{eq:Q-0}) and matching the result into 
(\ref{eq:Q-1}), we find 
the following
relation
\begin{equation}
C^{(1)}(q,\wbar q)=\left[\Omega(q)\right]^T C^{(0)}(q,\wbar q)\ \wbar
\Omega(\wbar{q})\,.
\lab{eq:C1-C0}
\end{equation}
The above matrix equation 
allows us to determine the matrices $C^{(0)}$ and $C^{(1)}$
and 
provides the quantization conditions for the integrals of
motion, $q_k$ and $\wbar q_k$ with $k=3,\ldots ,N$. 
Therefore, we can evaluate 
the eigenvalues of the Baxter $\mathbb{Q}-$operator,
Eq.~(\ref{eq:Q-R}). 
Formula (\ref{eq:C1-C0}) contains
$N^2$ equations with:
\begin{itemize}
\item $(N-1)$ $\alpha-$parameters inside the matrix $C^{(0)}$,
\item $2+(N-2)^2$ parameters $\beta_{1,2}$ 
and $\gamma_{m\wbar m}$ inside the matrix $C^{(1)}$,
\item $(N-2)$ integrals of motion $q_3,\ldots ,q_N$ where $\wbar
q_k=q_k^*$
\end{itemize}
Thus, we obtain
$(2N-3)$ nontrivial consistency conditions.

The solutions to the quantization conditions (\ref{eq:C1-C0}) 
will be presented in details in 
next Sections.

\chapter{Numerical results}

In this Chapter we present the spectra of the conformal charges
obtained by numerical calculations \ci{Derkachov:2002wz,Kotanski:2001iq}.
To this end we resum solutions (\ref{eq:power-series-0}) and 
(\ref{eq:v-series}) numerically, and using them we solve the quantization
conditions (\ref{eq:C1-C0}). 
First we discuss the numerical results
obtained in Refs. \ci{Kotanski:2001iq,Derkachov:2002wz}
for $N=3,4$ and for the ground states for $N=5,\ldots,8$.
Moreover, we present some results
which were never published before, i.e.
quantized values of $q_3$ for $N=3$ and $n_h>0$,
resemblant and winding
spectra of $q_3$, $q_4$ for $N=4$ and
corrections to the WKB approximation for $N=3$ and $N=4$.

\subsection{Trajectories}
Solving quantization conditions (\ref{eq:C1-C0})
we obtain continuous trajectories
in the space of conformal charges.
They are built of points, $(q_2(\nu_h),\ldots,q_N(\nu_h))$ which satisfy
(\ref{eq:C1-C0}) and depend on a continuous 
real parameter  $\nu_h$ entering $q_2$, (\ref{eq:q2}) and 
(\ref{eq:hpar}).
In order to label the trajectories 
we introduce the set of the integers 
\begin{equation}
\mybf{\ell}=\{\ell_1,\ell_2,\ldots,\ell_{2(N-2)}\}
\lab{eq:dell}
\end{equation}
which parameterize one specified point on each trajectory 
for given $h$.
Specific examples in the following Sections will further clarify
this point. 

Next we calculate the observables
along these trajectories, 
namely the energy (\ref{eq:E-fin}) 
and the quasimomentum (\ref{eq:parity-qc}). 
The quasimomentum is constant (\ref{eq:quask}) for all points 
situated on a given trajectory.
The minimum of the energy, which means the maximal intercept,
for almost all trajectories is located
at $\nu_h=0$.
It turns out that the energy behaves around $\nu_h=0$ like
\begin{equation}
E_N(\nu_h;\mybf{\ell}^{\rm ground})
=E_N^{\rm ground}+\sigma_N {\nu_h}^2+{\cal O}({\nu_h}^2)
\lab{eq:Enu}
\end{equation}
Thus, the ground state along its trajectory is gapless 
and the leading contribution
to the scattering amplitude around $\nu_h$ may be rewritten as
a series in the strong coupling constant:
\begin{equation}
{\cal A}(s,t) \sim -i s 
\sum _{N=2}^\infty (i \asbar)^N
\frac{s^{-\asbar E_N^{\rm ground}/4}}{(\asbar\sigma_N
\ln s)^{1/2}}\,\xi_{A,N}(t)
\xi_{B,N}(t)\,,
\lab{eq:amp}
\end{equation}
where  $\asbar=\alpha_s N_c/\pi$ 
and $\xi_{X,N}(t)$ are the impact factors
corresponding to the overlap between the wave-functions of scattered
particle with the wave-function of $N-$Reggeons, whereas 
$\sigma_N$ measures the dispersion of the energy on the
the trajectory around $\nu_h=0$.

On the other hand,
the energy along the trajectories 
grows with $\nu_h$ and for $|\nu_h| \rightarrow \infty$ 
and finally, we have 
$E_N(\nu_h;{\mybf \ell}) \sim \ln {\nu_h}^2$. 
These parts of the trajectory give the lowest contribution to the
scattering amplitude.

\subsection{Symmetries}
The spectrum of quantized charges $q_2,\ldots,q_N$ is degenerate.
This degeneration is caused by two symmetries:
\begin{equation}
q_k \leftrightarrow (-1)^k q_k
\lab{eq:qkmsym}
\end{equation}
which comes from
invariance of the Hamiltonian under mirror permutations of particles, 
(\ref{eq:PMsym}), and
\begin{equation}
q_k \leftrightarrow  \wbar q_k
\lab{eq:qkcsym}
\end{equation}
which is 
connected with
the symmetry under interchange of the $z-$ and 
$\wbar z-$sectors.
Therefore, the four points, $\{q_k\}$, $\{(-1)^k q_k\}$,
$\{{q_k}^\ast\}$ and $\{(-1)^k {q_k}^{\ast}\}$ with $k=2,\ldots,N$, 
are related and all of them satisfy the quantization conditions 
(\ref{eq:C1-C0})
and have the same energy.

\subsection{Descendent states}

Let us first discuss
the spectrum along the trajectories
with the highest conformal charge $q_N$
equal zero
for arbitrary $\nu_h \in \mathbb{R}$.
It turns out \ci{Vacca:2000bk,Lipatov:1998as,Bartels:1999yt,Derkachov:2002wz} 
that the wave-functions of these states
are built of $(N-1)-$particle states. Moreover, their energies
\ci{Korchemsky:1994um} are also equal to the energy of the ancestor
$(N-1)-$particle states: 
\begin{equation}
E_N(q_2,q_3,\ldots,q_N=0)=E_{N-1}(q_2,q_3,\ldots,q_{N-1})\,.
\lab{eq:Edes}
\end{equation}
Thus, we call them the descendent states
of the $(N-1)-$particle states. 

Generally, for odd $N$, the descendent state $\Psi_N^{(q_N=0)}$
with the minimal energy $E_N(q_N=0)=0$ 
has 
for $q_2=0$, i.e. for $h=0,1$,  the remaining 
integrals of motion 
$q_3=\ldots =q_N=0$ as well.
For $h=1+i \nu_h$, i.e. $q_2 \ne 0$,
the odd conformal charges $q_{2k+1}=0$ with $k=1,\ldots,(N-1)/2$
while the even ones $q_{2k}\ne 0$ and depend on $\nu_h$.

On the other hand, for even $N$, the eigenstate with the minimal energy 
$\Psi_N^{(q_N=0)}$  is the descendent state of the $(N-1)-$particle 
state 
which has minimal energy with $q_{N-1} \ne 0$.  Thus, 
${E_N^{min}(q_N=0)=E_{N-1}^{min}(q_{N-1}\ne0)>0}$.

Studying more exactly this problem one can obtain
\ci{Vacca:2000bk,Derkachov:2002wz}
a relation between the quasimomenum $\theta_N$ of the descendent state 
and the ancestor one  $\theta_{N-1}$, which takes the following form
\begin{equation}
\e^{i \theta_N} \bigg|_{q_{N}=0}=
-\e^{i \theta_{N-1}}=
(-1)^{N+1}\,.
\lab{eq:quasdes}
\end{equation} 

Additionally, one can define 
linear operator $\Delta$  
\ci{Vacca:2000bk,Derkachov:2002wz}
that maps 
the subspace $V_{N-1}^{(q_N-1)}$ of the $(N-1)-$particle 
ancestor eigenstates 
with the quasimomentum $\theta_{N-1}=\pi N$
into the $N-$particle descendent states with $q_N=0$
and $\theta_N=\pi (N+1)$ as
\begin{equation}
\Delta:
\quad
V_{N-1}^{(\theta_{N-1}=\pi N)} \to 
V_{N}^{(\theta_N=\pi (N+1))} \;.
\lab{eq:Delta}
\end{equation}
It turns out that this operator is nilpotent 
for the eigenstates 
which form trajectories
\ci{Vacca:2000bk}, i.e. $\Delta^2 \Psi=0$.
Thus, the descendent-state trajectory 
can not be ancestor one for $(N+1)-$particle states.
However, it is possible to built a single  state \ci{Derkachov:2002wz}
with $q_2=q_3=\ldots=q_N=0$, i.e. for only one point $\nu_h=0$,
that has $E_N=0$ and the eigenvalue of Baxter $Q-$operator
defined as
\begin{equation}
Q_N^{q=0}(u,\wbar u) \sim \frac{u-\wbar u}{\wbar u^N}\,,
\lab{eq:Qq0}
\end{equation}
where a normalization factor was omitted.
For $N=3$ this state corresponds to the wave-function defined in 
(\ref{eq:Psiq0}).

Additional 
examples of the descendent states for $N=3$ and $N=4$ will be 
described later in the next Sections.

\section{Quantum numbers of the $N=3$ states}

In this section we present the spectrum of $q_3$  
for three reggeized gluons. 
For the first time such solutions  for $\RRe[q_3]=0$
were obtained 
in \ci{Wosiek:1996bf}. 
The authors of Ref. \ci{Wosiek:1996bf}
used the method of the $\oq{3}$ eigenfunctions
described in Chapter 5. 
Similar quantization condition was also constructed in
\ci{Lipatov:1998as}.
Solutions with
$\RRe[q_3] \ne 0$ were found in \ci{Praszalowicz:1998pz,Kotanski:2001iq}.
Moreover, in the latter paper 
the solutions with $n_h \ne 0$ are described.

The first solution
using the Baxter $Q-$operator method
was described
in Ref.
\ci{Derkachov:2002wz}.
Later,  similar results were also obtained in
Ref. \ci{deVega:2002im}.
It turns out that the Baxter $Q-$operator method \ci{Derkachov:2002wz}
for $N=3$
is equivalent to the method of the $\oq{3}$ eigenfunctions
described in Chapter 5.
For higher $N>3$ only the Baxter $Q-$operator method was used to find
the quantization values of $q_N$ and the energy.

\subsection{Lattice structure}

Solving the quantization conditions (\ref{eq:C1-C0}) for $N=3$
and for $h=\frac{1+n_h}{2}$, i.e. with $\nu_h=0$,
we reconstruct the full spectrum of $q_3$.
It is convenient to show the spectrum in terms of $q_3^{1/3}$ rather 
than $q_3$. 
Since $q_3^{1/3}$  is a multi-valued function of complex $q_3$,
each eigenstate is represented on the complex
$q_3^{1/3}-$plane by $N=3$ different points.
Thus, the spectrum is symmetric under the
transformation
\begin{equation}
q_3^{1/3} \leftrightarrow \exp(2 \pi i k/3) q_3^{1/3},
\quad
\mbox{where}
\quad
0<k<3\,.
\end{equation} 
Additionally, symmetry (\ref{eq:qkmsym}) gives
a more regular structure.
For the total  $SL(2,\mathbb{C})$ spin of the system 
$h=1/2$, which means $n_h=0$ and $\nu_h=0$,
we present the spectrum in Figure \ref{fig:N3q0}.
\begin{figure}[h]
\centerline{
\begin{picture}(200,80)
\put(10,0){\epsfysize8.5cm \epsfbox{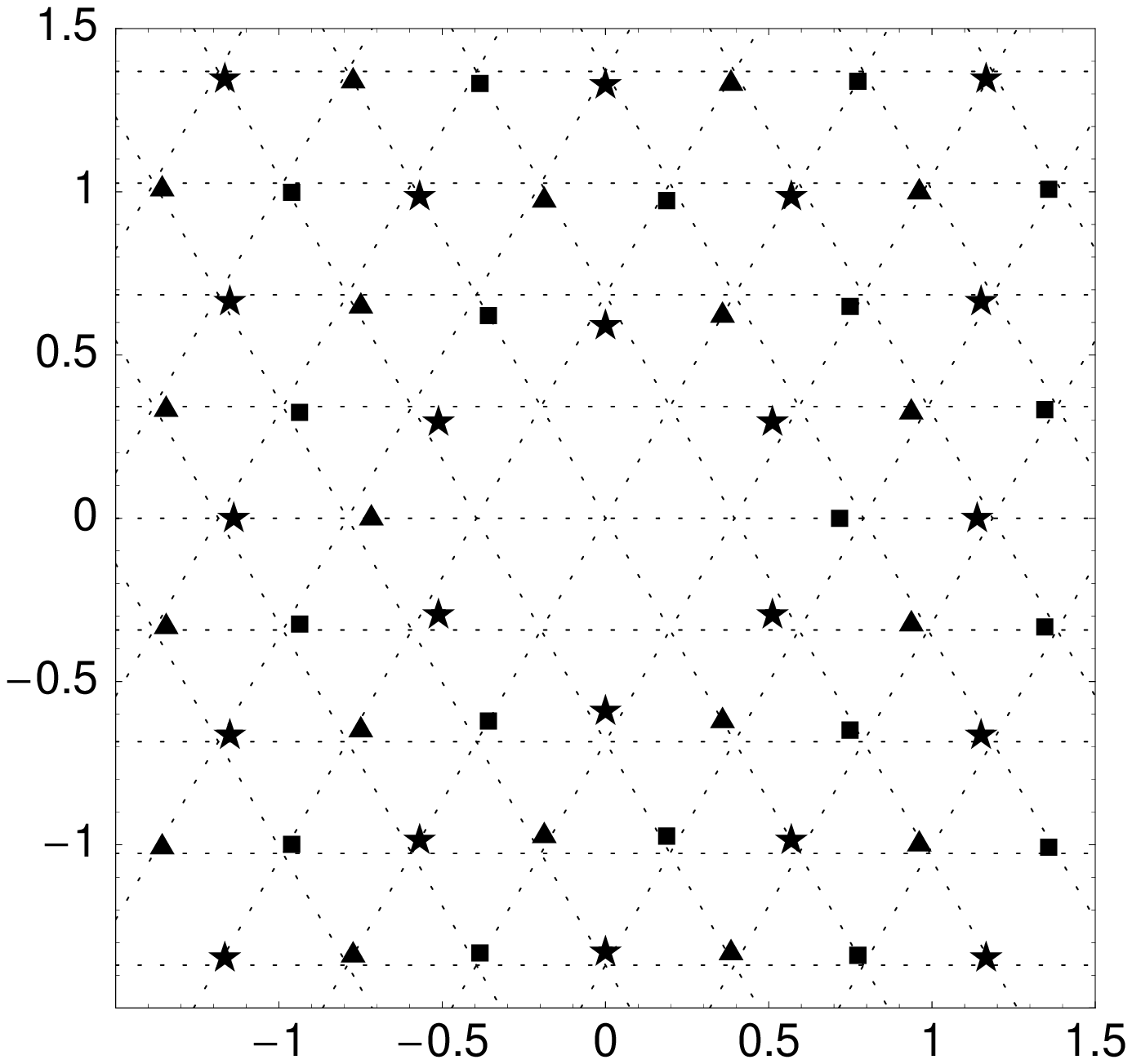}}
\put(105,0){\epsfysize8.5cm \epsfbox{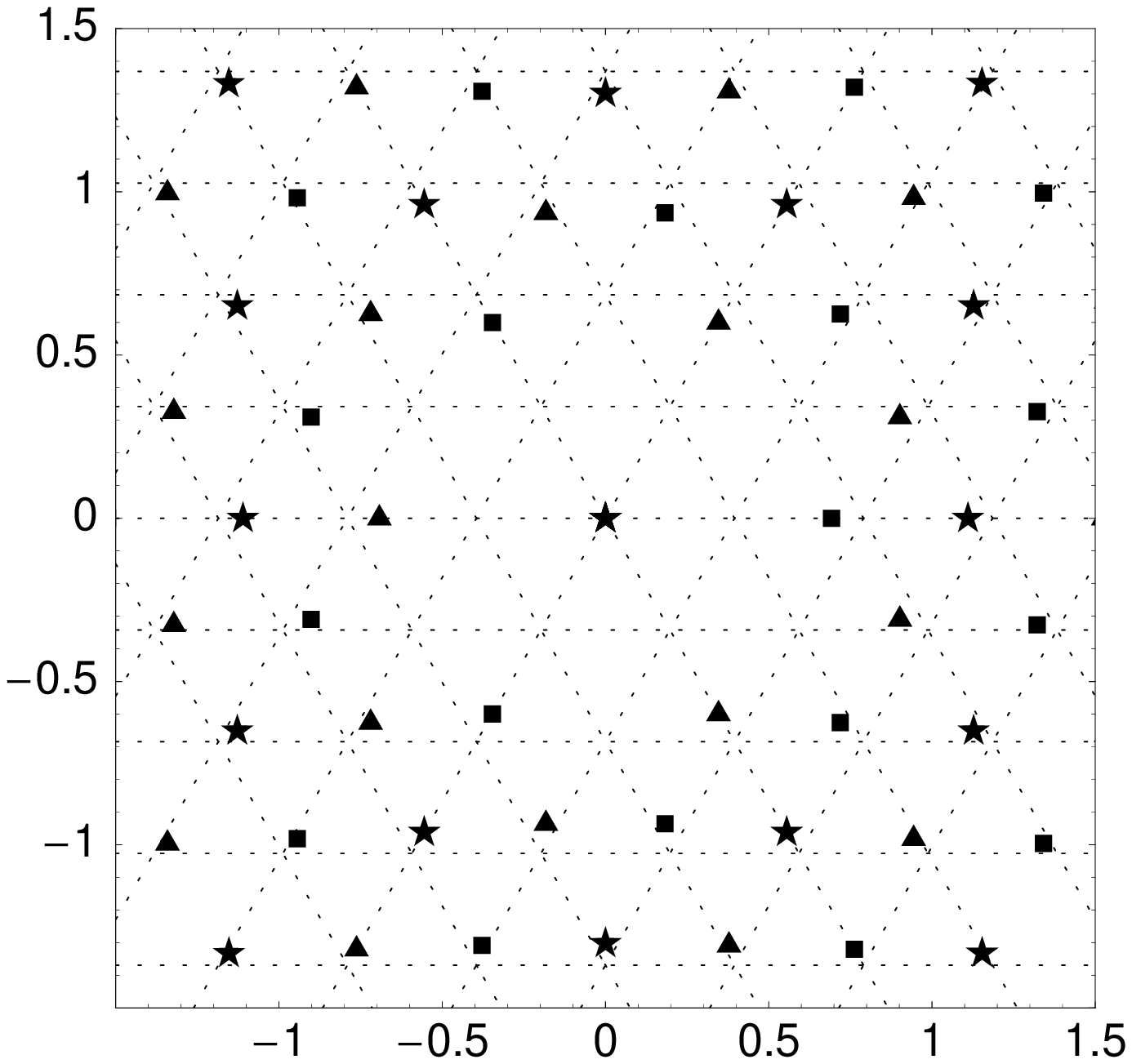}}
\put(48,-2){$\RRe[q_3^{1/3}]$}
\put(143,-2){$\RRe[q_3^{1/3}]$}
\put(9,37){\rotatebox{90}{$\IIm[q_3^{1/3}]$}}
\put(104,37){\rotatebox{90}{$\IIm[q_3^{1/3}]$}}
\end{picture}
}
\caption[The spectrum of $q_3^{1/3}$ for the 
system of $N=3$ particles for $h=\frac{1}{2}$ and $h=1$]
{The spectrum of quantized $q_3^{1/3}$ for the 
system of $N=3$ particles.
On the left 
the total $SL(2,\mathbb{C})$ spin of the system is equal to $h=\frac{1}{2}$,
while on the right $h=1$.
Different symbols stand for different quasimomenta $\theta_3$:
{\it stars} $\theta_3=0$
{\it boxes} $\theta_3=4 \pi/3$
{\it triangles} $\theta_3=2 \pi/3$.}
\lab{fig:N3q0}
\lab{fig:N3q1}
\end{figure}
One can easily notice that the spectrum has the structure
close to the equilateral triangle lattice
of the leading order WKB approximation (\ref{eq:qNqnt}). 
Indeed,
apart from a few points close to the origin, the quantized values of 
$q_3^{1/3}$ are located almost exactly at the vertices of the WKB
lattice. 
The WKB formula \ci{Derkachov:2002pb} gives
\begin{equation}
\left[q_{3}^{\rm WKB}(\ell_1,\ell_2)\right]^{1/3}
=\Delta_{N=3}\cdot
\left(\frac12\ell_1+i\frac{\sqrt{3}}2\ell_2\right),
\lab{eq:q3-WKB}
\end{equation}
where $\ell_1$ and $\ell_1$ are integers, 
such that their sum $\ell_1+\ell_1$ is
even. 
Here the lattice spacing is denoted by 
\begin{equation}
\Delta_{3}
=\left[\frac{3}{4^{1/3}\pi}
\int_{-\infty}^1\frac{dx}{\sqrt{1-x^3}}\right]^{-1}
=\frac{\Gamma^3(2/3)}{2\pi}=0.395175\ldots \,.
\end{equation}
The lattice of
$q_3^{1/3}$ 
extends on the whole complex plane except the interior of the
disk with the radius $\Delta_{3}$:
\begin{equation}
|q_3^{1/3}| > \Delta_{3}\,
\end{equation}
situated at $q_3=0$.

\begin{figure}[h]
\centerline{
\begin{picture}(200,80)
\put(10,0){\epsfysize8.5cm \epsfbox{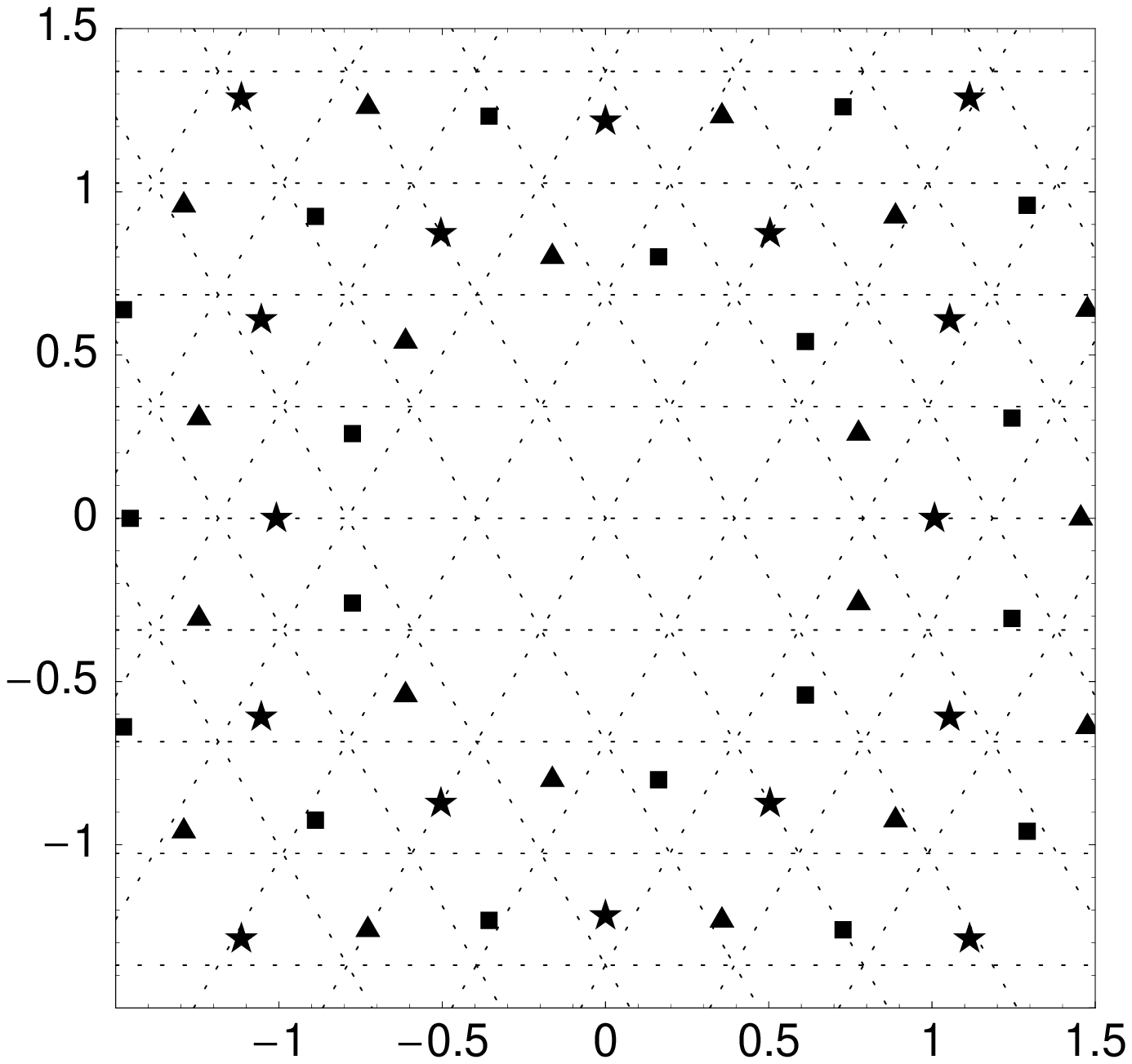}}
\put(105,0){\epsfysize8.5cm \epsfbox{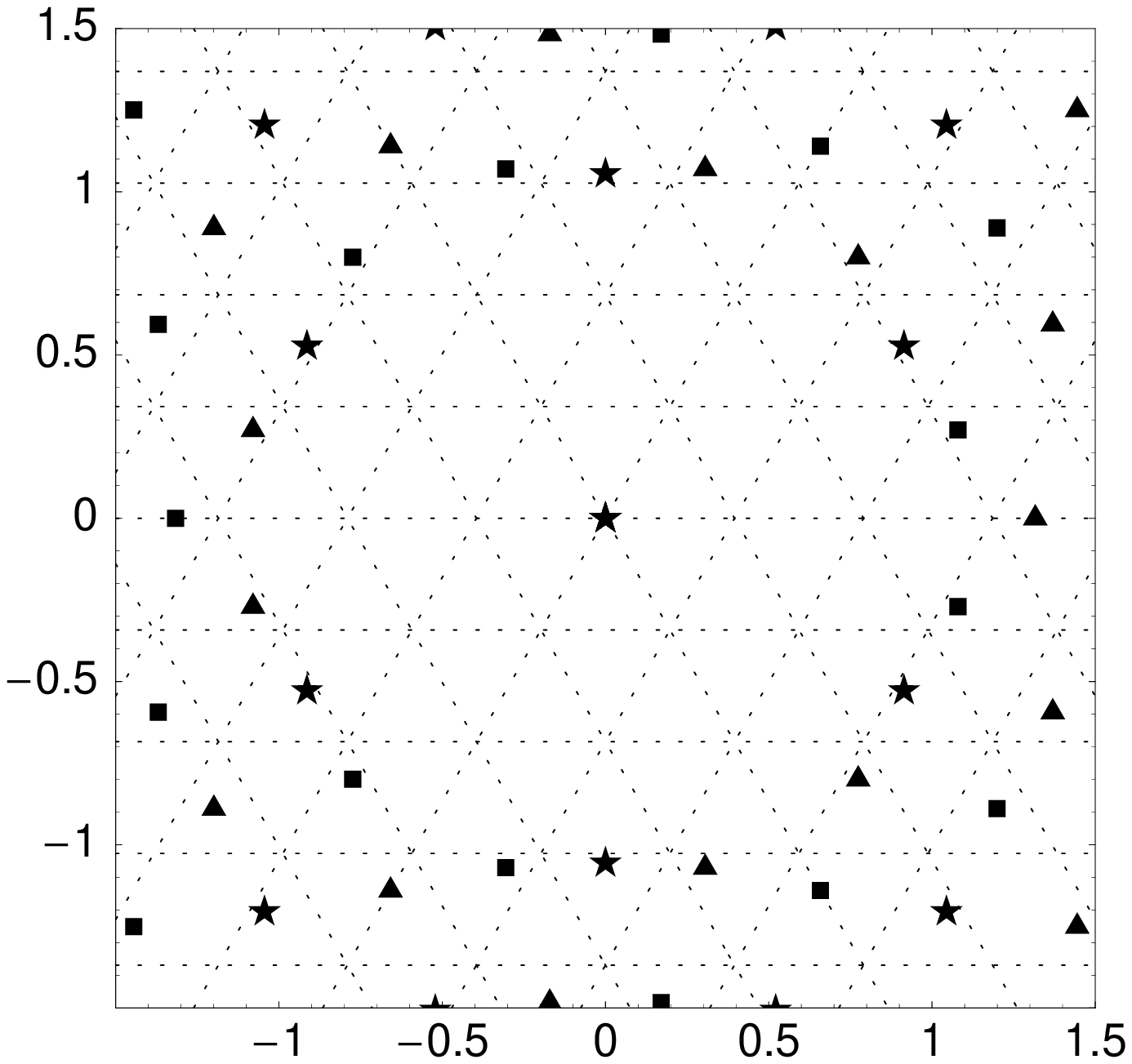}}
\put(48,-2){$\RRe[q_3^{1/3}]$}
\put(143,-2){$\RRe[q_3^{1/3}]$}
\put(9,37){\rotatebox{90}{$\IIm[q_3^{1/3}]$}}
\put(104,37){\rotatebox{90}{$\IIm[q_3^{1/3}]$}}
\end{picture}
}
\caption[The spectrum of $q_3^{1/3}$ for the 
system of $N=3$ particles for $h=\frac{3}{2}$ and $h=2$]
{The spectrum of quantized $q_3^{1/3}$ for the 
system of $N=3$ particles.
On the left 
the total $SL(2,\mathbb{C})$ spin of the system is equal to $h=\frac{3}{2}$,
while on the right $h=2$.
Different symbols stand for different quasimomenta $\theta_3$:
{\it stars} $\theta_3=0$
{\it boxes} $\theta_3=4 \pi/3$
{\it triangles} $\theta_3=2 \pi/3$.}
\lab{fig:N3q2}
\lab{fig:N3q3}
\end{figure}

In accordance with (\ref{eq:q3-WKB}),
a pair of integers $\ell_1$ and $\ell_2$
parameterize 
the quantized values of $q_3^{1/3}$. Going further,
one can calculate the quasimomentum as a function of 
$\ell_1$ and $\ell_2$. It has a following form
\begin{equation}
\theta_3(\ell_1,\ell_2)=\frac{2\pi}3\ell_1\quad ({\rm mod}~2\pi)
\,.
\lab{eq:quasi-3}
\end{equation}
Thus, as we can see states with the same value of 
$\RRe[q_3^{1/3}]$ have the same quasimomentum.
In Figure \ref{fig:N3q1}, different quasimomenta are
distinguished by {\it stars}, {\it boxes} and {\it triangles}.

The same lattice structure is exhibited by other spectra
with different $n_h$. 
However, they have different corrections to the 
leading order WKB approximation 
for $q_3^{1/3}$. These spectra are presented in 
Figures \ref{fig:N3q1}-\ref{fig:N3q3}. 
The corrections to the lattice structure 
depend on $q_2$ as seen in (\ref{eq:qNqnt}).
Since the WKB lattice is 
obtained 
in the leading order of 
the expansion
for large conformal charges, $1 \ll |q_2^{1/2}| \ll |q_3^{1/3}|$, 
the corrections are bigger for lower $|q_3^{1/3}|$.
Later, we shall discuss some other features of the corrections
to the WKB leading order approximation.

As we can see in Figs.\ \ref{fig:N3q1}-\ref{fig:N3q3} 
for $h \in \mathbb{Z}$ we have additionally
trajectories with  $q_3=0$. They are called
the descendent states because 
their spectra are related to the spectra for the $N-1=2$ Reggeon states. 
We discuss this point further in next Sections.

\subsection{Trajectories in $\nu_h$}

In the previous Section we considered the dependence of $q_3$ 
on $n_h$ for $\nu_h=0$.
However, the spectrum of conformal charges also depends 
on the continuous parameter $\nu_h$ with
$h=\frac{1+n_h}{2}+i \nu_h$. 
It turns out that 
the spectrum is built of trajectories 
parameterized by real parameter $\nu_h$.
Each trajectory
crosses one point ({\it star}, {\it box}, {\it triangle})
in Figs.\ \ref{fig:N3q1}-\ref{fig:N3q3}. 
An example of three trajectories is presented in Figure \ref{fig:traj-3D}.
They are numbered by $(\ell_1,\ell_2)=(0,2)$, $(2,2)$ and $(4,2)$ 
whereas
they quasimomentum $\theta_3(\ell_1,\ell_2)=0$, $4 \pi/3$ and $2 \pi/2$, 
respectively.

\begin{figure}[ht]
\vspace*{3mm}
\centerline{{\epsfysize6.5cm \epsfbox{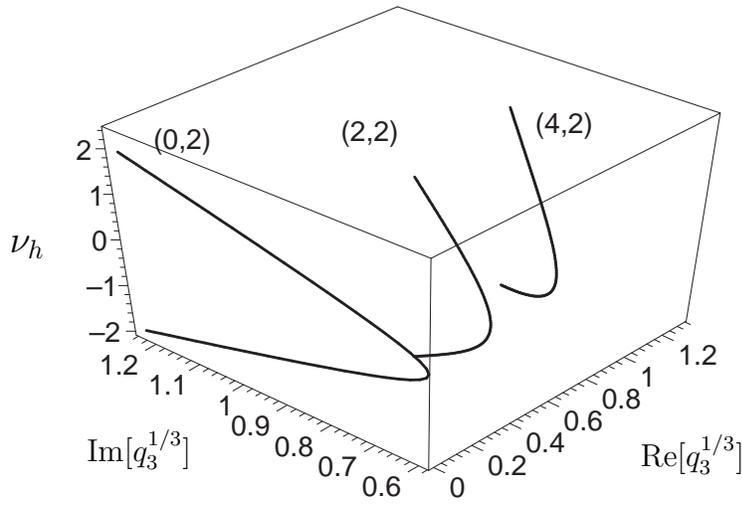}}}
\caption[The dependence of quantized $q_3(\nu_h;\ell_1,\ell_2)$ 
on the total spin $h=1/2+i\nu_h$.]{The dependence of 
quantized $q_3(\nu_h;\ell_1,\ell_2)$ on the total spin $h=1/2+i\nu_h$.
Three curves correspond to the trajectories 
with $(\ell_1,\ell_2)=(0,2)\,, (2,2)$
and $(4,2)$.}
\lab{fig:traj-3D}
\end{figure}

\begin{figure}[h]
\centerline{
\begin{picture}(200,100)
\put(10,0){\epsfysize8.5cm \epsfbox{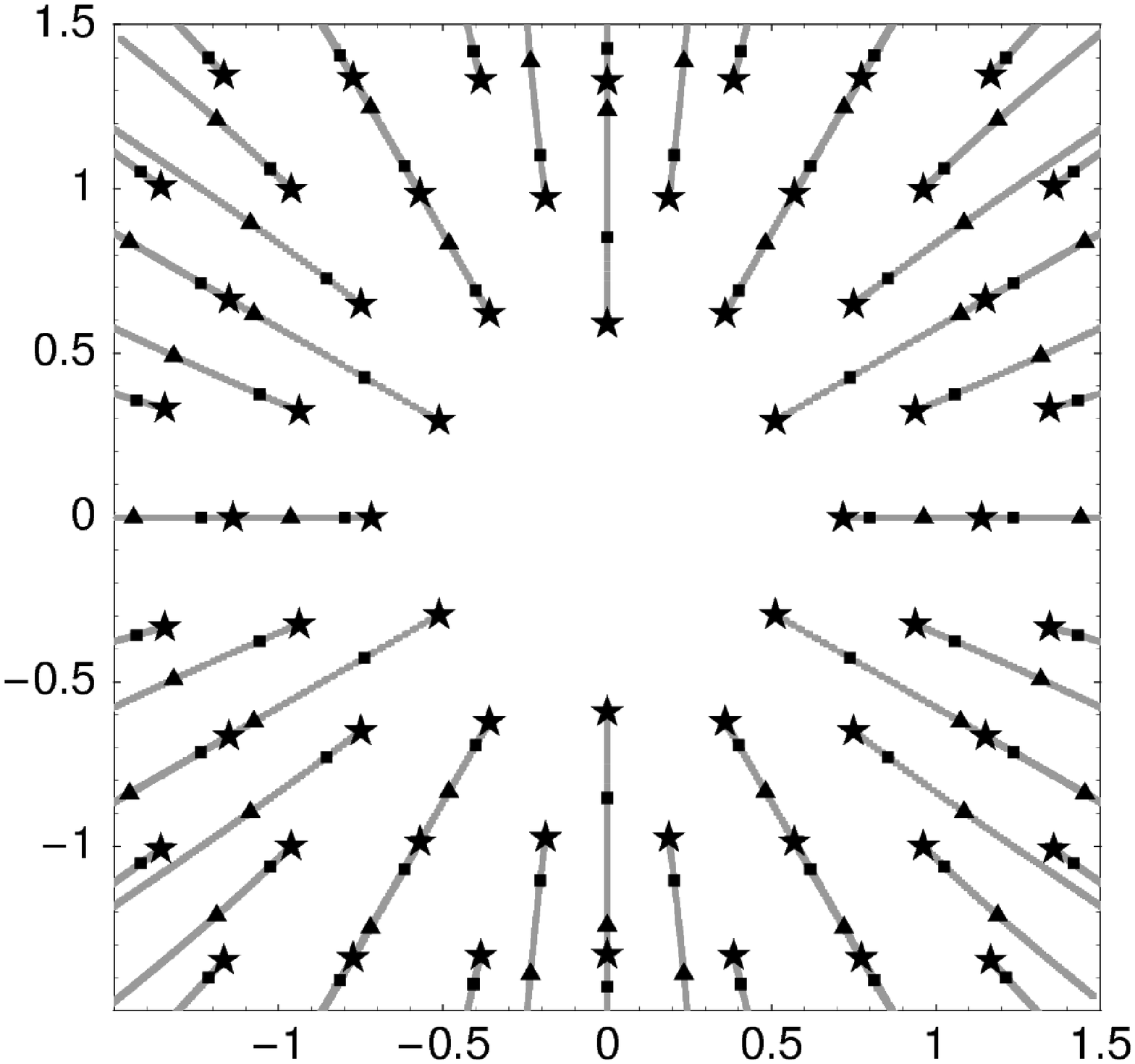}}
\put(105,0){\epsfysize8.5cm \epsfbox{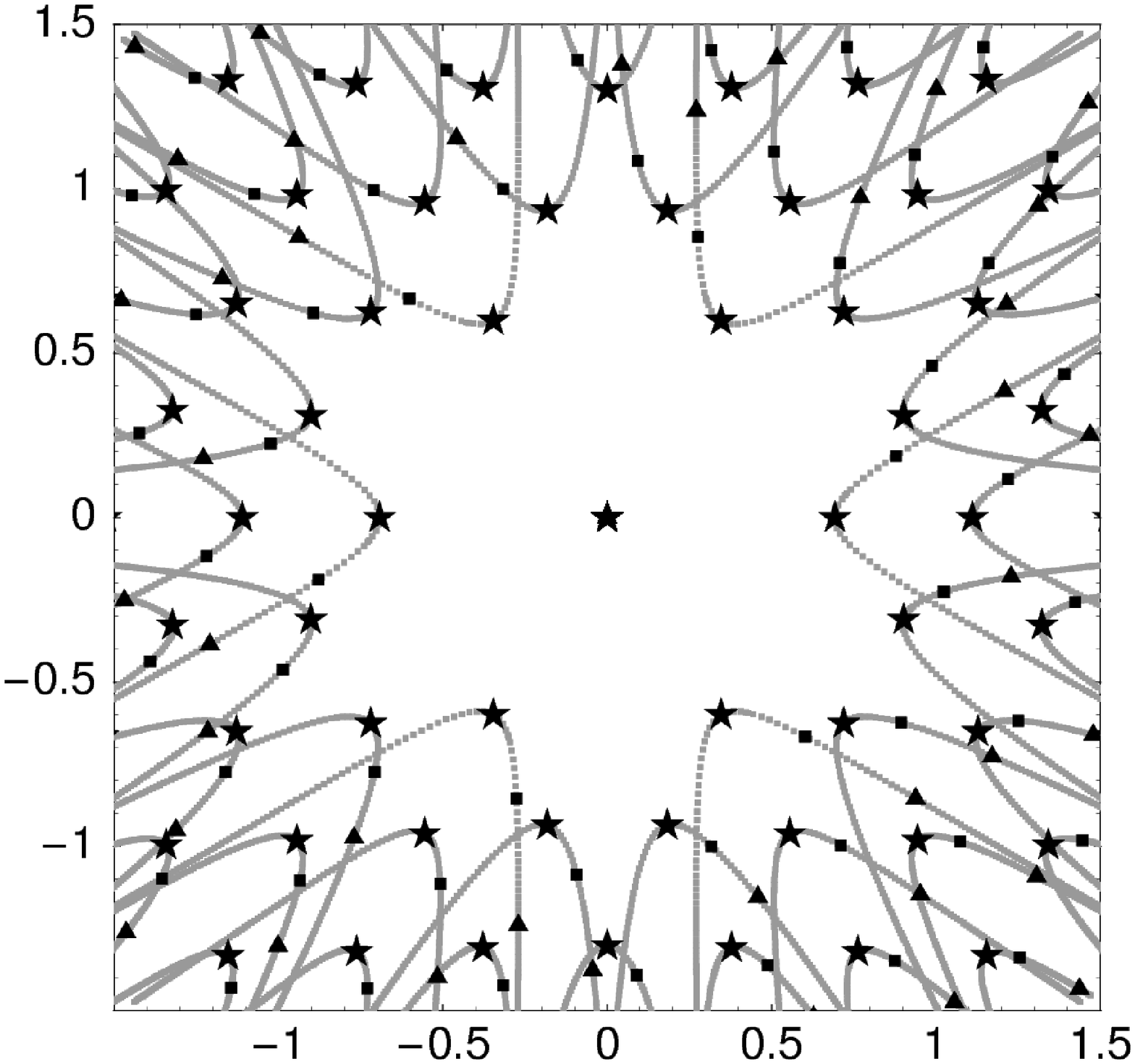}}
\put(48,-2){$\RRe[q_3^{1/3}]$}
\put(143,-2){$\RRe[q_3^{1/3}]$}
\put(9,37){\rotatebox{90}{$\IIm[q_3^{1/3}]$}}
\put(104,37){\rotatebox{90}{$\IIm[q_3^{1/3}]$}}
\end{picture}
}
\caption[The trajectories of $q_3^{1/3}$ projected on $\nu_h=0$
for $n_h=0$ and $n_h=1$.]
{The trajectories of $q_3^{1/3}$ projected on $\nu_h=0$.
On the left panel $h=\frac{1}{2}+ i \nu_h$, while on 
the right one $h=1+ i \nu_h$. {\it stars} denotes $\nu_h=0$,
{\it boxes} $\nu_h=1$ and {\it triangles} $\nu_h=2$.}
\lab{fig:prtraj}
\end{figure}

The trajectories cumulate at $\nu_h=0$. When we increase $\nu_h$, 
$q_3^{1/3}$ tends to infinity and the structure
of quantized charges 
starts to be less regular, especially for trajectories with lower 
$|q_3^{1/3}|$. 
We can see this in Figures \ref{fig:prtraj} where we project trajectories
with $h=\frac{1}{2}+i \nu_h$
on the $\nu_h=0$ plane. Here {\it stars} denote point with $\nu_h=0$,
{\it boxes} with $\nu_h=1$ and {\it circles} $\nu_h=2$.
Grey lines are drawn to show the projection of the trajectories for 
intermediate values of $\nu_h$.

\begin{figure}[h]
\centerline{
\begin{picture}(200,100)
\put(10,0){\epsfysize8.5cm \epsfbox{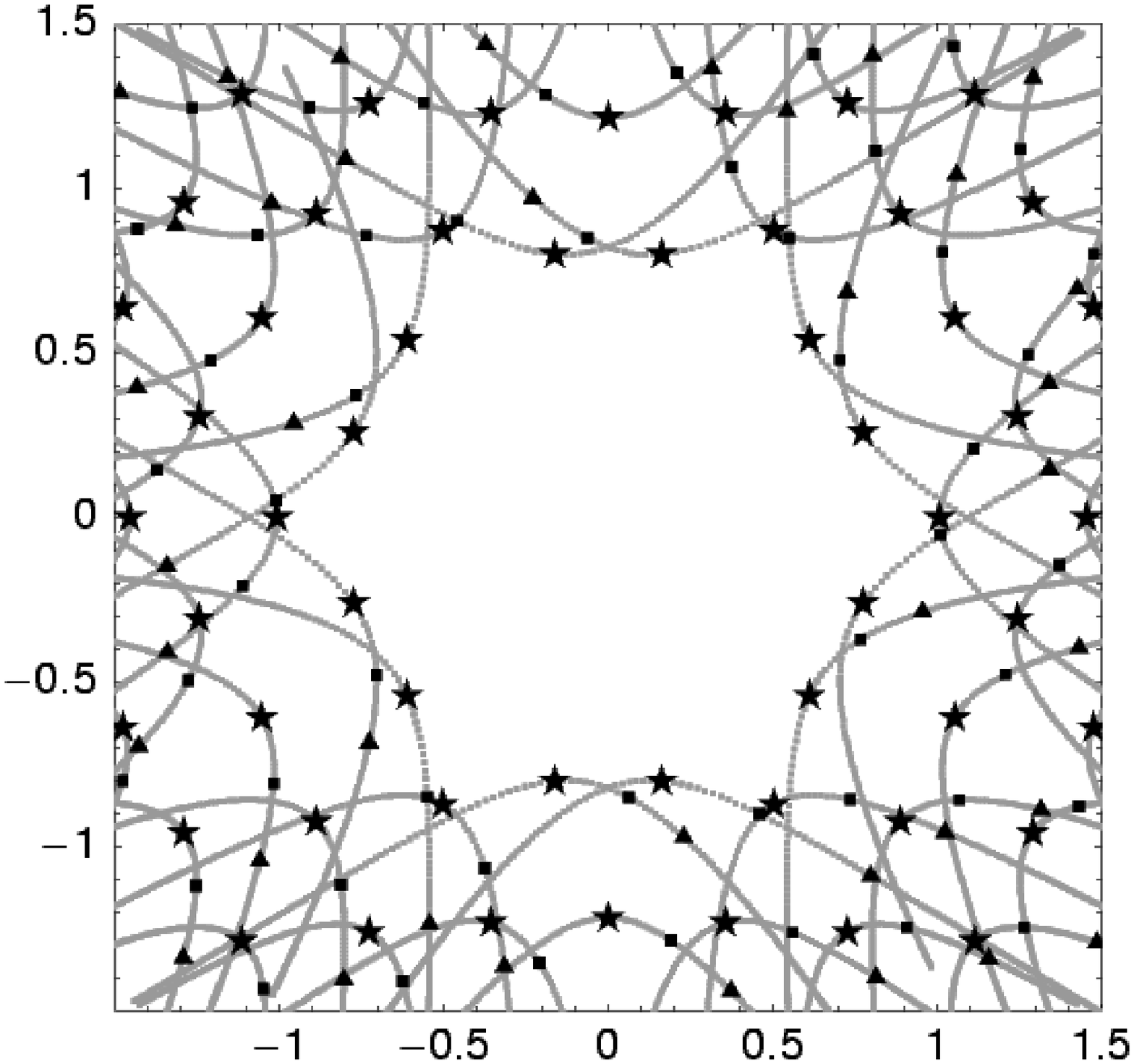}}
\put(105,0){\epsfysize8.5cm \epsfbox{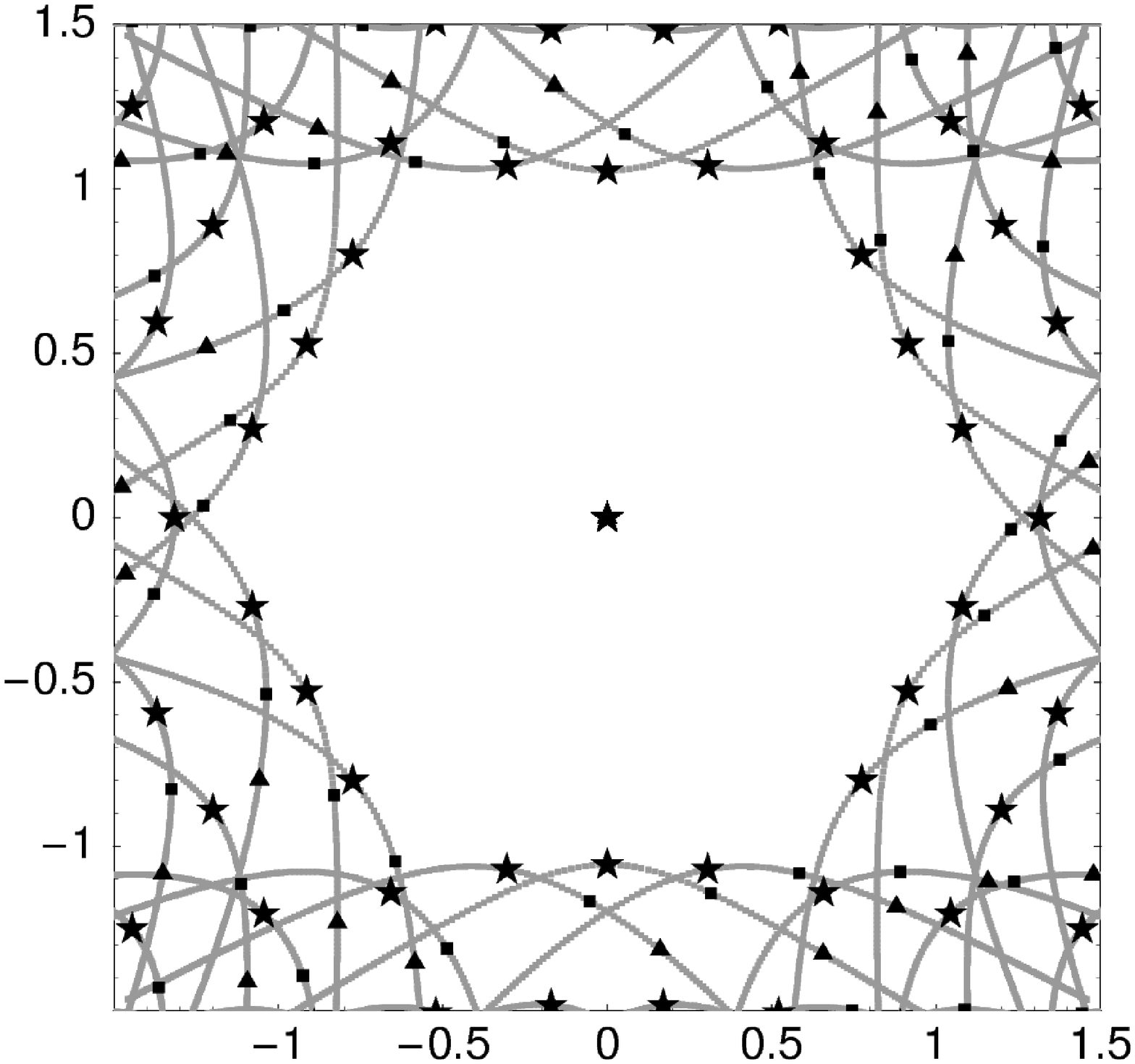}}
\put(48,-2){$\RRe[q_3^{1/3}]$}
\put(143,-2){$\RRe[q_3^{1/3}]$}
\put(9,37){\rotatebox{90}{$\IIm[q_3^{1/3}]$}}
\put(104,37){\rotatebox{90}{$\IIm[q_3^{1/3}]$}}
\end{picture}
}
\caption[The trajectories of $q_3^{1/3}$ projected on $\nu_h=0$
for $n_h=2$ and $n_h=3$.]
{The trajectories of $q_3^{1/3}$ projected on $\nu_h=0$.
On the left panel $h=\frac{3}{2}+ i \nu_h$, while on 
the right one $h=2+ i \nu_h$. {\it stars} denotes $\nu_h=0$,
{\it boxes} $\nu_h=1$ and {\it triangles} $\nu_h=2$.}
\lab{fig:prtraj2}
\end{figure}
For $n_h \ne 0$ we notice that the spectra start to rotate with $\nu_h$.
In Figures \ref{fig:prtraj} and \ref{fig:prtraj2}  
we present trajectories with positive $n_h=0,1,2,3$.
Due to the symmetry (\ref{eq:qkcsym}), which means 
$h\rightarrow 1-h^{\ast}$ or 
$n_h \rightarrow - n_h$ with $q_3 \rightarrow q_3^{\ast}$,
or equivalently,   
$\nu_h \rightarrow -\nu_h$ but $q_3 \rightarrow q_3$,
the spectrum for the negative $n_h$  is the
same as for the positive ones but
it rotates in the opposite direction with $\nu_h$.

Some of the results presented in this Section were found in earlier works
\ci{Korchemsky:1999is,Kotanski:2001iq}.
Trajectories with 
quasimomentum $\theta_3=0$ and
$n_h=0$ were obtained in Refs. \ci{Korchemsky:1999is,Kotanski:2001iq}.
The case for $h=2+i \nu_h$ with 
quasimomentum $\theta_3=0$ was discussed in Ref. \ci{Kotanski:2001iq}.
In this thesis we additionally analyse the spectra for 
$h=1+i\nu_h$ and 
$h=3/2+i\nu_h$.

\subsection{Energy and dispersion}

For all trajectories 
in $(q_2,q_3)-$space 
we can calculate the energy of the reggeized gluons
using Eq. (\ref{eq:E-fin}). 
Example of the energy spectrum for trajectories
from Figure \ref{fig:traj-3D} with $h=\frac{1}{2}+i \nu_h$ is shown in
Figure \ref{fig:energy3}.
\begin{figure}[ht]
\vspace*{3mm}
\centerline{{\epsfysize7cm \epsfbox{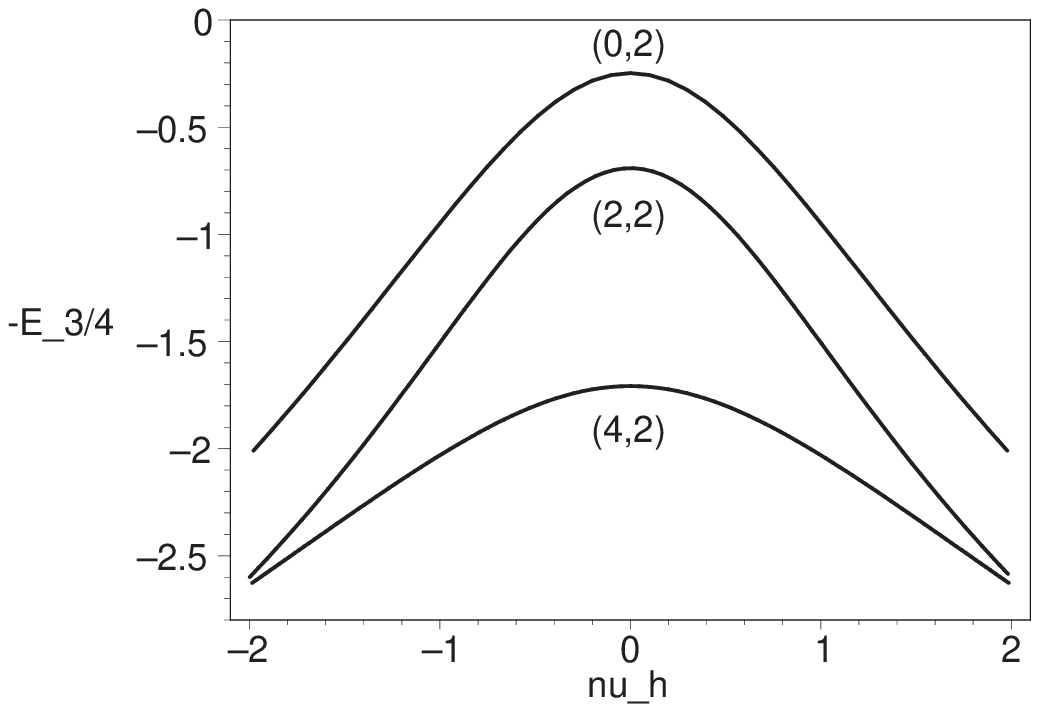}}}
\caption[The energy spectrum corresponding to three trajectories shown
in Figure~\ref{fig:traj-3D}]{The energy spectrum corresponding to three trajectories shown
in Figure~\ref{fig:traj-3D}.
The ground state is located on the $(0,2)-$trajectory at $\nu_h=0$.}
\lab{fig:energy3}
\end{figure}

The energy along the trajectories is a continuous gapless 
function of $\nu_h$.
As we can see the energy $E_3$ grows with rising $|\nu_h|$.
For $n_h=0$ it has a minimum value 
$\mbox{min}_{\nu_h} E_3(\nu_h;\ell_1,\ell_2)$ at $\nu_h=0$.
In the case $n_h \ne 0$, due the bending of the trajectories 
some minima of the energy are moved away from $\nu_h=0$ 
\ci{Kotanski:2001iq}. 
However, the ground state corresponds to the point(s) on the plane of 
$q_3^{1/3}$ (see Figure \ref{fig:N3q0}) closest to the origin.
For $N=3$ the ground state is located on the $(0,2)-$trajectory at 
$\nu_h=0$ and $n_h=0$  with quasimomentum equal $\theta_3=0$.
Due to the symmetry (\ref{eq:qkmsym})
it is doubly-degenerated and its conformal charge and energy 
take the following values:
\begin{equation}
iq_3^{\rm ground}=\pm 0.20526\ldots\,,\qquad E_3^{\rm ground}=0.98868\ldots\,.
\lab{eq:N3-ground}
\end{equation}
In the vicinity of $\nu_h=0$ 
the accumulation of the energy levels is described by the dispersion 
parameter $\sigma_3$ (\ref{eq:Enu}) 
given below in the Table \ref{tab:Summary}.

We show comparison the WKB result 
of Eq. (\ref{eq:q3-WKB}) with the exact expressions 
for $q_3$ at $h=1/2$
in Figure~\ref{fig:N3q0} and Table~\ref{tab:WKB}. 
One can find that the
expression (\ref{eq:q3-WKB}) describes the excited eigenstates 
with good accuracy. In the case where
the eigenstates have smaller $q_3$
agreement becomes less accurate. Thus,
for the ground state with $iq_3=0.20526\ldots $ the accuracy of (\ref{eq:q3-WKB})
is $\sim 20\%$.
Obviously, in the region where the WKB expansion
is valid, i.e. 
$|q_3^{1/3}|\gg |q_2^{1/2}|$, Eq.~(\ref{eq:q3-WKB}) can be systematically 
improved by including subleading WKB corrections. 

\begin{table}[h!]
\begin{center}
\begin{tabular}{|c||c|c|c|}
\hline
 $(\ell_1,\ell_2)$& $\left(q_3^{\rm \,exact}\right)^{1/3}$ & 
$\left(q_3^{\rm WKB}\right)^{1/3}$ & $-E_3/4$ 
\\
\hline
\hline

$(0,2)$ &  $0.590\,i$ & $0.684\,i$ & $-0.2472$ \\ \hline
$(2,2)$ & $0.358+0.621\,i$ & $0.395+0.684\,i$ & $-0.6910$ \\ \hline
$(4,2)$ & $0.749+0.649\,i$ & $0.790+0.684\,i$ & $-1.7080$ \\ \hline
$(6,2)$ & $1.150+0.664\,i$ & $1.186+0.684\,i$ & $-2.5847$ \\ \hline
$(8,2)$ & $1.551+0.672\,i$ & $1.581+0.684\,i$ & $-3.3073$ \\ \hline
$(10,2)$& $1.951+0.676\,i$ & $1.976+0.684\,i$ & $-3.9071$ \\ \hline

\end{tabular}
\end{center}
\caption[Comparison of the exact spectrum of $q_3^{1/3}$
with the 
WKB expression]{Comparison 
of the exact spectrum of $q_3^{1/3}$ at $h=1/2$ 
with the approximate
WKB expression (\ref{eq:q3-WKB}). 
The last line defines the corresponding energy
$E_3(0;\ell_1,\ell_2)$.}
\lab{tab:WKB}
\end{table}

\subsection{Descendent states for $N=3$}

One also can notice in Figures \ref{fig:prtraj} and \ref{fig:prtraj2}
that for odd $n_h$ we have states with 
$q_3=0$. For $n_h=0$ and $\nu_h=0$ it has the energy $E_3=0$, so
it is lower than (\ref{eq:N3-ground}). These states
are descendants 
of the states with two Reggeons
\ci{Lipatov:1998as,Bartels:1999yt,Bartels:2001hw,
Vacca:2000bk,Derkachov:2002wz}.
We constructed them using the $\oq{3}$ eigenfunction method.
The wave-functions of these states are described by 
(\ref{eq:Psib}),
(\ref{eq:BLV}),
(\ref{eq:Psiq0}) and
(\ref{eq:Psiq0b}).
These states have the same properties and the energy 
as the corresponding states with $N-1=2$ particles, 
$E_3(q_2,q_3=0)=E_2(q_2)$, with \ci{Balitsky:1978ic,Kuraev:1977fs}:
\begin{equation}
E_2(q_2)=4 \, \RRe[\psi(1-h)+\psi(h)-2 \psi(1)]
=8 \,\RRe\left[\psi\left(\frac{1+|n_h|}{2}+i \nu_h\right)-\psi(1)\right]\,,
\lab{eq:EN=2}
\end{equation}
where  $\psi(x)=\frac{d}{dx} \ln \Gamma(x)$ and $q_2=-h(h-1)$.
Moreover, their wave-functions are built of 
the two-Reggeon states \ci{Vacca:2000bk,Derkachov:2002wz}
and the quasimomentum $\theta_3=0$.
Contrary to the states with $q_3 \ne 0$,
the states with $q_3=0$ (\ref{eq:BLV}) couple to a point-like
hadronic impact factors 
\ci{Engel:1997cg,Czyzewski:1996bv,Bartels:2003zu},
like the one for the $\gamma^* \to \eta_c$ transition. 

\subsection{Corrections to WKB}

The WKB formula for the lattice structure of the conformal charge $q_{3}$ 
was derived
in paper \ci{Derkachov:2002pb}. This formula tells us that
for $q_{3}\rightarrow \infty $
\begin{equation}
q_{3}^{1/3}=\frac{{\Gamma ^{3}(2/3)}}{2\pi }
{\cal Q}(\mybf n)\left[1+\frac{b}{\left|{\cal Q}(\mybf n)\right|^{2}}
-\left(\frac{b}{\left|{\cal Q}(\mybf n)\right|^{2}}\right)^{2}+
\sum _{k=3}^{\infty }{a_{k}
\left(\frac{b}{\left|{\cal Q}(\mybf n)\right|^{2}}\right)^{k}}\right]\,,
\lab{eq:korq3}
\end{equation}
where 
\begin{equation}
{\cal Q}(\mybf n)=\frac{1}{2}(l_{1}+l_{2})+i\frac{{\sqrt{3}}}{2}(l_{1}-l_{2})
=\sum _{k=1}^{3}{n_{k}e^{i\pi (2k-1)/3}}
\lab{eq:Qn}
\end{equation}
and $l_{1}$, $l_{2}$, ${\mybf n}=\{n_1,\ldots,n_N\}$
are integers, 
while the coefficient
\begin{equation}
b=\frac{3\sqrt{3}}{2\pi }{q_{2}}^{\ast}\,,
\lab{eq:korb}
\end{equation}
where {\it star} denotes complex conjugation.

After numerical calculations we have noticed that 
better agreement with the exact results is obtained for
\begin{equation}
b=\frac{3\sqrt{3}}{2\pi}
\left({q_{2}}^{\ast}-\frac{2}{3}\right)\,.
\lab{eq:bcoef}
\end{equation}
In order to show this,
we calculated the values of the conformal charge $q_{3}$ for 
$h=\frac{1+n_{h}}{2}$ for $n_h=0,1,\ldots,19$.
We evaluated numerically $q_3$ with  $\IIm [q_3]=0$, i.e. $a^{(r)}_k$,
and separately with $\RRe [q_3]=0$, i.e. $a^{(i)}_k$
where the superscript $(i)$ and $(r)$ refers to 
imaginary and real parts, respectively.
Then
we fitted expansion coefficients $a^{(r,i)}_{k}$ for large $|q_{3}|$
in the range $0\ldots 5000$ with high numerical precision.
In order to save space  in Table \ref{tab:q3coefs} we present
results only for $n_h=0,1,2$ and $3$.


\begin{table}[h]
\begin{center}
$\begin{array}{|c|c||r|r|r|r|r|} 
    \hline
n_h & \mbox{coef.} & k=3 \qquad & k=4 \qquad & k=5 \quad & k=6 \quad & k=7 \quad
 \\ \hline \hline
0
&a^{(r)}_k           & 0.509799695633 & -10.065761318 & -76.722084 & -1508.927 & -44580. \\ 
&a^{(i)}_k           & 3.490200304367 &  20.065761318 & 104.722084 &  -600.068 & -40411. \\
&a^{(r)}_k+a^{(i)}_k & 4.000000000000 &  10.000000000 &  28.000000 & -2108.995 & -84991. \\
    \hline
1
&a^{(r)}_k           &-2.585231705744 & -34.383865187 & -209.366287 & -1828.065 & -26404. \\
&a^{(i)}_k           & 6.585231705744 &  44.383865187 &  237.366287 &   340.913 & -16641. \\
&a^{(r)}_k+a^{(i)}_k & 4.000000000000 &  10.000000000 &   28.000000 & -1487.152 & -43046. \\
    \hline
2
&a^{(r)}_k           & 1.336317342408 &  1.2054728742 & -3.08256385 & -31.51219 & -196.87 \\
&a^{(i)}_k           & 2.663682657592 &  8.7945271258 & 31.08256385 & 110.89046 &  377.64 \\
&a^{(r)}_k+a^{(i)}_k & 4.000000000000 & 10.0000000000 & 28.00000000 &  79.37827 &  180.77 \\
    \hline
3
&a^{(r)}_k           & 2.250754858908 &  6.7220206127 & 22.76106772 & 79.325249 & 268.13 \\
&a^{(i)}_k           & 1.749245141092 &  3.2779793873 &  5.23893228 & -0.430149 & -75.81 \\
&a^{(r)}_k+a^{(i)}_k & 4.000000000000 & 10.0000000000 & 28.00000000 & 78.895100 & 192.32 \\
    \hline

\end{array}$
\end{center}
\caption[The fitted coefficient to the series formula of $q_3^{1/3}$]
{The fitted coefficient to the series formula of $q_3^{1/3}$ 
(\ref{eq:korq3}) with  $n_h=0,1,2$ and $4$}
\lab{tab:q3coefs}
\end{table}

Coefficients $a_{k}$ for $k=0,\ldots ,2$ 
agree with formula (\ref{eq:korq3}),
i.e. $a_{0}=1$, $a_{1}=1$ and  $a_{2}=-1$, but as we previously mentioned
in (\ref{eq:korb}) and (\ref{eq:bcoef}), the expansion parameter 
$b/\left|{\cal Q}(\mybf n)\right|^{2}$ is different.
This difference comes from the fact that the series formula (\ref{eq:korq3}) 
with (\ref{eq:korb}) is derived in the limit
$1 \ll |q_2^{1/2}| \ll |q_3^{1/3}|$.
Since in (\ref{eq:bcoef}) the value of $q_2^{\ast}$ is much bigger
then $2/3$ the value of the parameter $b$ from (\ref{eq:bcoef}) 
in the above limit goes to (\ref{eq:korb}).
One may suppose that therefore the factor $2/3$ 
in Eq. (\ref{eq:bcoef}) as subleading was omitted in
the derivation presented in
Ref. \ci{Derkachov:2002pb}.

Secondly, we see that the coefficients $a_k$ with $k>2$ 
start to depend on $n_h$. Thus, to describe the behaviour of 
$q_3^{1/3}$ properly, we have to introduce a second 
expansion parameter, for example $q_2$. 
We can also notice that 
for $k>2$ for  
a few first coefficients
$a_k^{(i)}+a_k^{(r)} \in \mathbb{Z}$ and this sum
does not depend on $n_h$.

Moreover, we can see that after the numerical fitting we obtain two different 
sets of the expansion coefficient, $\{a_k^{(r)}\}$ and $\{a_k^{(i)}\}$, 
defined in (\ref{eq:korq3}), for real  and 
imaginary  $q_3^{1/3}$, respectively. Thus, in order to describe
full-complex values of $q_3^{1/3}$ in terms of
the series (\ref{eq:korq3})
we have to use both sets of coefficients,
one for real and one for imaginary part of $q_3^{1/3}$.
Alternatively, we can perform expansion with two small parameters, i.e.
$q^{\ast}_2/|{\cal Q}({\mybf n})|^2$ and $1/|{\cal Q}({\mybf n})|^2$.
Since the leading terms, i.e. with $a_0$, $a_1$ and $a_2$, 
for real and imaginary $q_3^{1/3}$ are equal and
known analytically,
good approximation is obtained using
Eq. (\ref{eq:korq3})
with (\ref{eq:bcoef}) 
and neglecting higher order terms with $a_{k\ge 3}$.


\section{Quantum numbers of the $N=4$ states}

In this Section we present the spectrum for four Reggeons. 
Earlier, some results for $N=4$ were only presented 
in \ci{Derkachov:2002wz} and some numerical results in Ref.
\ci{deVega:2002im}.
Here we show much more data and we present
more detailed analysis of this spectrum.

For four Reggeons the spectrum of the conformal charges 
is much more complicated than in the three-Reggeon case.
Indeed
we have here the space of three conformal charges $(q_2,q_3,q_4)$.
Thus, apart from the lattice structure in $q_4^{1/4}$ 
we have also respective lattice structures
in $q_3-$space.
Here we consider the case for $n_h=0$ so that $h=\frac{1}{2}+i \nu_h$. 
This spectrum includes
the ground states.
For clarity we split these spectra into several parts. 
We perform this separation by 
considering spectra with different quasimomenta 
$\theta_4(q,\wbar q)$ as well as
a different quantum number $\ell_3$,
which will be defined in the solution (\ref{eq:theta-4})--(\ref{eq:q3-quan})
of two quantization conditions (\ref{eq:ukcond}) for $N=4$ 
from Ref. \ci{Derkachov:2002pb}.


From the first quantization condition (\ref{eq:ukcond}) 
one can get the WKB approximation of
the charge $q_4$ as
\begin{equation}
q_4^{1/4}=\frac{\Gamma^2(3/4)}{4\sqrt{\pi}}\left[ \frac1{\sqrt {2}}
\ell_1+
\frac{i}{\sqrt {2}}\ell_2 \right]
\lab{eq:q4-quan}
\end{equation}
and the quasimomentum
 is equal to
\begin{equation}
\theta_4=-\frac{\pi}2\ell
=\frac{\pi}2 
(\ell_2+\ell_3-\ell_1)\qquad ({\rm mod}~ 2\pi)\,,
\lab{eq:theta-4}
\end{equation}
where $\ell_1$, $\ell_2$ and $\ell_3$ are even for even $\ell$ and
odd for odd $\ell$.
Thus, we have two kinds of lattices: with $\theta_4=0,\pi$ and
with $\theta_4=\pm \pi/2$. 
They are presented in Fig.\ \ref{fig:Q4}. In these pictures
gray
lines show the WKB lattice (\ref{eq:q4-quan})
with vertices at $\ell_1, \ell_2 \in \mathbb{Z}$.
To find  the leading approximation 
for the charge $q_3$, we apply the second relation in (\ref{eq:ukcond}) with
$m=2$  which gives
\begin{equation}
\Im \frac{q_3}{q_4^{1/2}} = (\ell_1-\ell_2-\ell)=\ell_3.
\lab{eq:q3-quan}
\end{equation}
Notice that the system (\ref{eq:q4-quan}) and (\ref{eq:q3-quan}) 
is underdetermined and 
it does not fix the charge $q_3$ completely \ci{Derkachov:2002pb}.

\medskip
\begin{figure}[h!]
\centerline{
\begin{picture}(200,80)
\put(10,0){\epsfysize8.5cm \epsfbox{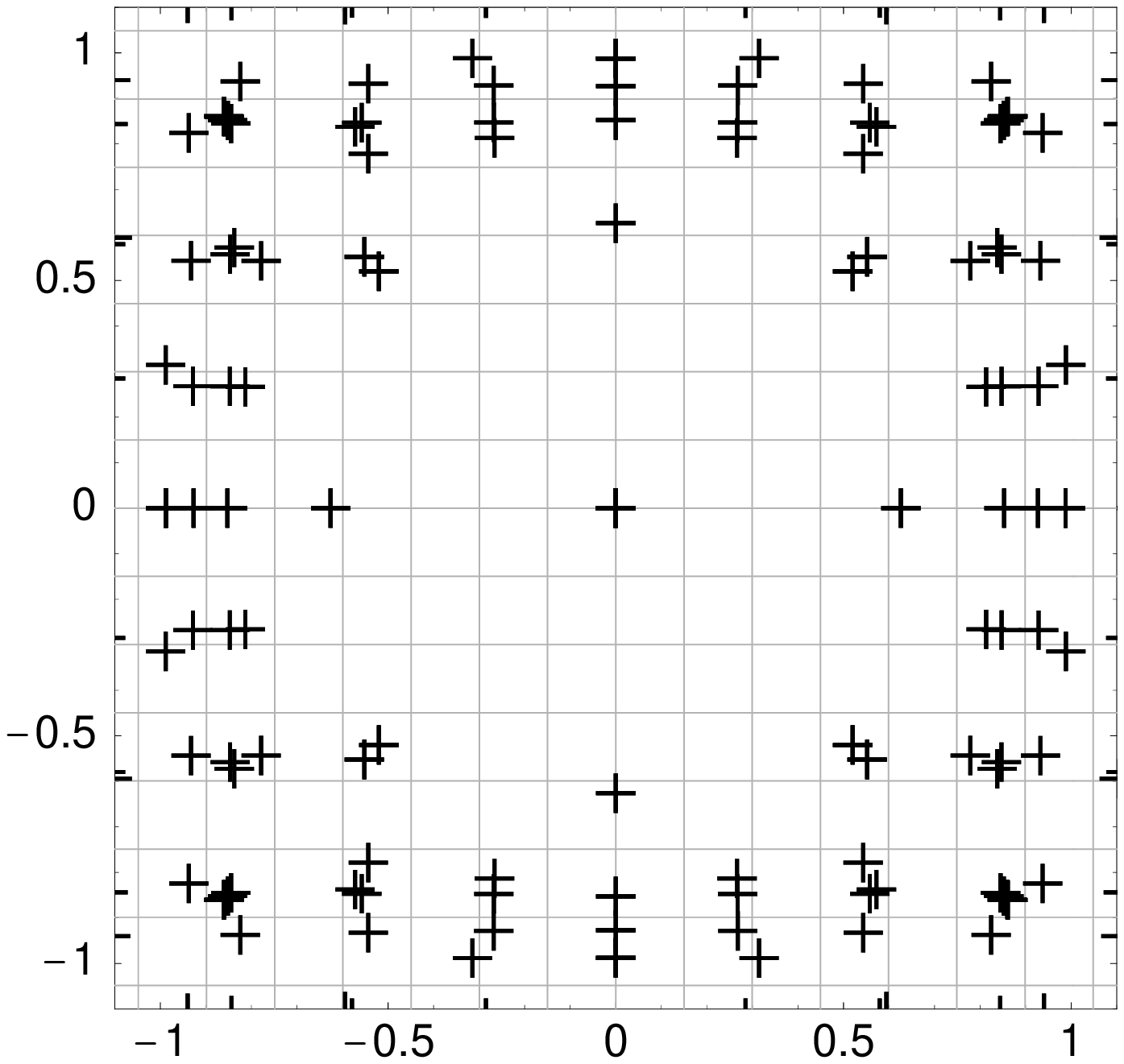}}
\put(105,0){\epsfysize8.5cm \epsfbox{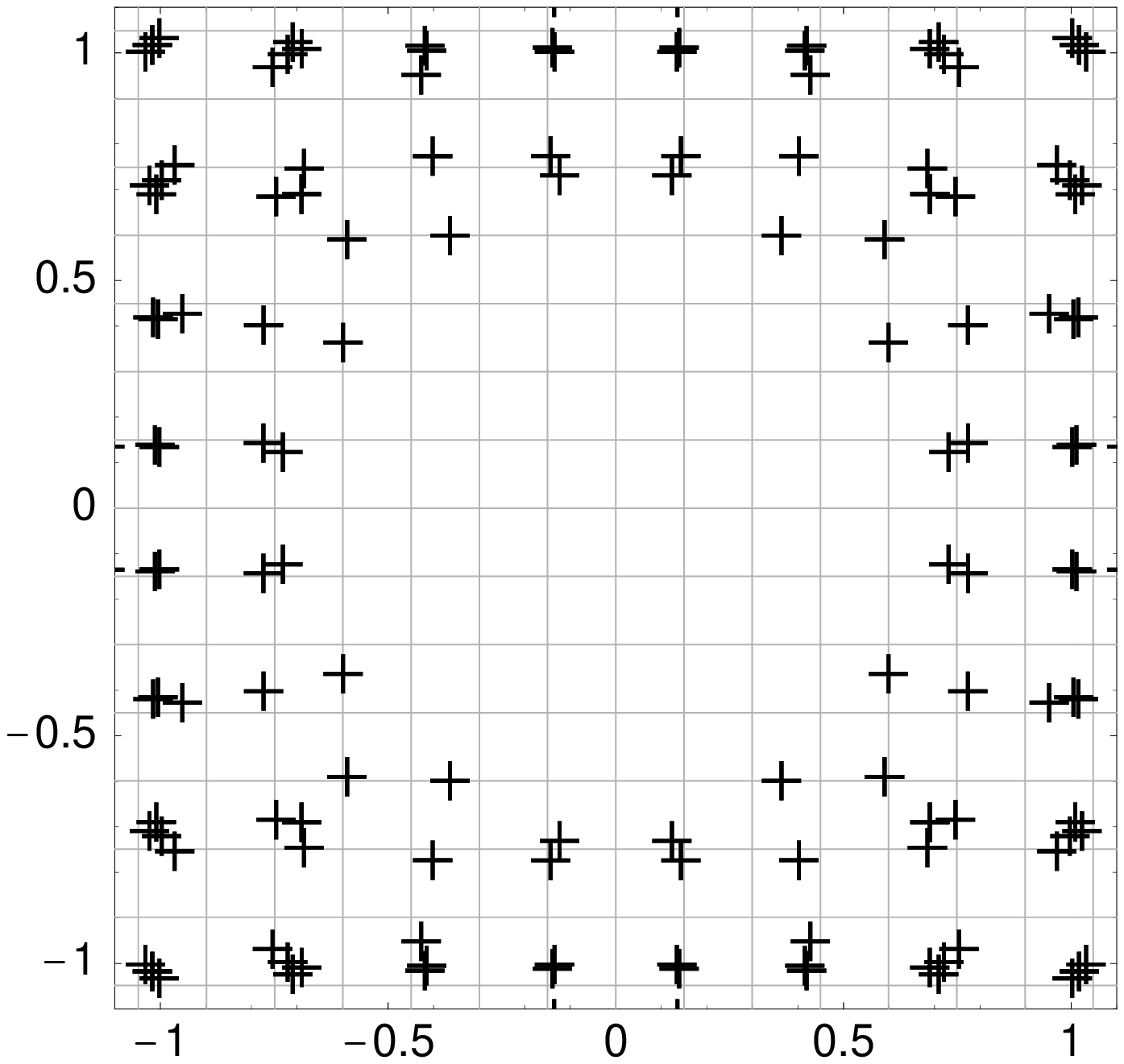}}
\put(48,-2){$\RRe[q_4^{1/4}]$}
\put(9,37){\rotatebox{90}{$\IIm[q_4^{1/4}]$}}
\put(143,-2){$\RRe[q_4^{1/4}]$}
\put(104,37){\rotatebox{90}{$\IIm[q_4^{1/4}]$}}
\end{picture}
}
\caption[The spectrum of  $q_4$ for $N=4$ 
and the total spin $h=1/2$.]{The spectrum of the integrals of 
motion $q_4$ for $N=4$ and the total spin $h=1/2$.
The left and right panels correspond to the eigenstates with different
quasimomenta $\e^{i\theta_4}=\pm 1$ and $\pm i$, respectively.}
\lab{fig:Q4}
\end{figure}

It turns out that after choosing one value of $\theta_4$,
the lattice in $q_4^{1/4}-$space 
is still spuriously degenerated\footnote{degeneration in the 
leading order of the WKB approximation} and this degeneration 
also corresponds to different lattices
in $q_3^{1/2}$. The parameter $\ell_3$ which is defined in (\ref{eq:q3-quan})
will be used to distinguish these different lattices.

\subsection{Descendent states for $N=4$}

One can notice that for $N=4$ and $n_h=0$ we have the descendent states.
They appear in sector
with the quasimomentum $\theta_4=\pi$, which agree with (\ref{eq:quasdes}).
The wave-functions of the descendent states
are built of three-particle eigenstates  with $\theta_3=0$.
These three-Reggeon states correspond to the 
lattice depicted by stars in Fig.\ \ref{fig:N3q0}. 
On the other hand, 
the lattice for the descendent state for $N=4$ is presented in
Fig.\ \ref{fig:WKB-N4} on the left panel.
Thus, one can compare the both lattices and notice that they 
are exactly the same.
Moreover, the energy of these descendent state and
the corresponding three-Reggeon states are also the same (\ref{eq:Edes}).

\medskip
\begin{figure}[h!]
\centerline{
\begin{picture}(200,80)
\put(10,0){\epsfysize8.5cm \epsfbox{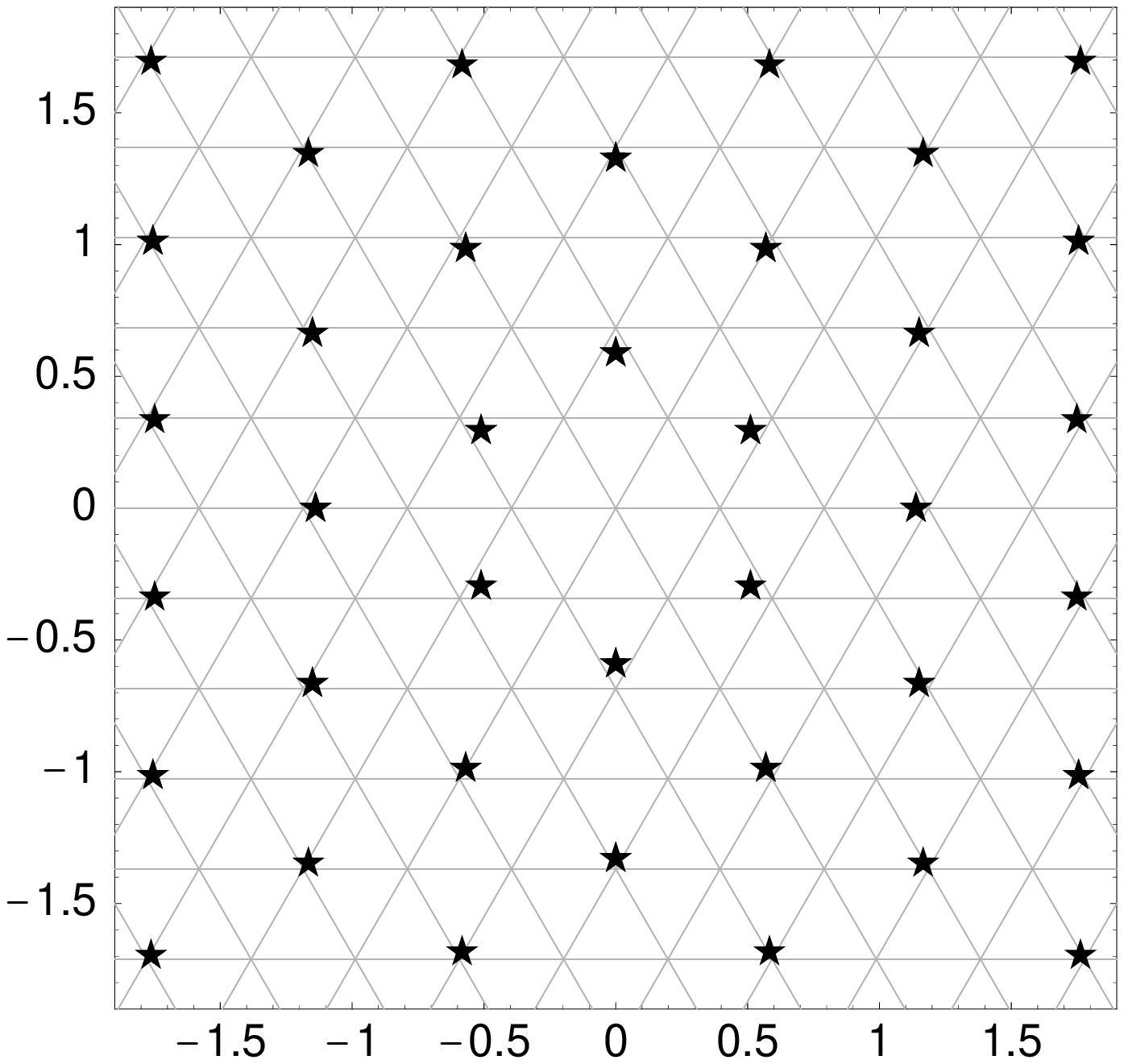}}
\put(105,0){\epsfysize8.5cm \epsfbox{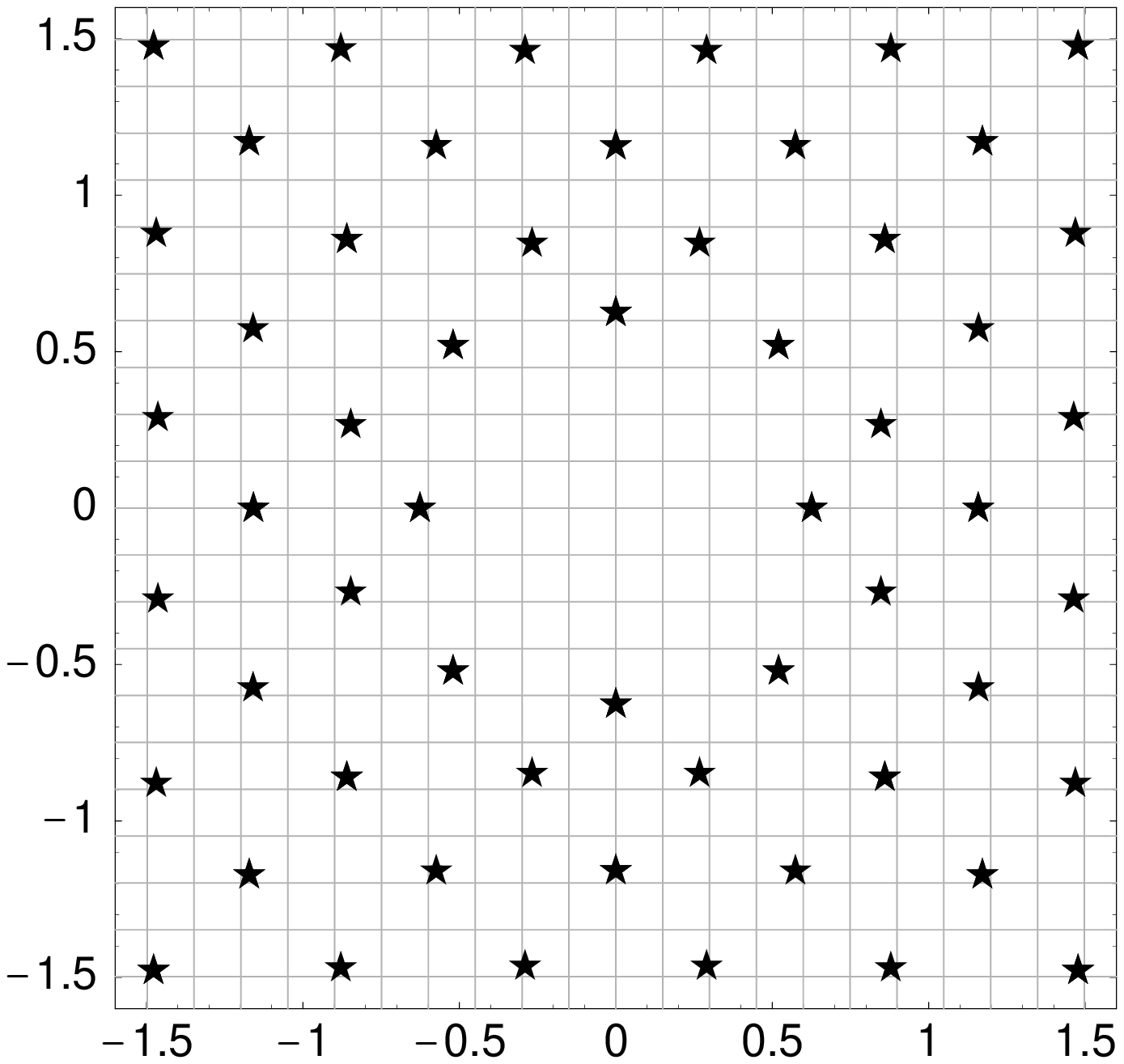}}
\put(48,-2){$\RRe[q_3^{1/3}]$}
\put(9,37){\rotatebox{90}{$\IIm[q_3^{1/3}]$}}
\put(143,-2){$\RRe[q_4^{1/4}]$}
\put(104,37){\rotatebox{90}{$\IIm[q_4^{1/4}]$}}
\end{picture}
}
\caption[The spectra of the conformal charges for $N=4$
]{
The spectra of the conformal charges for $N=4$
and comparison to the WKB expansion. On the left panel, the spectrum
of $q_3$ with $q_4=0$ corresponding
to the descendent states with $\theta_4=\pi$. On the right panel, 
the spectrum of $q_4$ for $h=1/2$ and $q_3=0$ with $\theta_4=0$. 
The WKB lattices are denoted by the 
gray
lines.}
\lab{fig:WKB-N4}
\end{figure}

\subsection{Lattice structure for $q_3=0$}

Let us consider the spectrum with $q_4 \ne 0$ and 
$q_3=0$, 
see the right panel of Fig \ref{fig:WKB-N4}.
In this case the quasimomentum $\theta_4=0$ and 
the lattice structure
include vertices that correspond to the ground state.

Similarly to the $N=3$ case we have in the $q_4^{1/4}-$space a lattice 
with a square-like structure described by (\ref{eq:q4-quan}). 
In this case even numbers $\ell_1$ and $\ell_2$ 
satisfy $\ell_1+\ell_2 \in 4 \mathbb{Z}$.
Thus, we have
the WKB formula
\begin{equation}
\left[q_4^{\rm WKB}(\ell_1,\ell_2)\right]^{1/4}
=\Delta_{N=4}\cdot\left(\frac{\ell_1}{2 \sqrt{2}}
+i\frac{\ell_2}{2 \sqrt{2}}\right),
\lab{eq:WKB-N4}
\end{equation}
where 
the vertices 
are placed outside a disk
around the origin of the radius
\begin{equation}
\Delta_{4}
=\left[\frac{4^{3/4}}{\pi}\int_{-1}^1\frac{dx}{\sqrt{1-x^4}}\right]^{-1}
=\frac{\Gamma^2(3/4)}{2\sqrt{\pi}}=0.423606\ldots \, .
\lab{eq:D4}
\end{equation}
As before, the leading-order WKB formula (\ref{eq:WKB-N4}) is valid only for
$|q_4^{1/4}|\gg |q_2^{1/2}|$. 

\begin{figure}[h!]
\vspace*{5mm}
\centerline{{\epsfysize6cm \epsfbox{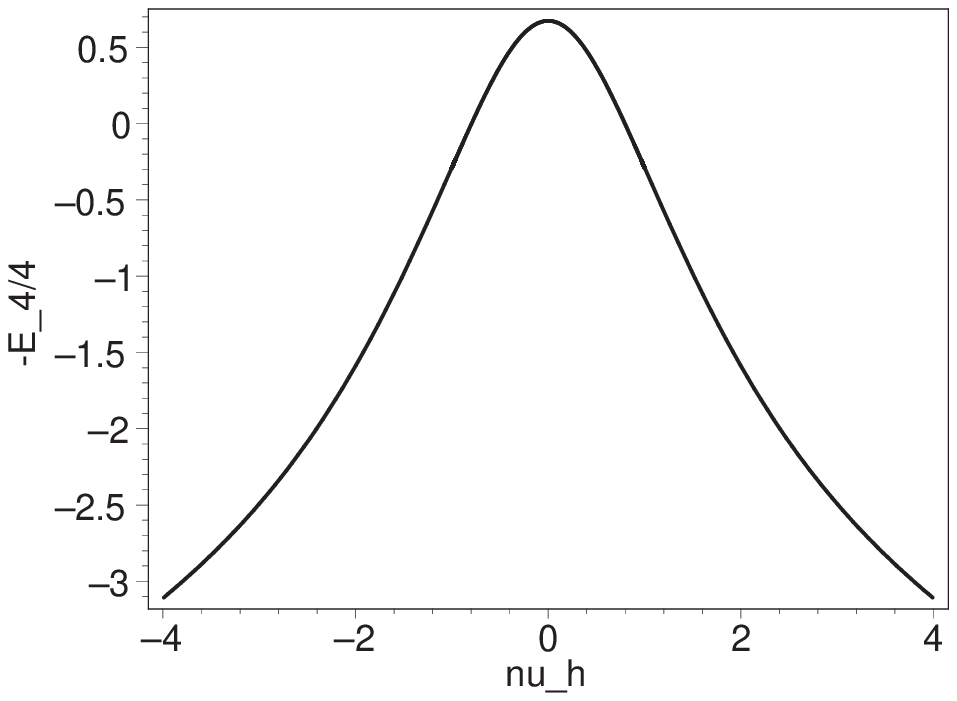}}\qquad{\epsfysize6cm 
\epsfbox{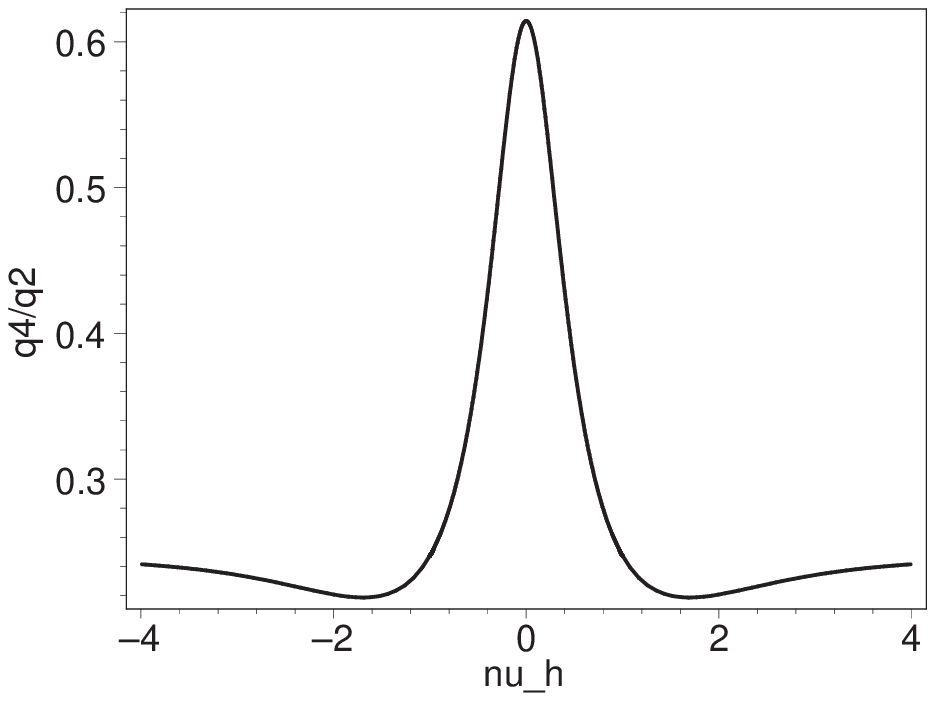}}}
\caption[The dependence of the energy, $-E_4/4$, and 
 $q_4/q_2$
with $q_2=1/4+\nu_h^2$]
{The dependence of the energy, $-E_4/4$, and the quantum number,
 $q_4/q_2$,
with $q_2=1/4+\nu_h^2$, on the total spin $h=1/2+i\nu_h$ along the ground state
trajectory for $N=4$.}
\lab{fig:N4-flow}
\end{figure}

The energy is lower for points  which are nearer to the origin. 
Similarly to the $N=3$ case, the spectrum is also built of trajectories which
extend  in the $(\nu_h,q_3,q_4)-$space.
Namely, each
point on the $q_4^{1/4}-$lattice belongs to one specific 
trajectory parameterized by
the set of integers $\{\ell_1,\ell_2,\ldots\}$. 

The ground state for $N=4$ is situated on a trajectory with
$(\ell_1,\ell_2)=(4,0)$.
We find that for this trajectory 
$q_3=\Im q_4=0$, 
whereas $\Re[q_4]$ and $E_4$ vary with $\nu_h$ as we show in
Figure~\ref{fig:N4-flow}. 
An accumulation of the energy levels 
in the vicinity of $\nu_h=0$ is described by Eq.~(\ref{eq:Enu}) with 
the dispersion parameter
$\sigma_4$ given below in Table~\ref{tab:Summary}.

On the $q_4^{1/4}-$plane the ground state is 
represented by four points with the
coordinates $(\ell_1,\ell_2)=(\pm 4,0)$ and $(0,\pm 4)$.
Due to a residual
symmetry $q_4^{1/4} \leftrightarrow \exp(ik \pi/2) q_4^{1/4}$, 
they describe a single eigenstate with  
\begin{equation}
q_3^{\rm ground}=0\,,\qquad q_4^{\rm ground}=0.153589\ldots\,,
\qquad E_4^{\rm ground}=-2.696640\ldots\,.
\end{equation}
with $h=1/2$.  
It has the quasimomentum $\theta_4=0$ and, in contrast to the
$N=3$ case, it is unique.

\begin{table}[h]
\begin{center}
\begin{tabular}{|c||c|c|c|c|c|c|}
\hline
 $(\ell_1/2,\ell_2/2)$&  $\left(q_4^{\rm \,exact}\right)^{1/4}$ & 
$\left(q_4^{\rm WKB}\right)^{1/4}$ & $-E_4/4$ \\
\hline
\hline
$(2,0)$ & $0.626 $ &  $0.599 $ & $0.6742$ \\ \hline
$(2,2)$ & $0.520+0.520\,i$ & $0.599+0.599\,i$ & $-1.3783$ \\ \hline
$(3,1)$ & $0.847+0.268\,i$ &  $0.899+0.299\,i$  & $-1.7919$ \\ \hline 
$(4,0)$ & $1.158$  &   $1.198$ & $-2.8356$ \\ \hline
$(3,3)$ & $0.860+0.860\,i$ & $0.899+0.899\,i$  & $-3.1410$  \\ \hline
$(4,2)$ & $1.159+0.574\,i$ & $1.198+0.599\,i$ & $-3.3487$ \\ \hline
\end{tabular}
\end{center}
\caption[Comparison of $q_4^{1/4}$ at $q_3=0$ 
and $h=1/2$ with the 
WKB expression]{Comparison of the exact spectrum of $q_4^{1/4}$ at $q_3=0$ 
and $h=1/2$ with the approximate
WKB expression \ref{eq:WKB-N4}. The last column shows the exact energy $E_4$.}
\lab{tab:WKB-4}
\end{table}

The comparison of (\ref{eq:WKB-N4}) with the 
exact
results for $q_4$ at $h=1/2$ is shown in Figure~\ref{fig:WKB-N4} and
Table~\ref{tab:WKB-4}. One can see that 
the WKB formula (\ref{eq:WKB-N4}) 
describes the spectrum with a good accuracy.

\subsection{Resemblant lattices with $\ell_3=0$}

In the previous Section we introduced the parameter $\ell_3$
which helps us to distinguish different lattices. This parameter
takes even values for $\theta_4=0,\pi$ and odd ones for 
$\theta_4=\pm \pi/2$. Let us take $\ell_3=0$. It turns out that 
in this case the spectrum lattice in the $q_3^{1/2}-$space is similar
to the corresponding lattice in 
the $q_4^{1/4}-$space, i.e.
considering only the 
leading order of the WKB approximation
the $q_3^{1/2}-$lattice in comparison
to the $q_4^{1/4}-$lattice 
is rescaled by some real number.
An example of such a lattice for
$\theta_4=0$
is shown
in Figure \ref{fig:l30k01}. 
One can notice that the non-leading corrections to the WKB approximation
cause the bending of the lattice structure in  Fig.\ \ref{fig:l30k01}: 
for $q_4^{1/4}$ concave whereas for $q_3^{1/2}$ convex. 
Moreover, we can see the $q_4^{1/4}-$lattice as well as
$q_3^{1/2}-$one have a similar structure to the lattice with $q_3=0$
presented in Fig.\ \ref{fig:WKB-N4}. 
These lattices also do not have 
the vertices inside the disk at the origin $q_3=q_4=0$.
\medskip
\begin{figure}[h!]
\centerline{
\begin{picture}(200,80)
\put(10,0){\epsfysize8.5cm \epsfbox{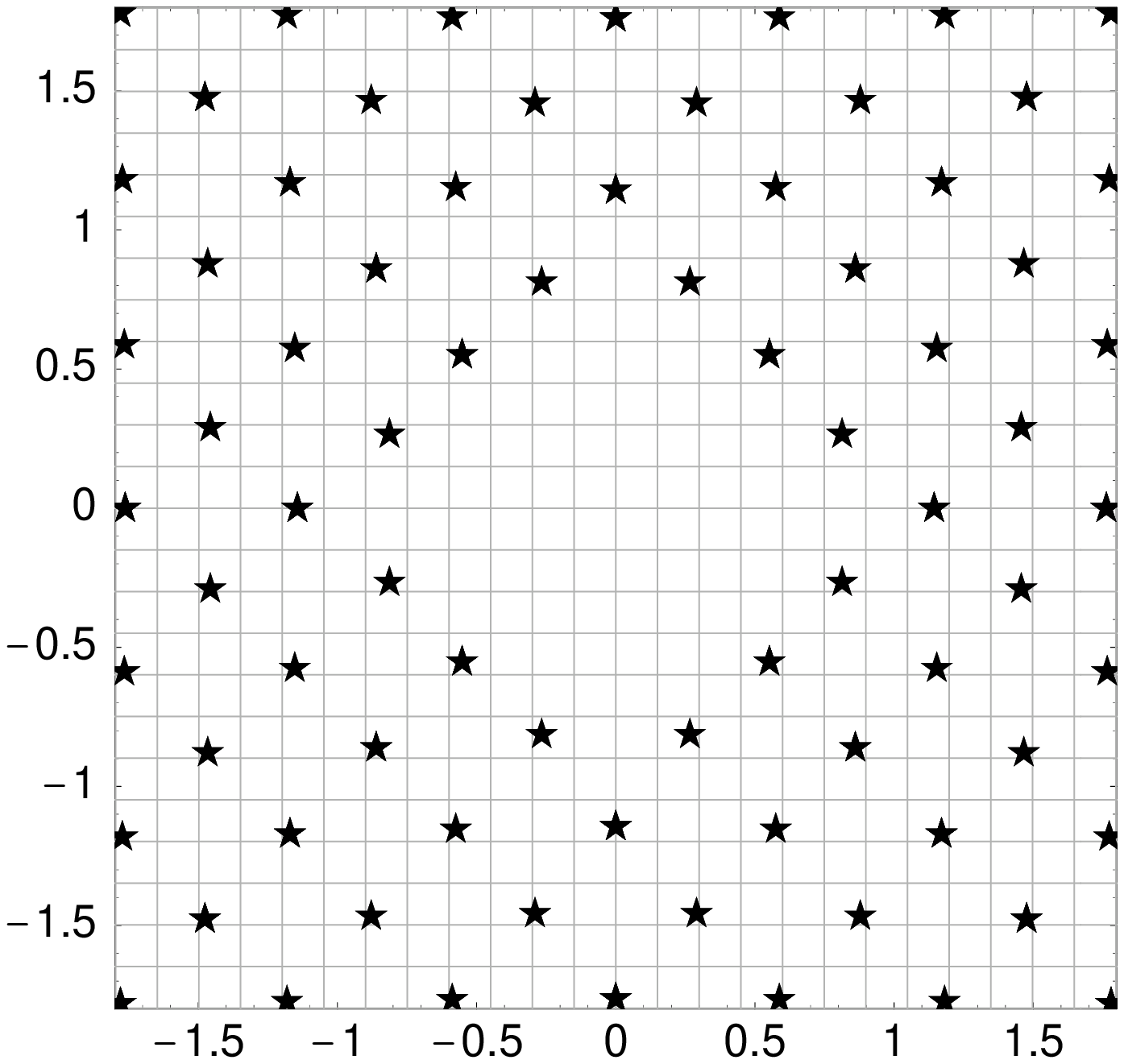}}
\put(105,0){\epsfysize8.5cm \epsfbox{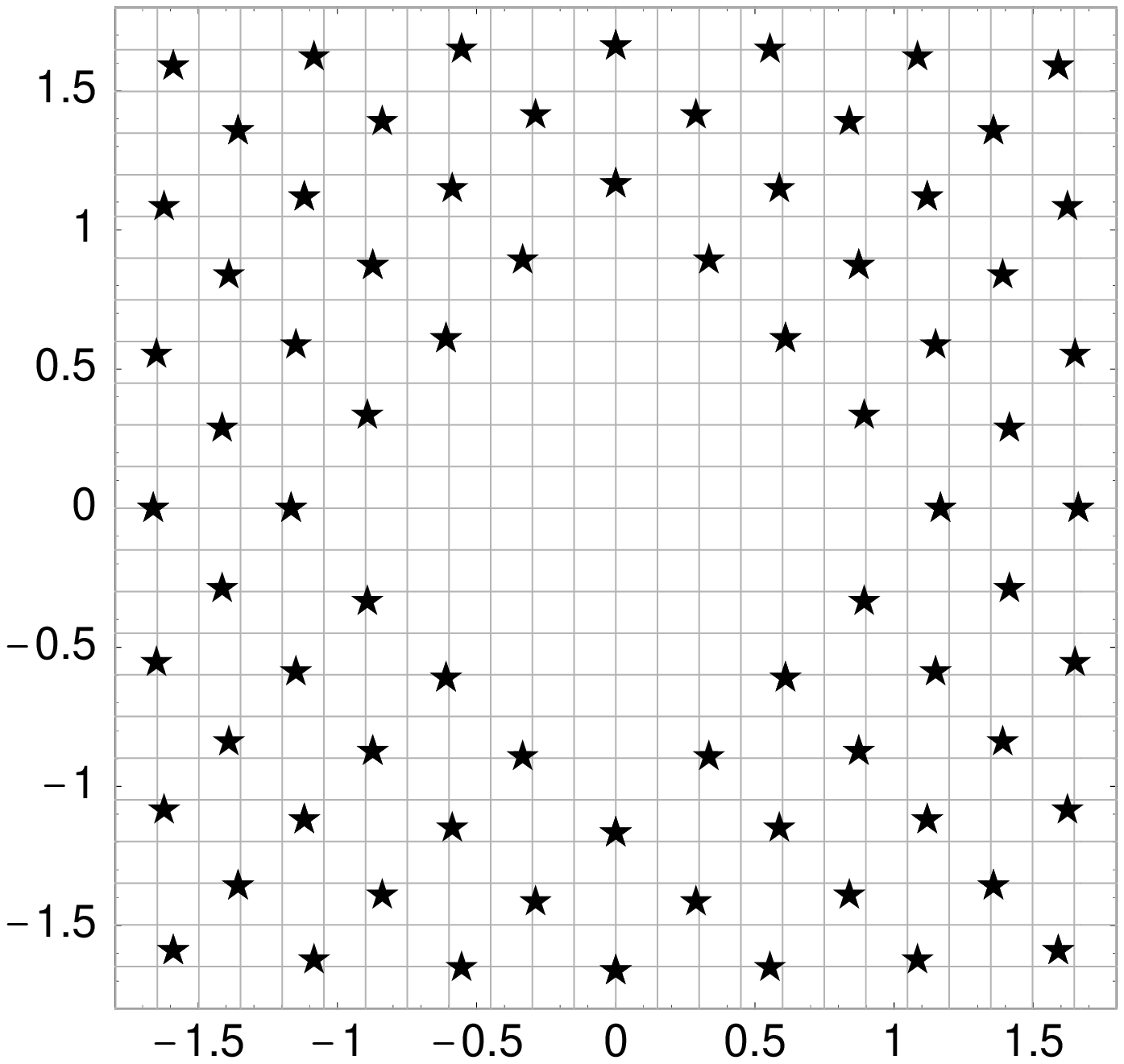}}
\put(48,-2){$\RRe[q_4^{1/4}]$}
\put(9,37){\rotatebox{90}{$\IIm[q_4^{1/4}]$}}
\put(143,-2){$\RRe[q_3^{1/2}]$}
\put(104,37){\rotatebox{90}{$\IIm[q_3^{1/2}]$}}
\end{picture}
}
\caption[The spectra of the conformal charges 
for $N=4$ with $\theta_4=0$,
$\ell_3=0$ and $\ell_4=1$.]{The  
spectra of the conformal charges for $N=4$ with $\theta_4=0$,
$\ell_3=0$ and $\ell_4=1$. On the left panel 
the spectrum of $q_4^{1/4}$, while on the right
panel the spectrum of $q_3^{1/2}$}
\lab{fig:l30k01}
\end{figure}


\medskip
\begin{figure}[h!]
\centerline{
\begin{picture}(200,80)
\put(10,0){\epsfysize8.5cm \epsfbox{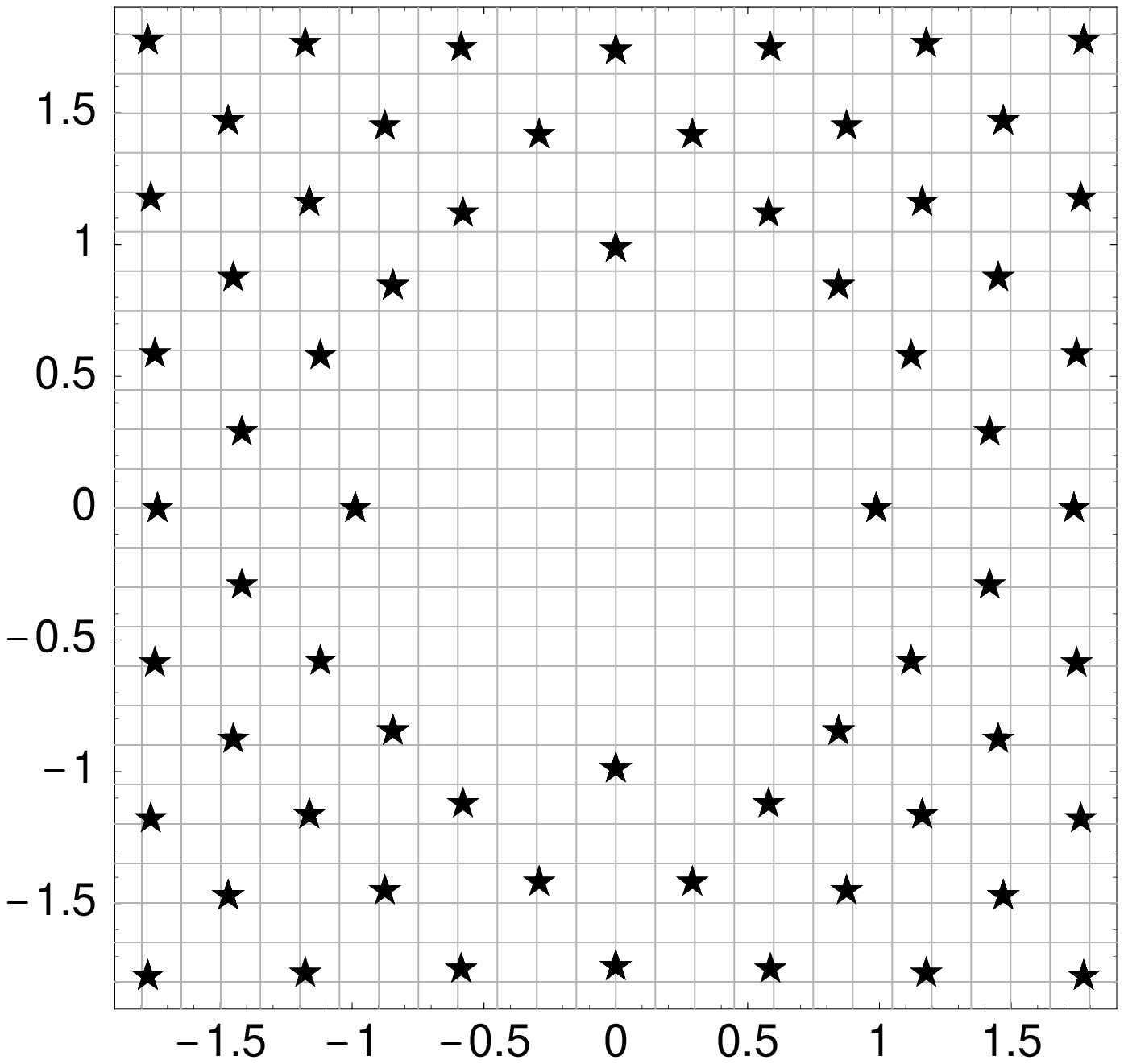}}
\put(107,2){\epsfysize8.cm \epsfbox{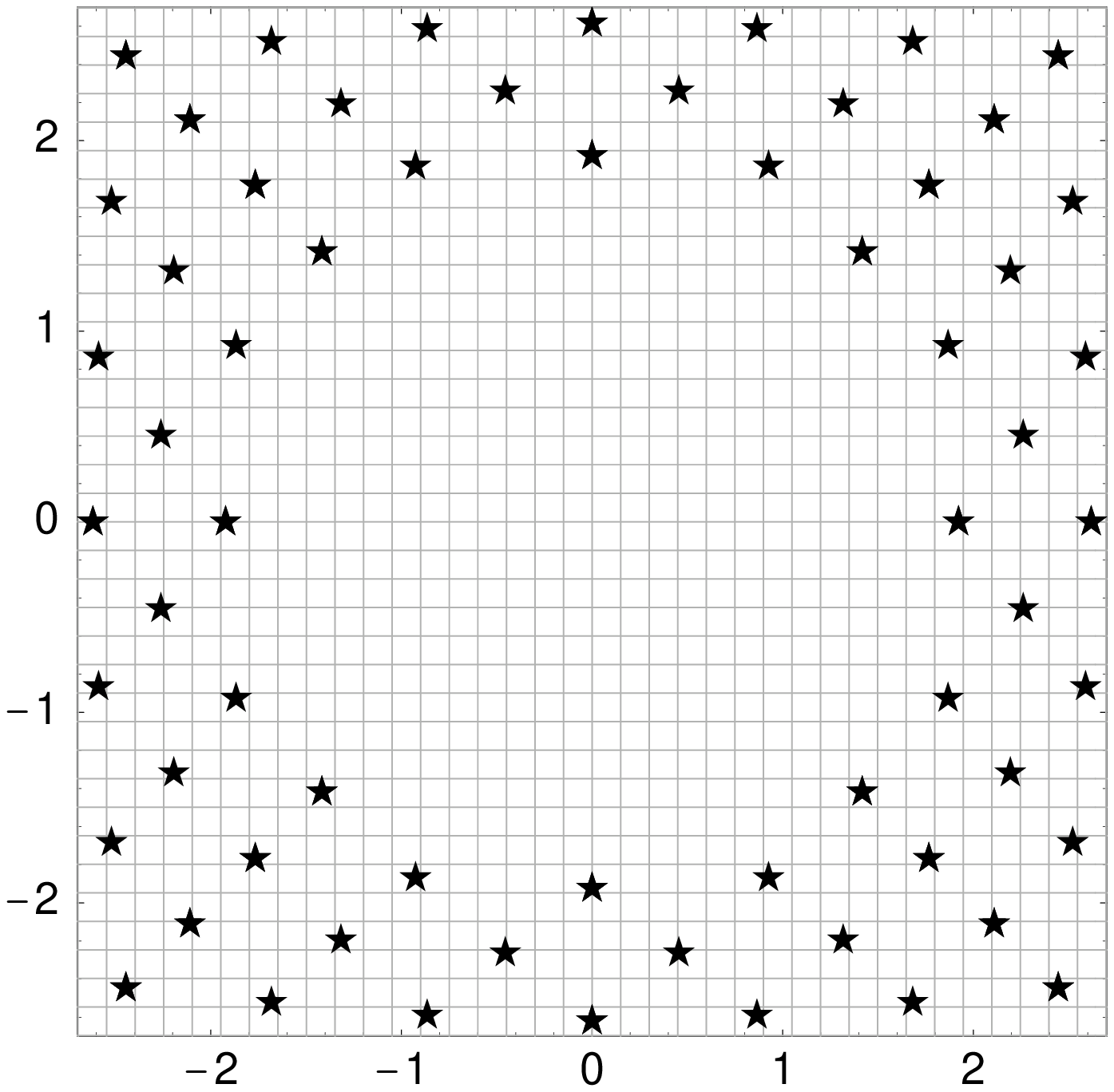}}
\put(48,-2){$\RRe[q_4^{1/4}]$}
\put(9,37){\rotatebox{90}{$\IIm[q_4^{1/4}]$}}
\put(143,-2){$\RRe[q_3^{1/2}]$}
\put(104,37){\rotatebox{90}{$\IIm[q_3^{1/2}]$}}
\end{picture}
}
\caption[The spectra of the conformal charges 
for $N=4$ with $\theta_4=0$,
$\ell_3=0$ and $\ell_4=2$.]{The spectra of 
the conformal charges for $N=4$ with 
$\theta_4=0$, $\ell_3=0$ and $\ell_4=2$. 
On the left panel the spectrum of $q_4^{1/4}$, while on the right
panel the spectrum of $q_3^{1/2}$}
\lab{fig:l30k02}
\end{figure}

Substituting 
\begin{equation}
q_3^{1/2}=r_3 \e^{i\phi_3}
\quad
\mbox{and}
\quad
q_4^{1/4}=r_4 \e^{i\phi_4}
\lab{eq:qrphi}
\end{equation}
into (\ref{eq:q3-quan})
we obtain a condition for the leading order of the WKB approximation
\begin{equation}
\ell_3=\left(\frac{r_3}{r_4}\right)^2 \sin \left(2 (\phi_3 -\phi_4) \right).
\lab{eq:q3-rphi}
\end{equation}
Thus, for $\ell_3=0$ and for a scale $\lambda=r_3/r_4 >0$ we have
$\phi_3=\phi_4$. This means that the vertices on the $q_3^{1/2}-$lattice
have the same angular coordinates as those from the $q_4^{1/4}-$lattice.
Looking at the numerical results in Figure \ref{fig:l30k01}
we notice that the missing quantization condition for $\ell_3=0$ in
the leading WKB order should 
have a form similar to
\begin{equation}
\RRe \frac{q_3}{q_4^{1/2}}={\lambda_{\ell_4}}^2\,,
\lab{eq:conlamb}
\end{equation}
where $\lambda_{\ell_4}=r_3/r_4 \in \mathbb{R}$ is a constant scale for
a given lattice $\ell_4$.

It turns out that
for a specified quasimomentum we have an infinite number of such lattices.
They differ from each other by a given scale $\lambda_{\ell_4}$.
For example for $\theta_4=0$ we have another lattice, shown 
in Fig \ref{fig:l30k02}. Its scale 
$\lambda_2$
differs from
the scale $\lambda_1$ of the lattice from Fig.\ \ref{fig:l30k01}.
The resemblant $q_4^{1/4}-$lattices for $\theta_4=0$ are described
by (\ref{eq:q4-quan}) with the integer parameters $\ell_1$ and $\ell_2$
satisfying $\ell_1+\ell_2 \in 4 \mathbb{Z}$.

The similar lattices also exist in the sector 
with the quasimomentum $\theta_4=\pi$.
Some of them are presented in Figures \ref{fig:l30k21}
and \ref{fig:l30k22}.
For $\theta_4=\pi$
the resemblant  $q_4^{1/4}-$lattices are also described 
by (\ref{eq:q4-quan}) but the integer parameters $\ell_1$ and $\ell_2$
satisfy $\ell_1+\ell_2 \in 4 \mathbb{Z}+2$.
\medskip
\begin{figure}[h!]
\centerline{
\begin{picture}(200,80)
\put(10,0){\epsfysize8.5cm \epsfbox{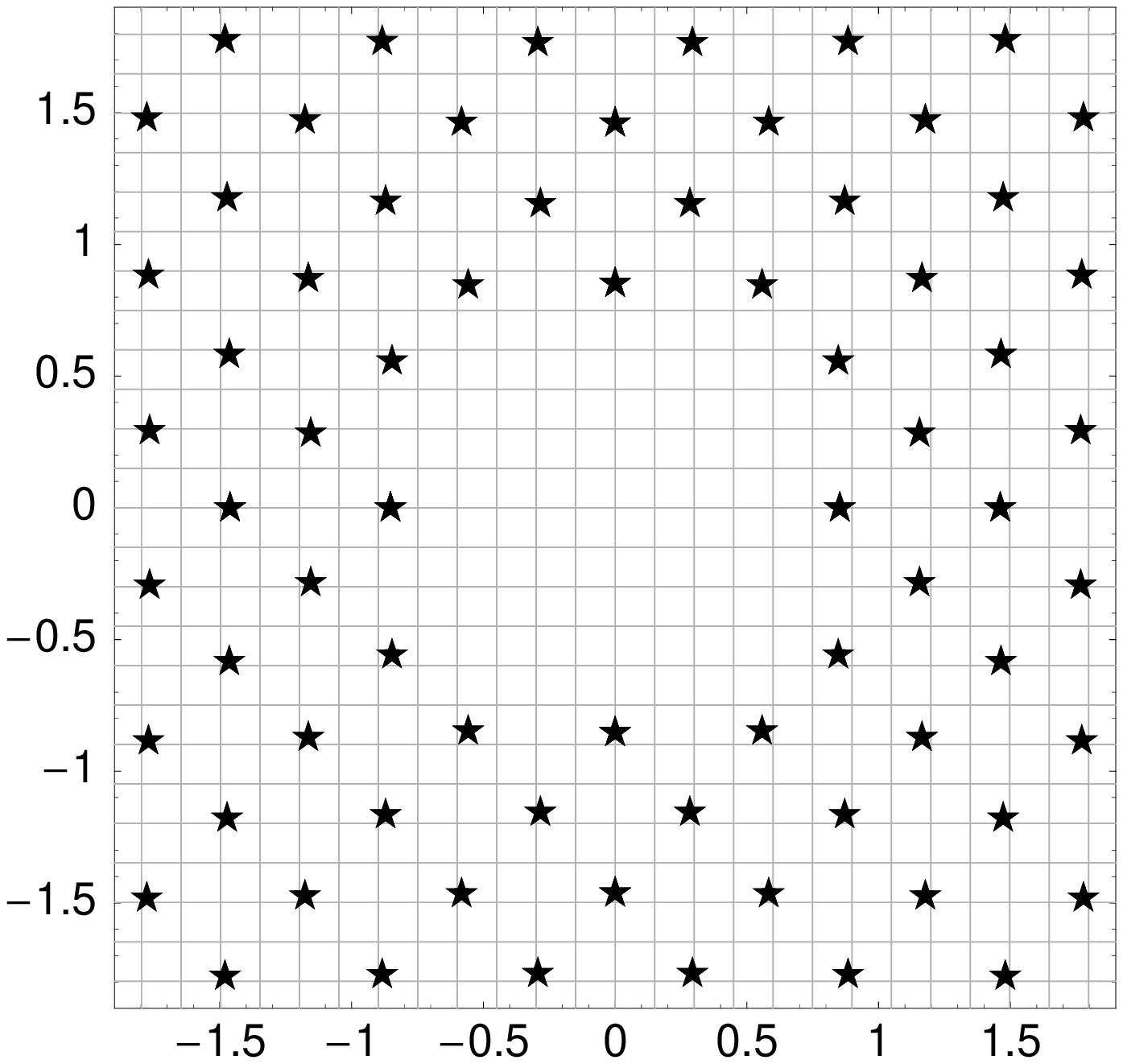}}
\put(105,0){\epsfysize8.5cm \epsfbox{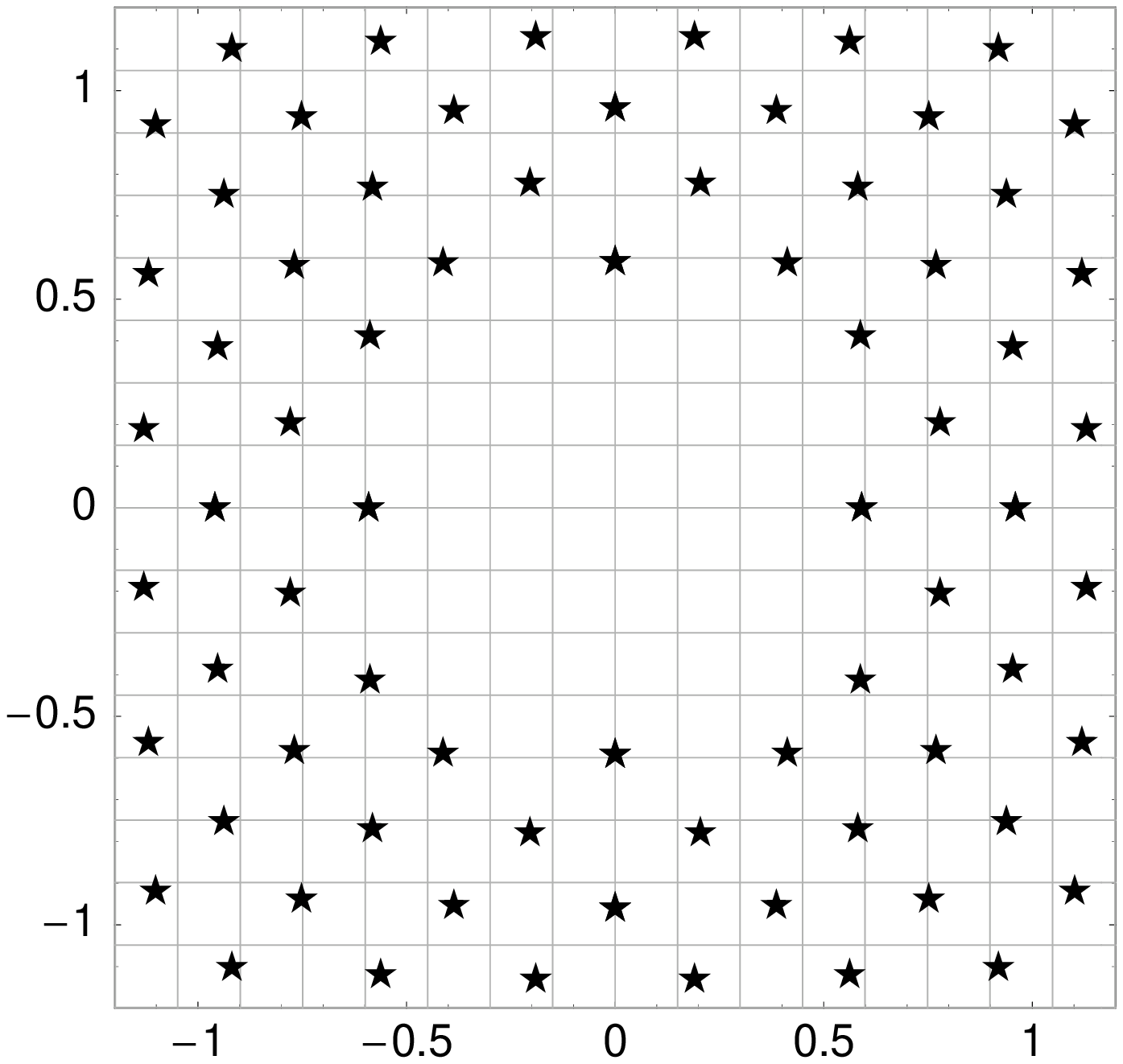}}
\put(48,-2){$\RRe[q_4^{1/4}]$}
\put(9,37){\rotatebox{90}{$\IIm[q_4^{1/4}]$}}
\put(143,-2){$\RRe[q_3^{1/2}]$}
\put(104,37){\rotatebox{90}{$\IIm[q_3^{1/2}]$}}
\end{picture}
}
\caption[The spectra of the conformal charges 
for $N=4$ with $\theta_4=\pi$ 
$\ell_3=0$ and $\ell_4=1$.]{The spectra of 
the conformal charges for $N=4$ with $\theta_4=\pi$,
$\ell_3=0$ and $\ell_4=1$. 
On the left panel the spectrum of $q_4^{1/4}$, while on the right
panel the spectrum of $q_3^{1/2}$}
\lab{fig:l30k21}
\end{figure}

\medskip
\begin{figure}[h!]
\centerline{
\begin{picture}(200,80)
\put(10,0){\epsfysize8.5cm \epsfbox{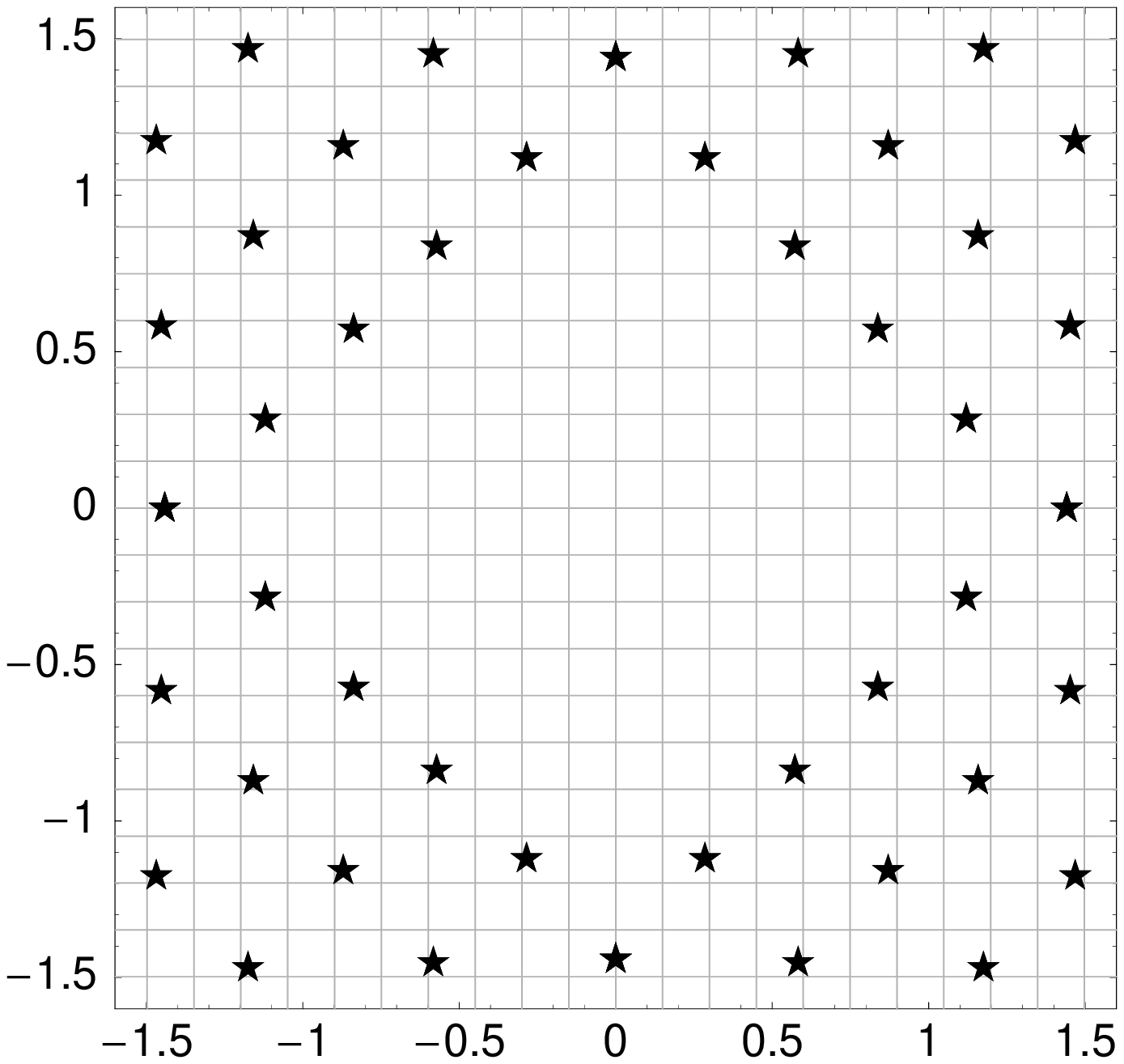}}
\put(105,0){\epsfysize8.5cm \epsfbox{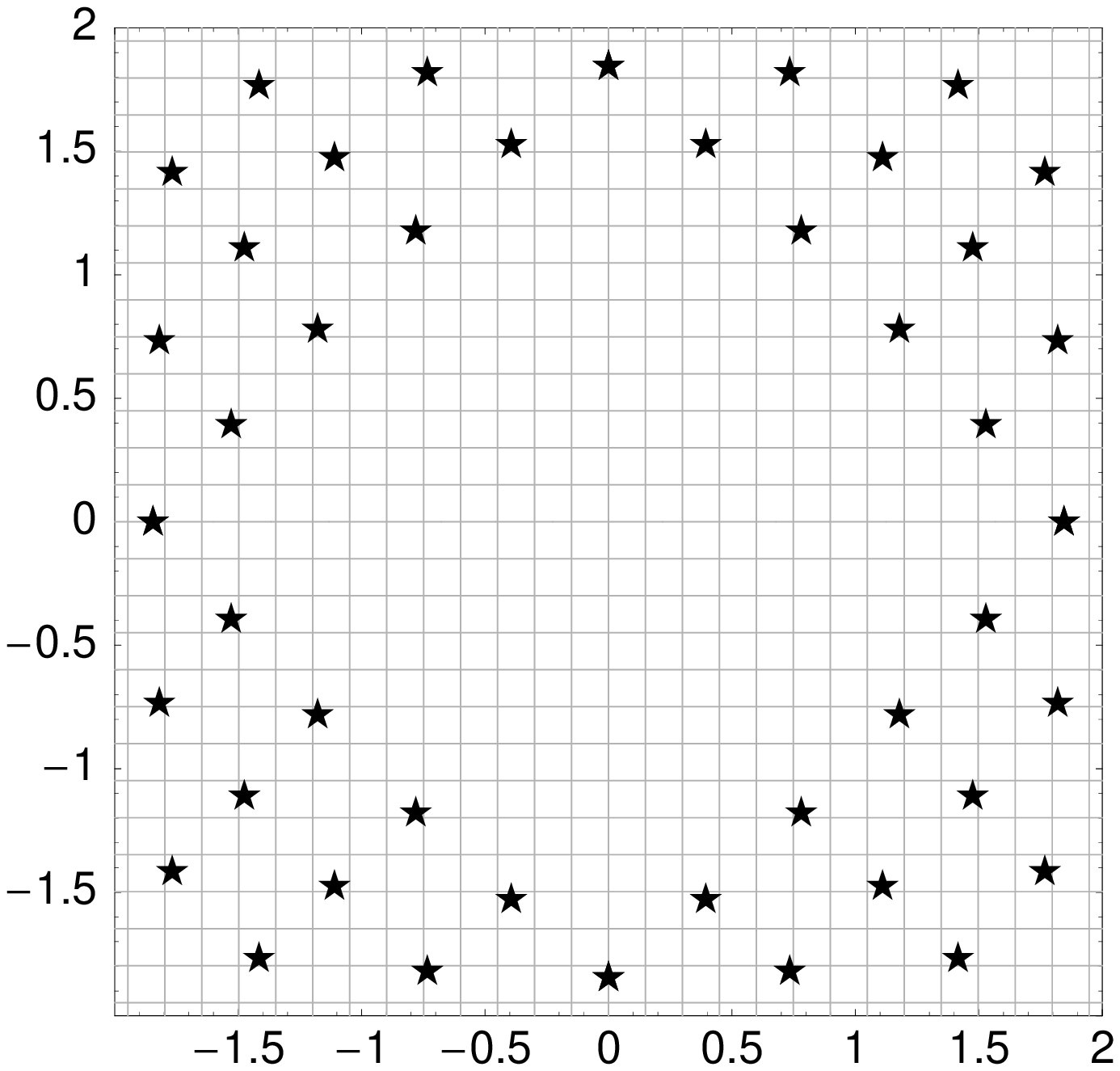}}
\put(48,-2){$\RRe[q_4^{1/4}]$}
\put(9,37){\rotatebox{90}{$\IIm[q_4^{1/4}]$}}
\put(143,-2){$\RRe[q_3^{1/2}]$}
\put(104,37){\rotatebox{90}{$\IIm[q_3^{1/2}]$}}
\end{picture}
}
\caption[The spectra of the conformal charges 
for $N=4$ with $\theta_4=\pi$, 
$\ell_3=0$ and $\ell_4=2$.]{The spectra of 
the conformal charges for $N=4$ with $\theta_4=\pi$,
$\ell_3=0$ and $\ell_4=2$.  
On the left panel the spectrum of $q_4^{1/4}$, while on the right
panel the spectrum of $q_3^{1/2}$}
\lab{fig:l30k22}
\end{figure}

\subsection{Winding lattices with $\ell_3 \ne 0$}

In the case with $\ell_3\ne0$ we have much more complicated 
situation than for $\ell_3=0$.
According to (\ref{eq:q3-rphi}) the angles $\phi_3$ and $\phi_4$
defined in (\ref{eq:qrphi}) are no more equal. Moreover,
they start to depend on the scale $\lambda=r_3/r_4$.

\medskip
\begin{figure}[h!]
\centerline{
\begin{picture}(200,80)
\put(12,2){\epsfysize8.2cm \epsfbox{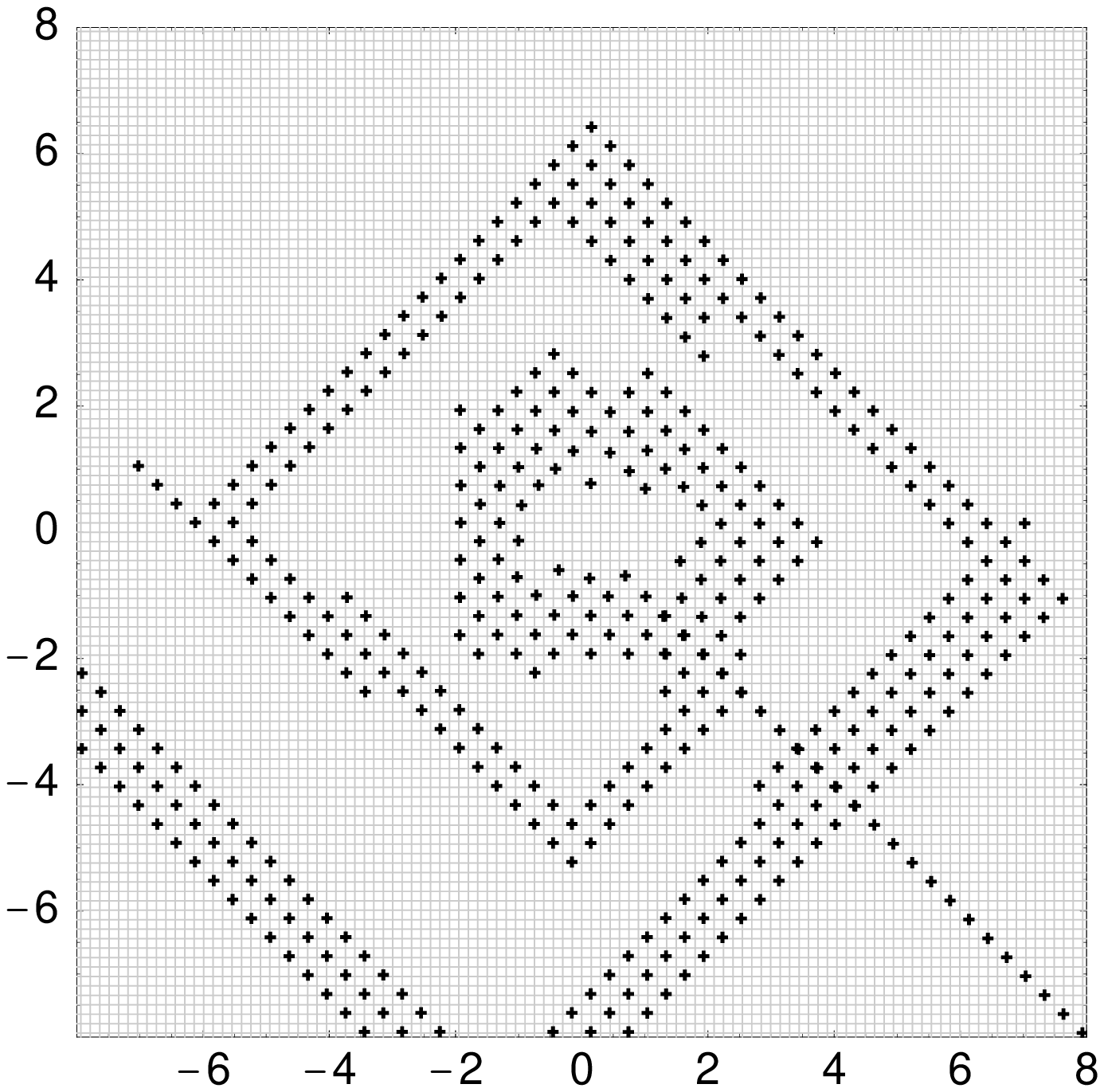}}
\put(105,0){\epsfysize8.5cm \epsfbox{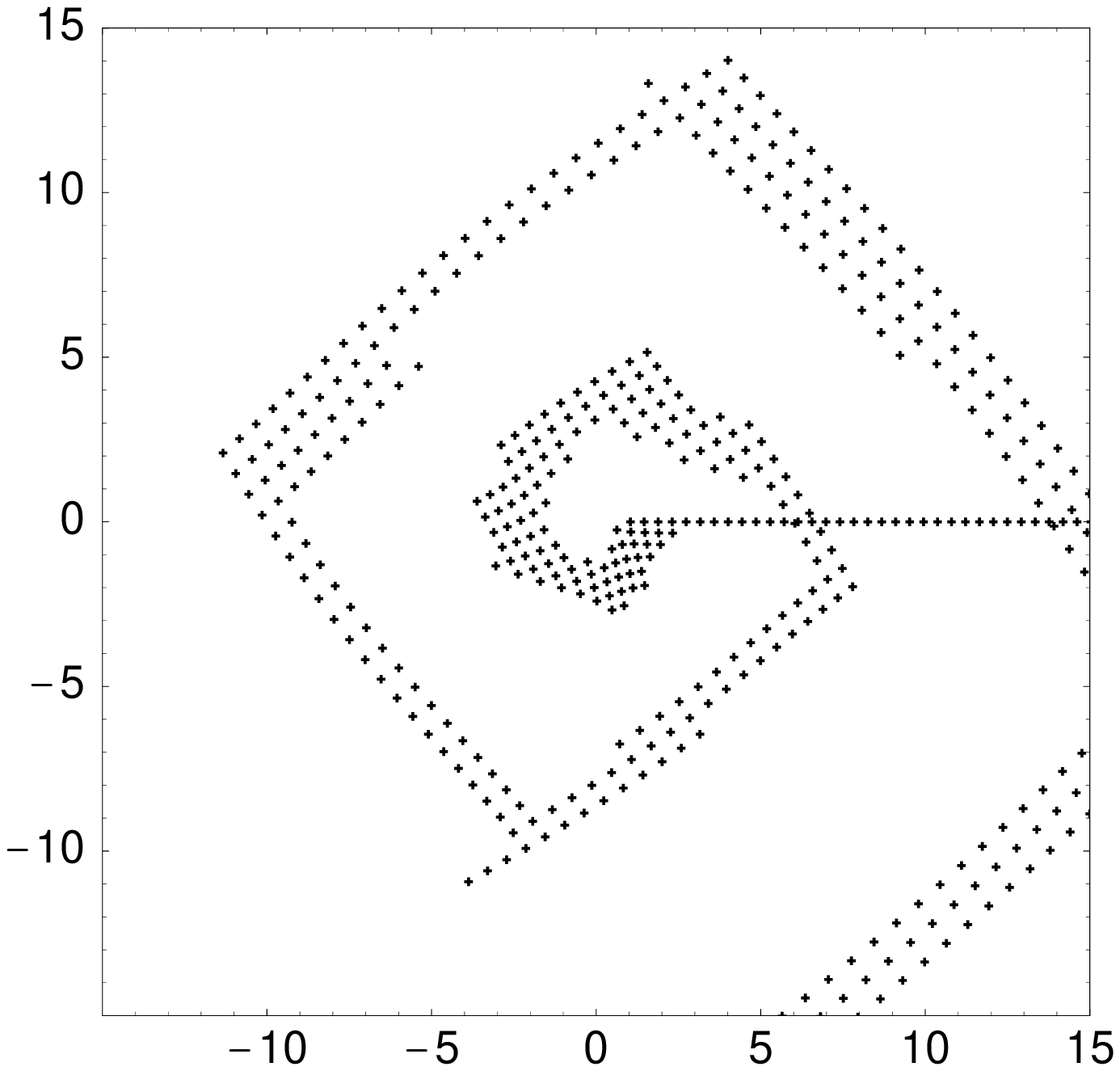}}
\put(48,-2){$\RRe[q_4^{1/4}]$}
\put(9,37){\rotatebox{90}{$\IIm[q_4^{1/4}]$}}
\put(143,-2){$\RRe[q_3^{1/2}]$}
\put(104,37){\rotatebox{90}{$\IIm[q_3^{1/2}]$}}
\end{picture}
}
\caption[The winding spectrum of $\{q_3,q_4\}$
for $N=4$ with $h=1/2$, $\theta_4=- \pi/2$ and 
$\ell_3=1$.]{The winding 
spectrum of the conformal charges for $N=4$ and $h=1/2$ 
with $\theta_4=-\pi/2$ and 
$\ell_3=1$. On the left panel the spectrum of $q_4^{1/4}$, while on the right
panel the spectrum of $q_3^{1/2}$}
\lab{fig:l31k31}
\end{figure}

An example of such a lattice, with $\ell_3=1$ and $\theta_4=-\pi/2$
we show in Figure \ref{fig:l31k31}.
For this case,
the $q_4^{1/4}-$lattice is defined by (\ref{eq:q4-quan})
with $\ell_1$ and $\ell_2$ odd integer numbers satisfying
$\ell_1+\ell_2 \in 4 \mathbb{Z}$.
In Figure \ref{fig:l31k31}, in order to present the correspondence 
between the $q_3^{1/2}-$ and $q_4^{1/4}-$lattice, we depict only some vertices
of the lattice which extends in the whole plane of the conformal charges
except the place nearby the origin $q_4=q_3=0$.
As we can see the $q_3^{1/2}-$lattice is still a square-like one
but it winds around the origin $q_3^{1/2}=0$.
Looking at Figure \ref{fig:l31k31}, let us start from
$\phi_3=0$ and $\phi_4=-\pi/4$ where $\phi_3$ and $\phi_4$ are defined
in (\ref{eq:qrphi}). 
Thus, the difference $\phi_3-\phi_4=\pi/4$
so that our scale at the beginning is $\lambda=\sqrt{2}$. 
In this region the vertices of the $q_3^{1/2}-$lattice 
are in the nearest place to the origin.
When we go clockwise 
around the origin of the lattices by decreasing $\phi_4$,
we notice that $\phi_3$ also decreases but much slower.
Thus, according to (\ref{eq:q3-rphi}) 
the difference $\phi_3-\phi_4$ changes. Moreover,
due to (\ref{eq:q3-rphi}), the scale $\lambda$ also continuously rises.
Therefore, the $q_3^{1/2}-$lattice winds in a different way 
than the $q_4^{1/4}-$lattice. After one revolution the vertices
of the $q_4^{1/4}-$lattices are at similar places as those 
with $\phi_4$ decreased by $2 \pi$. This provides
additional spurious degeneration in $q_4^{1/4}$. However, 
after revolution by the angle $2 \pi$, the vertices in $q_3^{1/2}-$space 
have completely different conformal charges $q_3$. The spurious degeneration
in the $q_3^{1/2}-$lattice does not appear.

We have to add that we obtain the second symmetric structure 
when we go in the opposite direction, i.e. anti-clockwise.
Moreover,
the winding lattice, like all other lattices,
extends to infinity on the $q_3^{1/2}-$
and $q_4^{1/4}-$plane and  does not have vertices 
in vicinity of the origin, $q_3=q_4=0$. However, 
the radii of these empty spaces grow with 
$\phi_{3,4}$.

Additionally, due to symmetry of the spectrum (\ref{eq:qkcsym})
we have a twin lattice with $q_k\rightarrow q_k^{\ast}$. 
Furthermore,
the second symmetry (\ref{eq:qkmsym}) produces another lattice
rotated in the $q_3^{1/2}-$space by an angle $\pi/2$. 
Notice that this symmetry (\ref{eq:qkmsym}) exchanges quasimomentum
${\theta_4 \leftrightarrow 2\pi - \theta_4 ({\rm mod}~ 2\pi)}$. 
Thus, the spectrum for $\theta_4=\pi/2$
is congruent with the spectrum of $\theta_4=-\pi/2$ but it is rotated
in the $q_3^{1/2}-$space by $\pi/2$.

\medskip
\begin{figure}[h!]
\centerline{
\begin{picture}(200,80)
\put(12,2){\epsfysize8.2cm \epsfbox{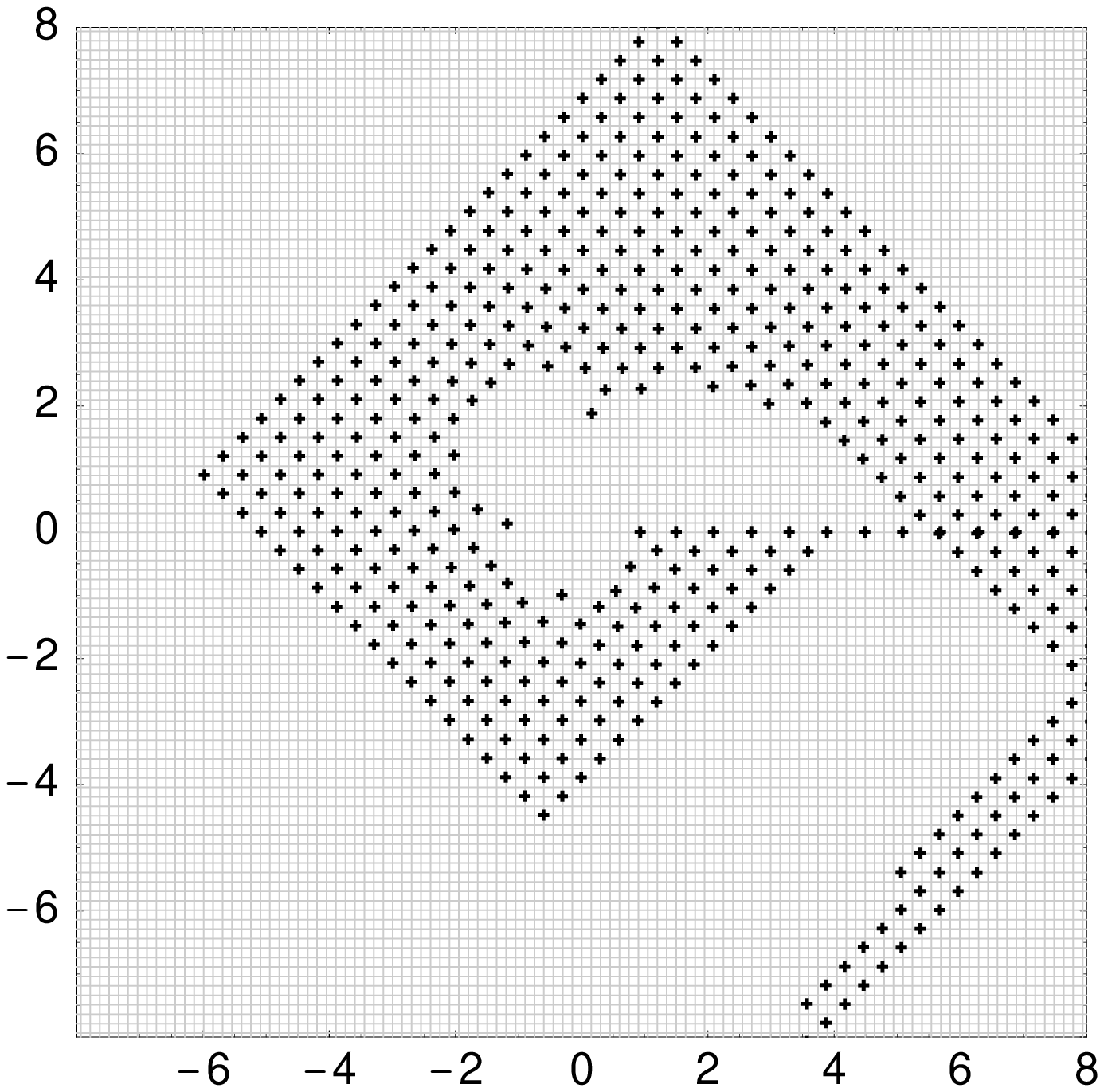}}
\put(105,0){\epsfysize8.5cm \epsfbox{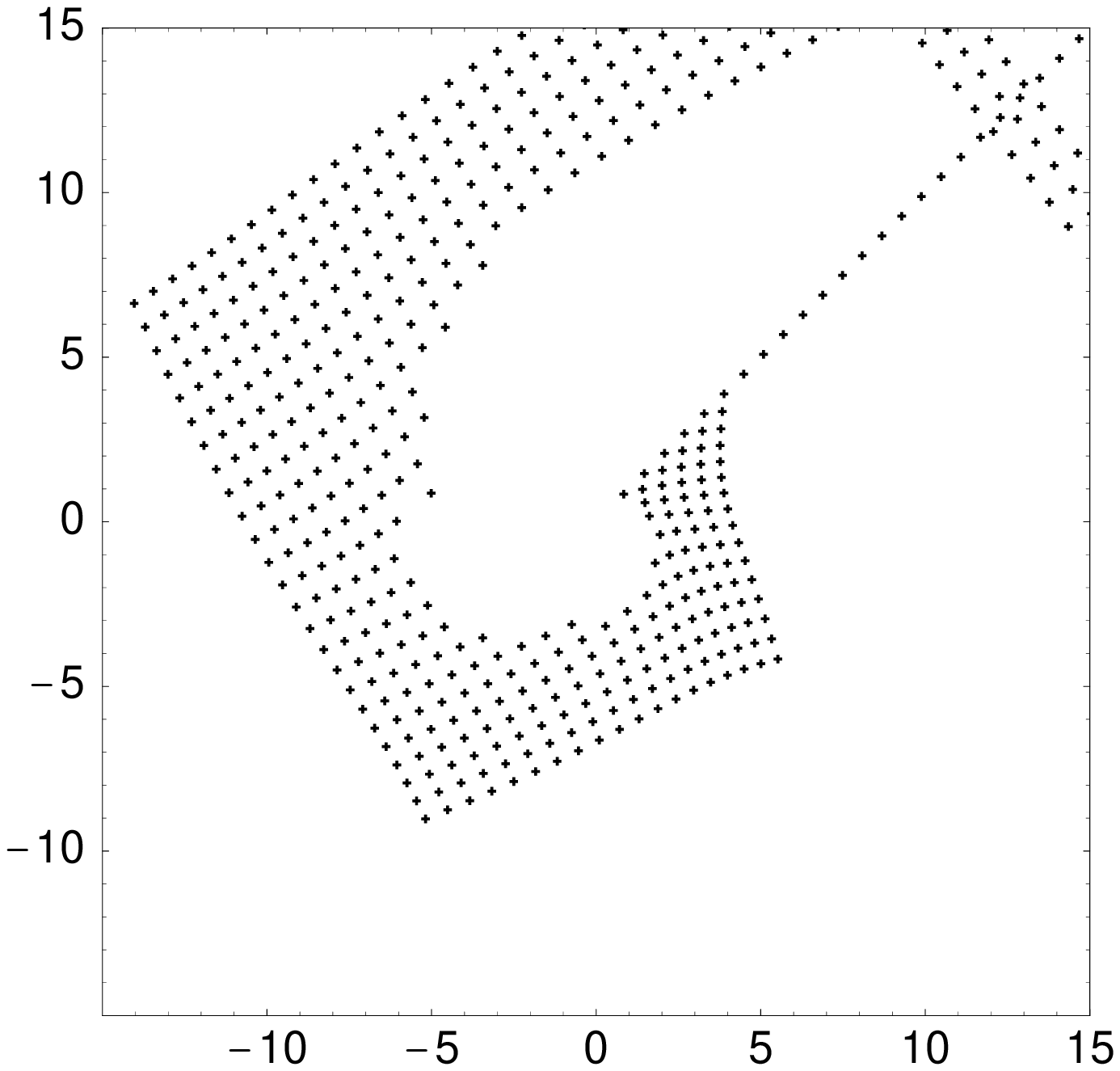}}
\put(48,-2){$\RRe[q_4^{1/4}]$}
\put(9,37){\rotatebox{90}{$\IIm[q_4^{1/4}]$}}
\put(143,-2){$\RRe[q_3^{1/2}]$}
\put(104,37){\rotatebox{90}{$\IIm[q_3^{1/2}]$}}
\end{picture}
}
\caption[The winding spectrum of $\{q_3,q_4\}$
for $N=4$ and $h=1/2$ with $\theta_4=0$ and 
$\ell_3=2$.]{The winding 
spectrum of the conformal charges for $N=4$ with $h=1/2$, $\theta_4=0$ and 
$\ell_3=2$. On the left panel the spectrum of $q_4^{1/4}$ while on the right
panel the spectrum of $q_3^{1/2}$}
\lab{fig:l32k0}
\end{figure}

As we said before 
for $\theta_4 \pm \pi/2$ we have winding spectra with odd $\ell_3$.
Similarly,
for even $\ell_3 \ne 0$ we have winding spectra with $\theta_4=0,\pi$,
thus, the lattice with the lowest non-zero $\ell_3$ corresponds to
$|\ell_3|=2$.
We present some points of this spectrum in Fig.\ \ref{fig:l32k0},
for which $\ell_1$, $\ell_2$ in 
(\ref{eq:q4-quan})
are even and 
$\ell_1+\ell_2 \in 4 \mathbb{Z}+2$.
In this case we start with $\phi_3=\pi/4$ and $\phi_4=0$ which implies
the beginning scale $\lambda=\sqrt{2}$. 
The spectra wind as in the previous case.

Similarly,
for $\theta_4=\pi$ we have also spectrum with the lowest  $|\ell_3|=2$.
It is defined by (\ref{eq:q4-quan}) with $\ell_1$, $\ell_2$ even
and $\ell_1+\ell_2 \in 4 \mathbb{Z}$.
In this case the angles $\phi_3=0$ and $\phi_4=-\pi/4$ so we also have
the beginning scale $\lambda=\sqrt{2}$, as depicted in Figure \ref{fig:l32k2}.
In order to describe the winding spectra better we may introduce
an integer parameter $\ell_4$ which helps us to number the 
overlapping winding planes
of the spectra and name spuriously-degenerated vertices
in the $q_4^{1/4}-$plane.

\begin{figure}[h!]
\centerline{
\begin{picture}(200,80)
\put(12,2){\epsfysize8.2cm \epsfbox{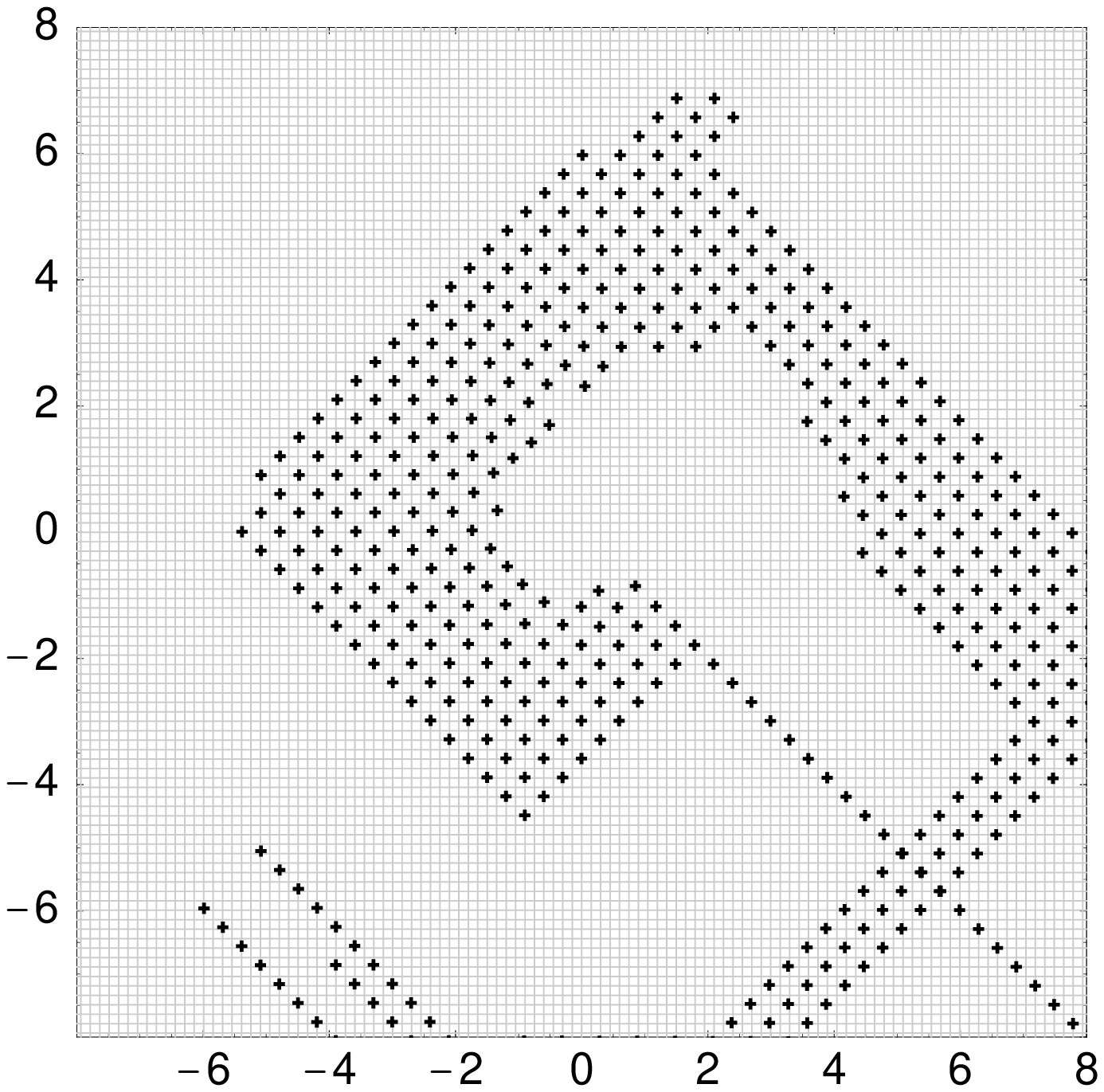}}
\put(105,0){\epsfysize8.5cm \epsfbox{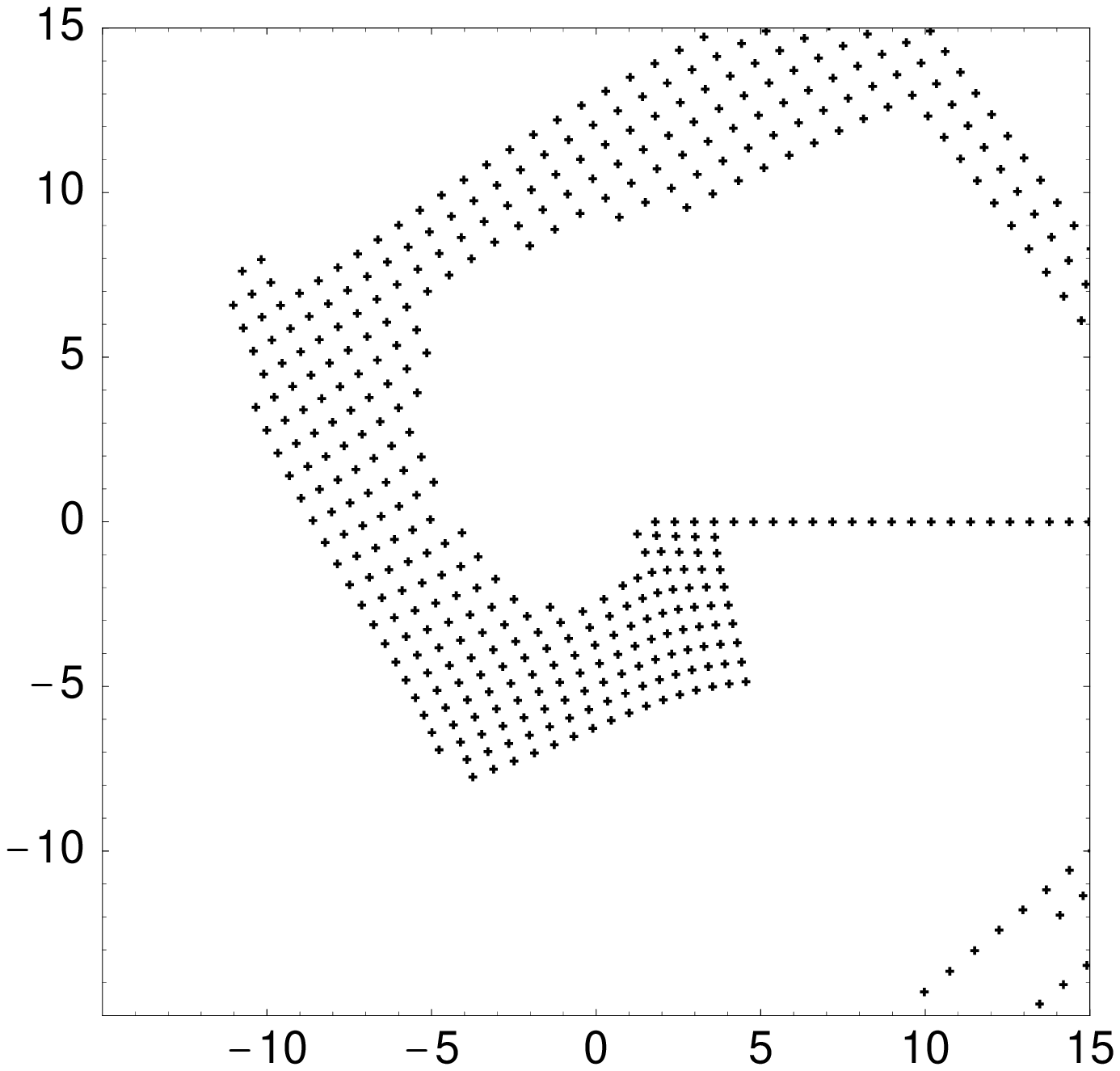}}
\put(48,-2){$\RRe[q_4^{1/4}]$}
\put(9,37){\rotatebox{90}{$\IIm[q_4^{1/4}]$}}
\put(143,-2){$\RRe[q_3^{1/2}]$}
\put(104,37){\rotatebox{90}{$\IIm[q_3^{1/2}]$}}
\end{picture}
}
\caption[The winding spectrum of $\{q_3,q_4\}$
for $N=4$ and $h=1/2$ with $\theta_4=\pi$ and 
$\ell_3=2$.]{The winding 
spectrum of the conformal charges for $N=4$
with $h=1/2$, $\theta_4=\pi$ and 
$\ell_3=2$. On the left panel the spectrum of $q_4^{1/4}$ while on the right
panel the spectrum of $q_3^{1/2}$}
\lab{fig:l32k2}
\end{figure}

To sum up, even for a given quasimomentum we have many lattices
which overlap, so that vertices of the lattices,
especially in $q_3^{1/2}-$space, make an impression
of being randomly distributed.
However, as we have shown
above, those spectra may be distinguished and finally described
by (\ref{eq:q4-quan}) and (\ref{eq:q3-quan}).
Still, there is a lack of one nontrivial WKB condition which
would uniquely explain the structure
of the resemblant and winding lattices.

\subsection{Corrections to WKB}

Let us consider the spectrum of 
the conformal charge $q_4$ for $N=4$ with $q_3=0$ and ${h=\frac{1+n_h}{2}}$.
It turns out that for $n_h \ne 0$ it has similar square-like
lattice structure like that with $n_h=0$,
see Fig.\ \ref{fig:WKB-N4} on the right panel.
Similarly to the case with three reggeized gluons
we have evaluated the conformal charges $q_4$
with even $n_h$ with high precision.
We have done it separately for $q_4$ with $\IIm[q_4]=0$
and $\RRe[q_4]=0$.
Next, we have fitted
coefficient expansion of the WKB series, $a^{(r)}_k$ and $a^{(i)}_k$, 
respectively.

In \ci{Derkachov:2002pb} the series formula for $q_4^{1/4}$
looks as follows
\begin{equation}
q_{4}^{1/4}=\frac{\pi^{3/2}}{2 \Gamma ^{2}(1/4)}
{\cal Q}(\mybf n)\left[1+\frac{b}{\left|{\cal Q}(\mybf n)\right|^{2}}+
\sum _{k=2}^{\infty }{a_{k}
\left(\frac{b}{\left|{\cal Q}(\mybf n)\right|^{2}}\right)^{k}}\right]\,,
\lab{eq:korq4}
\end{equation}
where 
\begin{equation}
{\cal Q}(\mybf n)
=\sum _{k=1}^{4}{n_{k}e^{i\pi (2k-1)/4}}
=\left( \frac{\ell_1}{\sqrt{2}} + i \frac{\ell_2}{\sqrt{2}}\right)
\lab{eq:Qn4}
\end{equation}
and $\ell_1$,$\ell_2$, ${\mybf n}=\{n_1,\ldots,n_N\}$ are integer.

\begin{table}[h]
\begin{center}
$\begin{array}{|c|c||r|r|r|r|r|} 
    \hline
n_h & \mbox{coef.} & k=2 \qquad & k=3 \qquad & k=4 \quad & k=5\quad
 \\ \hline \hline
0
&a^{(r)}_k           &     2.9910566246  & -24.021689 &   91.591 &  645.5 \\
&a^{(i)}_k           &    -4.9910566246  &  28.021689 & -148.830 & 1656.7 \\
&a^{(r)}_k+a^{(i)}_k &    -2.0000000000  &   4.000000 &  -57.239 & 2302.2 \\
    \hline
2
&a^{(r)}_k           &     -1.3991056625 &   4.674008 &   -4.516 &   -95.7 \\
&a^{(i)}_k           &     -0.6008943375 &  -0.674008 &   20.388 &  -200.8 \\
&a^{(r)}_k+a^{(i)}_k &     -2.0000000000 &   4.000000 &   15.872 &  -296.5 \\
    \hline
4
&a^{(r)}_k           &      -1.351212983 &   3.462323 &  -11.322 &  43.164 \\ 
&a^{(i)}_k           &      -0.648787017 &   0.537677 &   -0.096 &   0.611 \\
&a^{(r)}_k+a^{(i)}_k &      -2.000000000 &   4.000000 &  -11.418 &  43.775 \\  
    \hline
6
&a^{(r)}_k           &     -0.8882504145 & 1.883461 &  -5.4248 & 18.081 \\
&a^{(i)}_k           &     -1.1117495855 & 2.116539 &  -4.8117 & 12.091 \\
&a^{(r)}_k+a^{(i)}_k &     -2.0000000000 & 4.000000 & -10.2365 & 30.172 \\
    \hline
8
&a^{(r)}_k           &     -0.6326570719 &  1.197694 & -2.94579 &   8.3372 \\
&a^{(i)}_k           &     -1.3673429281 &  2.802305 & -5.96129 &  12.1238 \\
&a^{(r)}_k+a^{(i)}_k &     -2.0000000000 &  4.000000 & -8.90708 &  20.4610 \\
    \hline
\end{array}$
\end{center}
\caption[The fitted coefficient to the series formula of $q_4^{1/4}$]
{The fitted coefficient to the series formula of $q_4^{1/4}$ 
(\ref{eq:korq4}) with  $n_h=0,2,4,6$ and $8$}
\lab{tab:q4coefs}
\end{table}

Here for $N=4$, 
similarly to the $N=3$ case (\ref{eq:bcoef}), 
we have a different expansion parameter 
$b/\left|{\cal Q}(\mybf n)\right|^{2}$
and
in this case the parameter
\begin{equation}
b= \frac{4}{\pi}\left({q_{2}}^{\ast}-\frac{5}{4}\right)
\lab{eq:coef4}
\end{equation}
is decreased by $5/4$.
The expansion coefficients of the series (\ref{eq:korq4}) 
are shown in Table \ref{tab:q4coefs}.
Similarly to \ci{Derkachov:2002pb} 
the coefficients $a_0=1$ and $a_1=1$. 
The remaining coefficients 
depend on $n_h$. 
Moreover, the coefficients $a_k$ with $k>1$ are different for real
$q_4^{1/4}$ and for imaginary $q_4^{1/4}$
but one may notice that 
for $k=1,2$ the sum $a_k^{(r)}+a_k^{(i)} \in \mathbb{Z}$
and does not depend on $n_h$.
Thus, to describe 
the quantized values of $q_4^{1/4}$ 
more generally one has
to use  both sets of the coefficients, $a_k^{(r)}$ and $a_k^{(i)}$,
or perform the expansion with two small independent parameters, i.e.
$q^{\ast}_2/|{\cal Q}({\mybf n})|^2$ and $1/|{\cal Q}({\mybf n})|^2$.

Using the series (\ref{eq:korq4}) with (\ref{eq:coef4}) and coefficients from
Table \ref{tab:q4coefs} gives good approximation of the conformal charges 
$q_4$ with $q_3=0$. However, if someone wants to have a better precision
one has to introduce an additional expansion parameter.

\section{Quantum numbers of the states with higher $N$}

In the previous Sections we presented the spectra for $N=3$ 
and $N=4$ particles. One can notice that the latter spectrum is 
much more complicated than the one for $N=3$.
This complexity grows for larger $N$.
Thus, in this Section 
we discuss only the ground states for
$N=2,\ldots,8$
which we have obtained after solving numerically 
the quantization conditions (\ref{eq:C1-C0}) 
\ci{Korchemsky:2001nx,Derkachov:2002wz}.

For more than $N=8$ Reggeons numerical problems appear.
Firstly, the $\{q_k\}-$space becomes more dimensional,
i.e. it has $(2N-2)$ real dimensions. This means that the
method of solving non-linear equations \ci{Kotanski:2001iq}
for higher $N$ needs far more time. Secondly,
in the case $N>8$ the series 
(\ref{eq:power-series-0}) and (\ref{eq:v-series})
become less convergent and recurrence relations
for their coefficients, which include differences of
large numbers, start to require bigger precision.
Thus, in order to overcome this problem
one have to use multi-precision libraries.

\subsection{Properties of the ground states}

Performing the numerical calculations 
we found 
for each sector from $N=2$ to $8$
states which have the minimum energy.
It turns out that these states have common properties for even $N$.
However, the properties of states with even $N$ are different
than the ones with odd $N$.

For even $N$ 
the ground states are in sector where $n_h=0$ so that $h=\frac{1}{2}+i \nu_h$.
They are situated on the trajectories where 
the odd conformal charges vanish,
\begin{equation}
q_3=q_5=\ldots=q_{N-1}=0,
\lab{eq:qovan}
\end{equation}
while the even conformal charges are purely real,
\begin{equation}
\IIm[q_2]=\IIm[q_4]=\ldots=\IIm[q_{N}]=0.
\lab{eq:qeim}
\end{equation}
The minimal energies for these trajectories
are at $\nu_h=0$. 
The states are symmetric under
(\ref{eq:qkmsym}) and (\ref{eq:qkcsym}).
However,
due to (\ref{eq:qovan}) and (\ref{eq:qeim}) 
the ground states for even $N$ are not degenerated.
Moreover, the ground states with even $N$ 
have negative energies $E_N<0$.

For odd $N$ we have two types of the ground states: the descendent ones
and the others that have $q_N \ne 0$.

The descendent ground states are in the sector with $n_h=0$
so that $h=1+i \nu_h$ and they energies $E_N(q_N=0)=0$ vanish.
Along all trajectories $q_N=0$ so these states are compositions
of $(N-1)-$particle states.

On the other hand, the ground state trajectories for odd $N$
with $q_N \ne 0$ are in the sector with $n_h=0$ so $h=\frac{1}{2}+i \nu_h$.
For these trajectories even conformal charges are real,
\begin{equation}
\IIm[q_2]=\IIm[q_4]=\ldots=\IIm[q_{N-1}]=0
\lab{eq:qeim2}
\end{equation}
while the odd conformal charges are imaginary,
\begin{equation}
\RRe[q_3]=\RRe[q_5]=\ldots=\RRe[q_{N}]=0.
\lab{eq:qore}
\end{equation}
Similarly to the case with even $N$, 
these ground states are situated
at $\nu_h=0$.
However, they have positive energies
$E_N(q_N \ne 0)>0$.
Moreover, due to (\ref{eq:qkmsym}) and (\ref{eq:qkcsym}) the ground state
is double-degenerated.
We show exact values of the conformal charges and the energies for 
the ground states with $q_N \ne0$ in Table \ref{tab:Summary}.
Because of the degeneration for odd $N$ we have two equivalent sets 
of the ground states
with $q_k \leftrightarrow (-1)^k q_k$.

\begin{table}[h]
\begin{center}
\begin{tabular}{|c||c|c|c|c|c|c||r|r|}
\hline
  &   $\pm iq_3$ &  $q_4$ & $\pm iq_5$ &  $q_6$ & $\pm iq_7$ & $q_8$ & $-E_N/4$
  & $\sigma_N/4\,\,\,$
\\
\hline
${N=2}$ &            &           &           &    &&       & 2.77259 &16.829
\\
${N=3}$ & 0.205258 &           &           &    &&       & -0.24717 &0.908
\\
${N=4}$ &   0        & 0.153589 &           &     &&      &
\phantom{-}0.67416 & 1.318
\\
${N=5}$ & 0.267682 & 0.039452 & 0.060243 &     &&      & -0.12752 & 0.493
\\
${N=6}$ &          0 & 0.281825 & 0         & 0.070488 &&&
\phantom{-}0.39458 & 0.564
\\
$N=7$ & 0.313072 & 0.070993 & 0.128455 & 0.008494 & 0.019502 & & -0.08141
& 0.319
\\
$N=8$ & 0 & 0.391171 & 0 & 0.179077 & 0 & 0.030428 & 0.28099 & 0.341
\\
\hline
\end{tabular}
\end{center}
\caption[The exact quantum numbers, $q_N$, and the energy, $E_N$]{The exact quantum numbers, $q_N$, and the energy, $E_N$, 
of the ground state of $N$
reggeized gluons in multi-colour QCD. The dispersion parameter, $\sigma_N$,
defines the energy of the lowest excited states, Eq.~(\ref{eq:Enu}).}
\lab{tab:Summary}
\end{table}

\subsection{Energy for different $N$}

In order to analyse the dependence of the energy as a function of 
the number of particles $N$ we plot this dependence in Figure \ref{fig:E_N}.
As we can see we have two cases:
one for odd $N$ and one for even $N$.

\begin{figure}[h!]
\vspace*{3mm}
\centerline{\epsfysize7cm \epsfbox{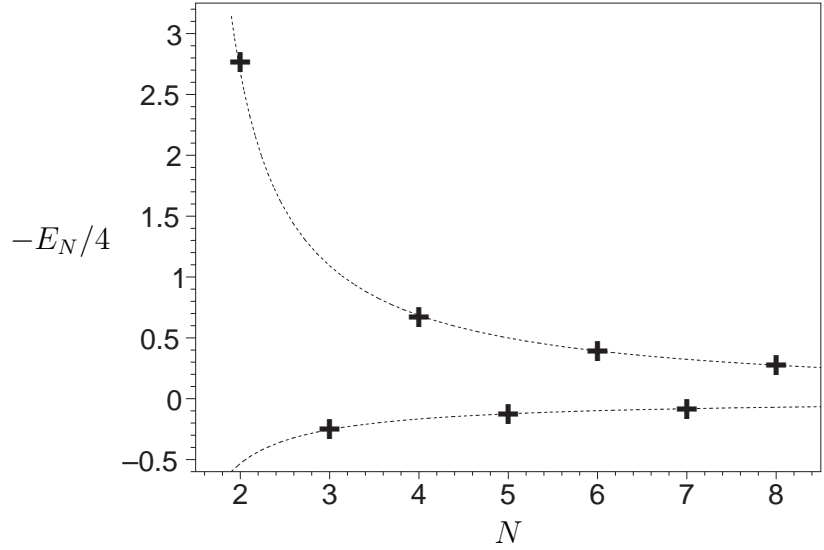}}
\caption[The dependence of  $-E_N/4$ on the number of
particles $N$.]{The dependence of the ground state energy, $-E_N/4$, 
on the number of
particles $N$. The exact values of the energy are denoted by crosses. The upper
and the lower dashed curves stand for the functions $1.8402/(N-1.3143)$ and
$-2.0594/(N-1.0877)$, respectively.}
\lab{fig:E_N}
\end{figure}

For even $N$ the lowest energy is for the ground state with $N=2$
and it grows with $N$ as
\begin{equation}
E_N^{\rm even} \sim \frac{1.8402}{N-1.3143}.
\lab{eq:ENev}
\end{equation}
Contrary to the even $N$ case, 
for odd $N$ we have the highest energy for the ground state with 
$N=3$ and it falls down with $N$ as
\begin{equation}
E_N^{\rm odd} \sim \frac{-2.0594}{N-1.0877}.
\lab{eq:ENod}
\end{equation}
In the both formulae for large $N$ we obtain the limit
of the ground-state energy as
\begin{equation}
|E_{N}| \sim \frac{1}{N} 
\stackrel{N \rightarrow \infty}{\longrightarrow} 0.
\lab{eq:ENlim}
\end{equation}

\subsection{Energy dependence on $\nu_h$}

The ground states are situated on the ground-state trajectories
which are parameterized by a continuous parameter $\nu_h$.
Let us consider the dependence of the energy $E_N(\nu_h)$ 
along these trajectories,
which are plotted in Figure \ref{fig:E-flow}.

\medskip
\begin{figure}[h!]
\centerline{
\begin{picture}(200,80)
\put(12,1){\epsfysize6.8cm \epsfbox{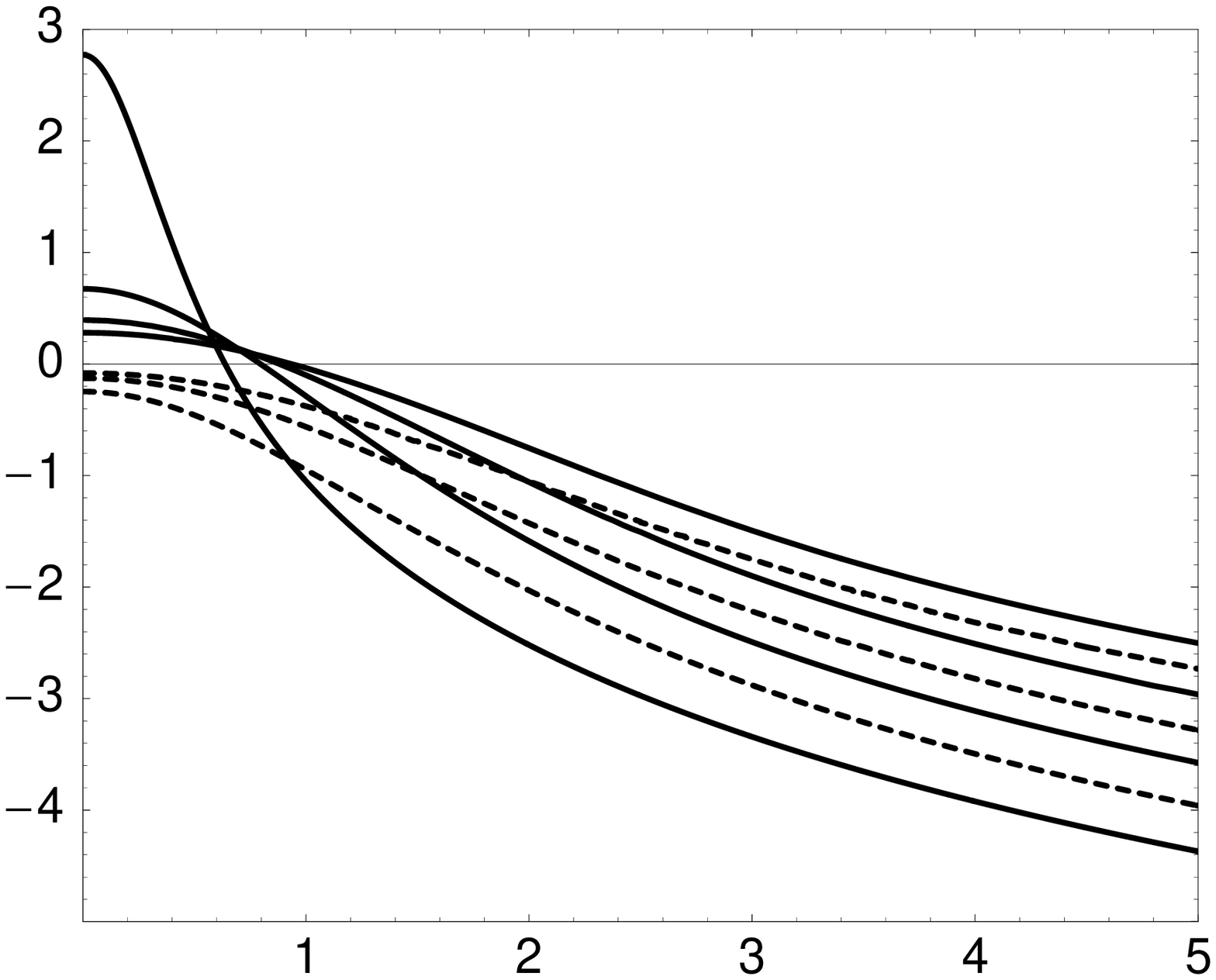}}
\put(102,0){\epsfysize6.8cm \epsfbox{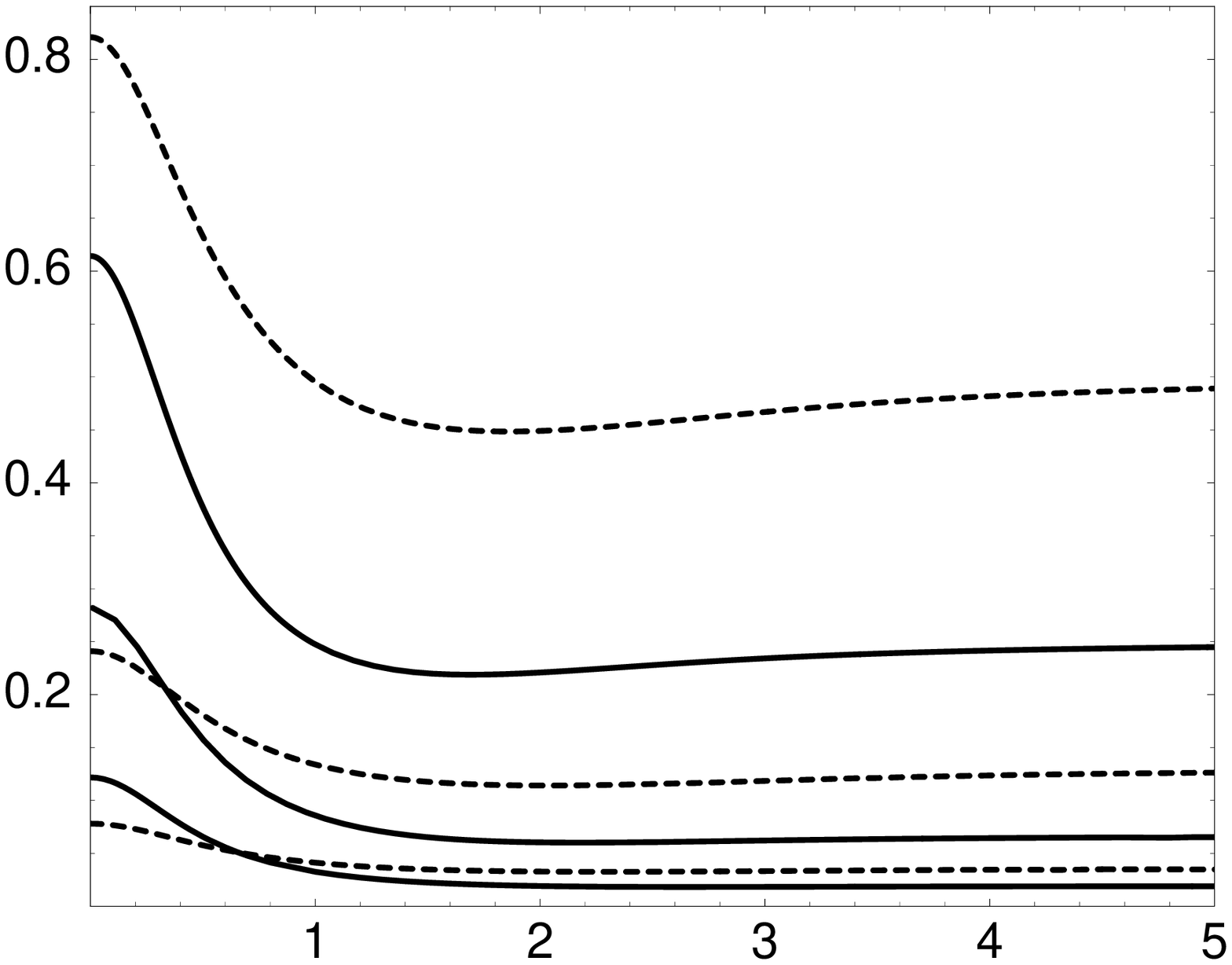}}
\put(54,-2){$\nu_h$}
\put(9,37){\rotatebox{90}{$-E_N/4$}}
\put(145,-2){$\nu_h$}
\put(97,37){\rotatebox{90}{$|q_N|/q_2$}}
\end{picture}
}
\caption[The dependence of the energy $-E_N(\nu_h)/4$ 
and  $|q_N|/q_2$  on $h=1/2+i\nu_h$.]{The dependence of the energy $-E_N(\nu_h)/4$ 
and the ``highest'' integral
of motion $|q_N|/q_2$ with $q_2=(1/4+\nu_h^2)$ on the total spin $h=1/2+i\nu_h$
along the ground state trajectory for different number of particles $2\le N\le
8$. At large $\nu_h$, $-E_8>\ldots>-E_3>-E_2$ 
on the left panel and $|q_8/q_2| <
\ldots < |q_3/q_2|$ on the right panel.
Dashed lines correspond to odd $N$ whereas solid lines to even $N$.}
\lab{fig:E-flow}
\end{figure}

We noticed in the previous Section that we have minima of the energy
at $\nu_h=0$ and these energies satisfy inequality
\begin{equation}
E_2 < E_4 < E_6 < E_8 < 0 < E_7 < E_5 < E_3.
\lab{eq:Eantifer}
\end{equation}
These values of $E_N(\nu_h=0)$ coincide with those
depicted by crosses in Figure~\ref{fig:E_N}.
Thus for $\nu_h$, the order of the energy levels
tell us
that we have anti-ferromagnetic system
\footnote{{In anti-ferromagnetic systems we have
two different sectors: with odd and even $N$ numbers of particles.
For example: in the simplest anti-ferromagnetic
periodic spin-chain with two possible spin 
orientations: {\it up} and {\it down}, 
for odd $N$ the lowest energy is when the neighbouring spins
have opposite orientations.
Since for odd $N$ 
we have always one pair of neighbouring spins with the same orientation
the ground states for odd $N$ has higher energies than the
the ground states for even $N$.
}}.
In the vicinity of $\nu_h=0$ the energies behave as 
(\ref{eq:Enu}) where $\sigma_N$ measure accumulation of the energies. 
The values of $\sigma_N$ are shown in Table \ref{tab:Summary}.

Going to the larger $\nu_h$ we notice that the energies grow. 
For $\nu_h \sim 1$ we have a quantum phase transition. The energies
of our system reorder and for the large $\nu_h$ the ground-state
energies of our system satisfy
\begin{equation}
0 < E_2 < E_3 < E_4 < E_5 < E_6 < E_7 < E_8.
\lab{eq:Efer}
\end{equation}
The reason for this behaviour is that for the total $SL(2,\mathbb{C})$ 
spin $h=1/2+i\nu_h$
with the large $\nu_h$ the system approaches a quasi-classical regime 
\ci{Korchemsky:1995be,Korchemsky:1996kh,Korchemsky:1997ve},
in which the energy $E_N(\nu_h)$ and the quantum numbers $q_N$ have a universal
scaling behaviour
\be
E_N(\nu_h)\sim 4 \,\Log |q_N| \,,\qquad |q_N| \sim  C_N\,\nu_h^2\,,
\ee
with $C_N$ decreasing with $N$. As can be seen from the right panels in
Figures~\ref{fig:N4-flow} and \ref{fig:E-flow}, this regime starts already 
at $\nu_h
\approx 2$ \ci{Derkachov:2002wz}.

\chapter{Other quantization conditions}

In this Chapter we show for completeness
another solution to the Baxter Equation
(\ref{eq:Baxeq}) that has been presented 
in Refs. \ci{DeVega:2001pu,deVega:2002im} by de Vega and Lipatov.
In their method two additional conditions
were assumed: the conformal charges should satisfy 
$\IIm[i^k q_k]=0$ and the holomorphic energy 
of each from $N$ linearly independent solutions
of the pertinent Baxter equation 
should be equal.
As was explained in Ref. \ci{Derkachov:2002wz} 
these conditions are too strong.

\section{De Vega and Lipatov's solution}

In the papers \ci{DeVega:2001pu,deVega:2002im}
the authors consider the Baxter equations (\ref{eq:Baxeq})
and (\ref{eq:Baxbeq}) with complex spins $(s=1,\wbar s=1)$.
The equal values of the complex spins 
cause that
the  Baxter equations
in the holomorphic and anti-holomorphic
sectors are identical. However, is this case
the complex spins $(s=1,\wbar s=1)$ do not satisfy (\ref{eq:ssbar})
so we have to use the different form of the scalar product  
given by (\ref{eq:norml}).

It has been shown in Ref. \ci{DeVega:2001pu} that
the simplest $N-$Reggeon solution to the Baxter
equation (\ref{eq:Baxeq}) 
may be conveniently written in a form of a sum over poles of 
orders $1$ up to $N-1$ situated in the upper semi-plane
of complex variable $u$:
\begin{equation}
Q^{(N-1)}(u;h,q)=\sum_{r=0}^{\infty} 
\frac{P_{r;h,q}^{(N-2)}(u)}{(u-ir)^{N-1}}
=\sum_{r=0}^{\infty} 
\left[
\frac{\tilde{a}_r(q)}{(u-ir)^{N-1}}+
\frac{\tilde{b}_r(q)}{(u-ir)^{N-2}}+
\ldots+
\frac{\tilde{z}_r(q)}{(u-ir)}
\right]\,,
\lab{eq:Qupol}
\end{equation}
where $q=\{q_2,q_3,\ldots,q_N\}$ and 
$P_{r;h,q}^{(N-2)}(u)$ are polynomials in $u$ of degree $N-2$, whereas
$\tilde{a}_r(q), \tilde{b}_r(q),\ldots$ are some residual coefficients.
Substituting (\ref{eq:Qupol}) to (\ref{eq:Baxeq}) we obtain the recurrence 
relations between the polynomials $P_{r;h,q}^{(N-2)}(u)$ which 
allows us to calculate them successively starting from 
$P_{0;h,q}^{(N-2)}(u)$.

The solution (\ref{eq:Qupol}) is normalized by the constraint
\begin{equation}
\lim_{u \to 0} P_{0;h,q}^{(N-2)}(u)=\tilde{a}_r(q)=1
\lab{eq:Qnorm}
\end{equation}
and the remaining coefficients of the polynomial 
$P_{0;h,q}^{(N-2)}(u)$, or equivalently the coefficients $b_r(q)$, $c_r(q)$,
$\ldots\,$, 
are calculated from the condition
\begin{equation}
\lim_{u \to \infty} Q^{(N-1)}(u;h,q) \sim  u^{h-N}
\lab{eq:Qasbah}
\end{equation}
which ensures \ci{DeVega:2001pu} 
that the solution (\ref{eq:Qupol}) 
satisfies the Baxter equation (\ref{eq:Baxeq})
for $u \to \infty$.
The condition (\ref{eq:Qasbah}) leads to 
\begin{equation}
\lim_{u \to \infty} 
u^{N-2}
\frac{P_{r;h,q}^{(N-2)}(u)}{(u-ir)^{N-1}}=0
\lab{eq:Qasc}
\end{equation}
which  gives $N-2$ equations and with (\ref{eq:Qnorm})
fixes  all the polynomial coefficients 
$P_{0;h,q}^{(N-2)}(u)$, or equivalently 
$\tilde{a}_0(q), \tilde{b}_0(q), \ldots$\,.

The second independent solution may be written in terms of the first one as
\begin{equation}
Q^{(0)}(u;h,q)=Q^{(N-1)}(-u;h,-q)=
\sum_{r=0}^{\infty}
\frac{P_{r;h,-q}^{(N-2)}(-u)}{(-u-ir)^{N-1}}\,,
\lab{eq:Q2up}
\end{equation}
where $-q=(q_2,-q_3,\ldots,(-)^N q_N)$. 

It turns out, that there is a set of Baxter 
functions $Q^{(t)}(u)$ with $t=0,1,\ldots,N-1$ which
have poles both in the upper and lower half$-u$ planes:
\begin{equation}
Q^{(t)}(u;h,q)=
\sum_{r=0}^{\infty}
\left[
\frac{P_{r;h,q}^{(t-1)}(u)}{(u-ir)^{t}}
+
\frac{P_{r;h,-q}^{(N-2-t)}(-u)}{(-u-ir)^{N-1-t}}
\right]\,,
\lab{eq:Qt}
\end{equation} 
where the polynomials $P_{r;h,q}^{(t-1)}(u)$
and $P_{r;h,-q}^{(N-2-t)}(-u)$ are fixed by 
the recurrence relations following from the Baxter equation
and from the condition that the solution should decrease at infinity
more rapidly than $u^{-N+2}$.

Using these functions in both holomorphic and anti-holomorphic sectors
it is possible to construct the Baxter function 
\begin{equation}
Q_{q,\wbar q}(u,\wbar u)= \sum_{t,l} C_{t,l} 
Q^{(t)}(u;h,q)
Q^{(l)}(\wbar u;\bar h,\wbar q)
\lab{eq:dlQuu}
\end{equation}
choosing the coefficients $C_{t,l}$,  appropriately.

In this method 
the coefficients $C_{t,l}$ are fixed by 
the normalization condition \ci{DeVega:2001pu}.
It leads to the requirement
that $Q_{q,\wbar q}(u,\wbar u)$
do not have poles at 
\begin{equation}
u=i m
\quad
\mbox{and} 
\quad
\wbar u=-i m 
\quad
\mbox{for} 
\quad
|m|>0\,. 
\lab{eq:dlpol}
\end{equation}
The pole at $u=\wbar u =0$ is
removed by the measure of the $(s=1,\wbar s=1)$ scalar product 
(\ref{eq:norml}).
Because in the bilinear solution we have products of the poles
$(u-ir)^{-s}$ and $(\wbar u- ir')^{-s'}$
we expand $Q_{q,\wbar q}(u, \wbar u)$ around the spurious
poles (\ref{eq:dlpol}) and we equate the residual expansion coefficients
to zero.
This results in the quantization conditions for the conformal charges 
$q$ and $\wbar q$.

The authors of  Refs. \ci{DeVega:2001pu,deVega:2002im}
claim that the same conditions are given by imposing equality
of holomorphic energies 
\begin{equation}
\epsilon_t=i \lim_{u\to i} \frac{\partial}{\partial u} \ln
\left[
u^N P_{1;h,q}^{(t-1)}(u)
\right]
\lab{eq:hole}
\end{equation}
for all solutions $Q^{(t)}(u)$.
However, our system is two-dimensional and the holomorphic energy is not a 
physical observable and it may depend on a calculation method. 
The real observable is the total energy defined 
as
\begin{equation}
E=2 i \lim_{u,\wbar u\to i} 
\frac{\partial}{\partial u}
\frac{\partial}{\partial \wbar u}
\ln \left[
(u-i)^{N-1}
(\wbar u-i)^{N-1}
u^{N}
{\wbar u}^{N}
Q_{q,\wbar q}(u,\wbar u)
\right]\,.
\lab{eq:Enlv}
\end{equation}

\subsection{Three Reggeon states}

For simplicity, let us consider the $N=3$ Reggeon case. The Baxter equation 
(\ref{eq:Baxeq}) with $s=1$ reduces to
\begin{eqnarray}
B_3(u;h,q_3)&\equiv&
\left[2 u^3-h(h-1) u + q_3
\right]Q(u;h,q_3)
\nonumber\\
& &-(u+i)^3 Q(u+i;h,q_3)
-(u-i)^3 Q(u-i;h,q_3)=0\,.
\lab{eq:Bax3}
\end{eqnarray}
Here the authors of Refs.  \ci{DeVega:2001pu,deVega:2002im}
claim that $i q_3$ should be real in order to 
obtain single-valued wave-functions. However, we have already shown
(\ref{eq:Psizz})--(\ref{eq:q3-WKB})
that it is possible to construct a single-valued function
without this assumption.

To find polynomial coefficients in $Q^{(t)}(u,\wbar u)$
we define auxiliary functions $f_r$, which look as follows
\begin{eqnarray}
f_2(u;h,q_3)&=&\sum_{l=0}^{\infty}
\left[
\frac{a_l(h,q_3)}{(-iu-l)^2}+
\frac{b_l(h,q_3)}{-iu-l}
\right]\,,
\nonumber\\
f_1(u;h,q)&=&\sum_{l=0}^{\infty}
\frac{a_l(h,q_3)}{-iu-l}\,.
\lab{eq:fr}
\end{eqnarray}
They help us to impose the conditions coming from behaviour 
of $Q(u;h,q_3)$ at $u\to \infty$.
Substituting (\ref{eq:fr}) to (\ref{eq:Bax3})
we obtain the recurrence relations for the residua of (\ref{eq:fr}):
\begin{eqnarray}
& &{\left(r+ 1 \right) }^3\,a_{r+1}(h,q_3) = 
\left[ 2\,r^3+h(h-1)\,r+i \,q_3  \right] \, a_r(h,q_3)
-{\left(r -1 \right) }^3\,a_{r-1}(h,q_3)
\nonumber\\
& & {\left(r+ 1 \right) }^3\,b_{r+1}(h,q_3) = 
\left[ 2\,r^3+h(h-1)\,r+i \,q_3  \right] \, b_r(h,q_3)
-{\left(r -1 \right) }^3\,b_{r-1}(h,q_3)
\nonumber\\ 
& &+\left[ 6 r^2 + h(h-1) \right] a_r(h,q_3)
-3 (r+1)^2 a_{r+1}(h,q_3)
-3 (r-1)^2 a_{r-11}(h,q_3)\,.
\lab{eq:recfr}
\end{eqnarray}
We can choose,
\begin{equation}
a_0(h,q_3)=1
\quad \mbox{and}
\quad
b_0(h,q_3)=0
\lab{eq:a0b0}
\end{equation}
what gives as
\begin{equation}
a_1(h,q_3)=i q_3
\,,
\qquad
b_1(h,q_3)=h(h-1)-3 i q_3\,.
\lab{eq:a1b1}
\end{equation}
Thus, we can calculate all the coefficients 
$a_r(h,q_3)$ and $b_r(h,q_3)$.

Now, we can construct $Q^{(t)}(u, \wbar u)$ as a linear combinations of
the auxiliary functions (\ref{eq:fr}) in a form
\begin{eqnarray}
Q^{(2)}(u;h,q_3)&=&
f_2(u;h,q)+
B(h,q_3) f_1(u;h,q)\,,
\nonumber\\
Q^{(1)}(u;h,q_3)&=&
f_1(u;h,q)+
C(h,q_3) f_1(-u;h,-q_3)\,,
\lab{eq:Q21}
\end{eqnarray}
where  $B(h,q_3)$ and $C(h,q_3)$ are coefficients we will
fix using the condition (\ref{eq:Qasc}) 
by demanding that 
the Baxter equation is satisfied at infinity.
Taking the limit $u \to \infty$ of (\ref{eq:Bax3}) with (\ref{eq:Q21})
we obtain
\begin{equation}
B_3(\infty;h,q_3)=\left[h(h-1)-2\right] \lim_{u \to \infty} 
\left[u\, Q(u;m,\mu)\right]\,.
\lab{eq:B3inf}
\end{equation}
Applying (\ref{eq:B3inf}) to (\ref{eq:Q21}) we get
the coefficients in a form
\begin{equation}
B(h,q_3)=-\frac{\sum_{r=0}^{\infty} b_r(h,q_3)}{
\sum_{r=0}^{\infty} a_r(h,q_3)}
\quad
\mbox{and}
\quad
C(h,q_3)=\frac{\sum_{r=0}^{\infty} a_r(h,q_3)}{
\sum_{r=0}^{\infty} a_r(h,-q_3)}\,.
\lab{eq:BC}
\end{equation}
Therefore, the solutions $Q^{(2)}(u;h,q_3)$ and
$Q^{(1)}(u;h,q_3)$ are completely determined.

Due to the symmetry (\ref{eq:qkmsym}) we can construct
the third solution 
\begin{equation}
Q^{(0)}(u;h,q_3)=Q^{(2)}(-u;h,-q_3)\,.
\lab{eq:Q0lv}
\end{equation}
Moreover, if we multiply
a solution of the Baxter equation by a periodic 
function of $i u$ with period 1 the result is also 
a solution of the Baxter equation (\ref{eq:Bax3}).
For instance 
\begin{equation}
Q^{(1)}(u;h,q_3) \pi \cot(i \pi u)
\lab{eq:Q1lv}
\end{equation}
is also the solution to (\ref{eq:Bax3}) and 
thanks to $\cot(i \pi u)$ it has second order poles
at $u=i m$ for $|m|>0$.

All these solutions are related by
the equation \ci{DeVega:2001pu,deVega:2002im} 
\begin{equation}
\left[X(h,q_3)-\coth (\pi u)\right]
Q^{(1)}(u;h,q_3)=
Q^{(2)}(u;h,q_3)
- C(h;q_3) Q^{(0)}(u;h,q_3)\,,
\lab{eq:QQQrel}
\end{equation}
where $X(h,q_3)$ can be obtained by performing the Laurent expansion
of (\ref{eq:QQQrel}) around the poles (\ref{eq:dlpol}).
In Refs. 
\ci{DeVega:2001pu,deVega:2002im} it is shown that
it is possible to construct the solution (\ref{eq:dlQuu})
with $h=\wbar h$ and $q_3=-\wbar q_3$ for which poles at (\ref{eq:dlpol})
disappear. It has the following form
\begin{equation}
Q_{q,\wbar q}(u,\wbar u)=
Q^{(2)}(u;h,q_3) Q^{(2)}(\wbar u;\wbar h,\wbar q_3)
-Q^{(0)}(u;h,q_3) Q^{(0)}(\wbar u;\wbar h,\wbar q_3)\,.
\lab{eq:QQ-QQ}
\end{equation}

Moreover, in Refs. \ci{DeVega:2001pu,deVega:2002im}
the authors assume that the condition $X(h,q_3)=0$
gives quantization of $q_3$.
Accidentally, for solutions with $X(h,q_3)=0$
the holomorphic energies (\ref{eq:hole})
of the solutions $Q^{(2)}(u;h,q_3)$ and
$\pi \cot(i u \pi) Q^{(1)}(u;h,q_3)$ 
are equal
what gives equivalent condition to $X(h,q_3)=0$
which looks like
\begin{equation}
B(h,q_3)-b_1(h,q_3)-\frac{1}{i q_3}
+\frac{1}{i q_3} \sum_{r=2}^{\infty}\frac{a_r(h,q_3)}{r-1}
+\frac{C(h,q_3)}{i q_3} \sum_{r=0}^{\infty}\frac{a_r(h,-q_3)}{r+1}=0\,.
\lab{eq:delee}
\end{equation}

However, the way of cancelling poles presented in 
\ci{DeVega:2001pu,deVega:2002im} is not unique.
Indeed, the condition (\ref{eq:delee}) is too strong and it is only
satisfied for some solutions with $\RRe \,q_3=0$. 
There are other solutions, for example with $\IIm \,q_3=0$,
for which the holomorphic energies are not equal. 

In Refs. \ci{DeVega:2001pu,deVega:2002im} 
for $N=4$ Reggeons 
the quantization conditions are also introduced by the equality of
the holomorphic energies. 
Similarly to the N=3 case, 
the numerical values of the conformal charges which result
from these conditions 
agree with Refs. \ci{Derkachov:2002wz} only for $h=\wbar h$, $\RRe \,q_3=0$ and
$\IIm\, q_4=0$.

Contrary to the solutions described in  \ci{Derkachov:2002wz}
the solutions presented in this Chapter have a simple pole structure,
so they look much simpler than those defined by
(\ref{eq:Q-R})--(\ref{eq:set-1}).
However, the series in (\ref{eq:fr}) are much slower convergent
than the series defined in (\ref{eq:power-series-0}) and (\ref{eq:v-series}).
Moreover, in \ci{Derkachov:2002wz} the quantisation conditions 
(\ref{eq:C1-C0}) 
come from single-valuedness and normalization of 
the Reggeon wave-functions
explicitly, whereas in \ci{DeVega:2001pu,deVega:2002im}
except for the normalization condition,
the quantization conditions (\ref{eq:delee}) 
follow from the
quality of the holomorphic energies (\ref{eq:hole}).

Thus, if one wants to find  a spectrum of the conformal charges
numerically
the method presented in \ci{Derkachov:2002wz}
is much more convenient.


\chapter{Anomalous dimensions}

In the preceding Chapters we used the reggeized gluon states
to describe the elastic scattering amplitude
of strongly interacting hadrons. It turns out
that the Reggeon states might be also  used to describe 
deep inelastic scattering of a virtual photon $\gamma^{\ast}(q_{\mu})$
off a (polarized) hadron with momentum $p_{\mu}$ 
\ci{Korchemsky:2003rc,deVega:2002im}.
In this case we also perform calculations in the limit of 
the low Bj\"orken $x$
but in the region where
\begin{equation}
M^2 \ll Q^2 \ll s^2=(p_{\mu}+q_{\mu})^2=\frac{Q^2(1-x)}{x}\,,
\lab{eq:DISlim}
\end{equation}
with
\begin{equation}
Q^2=-{q_{\mu}}^2 
\qquad
M^2={p_{\mu}}^2.
\lab{eq:defQM} 
\end{equation}
Notice that in the previous case (\ref{eq:rlim})
there was only one hard scale present, namely the mass of the
scattering particles
which
were assumed to be similar.

In the limit (\ref{eq:DISlim}), 
similarly to (\ref{eq:rlim}),
the moments of the structure function $F_2(x,Q^2)\equiv F(x,Q^2)$
may be rewritten as a power series in
the strong coupling constant $\alpha_s$:
\begin{equation}
\widetilde{F}(j,Q^2)
\equiv 
\int_{0}^{1} dx \, x^{j-2} F(x,Q^2)=
 \sum_{N=2}^{\infty} \bar\alpha_s^{N-2} \widetilde{F}_N(j,Q^2)\,,
\lab{eq:moments}
\end{equation}
where $\bar\alpha_s=\alpha_s N_c/\pi$.

The leading contribution comes from the Reggeon states  with
the intercept $j\to 1$ and in this limit
\begin{equation}
\widetilde{F}_N(j,Q^2)=\sum_{\mybf{q}} 
\frac1{j-1+\bar \alpha_s E_N(\Mybf{q})/4}
\beta_{\gamma^*}^{\,\mybf{q}}(Q)\,
 \beta_p^{\,\mybf{q}}(M)\,,
\lab{eq:moments2}
\end{equation}
where the conformal charges are denoted by 
$\mybf{q}=(q_2, \wbar q_2,q_3,\wbar q_3 ,\ldots,q_N,\wbar q_N)$
and $\beta_{\gamma^*}^{\,\mybf{q}}(Q)$, 
$\beta_p^{\,\mybf{q}}(M)$ are the impact factors
of virtual photon and scattered hadron, respectively,
while the energies $E_N$ are the eigenvalues of the Reggeon Hamiltonian
(\ref{eq:sepH}).

Performing the operator product expansion (OPE)
one may expand the moments of the structure function
$\widetilde{F}_N(x,Q^2)$ in inverse powers of hard scale $Q$
\begin{equation}
\widetilde F(j,Q^2)=\sum_{n=2,3,\ldots}
\frac1{Q^n} \sum_{a} C_n^a(j,\alpha_s(Q^2))\, 
\vev{p\,|\mathcal{O}^a_{n,j}(0)|p}\,,
\lab{eq:Fope}
\end{equation}
where the expansion coefficients are the forward matrix elements of Wilson
operators $\mathcal{O}^a_{n,j}(0)$ of increasing twist $n \ge 2$ 
and they satisfy renormalization group equations
\begin{equation}
Q^2\frac{d}{d Q^2}\vev{p\,|\mathcal{O}^a_{n,j}(0)|p}
=\gamma_n^a(j)\,\vev{p\,|\mathcal{O}^a_{n,j}(0)|p}
\lab{eq:Qren}
\end{equation}
with the anomalous dimensions of QCD $\gamma_n^a(j)$.
The parameter $a$ enumerates operators with the same twist $n$.
It also appears in (\ref{eq:Fope}) 
where $C_n^a(j,\alpha_s)$ denote
the coefficient functions.

The anomalous dimensions $\gamma_n^a(j)$
may be written as a power series
\begin{equation}
\gamma_n^a(j)=
\sum_{k=1}^\infty\gamma_{k,n}^a(j) \left(\alpha_s(Q^2)/\pi\right)^k.
\lab{eq:anod}
\end{equation}
For small Bj\"orken $x$ ($j \to 1$), the moments of the structure function
$\widetilde{F}_N(j,Q^2)$
take a form
\begin{equation}
\widetilde F(j,Q^2)=\frac{1}{Q^2} \sum_{n=2,3,\ldots}
\sum_{a} \tilde{C}_n^a(j,\alpha_s(Q^2))\ 
\left(\frac{M}{Q}\right)^{n-2-2\gamma_n^a(j)} \,.
\lab{eq:mFj1}
\end{equation}

It turns out that solving the Schr\"odinger equation (\ref{eq:Schr})
we are able to find poles which give the main contribution to 
(\ref{eq:moments2}).
Moreover, combining (\ref{eq:mFj1}) and 
(\ref{eq:moments2}) we are able to find the dependence
of $\gamma_{k,n}^a(j)$ on the strong coupling constant near $j \to 1$.
In this Chapter we show the way one can 
find  this dependence using techniques developed in the previous Chapters.

\section{Expansion for the large scale $Q$}

The moments of the structure function are defined in (\ref{eq:moments}).
In the limit $j \to 1$ they can be rewritten as (\ref{eq:moments2})
where the impact factors are given by 
\begin{equation}
\beta^{\mybf{q}}_{\gamma^*}(Q)=\int d^2 z_0\, 
\vev{\Psi_{\gamma^*}|\Psi_{\mybf{q}}(\vec z_0)}\,,
\qquad
\beta^{\mybf{q}}_p(M)=\int d^2 z_0\, \vev{\Psi_{\mybf{q}}(\vec z_0)|\Psi_p}
\lab{eq:imp-fac}
\end{equation}
and 
the functions $\Psi_{\mybf{q}}(\vec z_0)$
are orthonormal with respect to the scalar product 
(\ref{eq:norm}):
\begin{equation}
\vev{\Psi_{\mybf{q}}(\vec z_0)|\Psi_{\mybf{q}'}(\vec z_0')}\equiv\int
\prod_{k=1}^N d^2 z_k 
\Psi_{\mybf{q}}(\{\vec z\};\vec z_0) \lr{\Psi_{\mybf{q}'}(\{\vec z\};\vec
z_0')}^*= \delta^{(2)}(z_0-z_0')\,\delta_{\mybf{q}\mybf{q}'}\,,
\lab{eq:normd}
\end{equation}
where $\delta_{\mybf{q}\mybf{q}'}=\delta(\nu_h-\nu_h')
\delta_{n_h n_h'}\delta_{\mybf{\ell} \mybf{\ell'}}$
with $\mybf{\ell}=\{\ell_2,\ell_3,\ldots,\ell_{2(N-2)}\}$
defined in (\ref{eq:dell}).
Due to the scaling symmetry of the reggeized gluon states 
\begin{equation}
\Psi_{\mybf{q}}(\lambda \vec z_1,\lambda \vec z_2,\ldots,\lambda \vec z_N)=
\lambda^{2-h-\wbar h} \Psi_{\mybf{q}}(\vec z_1,\vec z_2,\ldots,\vec z_N)
\lab{eq:Psisc}
\end{equation}
we are
able to calculate dimensions of the impact factors (\ref{eq:imp-fac})
\begin{equation}
\beta^{\mybf{q}}_{\gamma^*}(Q)= C_{\gamma^*}^{\mybf{q}} Q^{-1-2i\nu_h}\,,
\qquad
\beta^{\mybf{q}}_p(M)=C_{p}^{\mybf{q}}\, M^{-1+2i\nu_h}\,,
\lab{eq:sc-imp}
\end{equation}
where $C_{\gamma^*}^{\mybf{q}}$ and $C_{p}^{\mybf{q}}$ are dimensionless.
Substituting (\ref{eq:sc-imp}) into (\ref{eq:moments2})
and expanding in powers of the ratio $(M/Q)$ 
\begin{equation}
\widetilde F_N(j,Q^2)=\frac{1}{Q^2}\sum_{\mybf{\ell}}\sum_{n_h\ge 0 }
\int_{-\infty}^\infty d\nu_h
\frac{C_{\gamma^*}^{\mybf{q}}\,C_{p^{\phantom{*}}}^{\mybf{q}}}{j-1
+\bar \alpha_s
E_N(\Mybf{q})/4} \lr{\frac{M}{Q}}^{-1+2i\nu_h}\,,
\lab{eq:F-fin}
\end{equation}
where $\mybf{q}=\mybf{q}(\nu_h;h_n,\mybf{\ell})$.
We shall calculate the integral in (\ref{eq:F-fin}) by performing analytical
continuation in $\nu_h$. Next, we shall close the integration contour in the 
$\nu_h-$complex plane and integrate by summing residua 
closed by the contour. The corresponding poles come from the denominator
of (\ref{eq:F-fin}) so they are placed at $\nu_h$ where 
\begin{equation}
j-1=\bar\alpha_s E_N(\Mybf{q}(\nu_h;n_h,\Mybf{\ell}))/4
\lab{eq:jN}
\end{equation}
is satisfied.
Here the parameters $n_h$ and 
$\Mybf{\ell}=\{\ell_1,\ell_2,\ldots,\ell_{2(N-2)}\}$
are integer numbers and they enumerate the quantizated energy levels.
The formula for (\ref{eq:F-fin}) is valid 
for $j>j_N=1 - \bar\alpha_s \min{}_{q}E_N(\Mybf{q})/4$.

In this way a contribution to the moments of $F(x,Q^2)$ 
coming from a given pole in $\nu_h$
takes the following form
\begin{equation}
\widetilde F_N(j,Q^2)\sim 
\frac{1}{Q^2} \left(\frac{M}{Q}\right)^{-1+2i\nu_h(j)}\,,
\lab{eq:Flead} 
\end{equation}
where we show the dependence on $j$ explicitly.
Comparing exponents in (\ref{eq:mFj1}) and (\ref{eq:Flead})
we obtain relation
\begin{equation}
\gamma_n(j)=(n-1)/2-i\nu_h(j)=[n-(h(j)+\bar h(j))]/2\,,
\lab{eq:expect1}
\end{equation}
where $h(j)$ and $\bar h(j)$ are the $SL(2,\mathbb{C})$ spins defined in 
(\ref{eq:hpar}) and $\gamma_n(j)$ is the anomalous dimension of the 
twist$-n$ operator (\ref{eq:Qren}).

Now we can calculate $\gamma_n(j)$ which 
with $\alpha_s(j)\to 0$ behave as $\gamma_n(j) \to 0$
\begin{equation}
\gamma_n(j)=\gamma_n^{(0)} \frac{\wbar \alpha_s}{j-1}+{\cal O}
({\wbar{\alpha}_s}^2).
\lab{eq:gambeh} 
\end{equation}
Substituting (\ref{eq:jN}) into (\ref{eq:gambeh})
one finds that $\gamma_n(j) \to 0$ 
corresponds to $E_N(\nu_h) \to \infty$.  
Thus, we can combine the expansion of the energy $E_N(\nu_h)$
around its poles in the complex $\nu_h-$plane:
\begin{equation}
E_N(\Mybf{q})=
-4\left[{\frac{c_{-1}}{\epsilon}+c_0+c_1\,\epsilon+\ldots}\right]
\lab{eq:Enpoles}
\end{equation}
with the small anomalous dimensions $\gamma_n(j)=-\epsilon$
where according to (\ref{eq:gambeh})
$i \nu_h=i\nu_h^{\rm pole}+\epsilon$.
Inverting the series (\ref{eq:Enpoles}) and using (\ref{eq:jN}) 
we obtain
\begin{equation}
\gamma_n(j)=-c_{{-1}}\left[\frac{\bar\alpha_s}{j-1}+c_{{0}}\,
\lr{\frac{\bar\alpha_s}{j-1}}^2+\left( c_{{1}}c_{{-1}}+c_{0}^{\,2}
\right)\lr{\frac{\bar\alpha_s}{j-1}}^3+\ldots\right]\,,
\lab{eq:gampol}
\end{equation}
where coefficients $c_k=c_k(n,n_h,\Mybf{\ell})$ in Eqs. (\ref{eq:gampol}) and 
(\ref{eq:Enpoles}) are identical.
Thus, evaluating these coefficient for the energy (\ref{eq:Enpoles})
we may calculate coefficients for the anomalous dimension expansion
in the coupling constant $\wbar \alpha_s$.

Moreover, we can notice that the positions of the energy poles
\begin{equation}
E_N(\Mybf{q})\sim 
\frac{\gamma_{n}^{(0)}}{i\nu_h-(n-1)/2}
\lab{eq:Enspol}
\end{equation} 
determines the twist $n$ of $\gamma_{n}^{(0)}$
\begin{equation}
i\nu_h=(n-1)/2 
\quad \mbox{with} \quad
n\ge N+n_h
\lab{eq:twist}
\end{equation}
with the number of Reggeons $N$ 
and the Lorentz spin $n_h$.

For $N=2$ formula for the energy \ci{Balitsky:1978ic,Fadin:1975cb}
is given by an analytical expression
\begin{equation}
E_2(\nu_h,n_h)=4\left(\psi\lr{\frac{1+n_h}2+i\nu_h}
+\psi\lr{\frac{1+n_h}2-i\nu_h}-2\psi(1)\right)\,,
\lab{eq:E2-anal}
\end{equation}
where $n_h$ are nonnegative integers.
In this case the anomalous dimensions for the leading twist $n=2$ 
have been calculated in \ci{Jaroszewicz:1982gr}:
\begin{equation}
\gamma_2(j)=\frac{\bar\alpha_s}{j-1}+2\zeta(3)\lr{\frac{\bar\alpha_s}{j-1}}^4
+2\zeta(5)\lr{\frac{\bar\alpha_s}{j-1}}^6 +\mathcal{O}(\bar\alpha_s^8)\,,
\lab{eq:sol-n}
\end{equation}
where $\zeta(k)$ is  the Riemann zeta-function.
The general formula for $N=2$ and arbitrary $n\ge 2$
and $n_h$
has been obtained in \ci{Lipatov:1985uk}:
\begin{equation}
\gamma_n(j)=
\frac{\bar\alpha_s}{j-1}
+(2 \psi(1) -\psi(1+|n_h|+n)-\psi(1+n))\lr{\frac{\bar\alpha_s}{j-1}}^2
+\mathcal{O}(\bar\alpha_s^3)\,.
\lab{eq:sol-2}
\end{equation}

\section{Analytical continuation}

The reggeized gluon functions $\Psi(\vec{z}_k)$ defined in (\ref{eq:Psip})
are normalized with respect to the scalar product (\ref{eq:normd}). 
Performing the analytical continuation is $\nu_h$ 
it turns out that the normalization condition (\ref{eq:sjdag}) is no 
longer satisfied.
The quantization conditions (\ref{eq:sjdag}) and (\ref{eq:C1-C0}) 
become relaxed. Thus,
conditions $\wbar q_k = q_k$ are not necessarily satisfied so that
$\wbar h \ne 1- h^{\ast}$.
However, we still have the conditions coming from the single-valuedness of 
the Reggeon wave-functions. The latter condition  (\ref{eq:C1-C0}) ensures 
that the 
two-dimensional integrals (\ref{eq:imp-fac}) are well defined.
In (\ref{eq:enQ2}) the energy $E_N(\nu_h)$ is a smooth function 
of real $\nu_h$.
After the analytical continuation into the complex $\nu_h-$plane
the energy spectrum exposes pole structure at $\RRe[\nu_h]=0$.

In Quantum Mechanics the problem of analytical continuation of 
energy $E(g)$ 
as a function of a coupling constant $g$
has been studied in various models 
\ci{Bender:1969si,Turbiner:1987kt,Bender:1992bk}.
In our case $\nu_h$ plays a role of such a coupling constant $g$.

Generally, the energy $E(g)$ is a multi-valued function of $g$.
It turns out that the number of energy levels for real $g$ 
is equal to the number of branches in the complex $g-$plane.
In order to analyse the global spectrum one may glue together different 
sheets corresponding to its branches and study  $E(g)$  as 
a single-valued function on the resulting Riemann surface.
Thus, the models may have a complicated structure of spectrum
in the complex $g-$plane. Additionally, these spectra may consist of
a few disconnected parts due to some additional symmetry \ci{Turbiner:1987kt}.

\subsection{Two level model}

In order to understand the analytical continuation more clearly
let us consider the simple model
with two energy levels $\epsilon_1$ and $\epsilon_2$:
\begin{equation}
\hat H=\left(
\begin{array}{cc}
\epsilon_1  & g\\
g & \epsilon_2\\
\end{array}
\right)\,,
\lab{eq:H2x2}
\end{equation}
where we add interaction by introducing 
a small coupling constant $g$.

The energy eigenvalues of the interacting Hamiltonian (\ref{eq:H2x2})
are given by
\begin{equation}
E_{\pm}(g)=\frac{\epsilon_1+\epsilon_2}{2}
\pm \sqrt{\left(\frac{\epsilon_1-\epsilon_2}{2}\right)^2+g^2}\,.
\lab{eq:Epm}
\end{equation}
Here we have two branches $E_+(g)$ and $E_-(g)$
which are shown in Fig \ref{fig:Epm}
where for illustration we have  chosen
with $\epsilon_1=0$ and $\epsilon_2=1$.
\begin{figure}[h]
\centerline{
\begin{picture}(200,80)
\put(15,5){\epsfysize7.5cm \epsfbox{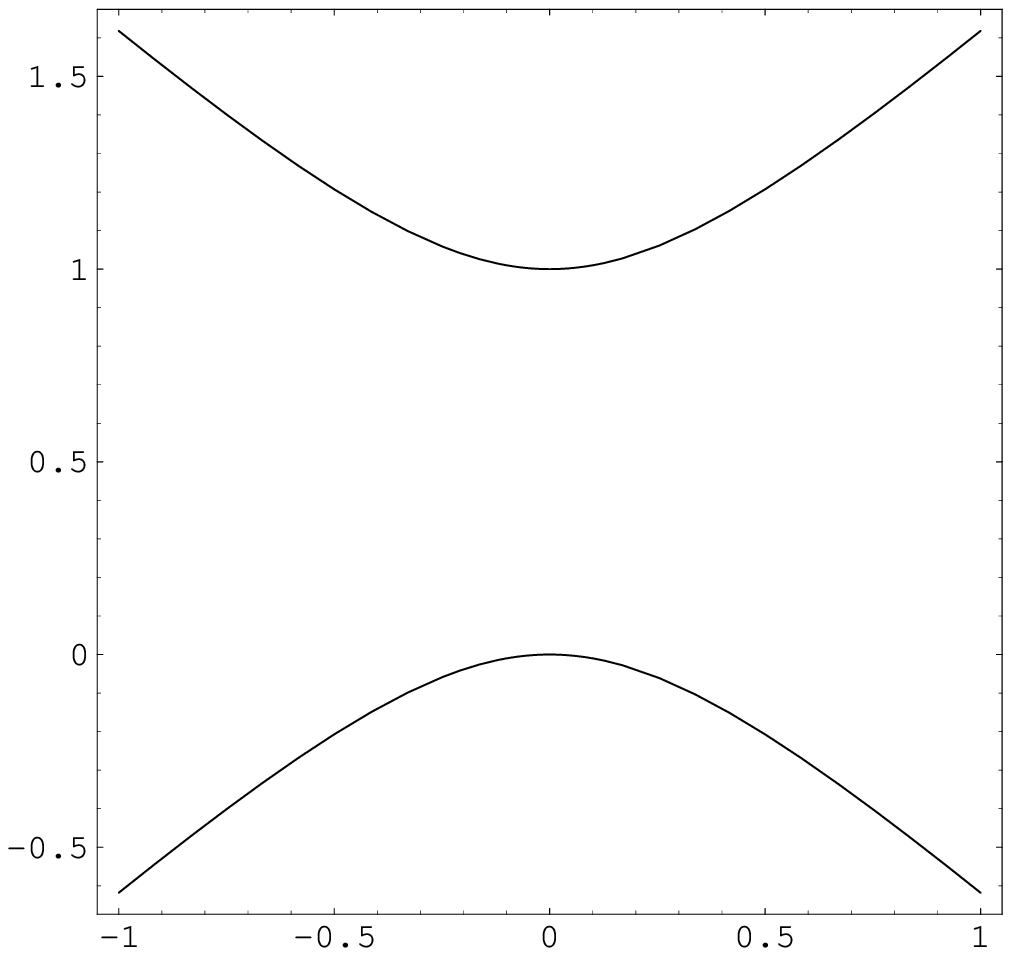}}
\put(53,3){$g$}
\put(9,37){\rotatebox{90}{$E_\pm(g)$}}
\put(105,0){\epsfysize8cm \epsfbox{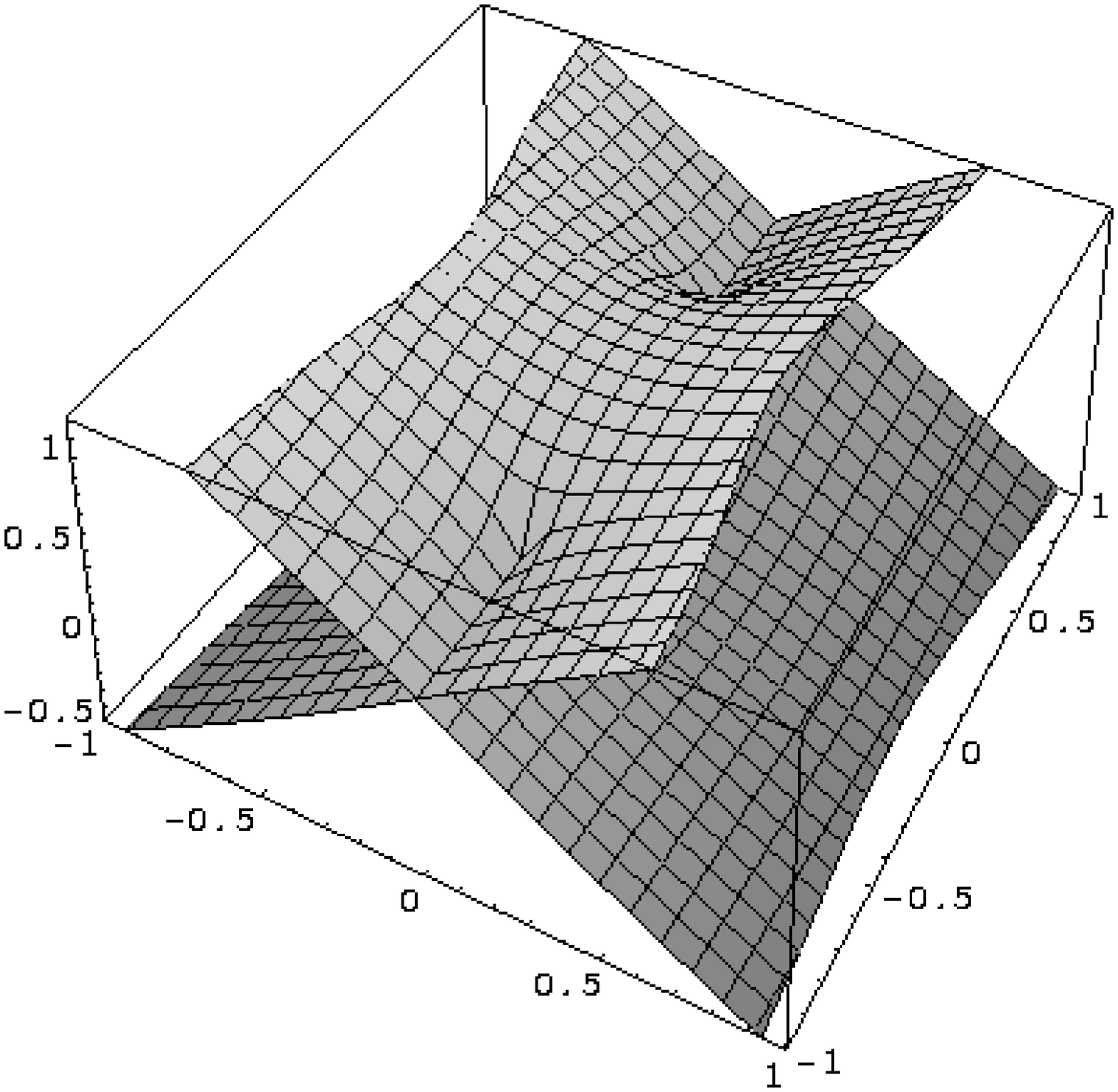}}
\put(123,13){$\RRe[g]$}
\put(178,23){$\IIm[g]$}
\put(99,37){\rotatebox{90}{$\RRe[E_\pm(g)]$}}
\end{picture}
}
\caption[Spectral surfaces for the two energy level model]
{Spectral surfaces for two energy level model.
On the left panel energy levels as a function of a real 
coupling constant $g$.
On the right panel real energy surfaces as function
of complex $g$}
\lab{fig:Epm}
\end{figure}
For the real $g$ the energy levels  $E_\pm(g)$ do not cross.
However, we have two branching points at 
$g_\pm=\pm\frac{i}{2}(\epsilon_1-\epsilon_2)$ for which
$E_+(g_{\pm})=E_-(g_{\pm})=\frac{\epsilon_1+\epsilon_2}{2}$.
The energy is a smooth two-valued function of complex $g$.
The spectral surfaces cross each other for $\RRe[g]=0$ where
$-i g >-i g_+$ or $-i g < -i g_-$.
Going around these branching points
we can pass from one surface to the other one.
In this case, 
the spectral curve, which is a function defined on 
one complicated Riemann surface, 
consists of  two branches $E_+(g)$ and $E_-(g)$.

One has to add that in this model the energy does not have any poles
which occur in our $SL(2,\mathbb{C})$ Heisenberg model.

\subsection{$SL(2,\mathbb{C})$ Heisenberg model for two Reggeons}

The energy for two Reggeons with nonnegative $n_h$ is 
defined by (\ref{eq:E2-anal}).
We notice that
analytical continuation of (\ref{eq:E2-anal})
exhibits the poles
at 
\begin{equation}
i \nu_h=\pm \frac{n-1}{2} \quad \mbox{with the twist} \quad n \ge 2+n_h.
\lab{eq:E2poles}
\end{equation}
The residua of these poles
are related to the anomalous dimensions given by (\ref{eq:sol-2}).
The leading twist $n=2$ corresponds to the pole at 
$\nu_h=-i/2$ with $n_h=0$.

The energy (\ref{eq:E2-anal}) does not have any branching points.
Thus, the spectral surfaces numbered by the Lorentz spin $n_h$ 
do not mix with each other in the complex $\nu_h-$plane.
The 
spectral surfaces are functions defined on 
trivial separated Riemann surfaces.

\subsection{$SL(2,\mathbb{C})$ Heisenberg model for more Reggeons}

For the case with more than two Reggeons 
the energy spectrum is much more complicated.
The energy $E_N(\nu_h;n_h,\mybf{\ell})$
depends not only on $\nu_h$ and $n_h$ but also
on $2(N-2)$ integer numbers 
$\mybf{\ell}=\{\ell_1,\ell_2,\ldots,\ell_{2(N-2)}\}$.
Similarly to the $N=2$ case the energy surfaces for $N>2$ possess poles 
for imaginary $\nu_h$.
Additionally, they also have branching points. 
Thus, the energy spectrum consists of
infinite number of spectral surfaces,
which are functions of complex $\nu_h$ with
complicated Riemann structure.

The energy in (\ref{eq:enQ2}) is a smooth function of
real $\nu_h$. In this region we have quantization conditions coming from
the single-valuedness of the Reggeon wave-function  as well as 
the normalization condition which ensure that Hamiltonian (\ref{eq:Ham})
is hermitian. The latter condition gives conjugation relations 
between the conformal charges 
\begin{equation}
q_k= {\wbar q_k}^{\ast} 
\quad
\mbox{with}
\quad
k=2,\ldots,N.
\lab{eq:qqb}
\end{equation}
Performing the analytical continuation into the complex $\nu_h-$plane
we have to relax the quantization conditions (\ref{eq:sjdag}).
The single-valuedness condition survives. However, the normalization condition 
is not valid anymore. Thus, the relations (\ref{eq:qqb}) in complex $\nu_h$ 
are not valid, so that, $q_k$ and $\wbar q_k$ may be independent.

We notice that the $N-$Reggeon spectrum contains the states 
with given $n_h$ and $\mybf{\ell}$ for which
\begin{equation}
\wbar q_k(\nu_h;n_h,\mybf{\ell})=\pm q_k(\nu_h;n_h,\mybf{\ell}).
\lab{eq:qqbpm}
\end{equation}
These states satisfy (\ref{eq:qqbpm}) on the whole complex $\nu_h-$plane.
They are of utmost interest because they 
give the higher order contribution 
to the structure function (\ref{eq:moments}).
Other interesting states are the descendent states.

For $N \ge 3$ the energy $E_N(\nu_h;n_h,\mybf{\ell})$ is 
a multi-valued function of $\nu_h$ parameterized by $n_h$ and $\mybf{\ell}$
which number the spectral surfaces. In the complex $\nu_h-$plane
the spectrum possesses poles at
\begin{equation}
i \nu_h=\pm \frac{n-1}{2} 
\quad 
\mbox{with the twist}
\quad 
n \ge N+ n_h\,.
\lab{eq:EpolN}
\end{equation}
It turns out that it also possesses the infinite set of
branching points which are square-like type similarly
to (\ref{eq:Epm}).
In order to construct the complex curves
$E_N(\nu;n_h,\mybf{\ell})$
we apply Eqs. (\ref{eq:enQ2})
developed in Refs. \ci{Derkachov:2001yn,Derkachov:2002wz}.
After analytical continuation we get
\begin{equation}
E_N=\left[\varepsilon(h,q)+\varepsilon(h,-q)+\lr{\varepsilon(1-\bar
h^*,\bar q^*)}^* +\lr{\varepsilon(1-\bar h^*,-\bar q^*)}^*\right],
\lab{eq:E-epsilon}
\end{equation}
where the $SL(2,\mathbb{C})$ spins $h$ and $\bar h$ are given by 
(\ref{eq:hpar}) and the conformal charges are denoted as 
$q=\{q_k\}$, $\wbar q=\{\wbar q_k\}$, 
$-q=\{(-1)^kq_k\}$ and $-\wbar q=\{(-1)^k\wbar q_k\}$.
As one can see $E_N(-q,-\wbar q)=E_N(q,\wbar q)$.
Since for real $\nu_h$ we have $h=1-{\wbar h}^{\ast}$ and
$q={\wbar q}^{\ast}$  the energy $E_N(q,\wbar q)$ takes real values on 
the real $\nu_h-$axis.

The function $\varepsilon(h,q)$ in (\ref{eq:E-epsilon})
is defined for arbitrary complex $h$ and $q$ by
\begin{equation}
\varepsilon(h,q)=i\frac{d}{d\epsilon}\ln \left[\epsilon^N
Q(i+\epsilon;h,q)\right]\bigg|_{\epsilon=0}\,,
\lab{eq:e}
\end{equation}
where $Q(u;h,q)$ has a form
\begin{equation}
Q(u;h,q)=
\int_0^1 {dz}\, z^{iu-1} Q_1(z)\,.
\lab{eq:ans}
\end{equation}

Going further, $Q_1(z)$
satisfies the differential equation (\ref{eq:Eq-1})
which comes from the Baxter equation (\ref{eq:Baxeq}).
Around $z=1$ the function $Q_1(z)$ corresponds to 
one of the solutions in 
(\ref{eq:set-1}) with asymptotics $Q_1(z)\sim (1-z)^{(-h-1)}$
and around $z=0$ from (\ref{eq:Q-0-h})
with $Q_1(z)\sim \ln^{N-1}z$.
It is convenient to normalize $Q(u;h,q)$ as
\begin{equation}
Q(i+\epsilon;h,q)=\frac1{\epsilon^N}-i\frac{\varepsilon(h,q)}{\epsilon^{N-1}}+
\mathcal{O}\left(\frac1{\epsilon^{N-2}}\right)\,.
\lab{eq:norm1}
\end{equation}
where the function  $\varepsilon(h,q)$  appears in (\ref{eq:E-epsilon}).

Thus, solving the quantization conditions (\ref{eq:C1-C0}) and 
using (\ref{eq:enQ2}) and (\ref{eq:E-epsilon}) 
we are able to evaluate the energy $E_N(\nu_h)$ on the whole complex
$\nu_h-$plane. Then, we find the poles which appear only for
$\RRe[\nu_h]=0$, and we evaluate expansion coefficients of $E_N(\nu_h)$
around its poles.
Next, using (\ref{eq:Enpoles}) and (\ref{eq:gampol}) we 
the calculate expansion coefficients of the anomalous dimensions (\ref{eq:gampol})
in the strong coupling constant $\alpha_s$.
These numerical results will be presented in the next sections.

\section{Spectral surfaces}

The two-Reggeon energy spectrum is described by the analytical formula 
(\ref{eq:E2-anal}). It is a meromorphic function of $\nu_h$ and depends
on the integer conformal Lorentz spin $n_h$. It possesses poles at 
(\ref{eq:E2poles}).

For higher $N\ge 3$
we do not have an analytical formula, so we have to calculate 
the spectrum numerically using Eqs. (\ref{eq:E-epsilon})--(\ref{eq:norm1}).
The energy $E_N(\nu_h,n_h,\ell)$ is a multi-valued function and its
branches are numerated by Lorentz spin $n_h$ and integer parameters 
$\mybf{\ell}=\{\ell_1,\ell_2,\ldots,\ell_{2(N-2)}\}$.
These spectra, comparing to the $N=2$ case,
except the  poles at (\ref{eq:EpolN}),
have additionally an infinite set of branching points at $\nu_h^{{\rm br},k}=\nu_h$ 
which allows us to glue some branches
along their cuts.
This causes that the spectrum consists of a set of
complicated spectral curves. Thus, we consider
the energy $E_N(\nu_h;n_h,\mybf{\ell})$ as a multi-valued function 
on complicated Riemann 
surfaces. The states from a given Riemann surface have the same quantum 
numbers, i.e. Lorentz spin $n_h$, $C-$parity, quasimomentum, Bose symmetry. 
These numbers impose additional rules for parameters $\mybf{\ell}$
which describe a single spectral curve on one Riemann surface.

\psfrag{E_N/4}[cc][cc]{$E_3/4$} \psfrag{Im_nu_h}[cc][bc]{$\mbox{\large$i\nu_h$}$}
\psfrag{Re_nu_h}[cc][bc]{$\mbox{\large$\nu_h$}$}
\begin{figure}[!h]
\vspace*{3mm} \centerline{{\epsfysize6cm \epsfbox{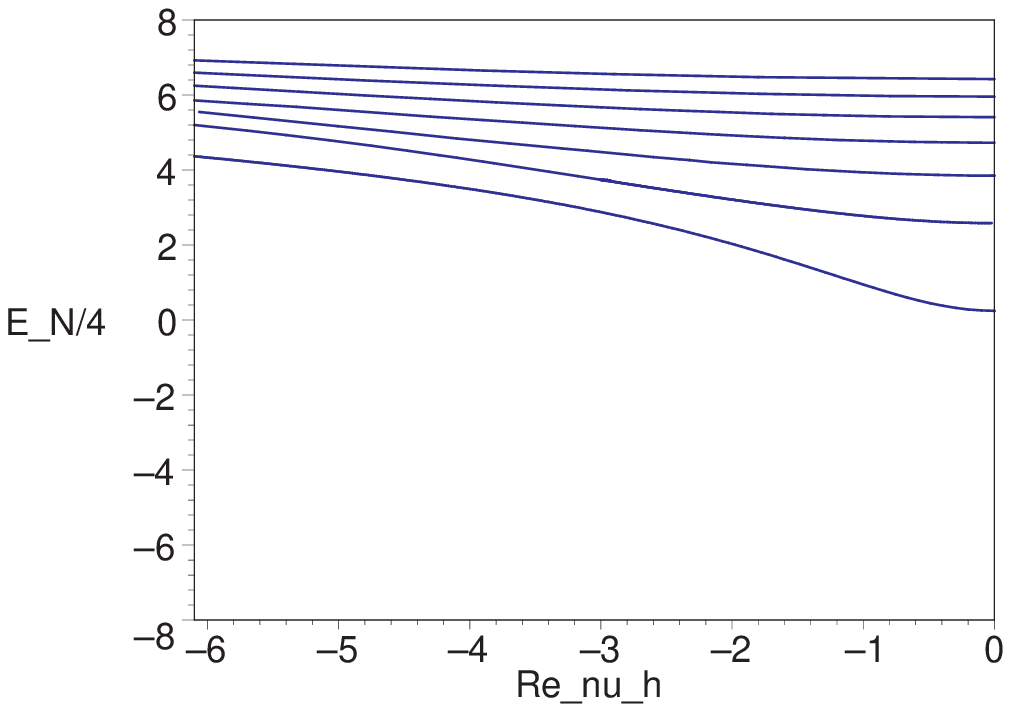}}{\epsfysize6cm
\epsfbox{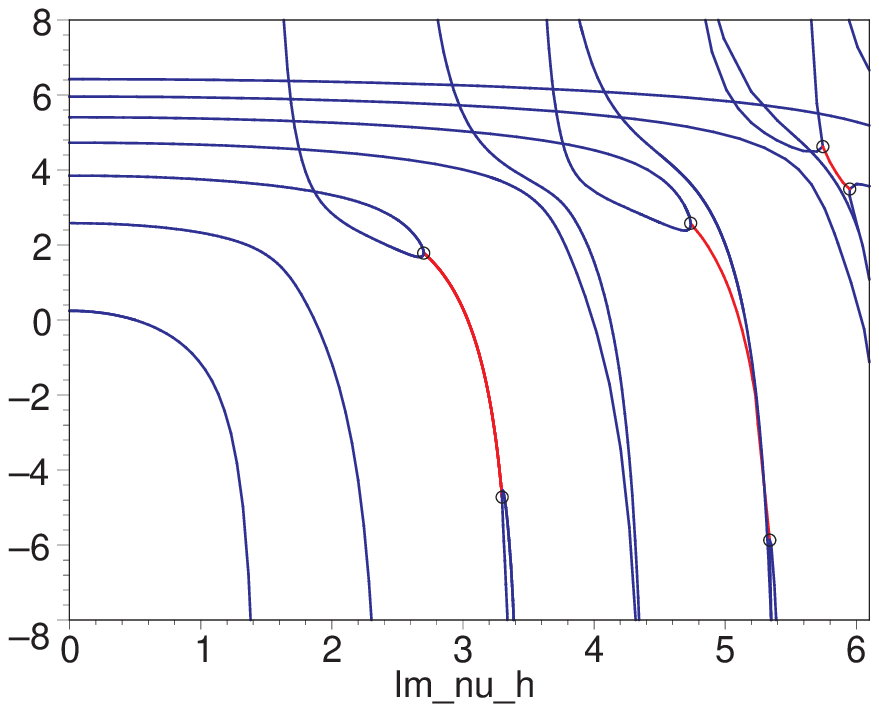}}}
%
%
%
\caption[The energy spectrum  of the $N=3$ Reggeon states $E_3(\nu_h;n_h,\Mybf{\ell})$
for $n_h=0$]
{The energy spectrum  of the $N=3$ Reggeon states $E_3(\nu_h;n_h,\Mybf{\ell})$
for $n_h=0$ and $\Mybf{\ell}=(0,\ell_2)$, with $\ell_2=2,\,4,\,\ldots,14$ from
the bottom to the top (on the left). Analytical continuation of the energy along
the imaginary $\nu_h-$axis  (on the right). The branching points are indicated by
open circles. The lines connecting the branching points represent $\Re E_3$.}
\lab{fig:E-comp}
\end{figure}

Let us consider the $N=3$ Reggeon state along the 
spectral curves labelled with 
$\ell_1=0$ and $\ell_2=$even positive numbers. This gives condition
$\wbar q_k+ q_k=0$. An example of these states is presented in Figure 
\ref{fig:E-comp}.
On the left panel we plot the energy
levels along the real $\nu_h-$axis whereas on the right panel
we continue them analytically. 
Here we show the states for imaginary $\nu_h$ only.

For real $\nu_h$ the energy is real and 
the energy levels are smooth functions of real $\nu_h$. 
Even though that we have only one slice of the spectral curve we
are able to observe that,
firstly,
we have poles at 
$i\nu_h^{\rm pole}=3/2,\, 5/2,\, 7/2,\,9/2,\,11/2$ and, 
secondly, different energy
levels collide at
$i\nu_h^{\rm br}=
    2.70
,\, 3.29
,\, 4.73
,\, 5.34
,\, 5.74
,\, 5.94
$.
which are denoted by circles in Fig.\ \ref{fig:E-comp}.
Moreover, we have also branching points at complex 
$i\nu_h^{\rm br}=1.723+i\,.248$ where the ground state
$\ell_2=2$ collide with the next state $\ell_2=4$. 
Thus, all these energy levels 
are glued in the complex $\nu_h-$plane and they form 
one multi-valued spectral surface.

The complex branching point values $\nu_h^{\rm br}$
imply that the contribution of the corresponding square-root cuts to the 
integral (\ref{eq:moments2}) scales as  
$1/[Q^{1+2i\nu_h^{\rm br}}\ln ^{3/2} Q]$ and,
therefore, it breaks the OPE expansion (\ref{eq:Fope}). 
Although such corrections
are present in \re{eq:F-fin} for given $n_h$ and $\Mybf{\ell}$, 
they cancel against
each other in the sum over all states \ci{Korchemsky:2003rc}.

Another interesting example of spectral surfaces are $N=3$
Reggeon states with the Lorentz spin $n_h=1$ and $q_3=\wbar q_3=0$.
They are descendent states and their energy levels coincide
with the levels for $N=2$ Reggeon states.
Performing analytical continuation we notice that 
$q_3(\nu_h)=\wbar q_3(\nu_h)=0$ and $E_{3,d}(\nu_h)=E_2(\nu_h,1)$
on the whole complex $\nu_h-$plane, where $E_2(\nu_h;1)$ is defined by 
(\ref{eq:E2-anal}).

Therefore, the energy of this state $E_{3,d}(\nu_h)$ 
is  a single-valued meromorphic function on the complex $\nu_h-$plane.
At $\nu_h=0$ the energy $E_{3,d}(\nu_h)$ vanishes and at this point we have 
the ``physical'' ground state for the system of $N=3$ Reggeons.
Similarly to the $N=2$ case, the descendent states have poles at 
(\ref{eq:E2poles}).

In the above example of the $N=3$ Reggeon states
we have presented two special states
$E_3(\nu_h=0,\nu_h=1)=0$ and
$E_3(\nu_h=0,\nu_h=0)/4=0.24717$, 
which are
considered as the ground states for the odderon in QCD
with the intercept
$j_\mathbb{O}=1-\bar\alpha_s E_3$~\cite{Janik:1998xj,Bartels:1999yt}.
We notice that the minimal twist corresponds to the position of the
energy pole in the complex
$\nu_h-$plane which is the nearest to the origin, $q_k=\wbar q_k=0$.

Thus, for $n_h=1$ we have the minimal twist $n_{min}=3$ whereas 
for $n_h=0$
we have the minimal twist $n_{min}=4$. Therefore, the leading contribution 
to the structure function (\ref{eq:moments}) 
with the intercepts $j_\mathbb{O}=1$ and
$j_\mathbb{O}<1$ scales at large $Q^2$ as $\sim 1/Q^3$ and $\sim 1/Q^4$,
respectively.

\section{Energy poles and the anomalous dimensions}

In order to evaluate the anomalous dimensions of $N$ Reggeon states 
around 
$\gamma_a(\alpha_s=0)=0$ we work out the Laurent expansion of the energy 
function around its poles (\ref{eq:Enpoles}).
As we said before the energy possesses the poles at 
$i \nu_h=(n-1)/2$ with the twist $n \ge N+n_h$
and the Lorentz spin $n_h$.

Firstly, let us consider the descendent states for $N=3$ Reggeons
with $n_h=1$. They are related to $N=2$ Reggeon states, so they have
equal energies $E_{3,d}(\nu_h)=E_2(\nu_h,1)$
defined in (\ref{eq:E2-anal}).
The pole which is the closest to the origin is located at $i \nu_h=1$.
Thus, $(h,\wbar h)=(2,1)$ and the twist $n=h+\wbar h=3$.
Expanding (\ref{eq:E2-anal}) around this pole we get
\begin{equation}
E_{3,\rm d}(2+\epsilon)=4\left(\frac1{\epsilon}
+1-\epsilon -\left( 2\zeta(3) -1
\right) {\epsilon}^{2}+\ldots\right)\,.
\lab{eq:E3d}
\end{equation}
Next, using (\ref{eq:E3d}) and (\ref{eq:Enpoles}) with (\ref{eq:gampol})
we find
the twist-3 anomalous dimension corresponding to the odderon state with the
intercept $j_\mathbb{O}=1$ as 
\begin{equation}
\gamma_{3}^{(N=3)}(j) = \frac{\bar\alpha_s}{j-1}
-\lr{\frac{\bar\alpha_s}{j-1}}^2
+(2\zeta(3)+1)\lr{\frac{\bar\alpha_s}{j-1}}^4+\ldots\,,
\lab{eq:gamma3}
\end{equation}
where the subscript and the superscript 
denote the twist and the number of Reggeons in the given state, respectively.
 
For the $N=3$ case with $q_3 \ne0$ we do not have analytical 
formula for the energy. However, applying 
(\ref{eq:E-epsilon})- (\ref{eq:norm1}) we are able 
to fit numerically the expansion coefficients of the energy around its poles.

Let us consider the spectral surfaces with $n_h=0$, which we show in Fig.\ 
\ref{fig:E-comp}. Applying (\ref{eq:E-epsilon})- (\ref{eq:norm1}),
we expand the energy in the vicinity of the poles at 
$h=1/2+i \nu_h=2,3,4,5$
with $E_3=E_3(h+\epsilon)$:
\begin{eqnarray}
E_3(2+\epsilon)&=& 4\left({\epsilon}^{-1} + \frac12 -\frac12\ \epsilon + 1.7021\,
\epsilon^2+\ldots\right)\,,
\nonumber
\end{eqnarray}
\begin{eqnarray}
E_3(3+\epsilon)&=& 4\left( {2}\,{\epsilon}^{-1} + {\frac {15}{8}} - 1.6172\ \epsilon +
 0.719\ {\epsilon}^{2}+\ldots \right)\,,
\nonumber 
\end{eqnarray}
\begin{eqnarray}
E_3^{\rm(a)}(4+\epsilon)&=& 4\left( {\epsilon}^{-1} + \frac{11}{12} - 0.6806\
\epsilon - 1.966\ \epsilon^2+\ldots \right)\,,
\nonumber 
\end{eqnarray}
\begin{eqnarray}
E_3^{\rm(b)}(4+\epsilon)&=& 4\left({2}\,{\epsilon}^{-1}+\frac{15}4-3.2187\ \epsilon
 + 3.430\ \epsilon^2+\ldots \right)\,,
\nonumber 
\end{eqnarray}
\begin{eqnarray}
E_3^{\rm(a)}(5+\epsilon)&=& 4\left(2\,{\epsilon}^{-1} + \frac{125}{48} - 2.0687\
\epsilon + 1.047\ \epsilon^2+\ldots \right)\,,
\nonumber 
\end{eqnarray}
\begin{eqnarray}
E_3^{\rm(b)}(5+\epsilon)&=& 4\left({2}\,{\epsilon}^{-1} + \frac{53}{12} - 2.4225\
\epsilon + 0.247\ {\epsilon}^{2}+\ldots \right)\,.
\lab{eq:res-N3}
\end{eqnarray}
Here $\epsilon \to 0$ and 
ellipses denote $\mathcal{O}(\epsilon^3)$ terms.
Since the energy is defined on complicated Riemann 
surfaces, therefore for one Riemann surface
at the same given $\nu_h$ we have many branches
and at some given $\nu_h$ we may have many poles.
Thus, we introduce the superscripts $(a)$ and $(b)$ to 
distinguish different branches of spectral curves 
as well as different poles at the same $\nu_h$.

In this case the nearest pole is at $\nu_h=3/2$.
Therefore, the minimal twist equals $n=h +\wbar h=4$.
Now applying 
(\ref{eq:Enpoles}) and (\ref{eq:gampol}) we find from the first 
relation in (\ref{eq:res-N3}) the anomalous dimension corresponding to
the odderon state with the intercept
$j_\mathbb{O}<1$ as
\begin{equation}
\gamma_4^{(N=3)}(j) =
\frac{\bar\alpha_s}{j-1}-\frac12\lr{\frac{\bar\alpha_s}{j-1}}^2
-\frac14\lr{\frac{\bar\alpha_s}{j-1}}^3-1.0771
\lr{\frac{\bar\alpha_s}{j-1}}^4+\ldots\,.
\lab{eq:gamma3-1}
\end{equation}
Similarly, we may calculate the other anomalous dimensions form 
(\ref{eq:res-N3}).
However, to save the space
we do not present them here.

All expansion series formulae for the energies around its poles may be 
written as 
\begin{equation}
E_3(h+\epsilon) =4\left(\frac{R}{\epsilon} + \gamma(h)+\mathcal{O}(\epsilon)
\right)\,,
\lab{eq:E3}
\end{equation}
where the parameter $R$ is integer and it is equal to $1$ or $2$. 
Moreover, the function $\gamma(h)$ is a rational number and 
is related to the energy of 
the Heisenberg $SL(2,\mathbb{R})$ magnet model.
More details 
concerning the properties of $\gamma(h)$
can be
found in Refs.
\ci{Braun:1999te,Belitsky:1999ru}. 
Because it is not the topic of this thesis
we omit this interesting subject.

Now, let us consider cases with $N\ge4$ Reggeon states. 
For even $N$ the main contributions comes from the sector  with 
$n_h=0$. The closest poles to the origin are located at 
$i \nu_h=(N-1)/2$ so the lowest twist $n=N$.
These states build Pomeron states.
For instance, the expansion series of the energy for $N=4$
and $N=6$ states which give the highest contribution look as follows
\begin{eqnarray}
E_4(2+\epsilon)&=& 4\left(\frac2{\epsilon} + 1 - \frac12\,\epsilon -
1.2021\,{\epsilon}^{ 2} + \ldots\right)\,,
\nonumber
\\
E_6(3+\epsilon)&=& 4\left(\frac4{\epsilon} + \frac32
-
\frac7{16}
\,\epsilon-0.238\,\epsilon^2+\ldots\right)\,.
\lab{eq:E46-pol}
\end{eqnarray}
Applying Eqs.~(\ref{eq:Enpoles}) and (\ref{eq:gampol}), 
we obtain the twist$-N$ anomalous dimensions of the $N-$Reggeon
states
\begin{eqnarray}
\gamma_4^{(N=4)}(j)&=&2\frac{\bar\alpha_s}{j-1}+
2\lr{\frac{\bar\alpha_s}{j-1}}^2
-13.6168\lr{\frac{\bar\alpha_s}{j-1}}^4+\ldots \nonumber\,,
\\
\gamma_6^{(N=6)}(j)&=&4\,\frac{\bar\alpha_s}{j-1}+
6\,\lr{\frac{\bar\alpha_s}{j-1}}^2 +2\,\lr{\frac{\bar\alpha_s}{j-1}}^3-
33.23\,\lr{\frac{\bar\alpha_s}{j-1}}^4+\ldots\,.
\lab{eq:gamma46}
\end{eqnarray}
The higher corrections are large and grow with $N$ so the expansion
series (\ref{eq:E46-pol}) has finite radius of convergence and 
its value decreases with $N$.
This agrees with the fact 
that the intercept of $N-$Reggeon state scales at large $N$ as
$|j_N-1|\sim 1/N$~\cite{Derkachov:2002wz}. 

In the case with odd $N$, we consider two interesting sectors: with 
$n_h=0$ and $n_h=1$. In the first sector the pole closest to the origin
is located at $i \nu_h=N/2$. These gives the minimal twist $n=N+1$.
For $N=3$ we have written expression for the energies 
and the anomalous dimensions
in (\ref{eq:res-N3}) and (\ref{eq:gamma3-1}).
For $N=5$ they look like
\begin{eqnarray}
E_5(3+\epsilon) &=&4\left( \frac3{\epsilon} +  \frac76  - 
\frac{101}{216}\ \epsilon - 0.1136 \
\epsilon^2 +\ldots\,.\right)\,,
\lab{E5-pol}
\\
\gamma_6^{(N=5)}(j) &=& 3\,\frac{\bar\alpha_s}{j-1}
+\frac72\lr{\frac{\bar\alpha_s}{j-1}}^2-
\frac18
\lr{\frac{\bar\alpha_s}{j-1}}^3
-13.032\lr{\frac{\bar\alpha_s}{j-1}}^4+\ldots\,.
\nonumber
\end{eqnarray}
These states possess the odderon quantum numbers and have intercepts 
$j_\mathbb{O}=1-\bar\alpha_s E^{\rm(br)}_5$ lower than $1$. 

In the second sector, with odd $N$ and $n_h=1$ the minimal twists
are smaller and they are equal to $n=N$.
It corresponds to the poles of the energy at $i \nu_h=(N-1)/2$.
At $N=5$ we have the Laurent expansions as follows
\begin{eqnarray}
&&E_{5,\rm d}(3+\epsilon)= 4 \left(
\frac{3+\sqrt{5}}{2\epsilon} + 1.36180 - 0.4349 \ \epsilon - 0.315\ \epsilon^2 +\ldots \right)\,,
\lab{eq:pole-des-5}
\\
&&\gamma_5^{(N=5)}(j)= \frac{3+\sqrt{5}}2\frac{\bar\alpha_s}{j-1}
+3.56524\lr{\frac{\bar\alpha_s}{j-1}}^2
+1.8743\lr{\frac{\bar\alpha_s}{j-1}}^3
-11.219\lr{\frac{\bar\alpha_s}{j-1}}^4+\ldots\,. \nonumber
\end{eqnarray}
For $N=3$ the results are shown in (\ref{eq:E3d})--(\ref{eq:gamma3}).
The poles belong to the same spectral surfaces that contain the ground states
for given odd number of particles $N$.
Because, these ground states are descendent ones,
they have minimum energy $E_N=0$, the intercept 
$j_\mathbb{O}=1$ and finally they have the odderon quantum numbers.

\chapter{Summary and Conclusions}

In this work we have considered the scattering processes in the Regge limit
(\ref{eq:rlim}) where the compound reggeized gluon states, i.e. Reggeons,
propagate in the $t-$channel and interact with each other.
We have performed calculations in the generalized leading logarithm
approximation (GLLA)
\ci{Bartels:1980pe,Kwiecinski:1980wb,Jaroszewicz:1980mq}, in which 
a number of Reggeons in the $t-$channel
is constant. In this approach the
reggeized gluon states appear in two different kinematical regions.
Thus, we have considered both of them, i.e.
the elastic scattering processes (\ref{eq:scatpr}) 
as well as the deep inelastic scattering processes (\ref{eq:DISlim}).
We attempted to find a scattering amplitude of the above processes 
with multi-Reggeon exchange. However, a structure of reggeized gluon states
as well as their properties have turned out to be so reach, 
complicated and interesting that 
in this work we have focused on
description of the Reggeon state properties as well as
on analysing the spectra of the energy and integrals of motion.

In order to simplify the problem we have applied the multi-colour limit 
 \ci{'tHooft:1973jz}, which makes the $N-$Reggeon system
$SL(2,\mathbb{C})$  (\ref{eq:HHam}) 
symmetric  (\ref{eq:trcoords}) and completely integrable. 
In this limit the equation for  the $N-$Reggeon wave-function
takes a form of Schr\"odinger equation (\ref{eq:Schr}) for
the non-compact XXX Heisenberg magnet model
of $SL(2,\mathbb{C})$ spins $s$
\ci{Takhtajan:1979iv,Faddeev:1979gh,KBI}. 
Its Hamiltonian describes the nearest neighbour interaction
of the Reggeons \ci{Lipatov:1993yb,Faddeev:1994zg} 
propagating in the two-dimensional transverse-coordinates space 
(\ref{eq:ki}).
The system has 
a hidden cyclic and mirror permutation symmetry (\ref{eq:PMsym}). 
It also possesses
the set of the $(N-1)$ of integrals of motion, which 
are eigenvalues of conformal charges  \ci{Derkachov:2001yn},
$\oq{k}$ and $\oqb{k}$,
(\ref{eq:qks0})--(\ref{eq:qks1}). Therefore, the operators of
conformal charges commute with each other and with
the Hamiltonian and they possess a common set of
the eigenstates.

Eigenvalues of the lowest conformal charge, $q_2$,
may be parameterized (\ref{eq:q2}) by the complex spins $s$
and the conformal weight $h$, where $h$ can be expressed by
the integer Lorentz spin $n_h$ and
the real parameter $\nu_h$ related to the scaling dimension (\ref{eq:hpar}).
Solving the eigenequation for $q_2$
we have derived ansatzes for $N-$Reggeons states with an arbitrary 
number of Reggeons $N$ as well as
arbitrary complex spins $s$ (\ref{eq:nNpsiz0}).
Since the $N=3$ Reggeon ansatz separates variables,
the $q_3-$eigenequation can be rewritten as a 
differential equation of a Fuchsian type
with three singular points (\ref{eq:ees0z32}) \ci{Janik:1998xj}.
We have solved this equation by a series method.
Gluing solutions for different singular points,
(\ref{eq:u-zero}), (\ref{eq:u-one}) and (\ref{eq:u-inf}),
and taking care for normalization and single-valuedness
of the Reggeon wave-function
(\ref{eq:Psizz}) we have obtained 
the quantization conditions 
(\ref{eq:azao})--(\ref{eq:aoai})
for integrals of motion 
$(q_2,\wbar q_2,q_3,\wbar q_3)$
which we have solved numerically \ci{Derkachov:2002wz,Kotanski:2001iq}.
For $q_3=\wbar q_3=0$ 
the series solutions have a simple form
and we are able to resum them.
Thus, we have obtained analytical expressions for the 
three-Reggeon wave-functions with $q_3=0$
(\ref{eq:Psib})--(\ref{eq:Psiq0b}).
For higher $N$ we obtained a set of $N-2$
non-separable differential equations
which are hard to solve even numerically.
We have presented explicitly such equations for $N=4$
(\ref{eq:N4eqt})--(\ref{eq:N4eqf}).

In order to solve the Reggeon problem for more than three particles
one can 
make use of the more sophisticated
technique, i.e. the Baxter $Q-$operator method 
\ci{Derkachov:2001yn}.
It relies on the existence of the operator $\mathbb{Q}(u,\wbar{u})$
depending on the pair of complex spectral parameters 
$u$ and $\wbar u$. The Baxter $Q-$operator 
has to commute with itself (\ref{eq:comQQ}) and with 
the conformal charges (\ref{eq:comQt}). It also has to satisfy
the Baxter equations
(\ref{eq:Baxeq})--(\ref{eq:Baxbeq}).
Furthermore, the $Q-$operator has
well defined analytical properties, i.e. known pole structure 
(\ref{eq:upoles}) 
and asymptotic behaviour at infinity (\ref{eq:Qanalb}).
The above conditions fix the $Q-$operator
uniquely and allow us to quantize the integrals $q_k$.
In turns out \ci{Derkachov:2001yn} 
that the Reggeon Hamiltonian 
can be rewritten in terms of 
Baxter $Q-$operator (\ref{eq:HNQQ}).
Therefore, combining together the solutions 
of the Baxter equations and the conditions for $q_k$
with the Schr\"odinger equation we can calculate the 
energy spectrum (\ref{eq:energy}). Moreover,
we are able to determine the quasimomentum of the
eigenstates (\ref{eq:quasQ}), i.e.
the observable which defines the properties of the
state with respect 
to the cyclic permutation symmetry (\ref{eq:PMsym}).

One can solve the Baxter equations (\ref{eq:Baxeq})--(\ref{eq:Baxbeq})
using the quasi-classical approach (\ref{eq:WKB}).
Following Refs. \ci{Derkachov:2002pb}
we have presented 
the WKB approximation 
for the quantized values of 
$q_k$ (\ref{eq:ukcond})
and explained the structure of their spectrum (\ref{eq:qNqnt}).
Applying this approximation, one introduces a small parameter $\eta$
and  performs expansion of the Baxter function for 
large values of $q_k$.
The single-valuedness requirement of the Baxter function
gives 
conditions for $q_k$ in a form 
of Bohr-Sommerfeld integrals (\ref{eq:sqcond}).
Adding the condition coming from the possible values
of the quasimomentum (\ref{eq:quasas}) one obtains
the relations between the quantized values of $q_k$ (\ref{eq:ukcond}).
However, not all quantization conditions can be obtained 
from (\ref{eq:ukcond}).
In turns
out that 
in the leading order of 
the WKB approximation
for a given value of $\nu_h$
the quantized values of 
$q_N^{1/N}$
correspond to vertices of equilateral lattices (\ref{eq:qNqnt}).
For lower conformal charges the conditions are underdetermined,
i.e. in the $N=4$ case one has only one of two required conditions 
(\ref{eq:q3-quan}).

In order to obtain the exact values of $q_k$
we have used the method \ci{Derkachov:2002pb,Derkachov:2002wz}
which makes it possible
to calculate numerically
the spectrum of the conformal charges as well as other
observables, i.e. the energy (\ref{eq:E-fin})
and the quasimomentum (\ref{eq:parity-qc}). 
The exact method consists in rewriting the
Baxter equations and the other conditions for 
$Q-$operator eigenvalues as the $N-$order
differential equation (\ref{eq:Eq-1}).
This equation can be solved similarly to 
the $q_3-$eigenequation (\ref{eq:ees0z32}). 
Thus, one obtains the conditions 
for quantized $q_k$ (\ref{eq:C1-C0}) and 
the formulae for the energy (\ref{eq:E-fin}) 
and the quasimomentum (\ref{eq:parity-qc}).

We have applied this exact method
for states with $N=2,\ldots,8$ particles.
The quantum values of $q_k$ form
continuous trajectories, Fig.\ \ref{fig:traj-3D}. This fact is related
to the dependence of $q_2$ on the real parameter $\nu_h$ (\ref{eq:q2}). 
The energy is a continuous function of $\nu_h$
so the energy spectrum is gapless. 
This energy continuity 
leads to additional logarithmic dependence of the scattering amplitude
(\ref{eq:amp}) on the total energy $s$.
The energy 
changes continuously along the trajectory (\ref{eq:Enu}) and
reaches the lowest value in the vicinity of $\nu_h=0$. 
Moreover, since the $q_k$ spectrum 
possesses symmetries, (\ref{eq:qkmsym}) and (\ref{eq:qkcsym}),
the energy for some cases is degenerated.

For $N=3$ we have calculated the behaviour of the 
$q_3-$spectrum
for the 
conformal Lorentz spins 
$n_h=0,1,2,3$
and  the scaling dimension $1+2 i \nu_h$.
Some results for $n_h>0$
were presented before in 
Ref. \ci{DeVega:2001pu} for $n_h=1$ and
Refs. \ci{Kotanski:2001iq} for $n_h=3$.
The quantized values of $q_3^{1/3}$ 
for given $n_h$ and fixed $\nu_h$
exhibit the WKB lattice structure (\ref{eq:qNqnt}), which
for N=3 takes a form an equilateral-triangle lattice (\ref{eq:q3-WKB}),
Figs.\ \ref{fig:N3q0}-\ref{fig:prtraj2}.
The non-leading WKB corrections
move the quantized value of $q_3$ away from the lattice and
cause that
that the quantized values of $q_3$
lie outside a disk located around
the origin of the lattice, 
i.e. near $q_3=0$.
However, for odd $n_h$ there are
states with $q_3=0$. They are called 
descendent states because their wave-function is 
effectively built of 
$N=2$ Reggeon states. 
The non-descendent state with the lowest energy
belongs to $n_h=0$ sector and its energy is positive
(\ref{eq:N3-ground}).
This state is double-degenerated and it
appears to be the nearest one to the origin of the $q_3^{1/3}$ lattice.
However, the ground state for $N=3$ is the descendent one with $n_h=1$
and energy $E_3=0$ (\ref{eq:EN=2}).
Having found the exact values of $q_3$
we are able to calculate corrections 
to the WKB approximation (\ref{eq:bcoef}),
Table \ref{tab:q3coefs}. 
These corrections differ from the corrections obtained earlier
in Ref. \ci{Derkachov:2002pb}.
The difference seems to be caused by
using only one expansion parameter $\eta$ for
two various conformal charges, $q_2$ and $q_3$.
The obtained corrections are subleading to
the WKB approximation \ci{Derkachov:2002pb}
which is an expansion for large values of
conformal charges, i.e. $1 \ll |q_2^{1/2}| \ll |q_3^{1/3}|$.

In the case with $N=4$ particles we have constructed
the spectrum for $n_h=0$
depicting complicated interplay between
the conformal charges: $q_3$ and $q_4$.
Such a complete analysis has been performed here for the first time. 
Earlier, for $N=4$ full spectrum of $q_4$ was shown in 
Ref. \ci{Derkachov:2002wz}
(however the corresponding $q_3$ spectrum was not discussed)
and some values of $q_4$ were found in Ref. \ci{DeVega:2001pu}.
The spectrum of $q_4^{1/4}$ has a structure of square-like lattice 
(\ref{eq:WKB-N4}), Fig.\ \ref{fig:Q4}.
In this case the spectrum of $q_3$ is more complicated. It turns
out that it has a few possible forms. 
Firstly, there are simple states with $q_3=0$, Fig.\ \ref{fig:WKB-N4}.
Moreover, we have found 
the $q_3^{1/2}-$lattices whose distribution of vertices
is similar to the distribution of vertices of 
$q_4^{1/4}-$lattice. We have called this structures as resemblant lattices,
Figs \ref{fig:l30k01}-\ref{fig:l30k22}.
The next $q_3^{1/2}-$lattices are called winding lattices.
They wind around the origin and in course of this winding the distance
between vertices increases and also the lattice goes away from the origin,
Figs.\ \ref{fig:l31k31}-\ref{fig:l32k2}.
The WKB approximation does not describe these structures (\ref{eq:q3-quan})
exactly because
one quantization condition is missing.
Finally, we have also considered the descendent states 
with $q_4=0$, Fig.\ \ref{fig:WKB-N4}.
They have quasimomentum $\theta_4=\pi$ and the structure
of their $q_3^{1/3}-$lattices is the same as for
$q_3^{1/3}-$lattices  (\ref{eq:q3-WKB})
in the three-Reggeon case with $\theta_4=0$. 
For the $N=4$ Reggeon states with $q_3=0$
we have calculated corrections to the WKB approximation 
(\ref{eq:korq4}), Table \ref{tab:q4coefs}.
Similarly to the $N=3$ case they come from the fact 
of using only one expansion parameter $\eta$ 
for two different conformal charges, $q_4$ and $q_2$.
One has to notice that these corrections are subleading
in comparison to the WKB limit \ci{Derkachov:2002pb}.

For $N>4$ we have more conformal charges so
their spectrum is  more complicated.
Therefore, in these cases we have been concentrating
only on the ground states, Table \ref{tab:Summary}.
Thus, we have shown the dependence of 
the ground state energy as a function of particle number
$N=2,\ldots,8$  (\ref{eq:ENod})--(\ref{eq:ENev}) \ci{Korchemsky:2001nx},
Fig.\ \ref{fig:E_N}.
When we enlarge number of particles $N$ the energy seems to go to zero 
as $1/N$. For even $N$ the energy of the ground state is negative
so the contribution of the even ground states to the scattering amplitude
increases with the total energy $s$. For odd $N$ we have to consider
separately the descendent
states with $q_N=0$ and non-descendent ones. 
For the latter states the energy is positive so
the contribution to the scattering amplitude decreases with $s$.
For the states with $q_N=0$
the lowest energy $E_N=0$. Therefore, the descendent states
include the ground states for odd $N$.
Additionally, we have calculated the energy along the 
ground state trajectories for $N=2,\ldots,8$ 
as a function of $\nu_h$, Fig.\ \ref{fig:E-flow}.
These functions are symmetric in $\nu_h$. The minimum energy is in 
$\nu_h=0$. The order of these energy levels tell us that the system for
$\nu_h=0$ is anti-ferromagnetic one. Increasing $\nu_h$
we have observed reordering of the energy levels which
makes the system for a large $\nu_h$ ferromagnetic.

Next we have made comments
on the approach of the Baxter $Q-$operator method presented 
in Ref. \ci{deVega:2002im,DeVega:2001pu}.
The authors of this approach solve the Baxter equation
by using the pole ansatz (\ref{eq:Qupol}) 
whereas the quantization conditions
come from the normalization requirement  
and condition
of equality of holomorphic energies  (\ref{eq:hole}) for different 
holomorphic solutions to the Baxter equation. 
We claim the the latter condition is too strong 
because the Reggeon system is two dimensional 
so the holomorphic energy is not a physical observable.

At the end, we have passed to 
the deep inelastic scattering region where
Reggeon states allow us to investigate
properties of the structure function (\ref{eq:moments})
of the scattered hadron \ci{Jaroszewicz:1982gr}.
Performing analytical continuation of the energy 
into complex $\nu_h$ we have calculated the anomalous dimensions 
as functions of
the strong coupling constant (\ref{eq:anod}) and the respective 
twists (\ref{eq:twist}) \ci{Korchemsky:2003rc}
which have been determined for each $N$ separately.
The case with $N=2$ was discussed in Refs. 
\ci{Jaroszewicz:1982gr,Lipatov:1985uk}.
It turns out that after the analytical continuation the energy for $N>2$
is multi-valued function defined on the complicated Riemann surfaces
and possesses branching points and poles 
\ci{Levin:2001cv,Bartels:1992ym,Bartels:1993ih,Korchemsky:2003rc},
Fig.\ \ref{fig:E-comp}.
The poles (\ref{eq:EpolN}) occur on imaginary axis of $\nu_h$ and
their positions correspond to the twist $n$ of the
structure functions while expansion coefficients of the energy
around the poles (\ref{eq:Enpoles}) are related to  expansion coefficients
of anomalous dimensions $\gamma(\wbar \alpha_s)$ in 
the strong coupling constant $\wbar \alpha_s$ (\ref{eq:gampol}).
Thus, we have found 
\ci{Korchemsky:2003rc} 
that for $N$ Reggeon system 
the minimal twist $n=N$. However, for even $N$
it is formed by the states belonging to the $n_h=0$ sector
while for odd $N$ it corresponds to the $n_h=1$ sector.


Thus, we have become convinced that 
although the $Q-$Baxter method
is complicated
it is a powerful tool.
It allows us to solve the reggeized gluon state problem
for more than $N=3$ particles. The above calculations
are of interest not only for perturbative QCD
but also to statistical physics 
as the $SL(2,\mathbb{C})$ non-compact XXX Heisenberg spin magnet model
\ci{Takhtajan:1979iv,Faddeev:1979gh,KBI}.

One has to mention that in this work we have not calculated 
all pieces of an elastic scattering amplitude but only
the part that is responsible for the dependence on the total energy $s$
(\ref{eq:amp}),
which is determined by the intercept $j=1-\wbar \alpha_s E_N/4$.
Similarly, in the deep inelastic scattering processes we have only
identified exponents of the operator product expansion (OPE)
series which are related to the twists and anomalous dimensions
of the structure function $F_2(x,Q^2)$ (\ref{eq:moments}). 
In order to find the full
scattering amplitude one has to calculate the impact factors 
(\ref{eq:imp-fac}), i.e.
overlaps between the Reggeon wave function and the wave functions
of the scattered particles.
These impact factors strongly depend on the scattered system
and they are hard to calculate due to a large number of 
integrations.
Therefore, with more than $N=2$ Reggeons
only overlaps of the simplest Reggeon states were calculated, i.e. 
in  the $\eta_c \to \gamma^*(q)$ diffractive process
\ci{Engel:1997cg,Czyzewski:1996bv,Bartels:2003zu},
the double-diffractive production of $J/\psi$ \ci{Schafer:1991na},
photon-photon processes 
\ci{Barakhovsky:1991ra,Ginzburg:1993gy,Ginzburg:1991hd,Motyka:1998kb} 
as well as the proton-proton 
scattering process \ci{Dosch:2002ai} and other similar processes.

One has to remember that apart from the GLLA method presented in this work
there are other techniques which also 
make use of the reggeized gluons.
Firstly, one may use the extended leading logarithm
approximation (EGLLA) \ci{Bartels:1993ke,Bartels:1994jj,Bartels:2004ef} 
where the number of gluons
in the $t-$channel
may fluctuate. Moreover, Balitsky and Kovchegov 
\ci{Balitsky:1995ub,Kovchegov:1999yj,Kovchegov:1999ua}
derived a non-linear differential (BK) equation
that
for the two-Reggeon Pomeron 
contains the linear BFKL equation and
additionally the non-linear terms which for higher energy $s$ describe
the saturation effect 
\ci{Kovchegov:1999ua,Levin:1999mw,Levin:2000mv,Levin:2001cv}.
Furthermore, one can
use the dipole model 
\ci{Mueller:1993rr,Mueller:1994jq,Mueller:1994gb,Chen:1995pa,
Kovchegov:2003dm}, 
the colour glass model 
\ci{McLerran:1993ka,McLerran:1993ni,McLerran:1994vd,Hatta:2005as}
or the Wilson loop operators model \ci{Balitsky:1998ya,Balitsky:2001gj}
in order to derive or 
explain diagrammaticly the BK equation. 
All these methods 
seem to be equivalent in the leading order 
to the BFKL.
However, they include
different non-leading corrections which 
cause that the different approaches
better describe the scattering processes
in different kinematical regions \ci{Kovchegov:2002qu}.

This work opens the way for further studies
related to the multi-Reggeon states.
Using the method presented in Refs. 
\ci{Derkachov:2002pb,Korchemsky:2001nx,Derkachov:2002wz} 
it will be interesting to calculate
the spectrum of the conformal charges 
for $N=4$ and $n_h \ne0$ and compare them to results
obtained in Refs. \ci{DeVega:2001pu,deVega:2002im}.
Additionally, increasing precision one my try to
find the energy of the ground states for $N>8$
and confirm the dependence $E_N\sim 1/N$.
It is also of interest to perform analytical
continuation of the energy for the states which
have not been considered here and calculate their
twist and anomalous dimensions.
Moreover, we have mentioned that there are
attempts to calculate the scattering  processes
with Reggeon exchange where a number
of exchanging reggeized gluons varies in the $t-$channel, 
i.e. Extended GLLA method \ci{Bartels:1993ke,Bartels:1994jj,Bartels:2004ef}.
These diagrams start to be important at higher energy $s$
\ci{Kovchegov:2002qu}
and describe saturation which might allow to
unitarize the scattering amplitude.
To complete this picture one should also include
the diagrams with $N>2$ Reggeon states and the diagrams
containing vertices of $N$ Reggeons going into $M$ 
Reggeons for any $N$ and $M$.
These diagrams  would allow to calculate the DIS processes
as well as the elastic scattering processes, i.e. 
onium-onium or proton-proton scattering.
Furthermore, we can go beyond the 
multi-colour approximation and try
to calculated NLO BKP diagrams.
The next-to-leading order corrections to the BFKL kernel
(NLO BFKL) was calculated by Fadin and Lipatov in Ref.
\ci{Fadin:1998py} and seems to play an important role 
for really high energy $s$. This calculation
needs further studies.
One has to notice the analysis of the
multi-colour limit seems to be interesting
for the anomalous dimensions
of the structure functions in DIS processes
and compare to Ref. \ci{Levin:1992mu}.
At the end we have to say that
there is a connection between the GLLA approach and
${\cal N}=2$ SUSY gauge theories presented in Ref. \ci{Gorsky:2002ju}.
It is very impressive and this topic also needs more research.

\chapter*{Acknowledgements}
I would like to warmly thank to
Micha{\l} Prasza{\l}owicz
for fruitful discussions and help during writing this dissertation.
I am very grateful to G.P. Korchemsky,
A.N.Manashov and S.\'E. Derkachov
I could work with them in early state of this project and without whom
this work couldn't be written.
I also thank to 
Jacek Wosiek
for illuminating discussions.
This work was supported by
KBN-PB-2-P03B-43-24 and
KBN-PB-0349-P03-2004-27.

\newpage

\newpage
\appendix
\newcommand{\LI}{\mbox{{\em L1}}}
\newcommand{\LII}{\mbox{{\em L2}}}
\newcommand{\LIII}{\mbox{{\em L3}}}
\newcommand{\NI}{\mbox{{\em N1}}}
\newcommand{\NII}{\mbox{{\em N2}}}
\newcommand{\NIII}{\mbox{{\em N3}}}
\newcommand{\M}{\mbox{\em M}}

\chapter{$SL(2,\mathbb{C})$ invariants and other variables}

Let us consider a difference of coordinates $(z_{1}-z_{2})$. It 
changes
under the $SL(2,\mathbb{C})$ transformation (\ref{eq:trcoords}) as
\begin{equation}
\begin{array}{ccc}
 (z_{1}^{\prime }-z_{2}^{\prime }) & = & 
\left(\frac{az_{1}+b}{cz_{1}+d}-\frac{az_{2}+b}{cz_{2}+d}\right)=
(cz_{1}+d)^{-1}(cz_{2}+d)^{-1}(z_{1}-z_{2})\end{array}\,.
\lab{eq:zdif}
\end{equation}
One can see that during the transformation (\ref{eq:zdif})
the factors $(cz_{i}+d)^{-1}$ appear
in front of the difference. 
These factors also exist in the transformation law of the wave-function
(\ref{eq:trpsiz0}). In order to cancel these factors one can construct
a fraction where the same additional factors appear 
in the denominator and numerator of the constructed fraction variable.

The fraction variable 
\begin{equation}
x=\frac{(z_{1}-z_{2})(z_{3}-z_{0})}{(z_{1}-z_{0})(z_{3}-z_{2})}\equiv (z_{1}z_{2}z_{3}z_{0})
\lab{eq:x}
\end{equation}
is invariant under the $SL(2,\mathbb{C})$ transformations (\ref{eq:trcoords}).
It is only one independent invariant for four coordinates 
\ci{CFT,Staruszkiewicz93}.
One can see that the fractions coming from the $SL(2,\mathbb{C})$
transformation simplified because:
\begin{itemize}
\item the variable is a function of the coordinate differences 
(\ref{eq:zdif}), 
\item the coordinates  
in the numerator and denominator of the variable $x$ are the same.
\end{itemize}
Therefore, simplifying the partial fractions we can obtain expression like $((az_{i}+b)(cz_{j}+d)-(az_{j}+b)(cz_{i}+d))$
which with making use of $ad-cb=1$ goes to $(z_{i}-z_{j})$. As we can
see, we have to build our invariants from differences of the coordinates.

It is easy to see that performing  permutations of coordinates we can construct
six different dependent invariants\begin{equation}
\begin{array}{cccccc}
 (z_{1}z_{2}z_{3}z_{0}) & = & x\mbox {,} & (z_{3}z_{2}z_{1}z_{0}) & = & 1/x\mbox {,}\\
 (z_{2}z_{3}z_{1}z_{0}) & = & 1/(1-x)\mbox {,} & (z_{1}z_{3}z_{2}z_{0}) & = & 1-x\mbox {,}\\
 (z_{3}z_{1}z_{2}z_{0}) & = & (x-1)/x\mbox {,} & (z_{2}z_{1}z_{3}z_{0}) & = & x/(x-1)\mbox {.}\end{array}\end{equation}

Let us take another product of $z_{ij}$
\begin{eqnarray}
w^{\prime } & = & 
\frac{(z_{1}^{\prime }-z_{2}^{\prime })}{
(z_{1}^{\prime }-z_{0}^{\prime })
(z_{2}^{\prime }-z_{0}^{\prime })}=
(cz_{0}+d)^{2}\frac{(z_{1}-z_{2})}{(z_{1}-z_{0})(z_{2}-z_{0})}
=(cz_{0}+d)^{2}w\,.
\lab{eq:wdef}
\end{eqnarray}
One can see that since in the denominator we have two additional $z_{0}$
variables after the transformation we 
obtain a multiplying factor $(cz_{0}+d)^{2}$
. 
Similarly for $w^h$ we obtain
\begin{equation}
w^{\prime h}=(cz_{0}+d)^{2h}w^{h}\,.
\end{equation}
Now one can compare this transformation to the transformation of
the $SL(2,\mathbb{C})$ wave-function (\ref{eq:trpsiz0}).

For $N=3$ the variables $w=\frac{(z_3-z_2)}{(z_3-z_0)(z_2-z0)}$ 
and $x=\frac{(z_{1}-z_{2})(z_{3}-z_{0})}{(z_{1}-z_{0})(z_{3}-z_{2})}$ 
transform under the cycling permutation 
(\ref{eq:PMsym}) as 
\begin{equation}
\begin{array}{c}
(x-1)\rightarrow\frac{(-x)}{(x-1)}\rightarrow\frac{1}{(-x)}
\rightarrow(x-1)\,,\\
(-x)\rightarrow\frac{1}{(x-1)}\rightarrow\frac{(x-1)}{(-x)}
\rightarrow(-x)\,,\\
w\rightarrow w(x-1)\rightarrow w(-x)\rightarrow w
\end{array}
\lab{eq:wxcyc}
\end{equation}
while under the mirror permutation:
\begin{equation}
\begin{array}{c}
(x-1)\rightarrow\frac{(x-1)}{(-x)}\rightarrow(x-1)\\
(-x)\rightarrow\frac{1}{(-x)}\rightarrow(-x)\\
w\rightarrow w\,x\rightarrow w\end{array}
\lab{eq:wxmir}
\end{equation}

For higher $N$ we have more invariants. All of them can be constructed 
\ci{Lipatov:1998as}
from
\begin{equation}
x_{r}=\frac{(z_{r-1}-z_{r})(z_{r+1}-z_{0})}{(z_{r-1}-z_{0})(z_{r+1}-z_{r})}\,.
;\qquad \prod _{r=1}^{N}x_{r}=(-1)^{N};\qquad \sum _{r=1}^{N}(-1)^{r}\prod _{k=r+1}^{N}x_{k}=0.
\lab{eq:xr}
\end{equation}
Variables $x_r$ are subject to the two conditions:
\begin{equation}
\qquad \prod _{r=1}^{N}x_{r}=(-1)^{N}
\lab{eq:xrc1}
\end{equation}
and 
\begin{equation}
\qquad \sum _{r=1}^{N}(-1)^{r}\prod _{k=r+1}^{N}x_{k}=0.
\lab{eq:xrc2}
\end{equation}
From (\ref{eq:xrc1}) we have
$x_{1}=(-1)^{N}/\prod _{r=2}^{N}x_{r}$. 
From (\ref{eq:xrc2}) we can
calculate 
\begin{equation}
x_{2}=\frac{\sum _{r=2}^{N}(-1)^{r}
\prod _{k=r+1}^{N}x_{k}}{\prod _{k=3}^{N}x_{k}}\,.
\end{equation}
Interchangeably,
one can derive from (\ref{eq:xrc1})
$x_{N}=(-1)^{N}/\prod _{r=1}^{N-1}x_{r}$ and next from
(\ref{eq:xrc2}) 
\begin{equation}
x_{N-1}=\frac{(-1)^{N}}{\sum _{r=1}^{N-1}(-1)^{r}
\prod _{k=r}^{N-2}x^{k}}\,.
\end{equation}
Thus, we can see that we have for $N$ particles $N-2$ independent
invariants built of the particle coordinates $z_i$.

\chapter{Solution of the $q_2$ eigenproblem }
 
\section{Case for $N=2$}

For two particles, i.e. $N=2$, we do not have any 
invariant variables $x_j$, (\ref{eq:xr}).
Substituting  (\ref{eq:zprod}) into the equation (\ref{eq:egq2}) 
we obtain
\begin{equation}
\oq{2}(z_{10})^{k_{10}}(z_{20})^{k_{20}}(z_{12})^{k_{12}}  =  
\left(-h(h-1)+s_{1}(s_{1}-1)+s_{2}(s_{2}-1)\right)
(z_{10})^{k_{10}}(z_{20})^{k_{20}}(z_{12})^{k_{12}}.
\lab{eq:eg2q2}
\end{equation}
Next, solving (\ref{eq:eg2q2}) for $k_{ij}$ we get the eigenstates:
\begin{enumerate}
\item $\psi _{1}(z_{10},z_{20})=
(z_{10})^{-h-s_{1}+s_{2}}(z_{20})^{-h+s_{1}-s_{2}}(z_{12})^{h-s_{1}-s_{2}}$\,,
\item $\psi _{2}(z_{10},z_{20})=
(z_{10})^{-1+h-s_{1}+s_{2}}(z_{20})^{-1+h+s_{1}-s_{2}}
(z_{12})^{1-h-s_{1}-s_{2}}$\,.
\end{enumerate}
The first solution transforms with the proper factor 
$(cz_{0}+d)^{2h}(cz_{1}+d)^{2s_{1}}(cz_{2}+d)^{2s_{2}}$.
The second one has $h\rightarrow 1-h$.

A full eigenstate of $\oq{2}$ with $N=2$
can be presented as a linear combination
of the difference-coordinate products. It comes from the fact that during
the $SL(2,\mathbb{C})$ transformation the factor $(cz_{i}+d)^{2s}$ have
to be obtained (\ref{eq:trpsiz0}). 
Thus, the eigenstate, up to some normalization factor, has a form
\ci{Balitsky:1978ic,Fadin:1975cb,Kuraev:1977fs}
\begin{equation}
\Psi(z_{10},z_{20})=(z_{10})^{-h-s_{1}+s_{2}}(z_{20})^{-h+s_{1}-s_{2}}
(z_{12})^{h-s_{1}-s_{2}}=
\left(\frac{z_{12}}{z_{20}z_{10}}\right)^{h}
\left(\frac{z_{20}}{z_{12}z_{10}}\right)^{s_{1}}
\left(\frac{z_{10}}{z_{12}z_{20}}\right)^{s_{2}}\,.
\lab{eq:psiq2f}
\end{equation}
For the homogeneous model, i.e. $s=s_{i}$,
\begin{equation}
\psi (z_{10},z_{20})=
(z_{10})^{-h}(z_{20})^{-h}(z_{12})^{h-2s}=
z_{12}^{-2s}\left(\frac{z_{12}}{z_{10}z_{20}}\right)^{h}\,.
\end{equation}
In the special case for $s=0$  it leads to
\begin{equation}
\psi (z_{10},z_{20})=
(z_{10})^{-h}(z_{20})^{-h}(z_{12})^{h}=
\left(\frac{z_{12}}{z_{10}z_{20}}\right)^{h}
\end{equation}
while for $s=1$ it gives
\begin{equation}
\psi (z_{10},z_{20})=
(z_{10})^{-h}(z_{20})^{-h}(z_{12})^{h-2}=
\frac{1}{z_{12}^{2}}\left(\frac{z_{12}}{z_{10}z_{20}}\right)^{h}=
\frac{1}{z_{10}^{2}z_{20}^{2}}\left(\frac{z_{12}}{z_{10}z_{20}}\right)^{h-2}\,.
\end{equation}

\section{Case for $N=3$}

In this case, the solutions with $z_{0}$ can be reduced 
to one independent function
\begin{equation}
\begin{array}{ccl}
 \psi (z_{10},z_{20},z_{30}) & = & 
(z_{10})^{-h+k_{32}-s_{1}+s_{2}+s_{3}}
(z_{12})^{k_{12}}(z_{20})^{-k_{32}-k_{12}-2s_{2}}
(z_{30})^{-h+k_{12}+s_{1}+s_{2}-s_{3}}\\
  &  & \times (z_{31})^{h-k_{32}-k_{12}-s_{1}-s_{2}-s_{3}}
(z_{32})^{k_{32}}=\\
  & = & (z_{10})^{-h-s_{1}+s_{2}+s_{3}}(z_{20})^{-2s_{2}}
(z_{30})^{-h+s_{1}+s_{2}-s_{3}}(z_{31})^{h-s_{1}-s_{2}-s_{3}}
\left(\frac{-x}{x-1}\right)^{k_{12}}\left(\frac{1}{x-1}\right)^{k_{32}}
\end{array}
\lab{eq:n3psiz0}
\end{equation}
and to the other one with $h\rightarrow 1-h$. It is invariant under the cyclic
permutation and contains two free parameters, $k_{12}$ and $k_{32}$. 

For $N=3$, there is one independent invariant variable (\ref{eq:x}), 
e.g. $x=x_{2}$.  
Next, we multiply our wave-function by an arbitrary function of $x$,
$F(x)$. There is only one pattern of the eigenfunction
which satisfy the eigenequation for $q_2$ and the 
$SL(2,\mathbb{C})$ transformation (\ref{eq:trpsiz0}) law. 
This eigenfunction looks like
\begin{equation}
\begin{array}{ccl}
 \Psi (z_{10},z_{20},z_{30}) & = & (z_{10})^{-h-s_{1}+s_{2}+s_{3}}
(z_{20})^{-2s_{2}}(z_{30})^{-h+s_{1}+s_{2}-s_{3}}
(z_{31})^{h-s_{1}-s_{2}-s_{3}}F(x)=\\
  & = & \frac{1}{(z_{10})^{2s_{1}}(z_{20})^{2s_{2}}(z_{30})^{2s_{3}}}
\left(\frac{z_{31}}{z_{10}z_{30}}\right)^{h-s_{1}-s_{2}-s_{3}} F(x).
\end{array}
\lab{eq:n3psiz0b2}
\end{equation}
For the homogeneous model ($s=s_{i}$) we have an ansatz
\begin{equation}
\Psi (z_{10},z_{20},z_{30})=(z_{10})^{-h+s}(z_{20})^{-2s}
(z_{30})^{-h+s}(z_{31})^{h-3s}F(x)
=\frac{1}{(z_{10}z_{20}z_{30})^{2s}}
\left(\frac{z_{31}}{z_{10}z_{30}}\right)^{h-3s}F(x)
\end{equation}
Substituting $s=0$ we have
\begin{equation}
\Psi (z_{10},z_{20},z_{30})=(z_{10})^{-h}(z_{30})^{-h}(z_{31})^{h}F(x)
=\left(\frac{z_{31}}{z_{10}z_{30}}\right)^{h}F(x)\,,
\end{equation}
whereas for $s=1$, it is
\begin{equation}
\Psi (z_{10},z_{20},z_{30})=(z_{10})^{-h+1}(z_{20})^{-2}
(z_{30})^{-h+1}(z_{31})^{h-3}F(x)
=\frac{1}{(z_{10}z_{20}z_{30})^{2}}
\left(\frac{z_{31}}{z_{10}z_{30}}\right)^{h-3}F(x).
\end{equation}

\section{Case for $N=4$}

Here, like in the previous cases, we have also four
$2\times 2$ solutions.
When we take solutions with $z_{0}$ they reduce into
two functions:
\begin{equation}
\begin{array}{ccl}
 \Psi (z_{10},z_{20},z_{30},z_{40})  & = & 
(z_{10})^{-k_{12}-k_{31}-k_{14}-2s_{1}}
(z_{12})^{k_{12}}(z_{20})^{-h+k_{31}+k_{34}+k_{14}+s_{1}-s_{2}+s_{3}+s_{4}}
(z_{24})^{k_{24}}\\
  &  & \times (z_{30})^{-h+k_{24}+k_{12}+k_{14}+s_{1}+s_{2}-s_{3}+s_{4}}
(z_{34})^{k_{34}}(k_{40})^{-k_{24}-k_{34}-k_{14}-2s_{4}}\\
  &  & \times (z_{14})^{k_{14}}(z_{31})^{k_{31}}
(z_{32})^{-h-k_{24}-k_{12}-k_{31}-k_{34}-k_{14}-s_{1}-s_{2}-s_{3}-s_{4}}
\end{array}
\lab{eq:n4psiz0}
\end{equation}
and the other one is with $h\rightarrow 1-h$. The function (\ref{eq:n4psiz0}) 
is invariant under cyclic
permutation and contains two free parameters. 

In the $N=4$ case we have $N-2=2$ independent invariant variables. 
Let us take 
$x_{1}=\frac{(z_{4}-z_{1})(z_{2}-z_{0})}{(z_{4}-z_{0})(z_{2}-z_{1})}$
and $x_{2}=\frac{(z_{1}-z_{2})(z_{3}-z_{0})}{(z_{1}-z_{0})(z_{3}-z_{2})}$.
We multiply our wave-function by an arbitrary function 
of these variables, 
$F(x_{1},x_{2})$.
Considering (\ref{eq:trpsiz0}), we obtain a solution for the 
$q_{2}$-eigenequation 
\begin{equation}
\begin{array}{ccl}
 \Psi (z_{10},z_{20},z_{30},z_{40})  & = & 
(z_{10})^{-k_{12}-k_{31}-k_{14}-2s_{1}}(z_{12})^{k_{12}}
(z_{20})^{-h+k_{31}+k_{34}+k_{14}+s_{1}-s_{2}+s_{3}+s_{4}}(z_{24})^{k_{24}}\\
  &  & \times (z_{30})^{-h+k_{24}+k_{12}+k_{14}+s_{1}+s_{2}-s_{3}+s_{4}}
(z_{34})^{k_{34}}(k_{40})^{-k_{24}-k_{34}-k_{14}-2s_{4}}\\
  &  & \times (z_{14})^{k_{14}}(z_{31})^{k_{31}}
(z_{32})^{-h-k_{24}-k_{12}-k_{31}-k_{34}-k_{14}-s_{1}-s_{2}-s_{3}-s_{4}}
f(x_1,x_2)=\\
  & = & \frac{1}{(z_{10})^{2s_{1}}(z_{20})^{2s_{2}}(z_{30})^{2s_{3}}
(z_{40})^{2s_{4}}}
\left(\frac{z_{31}}{z_{10}z_{30}}\right)^{h-s_{1}-s_{2}-s_{3}-s_{4}}
F(x_{1},x_{2})\,.
\end{array}
\end{equation}
For the homogeneous model ($s=s_{i}$) we have an ansatz
\begin{equation}
\Psi (z_{10},z_{20},z_{30},z_{40})=\frac{1}{(z_{10}z_{20}z_{30}z_{40})^{2s}}
\left(\frac{z_{31}}{z_{10}z_{30}}\right)^{h-4s}F(x_{1},x_{2})\,.
\end{equation}
 For $s=0$
\begin{equation}
\Psi (z_{10},z_{20},z_{30},z_{40}) =(z_{10})^{-h}(z_{30})^{-h}(z_{31})^{h}
F(x_{1},x_{2})=
\left(\frac{z_{31}}{z_{10}z_{30}}\right)^{h}F(x_{1},x_{2})
\end{equation}
 and for $s=1$
\begin{multline}
\Psi (z_{10},z_{20},z_{30},z_{40}) =(z_{10})^{-h+2}(z_{20})^{-2}
(z_{30})^{-h+2}(z_{40})^{-2}
(z_{31})^{h-4}F(x_{1},x_{2})=\\
=\frac{1}{(z_{10}z_{20}z_{30}z_{40})^{2}}
\left(\frac{z_{31}}{z_{10}z_{30}}\right)^{h-4}F(x_{1},x_{2})\,.
\end{multline}

\chapter{Solutions for $N=3$}
\section{Solutions for $s=0$ around $x=0^{+}$}
\subsection{Solutions for $q_3\ne0$ and $h\not\in\mathbb{Z}$ }

The eigenequation for $\oq{3}$ (\ref{eq:qF})
is a differential equation of the third order 
so around each singular point, $x=0,1,\infty$, it
has three independent solutions. 
Around $x=0^{+}$ we have an indicial
equation 
\begin{equation}
(h-n-r)(r+n-1)(n+r)=0\,,
\lab{eq:indx0}
\end{equation} 
so its solutions are 
given by $r_{1}=h$, $r_{2}=1$ and $r_{3}=0$. 
As we can see we have two cases when 
$h\not\in\mathbb{Z}$
(one solution with $\mbox{Log}(x)$) and $h\in\mathbb{Z}$ (one solution
with $\mbox{Log}(x)$ and one solution with $\mbox{Log}^{2}(x)$). As we
will see below we have to consider separately solutions with $q_{3}=0$.
In the first case\begin{equation}
\begin{array}{rcl}
u_{1}(x) & = & x^{r_{1}}\sum_{n=0}^{\infty}a_{n,r_{1}}x^{n}\,,\\
u_{2}(x) & = & x^{r_{2}}\sum_{n=0}^{\infty}a_{n,r_{2}}x^{n}\,,\\
u_{3}(x) & = & x^{r_{3}}\sum_{n=0}^{\infty}b_{n,r_{3}}x^{n}
+x^{r_{2}}\sum_{n=0}^{\infty}a_{n,r_{2}}x^{n}\mbox{Log}(x)\,,
\end{array}
\lab{eq:x0u123}
\end{equation}
where $a_{0,r}$ is arbitrary (e.g. equal to $1$)
\begin{equation}
a_{1,r}=\frac{(iq_{3}-(h-2r)(h-r)r)}{(h-1-r)r(1+r)}a_{0,r}
\end{equation}
and $m=n+r$
\begin{multline}
a_{n,r}=\frac{(h-m+1)(h-m+2)(m-2)}{(h-m)(m-1)m}a_{n-2,r}+\\
+\frac{(iq_{3}-(1+h-m)(m-1)(h-2(m-1)))}{(h-m)(m-1)m}a_{n-1,r}\,,
\end{multline}
whereas $b_{0,r_{3}}=\frac{(h-1)}{iq_{3}}a_{0,r_{2}}$ 
and
$b_{1,r_{3}}$ is arbitrary.
One can notice that coefficient 
$b_{0,r_{3}}$ is well defined only for $q_{3}\ne0$.
Moreover,
\begin{equation}
\begin{array}{rcl}
b_{2,r_{3}} & = & \frac{2+(h-3)h-iq_{3}}{2(2-h)}b_{1,r_{3}}
+\frac{3h-8}{2(2-h)}a_{1,r_{3}}+\frac{6+h(h-6)}{2(2-h)}a_{0,r_{3}}
\end{array}
\end{equation}
and
\begin{equation}
\begin{array}{rcl}
b_{n,r_{3}} & = & \frac{(h+1-m)(h-m+2)(m-2)}{(h-m)(m-1)m}b_{n-2,r_{3}}
+\frac{iq_{3}-(1+h-m)(m-1)(h-2(m-1))}{(h-m)(m-1)m}b_{n-1,r_{3}}+\\
 &  & +\frac{h^{2}+h(7-4m)+(m-2)(3m-4)}{(h-m)(m-1)m}a_{n-3,r_{3}}
-\frac{h^{2}-6h(m-1)+6(m-1)^{2}}{(h-m)(m-1)m}a_{n-2,r_{3}}+\\
 &  & -\frac{2m-3m^{2}+h(2m-1)}{(h-m)(m-1)m}a_{n-1,r_{3}}\,,
\end{array}
\end{equation}
where $m=r_{3}+n$.

\subsection{Solution with $q_{3}=0$}

In this case $a_{0,r}$ defined in (\ref{eq:x0u123})
is arbitrary (e.g. equal to $1$). 
For the first solution
with $r_{1}=h$ we have $a_{1,r_{1}}=0$ whereas for the third one $a_{1,r_{3}}$
is arbitrary (let us take $0$).
It turns out that we do not need $\mbox{Log}-$solutions.
The second solution is more complicated. One can derive an exact formula
for 
\begin{equation}
a_{n,r_{2}}=a_{0,r_{2}}\prod_{k=1}^{n}\frac{k-h}{k+1}
=\frac{\Gamma(1-h+n)}{\Gamma(1-h)\Gamma(n+2)}a_{0,r_{2}}
\end{equation}
and performing summations\begin{equation}
\begin{array}{rcl}
u_{1}(x) & = & x^{r_{1}}\sum_{n=0}^{\infty}a_{n,r_{1}}x^{n}=x^{h}a_{0,r_{1}}
\,,\\
u_{2}(x) & = & x^{r_{2}}\sum_{n=0}^{\infty}a_{n,r_{2}}x^{n}
=a_{0,r_{2}}x\sum_{n=0}^{\infty}
\frac{ \Gamma(1-h+n)}{\Gamma(1-h)\Gamma(n+2)}x^{n}
=-a_{0,r_{2}}\frac{1}{h}((1-x)^{h}-1)\,,\\
u_{3}(x) & = & x^{r_{3}}\sum_{n=0}^{\infty}a_{n,r_{3}}x^{n}
=a_{0,r_{3}}\,.
\end{array}
\end{equation}
Gathering together this all solutions we have
\begin{equation}
u(x)=A+B(-x)^{h}+C(x-1)^{h}
\end{equation}
where $A$, $B,$ $C$ are arbitrary. The above solution was presented
by Lipatov and Vacca in Refs\ . \ci{Lipatov:1998as,Vacca:2000bk}.

\subsection{Solution 
with $q_{3}\ne0$, $q_2=0$ and $h=1$ }

For $h=0$, i.e. $q_2=0$, and $q_3 \ne 0$ 
we have a different set of solutions. Here
we have three solutions of the indicial equation 
(\ref{eq:indx0})
which are integer $r_{1}=1$,
$r_{2}=1$, $r_{3}=0$ so the solutions are given by
\begin{equation}
\begin{array}{rcl}
u_{1}(x) & = & x^{r_{1}}\sum_{n=0}^{\infty}a_{n,r_{1}}x^{n}\,,\\
u_{2}(x) & = & x^{r_{2}}\sum_{n=0}^{\infty}b_{n,r_{2}}x^{n}
+x^{r_{1}}\sum_{n=0}^{\infty}a_{n,r_{1}}x^{n}\mbox{Log}(x)\,,\\
u_{3}(x) & = & x^{r_{3}}\sum_{n=0}^{\infty}c_{n,r_{3}}x^{n}
+2x^{r_{2}}\sum_{n=0}^{\infty}b_{n,r_{2}}x^{n}\mbox{Log}(x)
+x^{r_{1}}\sum_{n=0}^{\infty}a_{n,r_{1}}x^{n}\mbox{Log}^{2}(x)\,,
\end{array}
\end{equation}
where $a_{0,r}$ is arbitrary (e.g. equal $1$)
\begin{equation}
a_{1,r}=\frac{r+r^{2}(2r-3)}{r^{2}(1+r)-iq_{3}}a_{0,r}
\end{equation}
and $m=n+r$
\begin{equation}
a_{n,r}=-\frac{(m-3)(m-2)^{2}}{(m-1)^{2}m}a_{n-2,r}
+\frac{(m-2)(m-1)(2m-3)-iq_{3}}{(m-1)^{2}m}a_{n-1,r}\,,
\end{equation}
whereas 
\begin{equation}
b_{1,r_{2}}=\frac{1}{2}(-iq_{3}b_{0,r_{2}}+a_{0,r_{1}}-5a_{1,r_{1}})
\end{equation}
\begin{equation}
\begin{array}{rcl}
b_{n,r_{2}} & = & -\frac{(n-1)^{2}(n-2)}{n^{2}(n+1)}b_{n-2,r_{2}}
+\frac{(2m^{3}-3m^{2}+m-iq_{3})}{n^{2}(n+1)}b_{n-1,r_{2}}
+\frac{1+6n(n-1)}{n^{2}(n+1)}a_{n-1,r_{1}}+\\
 &  & +\frac{(n-1)(5-3n)}{n^{2}(n+1)}a_{n-2,r_{1}}
-\frac{(2+3n)}{n(n+1)}a_{n,r_{1}}
\end{array}
\end{equation}
while $c_{0,r_{3}}=\frac{-2}{i q_{3}}a_{0,r_{1}}$, $c_{1,r_{3}}$
is arbitrary
and
\begin{equation}
c_{2,r_{3}}=-\frac{1}{2}iq_{3}c_{1,r_{3}}+b_{0,r_{2}}-5b_{1,r_{2}}
+3a_{0,r_{1}}-4a_{1,r_{1}}\,,
\end{equation}
\begin{equation}
\begin{array}{rcl}
c_{n,r_{3}} & = & -\frac{(n-2)^{2}(n-3)}{n(n-1)^{2}}c_{n-2,r_{3}}
+\frac{-iq_{3}-6+n(13+n(2n-9))}{n(n-1)^{2}}c_{n-1,r_{3}}
-\frac{2(n-2)(3n-8)}{n(n-1)^{2}}b_{n-2,r_{2}}+\\
 &  & +\frac{26-36 n + 12 n^2}{n(n-1)^{2}}b_{n-1,r_{2}}
-\frac{6n^2-8n+2}{n(n-1)^{2}}b_{n,r_{2}}
+\frac{2(7-3n)}{n(n-1)^{2}}a_{n-3,r_{1}}
-\frac{6(3-2n)}{n(n-1)^{2}}a_{n-2,r_{1}}+\\
 &  & +\frac{2(2-3n)}{n(n-1)^{2}}a_{n-1,r_{1}}\,.
\end{array}
\end{equation}

\subsection{Solution 
with $q_2=q_{3}=0$ and $h=1$}

In this case $a_{0,r}$ is arbitrary (e.g. equal to $1$). 
For the first solution
with $r_{1}=(h=1)$ we have $a_{1,r_{1}}=0$ and for the third one
$a_{1,r_{3}}$ with $r_{3}=0$ is arbitrary (let us take $0$). The
second solution is more complicated. We need $\mbox{Log}-$solutions
\begin{equation}
u_{2}(x)=x^{r_{2}}\sum_{n=0}^{\infty}b_{n,r_{2}}x^{n}
+x^{r_{1}}\sum_{n=0}^{\infty}a_{n,r_{1}}x^{n}\mbox{Log}(x)\,.
\end{equation}
Using recurrence  relations with $b_{1,r_{2}}=\frac{1}{2}a_{0,r_{1}}$
(and $a_{n,r_{1}}=0$ for $n>0$) 
\begin{equation}
\begin{array}{rcl}
b_{2,r_{2}} & = & \frac{1}{3}b_{1,r_{2}}\,,\\
b_{n,r_{2}} & = & -\frac{(n-1)^{2}(n-2)}{n^{2}(n+1)}b_{n-2,r_{2}}
+\frac{(n-1)n(2n-1)}{n^{2}(n+1)}b_{n-1,r_{2}}
\end{array}
\end{equation}
one can derive an exact formula for $b_{n,r_{2}}=b_{1,r_{2}}\frac{2}{(n+1)n}$
and performing summations with arbitrary $b_{0,r_{2}}(=0)$ we have
$x\sum_{n=0}^{\infty}b_{n,r_{2}}x^{n}=2b_{1,r_{2}}(x+\mbox{Log}(1-x)
-x\mbox{Log}(1-x))$, so that
\begin{equation}
\begin{array}{rcl}
u_{1}(x) & = & x^{r_{1}}\sum_{n=0}^{\infty}a_{n,r_{1}}x^{n}=a_{0,r_{1}}
=xa_{0,r_{1}}\,,\\
u_{2}(x) & = & a_{0,r_{1}}(x+\mbox{Log}(1-x)-x\mbox{Log}(1-x)
+x\mbox{Log}(x))\,,\\
u_{3}(x) & = & x^{r_{3}}\sum_{n=0}^{\infty}a_{n,r_{3}}x^{n}
=a_{0,r_{3}}\,.
\end{array}
\end{equation}
Gathering together all this solutions we have
\begin{equation}
u(x)=A+B(-x)+C((x-1)\mbox{Log}(x-1)+(-x)\mbox{Log}(-x))\,,
\end{equation}
where $A$, $B,$ $C$ are arbitrary. See also Ref. \ci{DeVega:2001pu}.

\subsection{Solution 
with $q_{3}\ne0$, $q_2=0$ and $h=0$}

For $h=0$, i.e. $q_2=0$, and  arbitrary $q_3\ne 0$ 
we have three solutions 
\begin{equation}
\begin{array}{rcl}
u_{1}(x) & = & x^{r_{1}}\sum_{n=0}^{\infty}a_{n,r_{1}}x^{n}\\
u_{2}(x) & = & x^{r_{2}}\sum_{n=0}^{\infty}b_{n,r_{2}}x^{n}
+x^{r_{1}}\sum_{n=0}^{\infty}a_{n,r_{1}}x^{n}\mbox{Log}(x)\\
u_{3}(x) & = & x^{r_{3}}\sum_{n=0}^{\infty}c_{n,r_{3}}x^{n}
+2x^{r_{2}}\sum_{n=0}^{\infty}b_{n,r_{2}}x^{n}\mbox{Log}(x)
+x^{r_{1}}\sum_{n=0}^{\infty}a_{n,r_{1}}x^{n}\mbox{Log}^{2}(x)
\end{array}
\end{equation}
where $r_{1}=1$,
$r_{2}=0$ and $r_{3}=0$.
Here $a_{0,r}$ is arbitrary (e.g. $1$)
\begin{equation}
a_{1,r}=\frac{-iq_{3}+2r^{2}}{r(1+r)^{2}}a_{0,r}
\end{equation}
and $m=n+r$
\begin{equation}
a_{n,r}=-\frac{(m-2)^{2}}{m^{2}}a_{n-2,r}
+\frac{-iq_{3}+2(m-1)^{3}}{(m-1)m^{2}}a_{n-1,r}\,,
\end{equation}
whereas $b_{0,r_{2}}=-\frac{1}{iq_{3}}a_{0,r_{1}}$ and $b_{1,r_{2}}$
is arbitrary while
\begin{equation}
b_{2,r_{2}}=\frac{1}{4}((2-iq_{3})b_{1,r_{2}}+6a_{0,r_{1}}-8a_{1,r_{1}})\,,
\end{equation}
\begin{equation}
\begin{array}{rcl}
b_{n,r_{2}} & = & -\frac{(n-2)^{2}}{n^{2}}b_{n-2,r_{2}}
+\frac{-iq_{3}+2(n-1)^{3}}{n^{2}(n-1)}b_{n-1,r_{2}}
-\frac{(n-2)(3n-4)}{n^{2}(n-1)}a_{n-3,r_{1}}+\\
 &  & +\frac{6(n-1)}{n^{2}}a_{n-2,r_{1}}
+\frac{2-3n}{n^{2}(n-1)}a_{n-1,r_{1}}\,.
\end{array}
\end{equation}
The coefficient
$c_{0,r_{3}}=\frac{2}{iq_{3}}(2a_{0,r_{1}}+b_{1,r_{2}})$ and
$c_{1,r_{3}}$ is arbitrary and
\begin{equation}
c_{2,r_{3}}=\frac{1}{4}(2-iq_{3})c_{1,r_{3}}+3b_{1,r_{2}}
-4b_{2,r_{2}}+3a_{0,r_{1}}-\frac{5}{2}a_{1,r_{1}}\,,
\end{equation}
\begin{equation}
\begin{array}{rcl}
c_{n,r_{3}} & = & -\frac{(n-2)^{2}}{n^{2}}c_{n-2,r_{3}}
+\frac{-iq_{3}+2(n-1)^{3}}{n^{2}(n-1)}c_{n-1,r_{3}}
-\frac{2(n-2)(3n-4)}{n^{2}(n-1)}b_{n-2,r_{2}}+\\
 &  & +\frac{12(n-1)}{n^{2}}b_{n-1,r_{2}}+\frac{2(2-3n)}{n(n-1)}b_{n,r_{2}}
+\frac{2(5-3n)}{n^{2}(n-1)}a_{n-3,r_{1}}+\frac{12}{n^{2}}a_{n-2,r_{1}}+\\
 &  & +\frac{2(1-3n)}{n^{2}(n-1)}a_{n-1,r_{1}}\,.
\end{array}
\end{equation}

\subsection{Solution 
with $h=0$ and $q_2=q_{3}=0$}

In this case $a_{0,r}$ is arbitrary. For the first solution
with $r_{1}=1$ we have recurrence  relations with 
$a_{1,r_{1}}=\frac{1}{2}a_{0,r_{1}}$
\begin{equation}
\begin{array}{rcl}
a_{2,r_{1}} & = & \frac{1}{3}a_{1,r_{1}}\,,\\
a_{n,r_{1}} & = & -\frac{(n-1)^{2}}{n+1}a_{n-2,r_{1}}
+\frac{2n^{2}}{n+1}a_{n-1,r_{1}}\,.
\end{array}
\end{equation}
Summing series up we derive an exact formula for 
\begin{equation}
u_{1}(x)=x^{r_{1}}\sum_{n=0}^{\infty}a_{n,r_{1}}x^{n}
=x\sum_{n=0}^{\infty}x^{n}a_{0,r_{1}}\frac{1}{n+1}=-\mbox{Log}(1-x)\,.
\end{equation}

In the second solution, with $r_{2}=0$, we have arbitrary $a_{1,r_{2}}$
(let us take $0$) and solution is $u_{2}(x)=a_{0,r_{2}}$. The third
one is with $\mbox{Log}-$solutions 
\begin{equation}
u_{3}(x)=x^{r_{3}}\sum_{n=0}^{\infty}b_{n,r_{3}}x^{n}+x^{r_{2}}\sum_{n=0}^{\infty}a_{n,r_{2}}x^{n}\mbox{Log}(x)\,.
\end{equation}
Using recurrence relations with arbitrary $b_{0,r_{3}}$ 
\begin{equation}
\begin{array}{rcl}
b_{2,r_{3}} & = & \frac{1}{2}b_{1,r_{3}}\,\\
b_{n,r_{3}} & = & -\frac{(n-2)^{3}}{n^{2}}b_{n-2,r_{3}}
+\frac{2(n-1)^{2}}{n^{2}}b_{n-1,r_{3}}
\end{array}
\end{equation}
one can derive an exact formula for $b_{n,r_{3}}=b_{1,r_{3}}\frac{1}{n}$
and performing summations with arbitrary $b_{0,r_{3}}(=0)$ we have
$\sum_{n=0}^{\infty}b_{n,r_{3}}x^{n}=-b_{0,r_{3}}\mbox{Log}(1-x)$.
All these solutions look like 
\begin{equation}
\begin{array}{rcl}
u_{1}(x) & = & x^{r_{1}}\sum_{n=0}^{\infty}a_{n,r_{1}}x^{n}=-a_{1,r_{3}}\mbox{Log}(1-x)\,,\\
u_{2}(x) & = & x^{r_{2}}\sum_{n=0}^{\infty}a_{n,r_{2}}x^{n}=a_{0,r_{2}}\,,\\
u_{3}(x) & = & x^{r_{3}}\sum_{n=0}^{\infty}b_{n,r_{3}}x^{n}+a_{0,r_{2}}\mbox{Log}(x)=-b_{0,r_{3}}\mbox{Log}(1-x)+a_{0,r_{2}}\mbox{Log}(x)\,.
\end{array}
\end{equation}
Gathering together all this solutions we obtain
\begin{equation}
u(x)=A+B\mbox{Log}(-x)+C\mbox{Log}(x-1)\,,
\end{equation}
where $A$, $B,$ $C$ are arbitrary.
See also Ref.\ \ci{DeVega:2001pu}.

\subsection{Solution 
with $q_{3}\ne0$ and $h=2$}

For $h=2$ and $q_3\ne0$ 
we have three solutions 
\begin{equation}
\begin{array}{rcl}
u_{1}(x) & = & x^{r_{1}}\sum_{n=0}^{\infty}a_{n,r_{1}}x^{n}\,,\\
u_{2}(x) & = & x^{r_{2}}\sum_{n=0}^{\infty}b_{n,r_{2}}x^{n}+x^{r_{1}}\sum_{n=0}^{\infty}a_{n,r_{1}}x^{n}\mbox{Log}(x)\,,\\
u_{3}(x) & = & x^{r_{3}}\sum_{n=0}^{\infty}c_{n,r_{3}}x^{n}
+2x^{r_{2}}\sum_{n=0}^{\infty}b_{n,r_{2}}x^{n}\mbox{Log}(x)
+x^{r_{1}}\sum_{n=0}^{\infty}a_{n,r_{1}}x^{n}\mbox{Log}^{2}(x)
\end{array}
\end{equation}
with $r_{1}=2$,
$r_{2}=1$, $r_{3}=0$. 
Here $a_{0,r_{1}}$ is arbitrary (e.g. $1$)
\begin{equation}
a_{1,r_{1}}=\frac{-iq_{3}+2r(r-1)(r-2)}{r(r^{2}-1)}a_{0,r_{1}}
\end{equation}
and $m=n+r_{1}$\begin{equation}
a_{n,r_{1}}=-\frac{(m-3)(m-4)}{m(m-1)}a_{n-2,r_{1}}
+\frac{-iq_{3}+2(m-1)(m-2)(m-3)}{(m-2)(m-1)m}a_{n-1,r_{1}}\,,
\end{equation}
whereas $b_{0,r_{2}}=-\frac{2}{iq_{3}}a_{0,r_{1}}$ and $b_{1,r_{2}}$
is arbitrary while
\begin{equation}
b_{2,r_{2}}=\frac{1}{6}(-iq_{3}b_{1,r_{2}}+4a_{0,r_{1}}-11a_{1,r_{1}})\,,
\end{equation}
\begin{equation}
\begin{array}{rcl}
b_{n,r_{2}} & = & -\frac{(n-2)(n-3)}{n(n+1)}b_{n-2,r_{2}}
+\frac{-iq_{3}+2(n-2)(n-1)n}{n(n^{2}-1)}b_{n-1,r_{2}}
-\frac{11+3(n-4)n}{n(n^{2}-1)}a_{n-3,r_{1}}+\\
 &  & +\frac{6(n-2)n+4}{n(n^{2}-1)}a_{n-2,r_{1}}
+\frac{1-3n^2}{n(n^{2}-1)}a_{n-1,r_{1}}\,.
\end{array}
\end{equation}
Moreover, $c_{0,r_{3}}=\frac{2}{iq_{3}}b_{0,r_{2}}$ 
and $c_{1,r_{3}}=-\frac{4(b_{0,r_{2}}+b_{1,r_{2}})}{iq_{3}}
-\frac{6a_{0,r_{1}}}{iq_{3}}$
and $c_{2,r_{3}}$ is arbitrary while 
\begin{equation}
c_{3,r_{3}}=\frac{1}{6}(-iq_{3}c_{2,r_{3}}+2b_{0,r_{2}}
+8b_{1,r_{2}}-22b_{2,r_{2}}+12a_{0,r_{1}}-12a_{1,r_{1}})\,,
\end{equation}
\begin{equation}
\begin{array}{rcl}
c_{n,r_{3}} & = & -\frac{(n-3)(n-4)}{n(n-1)}c_{n-2,r_{3}}
+\frac{-iq_{3}+2(n-3)(n-2)(n-1)}{n(n-1)(n-2)}c_{n-1,r_{3}}
+\frac{2(3n(6-n)-26)}{n(n-1)(n-2)}b_{n-3,r_{2}}+\\
 &  & +\frac{4(11-12m+3m^{2})}{n(n-1)(n-2)}b_{n-2,r_{2}}
-\frac{2(2-6m+3m^{2})}{n(n-1)(n-2)}b_{n-1,r_{2}}
-\frac{6(n-3)}{n(n-1)(n-2)}a_{n-4,r_{1}}
+\frac{12}{n(n-1)}a_{n-3,r_{1}}+\\
 &  & -\frac{6}{n(n-2)}a_{n-2,r_{1}}\,.
\end{array}
\end{equation}

\subsection{Solution 
with $q_{3}=0$ and $h=2$}

In this case $a_{0,r}$ is arbitrary (e.g. equal to $1$). 
For the first solution
with $r_{1}=2$ we have remaining coefficients $a_{n>0,r_{1}}=0$.
The second solution, with $r_{2}=1$, has an arbitrary $a_{1,r_{2}}$
(we can take it as $0$), $a_{2,r_{2}}=0$ and third one with arbitrary
$a_{1,r_{3}}$, $a_{2,r_{3}}$
(we also set them to $0$). All these
solutions look like 
\begin{equation}
\begin{array}{rcl}
u_{1}(x) & = & x^{r_{1}}\sum_{n=0}^{\infty}a_{n,r_{1}}x^{n}
=x^{2}a_{0,r_{3}}\,,\\
u_{2}(x) & = & x^{r_{2}}\sum_{n=0}^{\infty}a_{n,r_{2}}x^{n}=xa_{0,r_{2}}\,,\\
u_{3}(x) & = & x^{r_{3}}\sum_{n=0}^{\infty}a_{n,r_{3}}x^{n}=a_{0,r_{3}}\,.
\end{array}
\end{equation}
Gathering together this all solutions we have
\begin{equation}
u(x)=A+B(-x)^{2}+C(x-1)^{2}\,,
\end{equation}
where $A$, $B,$ $C$ are arbitrary. As we can see this solution
corresponds to the solution with $q_{3}=0$ and arbitrary $h\not\in\{0,1\}$.
For other integer $h\not\in\{0,1\}$ we can observe the same correspondence.

\section{Solutions for $s=0$ around $x=1^{-}$}

\subsection{Solutions 
for $q_3\ne0$ and $h\not\in\mathbb{Z}$}

Similarly to the case $x=0$ for $x=1^{-}$ we have 
independent solutions 
where an indicial
equation 
has a form
\begin{equation} 
(h-n-r)(r+n-1)(n+r)=0\,.
\lab{eq:indx1}
\end{equation}
Its solutions have following values: $r_{1}=h$,
$r_{2}=1$ and $r_{3}=0$. 
As we can see we have two cases when $h\not\in\mathbb{Z}$
(one solution with $\mbox{Log}(x)$) and $h\in\mathbb{Z}$ (one solution
with $\mbox{Log}(x)$ and one solution with $\mbox{Log}^{2}(x)$). We
will see below that we have to consider solutions with $q_{3}=0$,
separately.

In this case, for $q_3 \ne 0$:
\begin{equation}
\begin{array}{rcl}
u_{1}(x) & = & (1-x)^{r_{1}}\sum_{n=0}^{\infty}a_{n,r_{1}}(1-x)^{n}\,,\\
u_{2}(x) & = & (1-x)^{r_{2}}\sum_{n=0}^{\infty}a_{n,r_{2}}(1-x)^{n}\,,\\
u_{3}(x) & = & (1-x)^{r_{3}}\sum_{n=0}^{\infty}b_{n,r_{3}}(1-x)^{n}
+(1-x)^{r_{2}}\sum_{n=0}^{\infty}a_{n,r_{2}}(1-x)^{n}\mbox{Log}(1-x)\,,
\end{array}
\end{equation}
where $a_{0,r}$ is arbitrary (e.g. $1$) and
\begin{equation}
a_{1,r}=\frac{(iq_{3}+(h-2r)(h-r)r)}{(h-1-r)r(1+r)}a_{0,r}
\end{equation}
and $m=n+r$
\begin{equation}
a_{n,r}=\frac{(h-m+1)(h-m+2)(m-2)}{(h-m)(m-1)m}a_{n-2,r}
-\frac{(iq_{3}+(1+h-m)(m-1)(h-2(m-1)))}{(h-m)(m-1)m}a_{n-1,r}\,,
\end{equation}
whereas $b_{0,r_{3}}=\frac{(1-h)}{iq_{3}}a_{0,r_{2}}$ (one can notice
that coefficient $b_{0,r_{3}}$ is valid only for $q_{3}\ne0$) and
$b_{1,r_{3}}$ is arbitrary
\begin{equation}
\begin{array}{rcl}
b_{2,r_{3}} & = & \frac{iq_{3}+(h-2)(h-1)}{2(2-h)}b_{1,r_{3}}
-\frac{8-3h}{2(2-h)}a_{1,r_{3}}
+\frac{6+h(h-6)}{2(2-h)}a_{0,r_{3}}
\end{array}
\end{equation}
and\begin{equation}
\begin{array}{rcl}
b_{n,r_{3}} & = & \frac{(h+1-m)(h-m+2)(m-2)}{(h-m)(m-1)m}b_{n-2,r_{3}}
-\frac{iq_{3}+(1+h-m)(m-1)(h-2(m-1))}{(h-m)(m-1)m}b_{n-1,r_{3}}+\\
 &  & +\frac{h^{2}+h(7-4m)+(m-2)(3m-4)}{(h-m)(m-1)m}a_{n-3,r_{3}}
-\frac{h^{2}-6h(m-1)+6(m-1)^{2}}{(h-m)(m-1)m}a_{n-2,r_{3}}+\\
 &  & -\frac{2m-3m^{2}+h(2m-1)}{(h-m)(m-1)m}a_{n-1,r_{3}}
\end{array}
\end{equation}
where $m=r_{3}+n$.

We can easy see that above coefficients correspond to coefficients
for solution around $x=0^{-}$ with $q_{3}\rightarrow-q_{3}$.

\subsection{Solution 
with $q_{3}=0$}


Here $a_{0,r}$ is arbitrary (e.g. equal to  $1$). 
For the first solution with
$r_{1}=h$ we have $a_{1,r_{1}}=0$ and for the third one $a_{1,r_{3}}$
is arbitrary (let us take $0$) . We do not need $\mbox{Log}$ solutions.
The second solution is more complicated. One can derive an exact formula
for 
\begin{equation}
a_{n,r_{2}}=a_{0,r_{2}}\prod_{k=1}^{n}\frac{k-h}{k+1}
=\frac{\Gamma(1-h+n)}{\Gamma(1-h)\Gamma(n+2)}a_{0,r_{2}}
\end{equation}
and performing summations
\begin{equation}
\begin{array}{rcl}
u_{1}(x) & = & (1-x)^{r_{1}}\sum_{n=0}^{\infty}a_{n,r_{1}}(1-x)^{n}
=(1-x)^{h}a_{0,r_{1}}\,,\\
u_{2}(x) & = & (1-x)^{r_{2}}\sum_{n=0}^{\infty}a_{n,r_{2}}(1-x)^{n}
 =  a_{0,r_{2}}(1-x)\sum_{n=0}^{\infty}
\frac{\Gamma(1-h+n)}{\Gamma(1-h)\Gamma(n+2)}(1-x)^{n}\\
& = & -a_{0,r_{2}}\frac{1}{h}(x^{h}-1)\,,\\
u_{3}(x) & = & (1-x)^{r_{3}}\sum_{n=0}^{\infty}a_{n,r_{3}}(1-x)^{n}
=a_{0,r_{3}}\,.
\end{array}
\end{equation}
Gathering together all this solutions we have
\begin{equation}
u(x)=A+B(-x)^{h}+C(x-1)^{h}\,,
\end{equation}
where $A$, $B,$ $C$ are arbitrary. The above solution was presented
by Lipatov and Vacca in Refs.\ \ci{Bartels:1999yt,Lipatov:1998as}.

\subsection{Solution 
with $q_{3}\ne0$, $q_2=0$ and $h=1$}

For $h=1$, i.e. $q_2=0$, and $q_3\ne 0$ we have three solutions 
to the indicial equation 
(\ref{eq:indx1})
which are integer $r_{1}=1$,
$r_{2}=1$, $r_{3}=0$ so solutions are
\begin{equation}
\begin{array}{rcl}
u_{1}(x) & = & (1-x)^{r_{1}}\sum_{n=0}^{\infty}a_{n,r_{1}}(1-x)^{n}\,,\\
u_{2}(x) & = & (1-x)^{r_{2}}\sum_{n=0}^{\infty}b_{n,r_{2}}(1-x)^{n}
+x^{r_{1}}\sum_{n=0}^{\infty}a_{n,r_{1}}(1-x)^{n}\mbox{Log}(1-x)\,,\\
u_{3}(x) & = & (1-x)^{r_{3}}\sum_{n=0}^{\infty}c_{n,r_{3}}(1-x)^{n}
+2(1-x)^{r_{2}}\sum_{n=0}^{\infty}b_{n,r_{2}}(1-x)^{n}\mbox{Log}(1-x)+\\
 & & +(1-x)^{r_{1}}\sum_{n=0}^{\infty}
a_{n,r_{1}}(1-x)^{n}\mbox{Log}^{2}(1-x)\,,
\end{array}
\end{equation}
where $a_{0,r}$ is arbitrary (e.g. equal to  $1$)
\begin{equation}
a_{1,r}=\frac{iq_{3}+r+r^{2}(2r-3)}{r^{2}(1+r)}a_{0,r}
\end{equation}
and $m=n+r$\begin{equation}
a_{n,r}=-\frac{(m-3)(m-2)^{2}}{(m-1)^{2}m}a_{n-2,r}
+\frac{(iq_{3}+(m-2)(m-1)(2m-3))}{(m-1)^{2}m}a_{n-1,r}\,,
\end{equation}
whereas 
\begin{equation}
b_{1,r_{2}}=\frac{1}{2}(iq_{3}b_{0,r_{2}}+a_{0,r_{1}}-5a_{1,r_{1}})\,,
\end{equation}
\begin{equation}
\begin{array}{rcl}
b_{n,r_{2}} & = & -\frac{(n-1)^{2}(n-2)}{n^{2}(n+1)}b_{n-2,r_{2}}
+\frac{(iq_{3}+2n^{3}-3n^{2}+n)}{n^{2}(n+1)}b_{n-1,r_{2}}
+\frac{1+6n(n-1)}{n^{2}(n+1)}a_{n-1,r_{1}}+\\
 &  & +\frac{(n-1)(5-3n)}{n^{2}(n+1)}a_{n-2,r_{1}}
-\frac{(2+3n)}{n(n+1)}a_{n,r_{1}}\,.
\end{array}
\end{equation}
Moreover, $c_{0,r_{3}}=\frac{2i}{q_{3}}a_{0,r_{1}}$ $c_{1,r_{3}}$ 
is arbitrary
and
\begin{equation}
c_{2,r_{3}}=\frac{1}{2}iq_{3}c_{1,r_{3}}+b_{0,r_{2}}-5b_{1,r_{2}}
+3a_{0,r_{1}}-4a_{1,r_{1}}\,,
\end{equation}
\begin{equation}
\begin{array}{rcl}
c_{n,r_{3}} & = & -\frac{(n-2)^{2}(n-3)}{n(n-1)^{2}}c_{n-2,r_{3}}
+\frac{iq_{3}-6+n(13+n(2n-9))}{n(n-1)^{2}}c_{n-1,r_{3}}
-\frac{2(n-2)(3n-8)}{n(n-1)^{2}}b_{n-2,r_{2}}+\\
 &  & +\frac{2(13+6(n-3)n)}{n(n-1)^{2}}b_{n-1,r_{2}}
-\frac{2(1+n(3n-4)}{n(n-1)^{2}}b_{n,r_{2}}
+\frac{2(7-3n)}{n(n-1)^{2}}a_{n-3,r_{1}}
-\frac{6(3-2n)}{n(n-1)^{2}}a_{n-2,r_{1}}+\\
 &  & +\frac{2(2-3n)}{n(n-1)^{2}}a_{n-1,r_{1}}\,.
\end{array}
\end{equation}

\subsection{Solution 
with $q_2=q_{3}=0$ and $h=1$}

In this case $a_{0,r}$ is arbitrary (e.g. $1$). For the first solution
with $r_{1}=(h=1)$ we have $a_{1,r_{1}}=0$ and for the third one
$a_{1,r_{3}}$ with $r_{3}=0$ is arbitrary (let us take $0$). The
second solution is more complicated. We need $\mbox{Log}-$solutions
\begin{equation}
u_{2}(x)=(1-x)^{r_{2}}\,,
\sum_{n=0}^{\infty}b_{n,r_{2}}(1-x)^{n}+x^{r_{1}}\,,
\sum_{n=0}^{\infty}a_{n,r_{1}}(1-x)^{n}\mbox{Log}(1-x)\,.
\end{equation}
Using recursive relations with $b_{1,r_{2}}=\frac{1}{2}a_{0,r_{1}}$
(and $a_{n,r_{1}}=0$ for $n>0$) 
\begin{equation}
\begin{array}{rcl}
b_{2,r_{2}} & = & \frac{1}{3}b_{1,r_{2}}\,,\\
b_{n,r_{2}} & = & -\frac{(n-1)^{2}(n-2)}{n^{2}(n+1)}b_{n-2,r_{2}}
+\frac{(n-1)n(2n-1)}{n^{2}(n+1)}b_{n-1,r_{2}}
\end{array}
\end{equation}
one can derive an exact formula for $b_{n,r_{2}}=b_{1,r_{2}}\frac{2}{(n+1)n}$
and performing summations with arbitrary $b_{0,r_{2}}(=0)$
we have
\begin{equation}
x\sum_{n=0}^{\infty}b_{n,r_{2}}(1-x)^{n}=2b_{1,r_{2}}((1-x)
+\mbox{Log}(x)-(1-x)\mbox{Log}(x))
\end{equation}
what gives us 
\begin{equation}
\begin{array}{rcl}
u_{1}(x) & = & (1-x)^{r_{1}}\sum_{n=0}^{\infty}a_{n,r_{1}}(1-x)^{n}
=a_{0,r_{1}}=(1-x)a_{0,r_{1}}\,,\\
u_{2}(x) & = & a_{0,r_{1}}((1-x)+\mbox{Log}(x)-(1-x)\mbox{Log}(x)
+(1-x)\mbox{Log}(1-x))\,,\\
u_{3}(x) & = & (1-x)^{r_{3}}\sum_{n=0}^{\infty}a_{n,r_{3}}(1-x)^{n}
=a_{0,r_{3}}\,.
\end{array}
\end{equation}
Gathering together all this solutions we have
\begin{equation}
u(x)=A+B(-x)+C((x-1)\mbox{Log}(x-1)+(-x)\mbox{Log}(-x))\,,
\end{equation}
where $A$, $B,$ $C$ are arbitrary. Here we have also obtained 
the consistent solution.

\subsection{Solution 
with  $q_{3}\ne0$, $q_2=0$ and $h=0$}

For $h=0$, i.e. $q_2=0$, and $q_3 \ne 0$ we have 
three solutions 
\begin{equation}
\begin{array}{rcl}
u_{1}(x) & = & (1-x)^{r_{1}}\sum_{n=0}^{\infty}a_{n,r_{1}}(1-x)^{n}\,,\\
u_{2}(x) & = & (1-x)^{r_{2}}\sum_{n=0}^{\infty}b_{n,r_{2}}(1-x)^{n}
+(1-x)^{r_{1}}\sum_{n=0}^{\infty}a_{n,r_{1}}(1-x)^{n}\mbox{Log}(1-x)\,,\\
u_{3}(x) & = & (1-x)^{r_{3}}\sum_{n=0}^{\infty}c_{n,r_{3}}(1-x)^{n}
+2(1-x)^{r_{2}}\sum_{n=0}^{\infty}b_{n,r_{2}}(1-x)^{n}\mbox{Log}(1-x)+\\
 &  & +(1-x)^{r_{1}}\sum_{n=0}^{\infty}
a_{n,r_{1}}(1-x)^{n}\mbox{Log}^{2}(1-x)\,,
\end{array}
\end{equation}
where coefficients 
$r_{1}=1$,
$r_{2}=0$, $r_{3}=0$ are integer
while $a_{0,r}$ is arbitrary (e.g. $1$)
\begin{equation}
a_{1,r}=\frac{iq_{3}+2r^{2}}{r(1+r)^{2}}a_{0,r}
\end{equation}
and $m=n+r$
\begin{equation}
a_{n,r}=-\frac{(m-2)^{2}}{m^{2}}a_{n-2,r}+\frac{iq_{3}
+2(m-1)^{3}}{(m-1)m^{2}}a_{n-1,r}\,,
\end{equation}
whereas $b_{0,r_{2}}=\frac{1}{iq_{3}}a_{0,r_{1}}$ and $b_{1,r_{2}}$
is arbitrary while
\begin{equation}
b_{2,r_{2}}=\frac{1}{4}((2+iq_{3})b_{1,r_{2}}+6a_{0,r_{1}}-8a_{1,r_{1}})\,,
\end{equation}
\begin{equation}
\begin{array}{rcl}
b_{n,r_{2}} & = & -\frac{(n-2)^{2}}{n^{2}}b_{n-2,r_{2}}
+\frac{iq_{3}+2(n-1)^{3}}{n^{2}(n-1)}b_{n-1,r_{2}}
-\frac{(n-2)(3n-4)}{n^{2}(n-1)}a_{n-3,r_{1}}+\\
 &  & +\frac{6(n-1)}{n^{2}}a_{n-2,r_{1}}
+\frac{2-3n}{n(n-1)}a_{n-1,r_{1}}\,.
\end{array}
\end{equation}
Moreover, $c_{0,r_{3}}=\frac{2}{iq_{3}}(2a_{0,r_{1}}+b_{1,r_{2}})$,
$c_{1,r_{3}}$
is arbitrary and
\begin{equation}
c_{2,r_{3}}=\frac{1}{4}(2+iq_{3})c_{1,r_{3}}+3b_{1,r_{2}}-4b_{2,r_{2}}
+3a_{0,r_{1}}-\frac{5}{2}a_{1,r_{1}}\,,
\end{equation}
\begin{equation}
\begin{array}{rcl}
c_{n,r_{3}} & = & -\frac{(n-2)^{2}}{n^{2}}c_{n-2,r_{3}}
+\frac{iq_{3}+2(n-1)^{3}}{n^{2}(n-1)}c_{n-1,r_{3}}
-\frac{2(n-2)(3n-4)}{n^{2}(n-1)}b_{n-2,r_{2}}+\\
 &  & +\frac{12(n-1)}{n^{2}}b_{n-1,r_{2}}+\frac{2(2-3n)}{n(n-1)}b_{n,r_{2}}
+\frac{2(5-3n)}{n^{2}(n-1)}a_{n-3,r_{1}}+\frac{12}{n^{2}}a_{n-2,r_{1}}+\\
 &  & +\frac{2(1-3n)}{n^{2}(n-1)}a_{n-1,r_{1}}\,.
\end{array}
\end{equation}

\subsection{Solution 
with $q_2=q_{3}=0$ and $h=0$}

In this case $a_{0,r}$ is arbitrary (e.g. $1$). For the first solution
with $r_{1}=1$ we have recurrence  relations with 
$a_{1,r_{1}}=\frac{1}{2}a_{0,r_{1}}$
\begin{equation}
\begin{array}{rcl}
a_{2,r_{1}} & = & \frac{1}{3}a_{1,r_{1}}\,,\\
a_{n,r_{1}} & = & -\frac{(n-1)^{2}}{n+1}a_{n-2,r_{1}}
+\frac{2n^{2}}{n+1}a_{n-1,r_{1}}\,.
\end{array}
\end{equation}
Summing series up we derive an exact formula for 
\begin{equation}
u_{1}(x)=(1-x)^{r_{1}}\sum_{n=0}^{\infty}a_{n,r_{1}}(1-x)^{n}
=(1-x)\sum_{n=0}^{\infty}(1-x)^{n}a_{0,r_{1}}\frac{1}{n+1}=-\mbox{Log}(x)\,.
\end{equation}

In the second solution, with $r_{2}=0$, we have an arbitrary $a_{1,r_{2}}$
(let us take $0$) so the solution is $u_{2}(x)=a_{0,r_{2}}$. The third
one is with $\mbox{Log}$ solutions 
\begin{equation}
u_{3}(x)=(1-x)^{r_{3}}\sum_{n=0}^{\infty}b_{n,r_{3}}(1-x)^{n}
+(1-x)^{r_{2}}\sum_{n=0}^{\infty}a_{n,r_{2}}(1-x)^{n}\mbox{Log}(1-x)\,.
\end{equation}
Using recurrence  relations with arbitrary $b_{0,r_{3}}$ 
\begin{equation}
\begin{array}{rcl}
b_{2,r_{3}} & = & \frac{1}{2}b_{1,r_{3}}\,,\\
b_{n,r_{3}} & = & -\frac{(n-2)^{3}}{n^{2}}b_{n-2,r_{3}}
+\frac{2(n-1)^{2}}{n^{2}}b_{n-1,r_{3}}\end{array}
\end{equation}
one can derive an exact formula for $b_{n,r_{3}}=b_{1,r_{3}}\frac{1}{n}$
and performing summations with arbitrary $b_{0,r_{3}}(=0)$we have
$\sum_{n=0}^{\infty}b_{n,r_{3}}(1-x)^{n}=-b_{0,r_{3}}\mbox{Log}(x)$.
All these solutions look like 
\begin{equation}
\begin{array}{rcl}
u_{1}(x) & = & (1-x)^{r_{1}}\sum_{n=0}^{\infty}a_{n,r_{1}}(1-x)^{n}
=-a_{1,r_{3}}\mbox{Log}(x)\,,\\
u_{2}(x) & = & (1-x)^{r_{2}}\sum_{n=0}^{\infty}a_{n,r_{2}}(1-x)^{n}
=a_{0,r_{2}}\,,\\
u_{3}(x) & = & (1-x)^{r_{3}}\sum_{n=0}^{\infty}b_{n,r_{3}}(1-x)^{n}
+a_{0,r_{2}}\mbox{Log}(1-x)=-b_{0,r_{3}}\mbox{Log}(x)+a_{0,r_{2}}
\mbox{Log}(1-x)\,.
\end{array}
\end{equation}
Gathering together all this solutions we have
\begin{equation}
u(x)=A+B\mbox{Log}(-x)+C\mbox{Log}(x-1)\,,
\end{equation}
where $A$, $B,$ $C$ are arbitrary.

\subsection{Solution 
with $q_{3}\ne0$ and $h=2$}

For integer $h=2$ and $q_3\ne 0$ 
we have three solutions to the indicial equation 
(\ref{eq:indx1})
which are integer $r_{1}=2$,
$r_{2}=1$, $r_{3}=0$ so solutions are
\begin{equation}
\begin{array}{rcl}
u_{1}(x) & = & (1-x)^{r_{1}}\sum_{n=0}^{\infty}a_{n,r_{1}}(1-x)^{n}\,,\\
u_{2}(x) & = & (1-x)^{r_{2}}\sum_{n=0}^{\infty}b_{n,r_{2}}(1-x)^{n}
+(1-x)^{r_{1}}\sum_{n=0}^{\infty}a_{n,r_{1}}(1-x)^{n}\mbox{Log}(1-x)\,,\\
u_{3}(x) & = & (1-x)^{r_{3}}\sum_{n=0}^{\infty}c_{n,r_{3}}(1-x)^{n}
+2(1-x)^{r_{2}}\sum_{n=0}^{\infty}b_{n,r_{2}}(1-x)^{n}\mbox{Log}(1-x)+\\
 &  & +(1-x)^{r_{1}}\sum_{n=0}^{\infty}
a_{n,r_{1}}(1-x)^{n}\mbox{Log}^{2}(1-x)\,,
\end{array}
\end{equation}
where $a_{0,r_{1}}$ is arbitrary (e.g. $1$)
\begin{equation}
a_{1,r_{1}}=\frac{iq_{3}+2r(r-1)(r-2)}{r(r^{2}-1)}a_{0,r_{1}}
\end{equation}
and $m=n+r_{1}$
\begin{equation}
a_{n,r_{1}}=-\frac{(m-3)(m-4)}{m(m-1)}a_{n-2,r_{1}}
+\frac{iq_{3}+2(m-1)(m-2)(m-3)}{(m-2)(m-1)m}a_{n-1,r_{1}}\,,
\end{equation}
whereas $b_{0,r_{2}}=\frac{2}{iq_{3}}a_{0,r_{1}}$ and $b_{1,r_{2}}$
is arbitrary while
\begin{equation}
b_{2,r_{2}}=\frac{1}{6}(iq_{3}b_{1,r_{2}}+4a_{0,r_{1}}-11a_{1,r_{1}})\,,
\end{equation}
\begin{equation}
\begin{array}{rcl}
b_{n,r_{2}} & = & -\frac{(n-2)(n-3)}{n(n+1)}b_{n-2,r_{2}}
+\frac{iq_{3}+2(n-2)(n-1)n}{n(n^{2}-1)}b_{n-1,r_{2}}
-\frac{11+3(n-4)n}{n(n^{2}-1)}a_{n-3,r_{1}}+\\
 &  & +\frac{6(n-2)n+4}{n(n^{2}-1)}a_{n-2,r_{1}}
+\frac{1-3n}{n(n^{2}-1)}a_{n-1,r_{1}}\,.
\end{array}
\end{equation}
Therefore, $c_{0,r_{3}}=-\frac{2}{iq_{3}}b_{0,r_{2}}$ and $c_{1,r_{3}}
=\frac{4(b_{0,r_{2}}+b_{1,r_{2}})}{iq_{3}}+\frac{6a_{0,r_{1}}}{iq_{3}}$
and $c_{2,r_{3}}$is arbitrary while
\begin{equation}
c_{3,r_{3}}=\frac{1}{6}(iq_{3}c_{2,r_{3}}+2b_{0,r_{2}}+8b_{1,r_{2}}
-22b_{2,r_{2}}-12a_{0,r_{1}}+12a_{1,r_{1}})\,,
\end{equation}
\begin{equation}
\begin{array}{rcl}
c_{n,r_{3}} & = & -\frac{(n-3)(n-4)}{n(n-1)}c_{n-2,r_{3}}
+\frac{iq_{3}+2(n-3)(n-2)(n-1)}{n(n-1)(n-2)}c_{n-1,r_{3}}
+\frac{2(3n(6-n)-26)}{n(n-1)(n-2)}b_{n-3,r_{2}}+\\
 &  & +\frac{4(11-12m+3m^{2})}{n(n-1)(n-2)}b_{n-2,r_{2}}
-\frac{2(2-6m+3m^{2})}{n(n-1)(n-2)}b_{n-1,r_{2}}
-\frac{6(n-3)}{n(n-1)(n-2)}a_{n-4,r_{1}}+\frac{12}{n(n-1)}a_{n-3,r_{1}}+\\
 &  & -\frac{6}{n(n-2)}a_{n-2,r_{1}}\,.
\end{array}
\end{equation}

\subsection{Solution 
with $q_{3}=0$ and $h=2$}

In this case $a_{0,r}$ is arbitrary (e.g. $1$). For the first solution
with $r_{1}=2$ we have other coefficients $a_{n>0,r_{1}}=0$.
The second solution, with $r_{2}=1$, has arbitrary $a_{1,r_{2}}$
(we can take it as $0$) and $a_{2,r_{2}}=0$. The third one is with arbitrary
$a_{1,r_{3}}$,$a_{2,r_{3}}$ (we also set them to $0$). All these
solutions look like 
\begin{equation}
\begin{array}{rcl}
u_{1}(x) & = & (1-x)^{r_{1}}\sum_{n=0}^{\infty}a_{n,r_{1}}(1-x)^{n}
=(1-x)^{2}a_{0,r_{3}}\,,\\
u_{2}(x) & = & (1-x)^{r_{2}}\sum_{n=0}^{\infty}a_{n,r_{2}}(1-x)^{n}
=(1-x)a_{0,r_{2}}\,,\\
u_{3}(x) & = & (1-x)^{r_{3}}\sum_{n=0}^{\infty}a_{n,r_{3}}(1-x)^{n}
=a_{0,r_{3}}\,.
\end{array}
\end{equation}
Gathering together all this solutions we have
\begin{equation}
u(x)=A+B(-x)^{2}+C(x-1)^{2}\,,
\end{equation}
where $A$, $B,$ $C$ are arbitrary. As we can see this solution
corresponds to the solution with $q_{3}=0$ and arbitrary $h\not\in\{0,1\}$.
For other integer $h\not\in\{0,1\}$ we can observe the same correspondence.

\section{Solutions for $s=0$ around $x=\infty^{-}$}

\subsection{Solutions 
for $q_3\ne0$ and $h\not\in\mathbb{Z}$}

When $x=\infty^{+}$ we have an indicial
equation 
\begin{equation}
(h+n+r)(h-1+r+n)(n+r)=0
\lab{eq:indxi}
\end{equation} 
so its solutions are $r_{1}=0$,
$r_{2}=1-h$ and $r_{3}=h$. As we can see we have to cases 
when $h\not\in\mathbb{Z}$
(one solution with $\mbox{Log}(x)$) and $h\in\mathbb{Z}$ (one solution
with $\mbox{Log}(x)$ and one solution with $\mbox{Log}^{2}(x)$). As we
will see below we have to consider solutions with $q_{3}=0$, separately.
In the first case\begin{equation}
\begin{array}{rcl}
u_{1}(x) & = & (1/x)^{r_{1}}\sum_{n=0}^{\infty}a_{n,r_{1}}(1/x)^{n}\,,\\
u_{2}(x) & = & (1/x)^{r_{2}}\sum_{n=0}^{\infty}a_{n,r_{2}}(1/x)^{n}\,,\\
u_{3}(x) & = & (1/x)^{r_{3}}\sum_{n=0}^{\infty}b_{n,r_{3}}(1/x)^{n}
+(1/x)^{r_{2}}\sum_{n=0}^{\infty}a_{n,r_{2}}(1/x)^{n}\mbox{Log}(1/x)\,,
\end{array}
\end{equation}
where $a_{0,r}$ is arbitrary (e.g. $1$)
\begin{equation}
a_{1,r}=\frac{(iq_{3}+r(h+r)(h+2r))}{(1+r)(h+r)(h+1+r)}a_{0,r}
\end{equation}
and $m=n+r$\begin{equation}
a_{n,r}=\frac{(m-2)(m-1)(2-h-m)}{(h+m)(h+m-1)m}a_{n-2,r}
+\frac{iq_{3}+(h-1+m)(m-1)(h+2(m-1))}{(h+m)(h+m-1)m}a_{n-1,r}\,,
\end{equation}
whereas $b_{0,r_{3}}=\frac{(1-h)}{iq_{3}}a_{0,r_{2}}$ (one can notice
that coefficient $b_{0,r_{3}}$ is valid only for $q_{3}\ne0$) and
$b_{1,r_{3}}$ is arbitrary
\begin{equation}
\begin{array}{rcl}
b_{2,r_{3}} & = & \frac{iq_{3}+2+(h-3)h}{2(2-h)}b_{1,r_{3}}
-\frac{8-3h}{2(2-h)}a_{1,r_{3}}+\frac{6+h(h-6)}{2(2-h)}a_{0,r_{3}}
\end{array}
\end{equation}
and\begin{equation}
\begin{array}{rcl}
b_{n,r_{3}} & = & \frac{(h+1-m)(h-m+2)(m-2)}{(h-m)(m-1)m}b_{n-2,r_{3}}
-\frac{iq_{3}+(1+h-m)(m-1)(h-2(m-1))}{(h-m)(m-1)m}b_{n-1,r_{3}}+\\
 &  & +\frac{h^{2}+h(7-4m)+(m-2)(3m-4)}{(h-m)(m-1)m}a_{n-3,r_{3}}
-\frac{h^{2}-6h(m-1)+6(m-1)^{2}}{(h-m)(m-1)m}a_{n-2,r_{3}}+\\
 &  & -\frac{2m-3m^{2}+h(2m-1)}{(h-m)(m-1)m}a_{n-1,r_{3}}\,,
\end{array}
\end{equation}
where $m=r_{3}+n$.

\subsection{Solution 
with $q_{3}=0$}

In this case $a_{0,r}$ is arbitrary (e.g. $1$). For the first solution
with $r_{1}=0$ we have $a_{1,r_{1}}=0$ and for the third one $a_{1,r_{3}}$
is arbitrary (let us take $0$) . 
Here it also turns out that we do not need $\mbox{Log}$ solutions.
The second solution is more complicated. One can derive exact formula
for 
\begin{equation}
a_{n,r_{2}}=a_{0,r_{2}}\prod_{k=1}^{n}\frac{k-h}{k+1}
=\frac{\Gamma(1-h+n)}{\Gamma(1-h)\Gamma(n+2)}a_{0,r_{2}}
\end{equation}
and performing summations
\begin{equation}
\begin{array}{rcl}
u_{1}(x) & = & (1/x)^{r_{1}}\sum_{n=0}^{\infty} a_{n,r_{1}}(1/x)^{n}
=a_{0,r_{1}}\,,\\
u_{2}(x) & = & (1/x)^{r_{2}}\sum_{n=0}^{\infty} a_{n,r_{2}}(1/x)^{n}
=a_{0,r_{2}}(1/x)^{1-h}\sum_{n=0}^{\infty}
\frac{\Gamma(1-h+n)}{\Gamma(1-h)\Gamma(n+2)}(1/x)^{n}\\
& = & -a_{0,r_{2}}\frac{1}{h}((x-1)^{h}-x^{h})\,,\\
u_{3}(x) & = & (1/x)^{r_{3}}
\sum_{n=0}^{\infty}a_{n,r_{3}}(1/x)^{n}=a_{0,r_{3}}(1/x)^{-h}\,.
\end{array}
\end{equation}
Gathering together all this solutions we have
\begin{equation}
u(x)=A+B(-x)^{h}+C(x-1)^{h}\,,
\end{equation}
where $A$, $B,$ $C$ are arbitrary. The above solution was presented
by Lipatov and Vacca in Refs.\ \ci{Bartels:1999yt,Lipatov:1998as}.

\subsection{Solution 
with $q_{3}\ne0$, $q_2=0$ and $h=1$}

For integer $h=1$, i.e. $q_2=0$, and $q_3 \ne 0$ 
we have three solutions to the indicial equation 
(\ref{eq:indxi})
which are integer $r_{1}=0$,
$r_{2}=0$, $r_{3}=-1$ so solutions are
\begin{equation}
\begin{array}{rcl}
u_{1}(x) & = & (1/x)^{r_{1}}\sum_{n=0}^{\infty}a_{n,r_{1}}(1/x)^{n}\\
u_{2}(x) & = & (1/x)^{r_{2}}\sum_{n=0}^{\infty}b_{n,r_{2}}(1/x)^{n}
+(1/x)^{r_{1}}\sum_{n=0}^{\infty}a_{n,r_{1}}(1/x)^{n}\mbox{Log}(1/x)\\
u_{3}(x) & = & (1/x)^{r_{3}}\sum_{n=0}^{\infty}c_{n,r_{3}}(1/x)^{n}
+2(1/x)^{r_{2}}\sum_{n=0}^{\infty}b_{n,r_{2}}(1/x)^{n}\mbox{Log}(1/x)+\\
 &  & +(1/x)^{r_{1}}\sum_{n=0}^{\infty}a_{n,r_{1}}(1/x)^{n}
\mbox{Log}^{2}(1/x)\,,
\end{array}
\end{equation}
where $a_{0,r}$ is arbitrary (e.g. $1$)
\begin{equation}
a_{1,r}=\frac{iq_{3}+r(1+r)(1+2r)}{(1+r)^{2}(2+r)}a_{0,r}
\end{equation}
and $m=n+r$
\begin{equation}
a_{n,r}=-\frac{(m-2)(m-1)^{2}}{(m+1)m^{2}}a_{n-2,r}
+\frac{iq_{3}+(m-1)m(2m-1)}{(m+1)m^{2}}a_{n-1,r}\,,
\end{equation}
whereas 
\begin{equation}
b_{1,r_{2}}=\frac{1}{2}(iq_{3}b_{0,r_{2}}+a_{0,r_{1}}-5a_{1,r_{1}})\,,
\end{equation}
\begin{equation}
\begin{array}{rcl}
b_{n,r_{2}} & = & -\frac{(n-1)^{2}(n-2)}{n^{2}(n+1)}b_{n-2,r_{2}}+
\frac{(iq_{3}+2m^{3}-3m^{2}+m)}{n^{2}(n+1)}b_{n-1,r_{2}}
+\frac{1+6n(n-1)}{n^{2}(n+1)}a_{n-1,r_{1}}+\\
 &  & +\frac{(n-1)(5-3n)}{n^{2}(n+1)}a_{n-2,r_{1}}
-\frac{(2+3n)}{n(n+1)}a_{n,r_{1}}
\end{array}
\end{equation}
and $c_{0,r_{3}}=\frac{2}{i q_{3}}a_{0,r_{1}}$ $c_{1,r_{3}}$ is arbitrary
while
\begin{equation}
c_{2,r_{3}}=\frac{1}{2}iq_{3}c_{1,r_{3}}+b_{0,r_{2}}-5b_{1,r_{2}}
+3a_{0,r_{1}}-4a_{1,r_{1}}\,,
\end{equation}
\begin{equation}
\begin{array}{rcl}
c_{n,r_{3}} & = & -\frac{(n-2)^{2}(n-3)}{n(n-1)^{2}}c_{n-2,r_{3}}
+\frac{-6+n(13+n(2n-9))-iq_{3}}{n(n-1)^{2}}c_{n-1,r_{3}}
-\frac{2(n-2)(3n-8)}{n(n-1)^{2}}b_{n-2,r_{2}}+\\
 &  & +\frac{2(13+6(n-3)n)}{n(n-1)^{2}}b_{n-1,r_{2}}
-\frac{2(3n-1)}{n(n-1)}b_{n,r_{2}}+\frac{2(7-3n)}{n(n-1)^{2}}a_{n-3,r_{1}}
-\frac{6(3-2n)}{n(n-1)^{2}}a_{n-2,r_{1}}+\\
 &  & -\frac{2(3n-2)}{n(n-1)}a_{n-1,r_{1}}\,.
\end{array}
\end{equation}

\subsection{Solution 
with $q_2=q_{3}=0$ and $h=1$}

In this case $a_{0,r}$ is arbitrary (e.g. $1$). For the first solution
with $r_{1}=0$ we have $a_{1,r_{1}}=0$ and for the third one $a_{1,r_{3}}$
with $r_{3}=0$ is arbitrary (let us take $0$). The second solution
is more complicated. We need $\mbox{Log}-$solutions 
\begin{equation}
u_{2}(x)=(1/x)^{r_{2}}\sum_{n=0}^{\infty}b_{n,r_{2}}(1/x)^{n}
+(1/x)^{r_{1}}\sum_{n=0}^{\infty}a_{n,r_{1}}(1/x)^{n}\mbox{Log}(1/x)\,.
\end{equation}
Using recurrence  relations with $b_{1,r_{2}}=\frac{1}{2}a_{0,r_{1}}$
(and $a_{n,r_{1}}=0$ for $n>0$) 
\begin{equation}
\begin{array}{rcl}
b_{2,r_{2}} & = & \frac{1}{3}b_{1,r_{2}}\,,\\
b_{n,r_{2}} & = & -\frac{(n-1)^{2}(n-2)}{n^{2}(n+1)}b_{n-2,r_{2}}
+\frac{(n-1)n(2n-1)}{n^{2}(n+1)}b_{n-1,r_{2}}
\end{array}
\end{equation}
one can derive an exact formula for $b_{n,r_{2}}=b_{1,r_{2}}\frac{2}{(n+1)n}$
and performing summations with arbitrary $b_{0,r_{2}}(=0)$ we have
$\sum_{n=0}^{\infty}b_{n,r_{2}}(1/x)^{n}=2b_{1,r_{2}}(1-\mbox{Log}(1-1/x)
+x\mbox{Log}(1-1/x))$, so that
\begin{equation}
\begin{array}{rcl}
u_{1}(x) & = & (1/x)^{r_{1}}\sum_{n=0}^{\infty}a_{n,r_{1}}(1/x)^{n}
=a_{0,r_{1}}\,,\\
u_{2}(x) & = & a_{0,r_{1}}(1-\mbox{Log}(1-1/x)+x\mbox{Log}(1-1/x)
+\mbox{Log}(1/x))\,,\\
u_{3}(x) & = & (1/x)^{r_{3}}\sum_{n=0}^{\infty}a_{n,r_{3}}(1/x)^{n}
=(1/x)^{-1}a_{0,r_{3}}\,.
\end{array}
\end{equation}
Gathering together all this solutions we have
\begin{equation}
u(x)=A+B(-x)+C((x-1)\mbox{Log}(x-1)+(-x)\mbox{Log}(-x))\,,
\end{equation}
where $A$, $B,$ $C$ are arbitrary.

\subsection{Solution 
with $q_{3}\ne0$, $q_2=0$ and $h=0$}

For $h=0$, i.e. $q_2=0$, and $q_3 \ne 0$ 
we have 
three solutions of the indicial equation 
(\ref{eq:indxi})
which are integer $r_{1}=1$,
$r_{2}=0$, $r_{3}=0$ so solutions are
\begin{equation}
\begin{array}{rcl}
u_{1}(x) & = & (1/x)^{r_{1}}\sum_{n=0}^{\infty}a_{n,r_{1}}(1/x)^{n}\,,\\
u_{2}(x) & = & (1/x)^{r_{2}}\sum_{n=0}^{\infty}b_{n,r_{2}}(1/x)^{n}
+(1/x)^{r_{1}}\sum_{n=0}^{\infty}a_{n,r_{1}}(1/x)^{n}\mbox{Log}(1/x)\,,\\
u_{3}(x) & = & (1/x)^{r_{3}}\sum_{n=0}^{\infty}c_{n,r_{3}}(1/x)^{n}
+2(1/x)^{r_{2}}\sum_{n=0}^{\infty}b_{n,r_{2}}(1/x)^{n}\mbox{Log}(1/x)+\\
 &  & +(1/x)^{r_{1}}\sum_{n=0}^{\infty}a_{n,r_{1}}(1/x)^{n}
\mbox{Log}^{2}(1/x)\,,
\end{array}
\end{equation}
where $a_{0,r}$ is arbitrary (e.g. $1$)
\begin{equation}
a_{1,r}=\frac{iq_{3}+2r^{2}}{r(1+r)^{2}}a_{0,r}
\end{equation}
and $m=n+r$
\begin{equation}
a_{n,r}=-\frac{(m-2)^{2}}{m^{2}}a_{n-2,r}
+\frac{iq_{3}+2(m-1)^{3}}{(m-1)m^{2}}a_{n-1,r}\,,
\end{equation}
whereas $b_{0,r_{2}}=\frac{1}{iq_{3}}a_{0,r_{1}}$ and $b_{1,r_{2}}$
is arbitrary while
\begin{equation}
b_{2,r_{2}}=\frac{1}{4}((2+iq_{3})b_{1,r_{2}}+6a_{0,r_{1}}-8a_{1,r_{1}})\,,
\end{equation}
\begin{equation}
\begin{array}{rcl}
b_{n,r_{2}} & = & -\frac{(n-2)^{2}}{n^{2}}b_{n-2,r_{2}}+\frac{iq_{3}
+2(n-1)^{3}}{n^{2}(n-1)}b_{n-1,r_{2}}
-\frac{(n-2)(3n-4)}{n^{2}(n-1)}a_{n-3,r_{1}}+\\
 &  & +\frac{6(n-1)}{n^{2}}a_{n-2,r_{1}}+\frac{2-3n}{n(n-1)}a_{n-1,r_{1}}\,.
\end{array}
\end{equation}
Furthermore, $c_{0,r_{3}}=\frac{2}{iq_{3}}(2a_{0,r_{1}}+b_{1,r_{2}})$ and
$c_{1,r_{3}}$ is arbitrary and 
\begin{equation}
c_{2,r_{3}}=\frac{1}{4}(2+iq_{3})c_{1,r_{3}}+3b_{1,r_{2}}-4b_{2,r_{2}}
+3a_{0,r_{1}}-\frac{5}{2}a_{1,r_{1}}\,,
\end{equation}
\begin{equation}
\begin{array}{rcl}
c_{n,r_{3}} & = & -\frac{(n-2)^{2}}{n^{2}}c_{n-2,r_{3}}
+\frac{iq_{3}+2(n-1)^{3}}{n^{2}(n-1)}c_{n-1,r_{3}}
-\frac{2(n-2)(3n-4)}{n^{2}(n-1)}b_{n-2,r_{2}}+\\
 &  & +\frac{12(n-1)}{n^{2}}b_{n-1,r_{2}}+\frac{2(2-3n)}{n(n-1)}b_{n,r_{2}}
+\frac{2(5-3n)}{n^{2}(n-1)}a_{n-3,r_{1}}+\frac{12}{n^{2}}a_{n-2,r_{1}}+\\
 &  & +\frac{2(1-3n)}{n^{2}(n-1)}a_{n-1,r_{1}}\,.
\end{array}
\end{equation}

\subsection{Solution 
with $q_2=q_{3}=0$ and $h=0$}

In this case $a_{0,r}$ is arbitrary (e.g. $1$). For the first solution
with $r_{1}=1$ we have recurrence  relations with 
$a_{1,r_{1}}=\frac{1}{2}a_{0,r_{1}}$
\begin{equation}
\begin{array}{rcl}
a_{2,r_{1}} & = & \frac{1}{3}a_{1,r_{1}}\,,\\
a_{n,r_{1}} & = & -\frac{(n-1)^{2}}{n+1}a_{n-2,r_{1}}
+\frac{2n^{2}}{n+1}a_{n-1,r_{1}}\,.
\end{array}
\end{equation}
Summing series up we derive an exact formula for 
\begin{equation}
u_{1}(x)=(1/x)^{r_{1}}\sum_{n=0}^{\infty}a_{n,r_{1}}(1/x)^{n}
=(1/x)\sum_{n=0}^{\infty}(1/x)^{n}a_{0,r_{1}}\frac{1}{n+1}
=-\mbox{Log}(1-1/x)\,.
\end{equation}

In the second solution, with $r_{2}=0$, we have arbitrary $a_{1,r_{2}}$
(let us take $0$) so the second solution is $u_{2}(x)=a_{0,r_{2}}$. The third
one is with $\mbox{Log}$ solutions 
\begin{equation}
u_{3}(x)=(1/x)^{r_{3}}\sum_{n=0}^{\infty}b_{n,r_{3}}(1/x)^{n}+
(1/x)^{r_{2}}\sum_{n=0}^{\infty}a_{n,r_{2}}(1/x)^{n}\mbox{Log}(1/x)\,.
\end{equation}
Using recurrence relations with arbitrary $b_{0,r_{3}}$ 
\begin{equation}
\begin{array}{rcl}
b_{2,r_{3}} & = & \frac{1}{2}b_{1,r_{3}}\,,\\
b_{n,r_{3}} & = & -\frac{(n-2)^{3}}{n^{2}}b_{n-2,r_{3}}
+\frac{2(n-1)^{2}}{n^{2}}b_{n-1,r_{3}}
\end{array}
\end{equation}
one can derive an exact formula for $b_{n,r_{3}}=b_{1,r_{3}}\frac{1}{n}$
and performing summations with arbitrary $b_{0,r_{3}}(=0)$ we have
$\sum_{n=0}^{\infty}b_{n,r_{3}}(1/x)^{n}=-b_{0,r_{3}}\mbox{Log}(1-(1/x))$.
All these solutions look like 
\begin{equation}
\begin{array}{rcl}
u_{1}(x) & = & (1/x)^{r_{1}}\sum_{n=0}^{\infty}a_{n,r_{1}}(1/x)^{n}
=-a_{1,r_{3}}\mbox{Log}(1-1/x)\,,\\
u_{2}(x) & = & (1/x)^{r_{2}}\sum_{n=0}^{\infty}a_{n,r_{2}}x^{n}
=a_{0,r_{2}}\,,\\
u_{3}(x) & = & (1/x)^{r_{3}}\sum_{n=0}^{\infty}b_{n,r_{3}}(1/x)^{n}
+a_{0,r_{2}}\mbox{Log}(1/x)=-b_{0,r_{3}}\mbox{Log}(1-1/x)
+a_{0,r_{2}}\mbox{Log}(1/x)\,.
\end{array}
\end{equation}
Gathering together this all solutions we have
\begin{equation}
u(x)=A+B\mbox{Log}(-x)+C\mbox{Log}(x-1)\,,
\end{equation}
where $A$, $B,$ $C$ are arbitrary.

\subsection{Solution 
with $q_{3}\ne0$ and $h=2$} 

For $h=2$ and $q_{3}\ne0$ we have 
three solutions to the indicial equation 
(\ref{eq:indxi})
which are integer $r_{1}=0$,
$r_{2}=-1$, $r_{3}=-2$ so solutions are
\begin{equation}
\begin{array}{rcl}
u_{1}(x) & = & (1/x)^{r_{1}}\sum_{n=0}^{\infty}a_{n,r_{1}}(1/x)^{n}\,,\\
u_{2}(x) & = & (1/x)^{r_{2}}\sum_{n=0}^{\infty}b_{n,r_{2}}(1/x)^{n}
+(1/x)^{r_{1}}\sum_{n=0}^{\infty}a_{n,r_{1}}(1/x)^{n}\mbox{Log}(1/x)\,,\\
u_{3}(x) & = & (1/x)^{r_{3}}\sum_{n=0}^{\infty}c_{n,r_{3}}(1/x)^{n}
+2(1/x)^{r_{2}}\sum_{n=0}^{\infty}b_{n,r_{2}}(1/x)^{n}\mbox{Log}(1/x)+\\
 &  & +(1/x)^{r_{1}}\sum_{n=0}^{\infty}a_{n,r_{1}}x^{n}\mbox{Log}^{2}(1/x)\,,
\end{array}
\end{equation}
where $a_{0,r_{1}}$ is arbitrary (e.g. $1$)
\begin{equation}
a_{1,r_{1}}=\frac{iq_{3}+2r(r+1)(r+2)}{(r+1)(r+2)(r+3)}a_{0,r_{1}}
\end{equation}
and $m=n+r_{1}$\begin{equation}
a_{n,r_{1}}=-\frac{(m-1)(m-2)}{(m+1)(m+2)}a_{n-2,r_{1}}
+\frac{iq_{3}+2m(m^{2}-1)}{(m+2)(m+1)m}a_{n-1,r_{1}}\,,
\end{equation}
whereas $b_{0,r_{2}}=\frac{2}{iq_{3}}a_{0,r_{1}}$ and $b_{1,r_{2}}$
is arbitrary while
\begin{equation}
b_{2,r_{2}}=\frac{1}{6}(iq_{3}b_{1,r_{2}}+4a_{0,r_{1}}-11a_{1,r_{1}})\,,
\end{equation}
\begin{equation}
\begin{array}{rcl}
b_{n,r_{2}} & = & -\frac{(n-2)(n-3)}{n(n+1)}b_{n-2,r_{2}}
+\frac{iq_{3}+2(n-2)(n-1)n}{n(n^{2}-1)}b_{n-1,r_{2}}
-\frac{11+3(n-4)n}{n(n^{2}-1)}a_{n-3,r_{1}}+\\
 &  & +\frac{6(n-2)n+4}{n(n^{2}-1)}a_{n-2,r_{1}}
+\frac{1-3n}{n(n^{2}-1)}a_{n-1,r_{1}}\,.
\end{array}
\end{equation}
Moreover, $c_{0,r_{3}}=-\frac{2}{iq_{3}}b_{0,r_{2}}$ and $c_{1,r_{3}}
=\frac{4(b_{0,r_{2}}+b_{1,r_{2}})}{iq_{3}}+\frac{6a_{0,r_{1}}}{iq_{3}}$
and $c_{2,r_{3}}$ is arbitrary while
\begin{equation}
c_{3,r_{3}}=\frac{1}{6}(iq_{3}c_{2,r_{3}}+2b_{0,r_{2}}+8b_{1,r_{2}}
-22b_{2,r_{2}}-12a_{0,r_{1}}+12a_{1,r_{1}})\,,
\end{equation}
\begin{equation}
\begin{array}{rcl}
c_{n,r_{3}} & = & -\frac{(n-3)(n-4)}{n(n-1)}c_{n-2,r_{3}}
+\frac{iq_{3}+2(n-3)(n-2)(n-1)}{n(n-1)(n-2)}c_{n-1,r_{3}}
+\frac{2(3n(6-n)-26)}{n(n-1)(n-2)}b_{n-3,r_{2}}+\\
 &  & +\frac{4(11-12m+3m^{2})}{n(n-1)(n-2)}b_{n-2,r_{2}}
-\frac{2(2-6m+3m^{2})}{n(n-1)(n-2)}b_{n-1,r_{2}}
-\frac{6(n-3)}{n(n-1)(n-2)}a_{n-4,r_{1}}+\frac{12}{n(n-1)}a_{n-3,r_{1}}+\\
 &  & -\frac{6}{n(n-2)}a_{n-2,r_{1}}\,.
\end{array}
\end{equation}

\subsection{Solution 
with $q_{3}=0$ and $h=2$}

In this case $a_{0,r}$ is arbitrary (e.g. $1$). For the first solution
with $r_{1}=0$ we have other coefficients $a_{n>0,r_{1}}=0$. The
second solution, with $r_{2}=-1$, has arbitrary $a_{1,r_{2}}$(we
can take it as $0$) and $a_{2,r_{2}}=0$. and third one with arbitrary
$a_{1,r_{3}}$, $a_{2,r_{3}}$ (we also set them to $0$). All these
solutions look like 
\begin{equation}
\begin{array}{rcl}
u_{1}(x) & = & (1/x)^{r_{1}}\sum_{n=0}^{\infty}a_{n,r_{1}}(1/x)^{n}
=a_{0,r_{3}}\,,\\
u_{2}(x) & = & (1/x)^{r_{2}}\sum_{n=0}^{\infty}a_{n,r_{2}}(1/x)^{n}
=xa_{0,r_{2}}\,,\\
u_{3}(x) & = & (1/x)^{r_{3}}\sum_{n=0}^{\infty}a_{n,r_{3}}(1/x)^{n}
=x^{2}a_{0,r_{3}}\,.
\end{array}
\end{equation}
Gathering together all this solutions we have
\begin{equation}
u(x)=A+B(-x)^{2}+C(x-1)^{2}\,,
\end{equation}
where $A$, $B,$ $C$ are arbitrary. As we can see this solution
corresponds to the solution with $q_{3}=0$ and arbitrary $h\not\in\{0,1\}$.
For other integer $h\not\in\{0,1\}$ we can observe the same correspondence.

\chapter{Coefficients in the eigenequations for $N=4$}
Coefficients for the eigenequations 
of $\oq{3}$ and $\oq{4}$ for $N=4$ presented in 
(\ref{eq:N4eqt}) and (\ref{eq:N4eqf})
look like
\begin{equation}
\begin{array}{rcl}
t_{0,0} & = & \frac{1}{x_{2}-1}(i(-h+s_{1}+s_{2}+s_{3}+s_{3})
((s_{2}-s_{4})(1-h+s_{1}+s_{2}-s_{3}+s_{4})+\\
& &+({s_{2}}^{2}-s_{2}(-1+h+s_{1}-s_{3}+2s_{4}-2s_{4}x_{1})+\\
& & +s_{4}(1+s_{1}-s_{3}+s_{4}
-2(1+s_{4})x_{1}+h(-1+2x_{1})))x_{2}))-q_{3}\,,
\end{array}
\end{equation}
\begin{equation}
\begin{array}{rcl}
t_{1,0} & = & -i((s_{1}+s_{2})(1-h+s_{1}+s_{2}-s_{3}+s_{4})
-(2+h-h^{2}+s_{1}(3+s_{1})-s_{2} +\\
& &+4hs_{2}-3{s_{2}}^{2}-3s_{3}-2s_{2}s_{3}+{s_{3}}^{2}+s_{4}
+{s_{4}}^{2}-2s_{1}(s_{2}+s_{3}+s_{4}(x_{1}-1))+\\
 &  & -2s_{2}s_{4}x_{1})x_{2}
-(s_{2}+s_{3}-2)(1+h+s_{1}-s_{2}-s_{3}+s_{4}-2s_{4}x_{1}){x_{2}}^{2})\,,
\end{array}
\end{equation}
\begin{equation}
\begin{array}{rcl}
t_{0,1} & = & \frac{1}{(x_{2}-1)x_{2}}(i((h-s_{1}-s_{2}+s_{3}-s_{4})(s_{1}
+s_{4}+(s_{1}+s_{2}-2)x_{1})+\\
& & +((s_{1}+s_{4})(1-2h+2s_{1}-2s_{3}+
2s_{4})-2(1+h^{2}+s_{2}+{s_{2}}^{2}-s_{3}+s_{2}s_{3}+\\
& & -{s_{3}}^{2}+(4+s_{2})s_{4}+2s_{4}+s_{1}(2+s_{2}+s_{3}+2s_{4})
-h(2+s_{1}+2s_{2}+3s_{4}))x_{1}+\\
& &-(s_{1}+s_{2}-2)(1+2s_{4}){x_{1}}^{2})
x_{2}+(-(s_{1}+s_{4})(1-h+s_{1}-s_{2}-s_{3}+s_{4})+\\ 
& &+(2+h^{2}+{s_{1}}^{2}+s_{2}+s_{3}-2(s_{2}+s_{3})^{2}
+h(-3-2s_{1}+s_{2}+s_{3}-6s_{4})+\\
 &  & +7s_{4}+(s_{3}-s_{2})s_{4}+
5{s_{4}}^{2}+s_{1}(3-s_{2}+s_{3}+6s_{4}))x_{1}+\\
& &+(-2-h^{2}-2s_{2}+{s_{2}}^{2}-3s_{3}+s_{2}s_{3}
+s_{1}(-2+s_{2}-3s_{4})+\\
& &-(7+2s_{2}+5s_{3})s_{4}-5{s_{4}}^{2}
+h(3+s_{1}+s_{3}+6s_{4})){x_{1}}^{2}){x_{2}}^{2}))\,,
\end{array}
\end{equation}
\begin{equation}
\begin{array}{rcl}
t_{2,0} & = & -\frac{1}{x_{2}(x_{2}-1)}(i(x_{1}-1)x_{1}(h-s_{1}
-s_{2}+s_{3}-s_{4}+(2-2h+3s_{1}-2s_{3}+3s_{4}+\\ 
& &+ (-4+s_{1}+s_{2}-s_{4})x_{1})x_{2}+(-2+h-2s_{1}+s_{2}
+s_{3}-2s_{4}-2hx_{1}+\\
& & +(4+s_{1}+s_{2}+2s_{3}+4s_{4})x_{1}){x_{2}}^{2}))\,,
\end{array}
\end{equation}
\begin{equation}
\begin{array}{rcl}
t_{0,2} & = & -i(x_{2}-1)x_{2}(h-2-2s_{1}-2s_{2}+s_{3}-s_{4}+\\
& &+(4+h+s_{1}-2s_{2}-2s_{3}+s_{4}-2s_{4}x_{1})x_{2})\,,
\end{array}
\end{equation}
\begin{equation}
\begin{array}{rcl}
t_{1,1} & = & -i(x_{1}(2h-3s_{2}+(2h-4(s_{2}+s_{3})+(s_{2}-2)x_{1})x_{2}+\\
 &  & +(s_{2}+s_{3}-2)(x_{1}-1){x_{2}}^{2}
+2(s_{3}+x_{2}))+s_{1}(x_{2}-1+x_{1}(x_{1}x_{2}-3))
+\\
& & +s_{4}(x_{2}-1-2x_{1}(1+2x_{1}x_{2})))\,,
\end{array}
\end{equation}

\begin{equation}
\begin{array}{rcl}
t_{3,0} & = & -i(x_{1}-1)^{2}{x_{1}}^{2}\,,
\end{array}
\end{equation}
\begin{equation}
\begin{array}{rcl}
t_{0,3} & = & -i(x_{2}-1)^{2}{x_{2}}^{2}\,,
\end{array}
\end{equation}
\begin{equation}
\begin{array}{rcl}
t_{2,1} & = & i(x_{1}-1)x_{1}(x_{2}-1+2x_{1}x_{2})\,,
\end{array}
\end{equation}
\begin{equation}
\begin{array}{rcl}
t_{1,2} & = & ix_{1}(x_{2}-1)x_{2}((x_{1}-1)x_{2}-2)
\end{array}
\end{equation}
and 
where
\begin{equation}
\begin{array}{rcl}
f_{1,0} & = & \frac{1}{(x_{2}-1)^{2}x_{2}}((-h+s_{1}+s_{2}-s_{3}+s_{4})
((s_{2}-1)(s_{1}+s_{4})-(-2+s_{1}+
\\&&+s_{2})(1+s_{4})x_{1}) 
+(-(s_{1}+s_{4})(-h(-2+s_{2})+s_{1}(s_{2}-2)+3s_{3}+\\
& &-2s_{4}+s_{2}(-2+s_{2}-3s_{3}+s_{4})) 
+(-5s_{2}+3{s_{1}}^{2}s_{2}+{s_{2}}^{2}+{s_{3}}^{3}+5s_{3}+\\
& &-2s_{2}s_{3}
+{s_{3}}^{2}-s_{2}{s_{3}}^{2}+h^{2}(-1+s_{2}-s_{4})+(-5+4{s_{2}}^{2}
+s_{2}(5+s_{3})+\\
 &  & -s_{2}(7+3s_{3}))s_{4}
+2(s_{2}-4){s_{4}}^{2}-{s_{4}}^{3}+h(5+s_{1}-4s_{1}s_{2}-2{s_{2}}^{2}+\\
 &  & +(9+s_{1}-3s_{2})s_{4}+2{s_{4}}^{2})+s_{1}(-5+s_{2}
+4{s_{2}}^{2}+6s_{2}s_{4} -3s_{3}(1+s_{4})+\\
 &  & -s_{4}(8+s_{4})))x_{1}-(-2+s_{1}+s_{2})(-h+s_{1}+s_{2}+s_{4})
(1+2s_{4}){x_{1}}^{2})x_{2}+\\
 &  & +(-(s_{1}+s_{4})(s_{1}+s_{2}-h(1+s_{2})-3s_{3}+s_{4}+s_{2}(s_{1}
+s_{2}+3s_{3}+s_{4}))+\\ & &
+(3s_{2}-5s_{3}+2({s_{2}}^{2}+s_{2}(1-s_{2})s_{3}
+(-1+s_{2}){s_{3}}^{2})+3s_{4}+2({s_{2}}^{2}+\\
& &+4s_{2}(1+s_{3})-s_{3}(4+s_{3}))s_{4}
+(6+3s_{2}-s_{3}){s_{4}}^{2}+{s_{4}}^{3}+h^{2}(1+s_{4})+\\
 &  & 
+{s_{1}}^{2}(1-2s_{2}+s_{4})+s_{1}(3-2{s_{2}}^{2}+s_{3}
+(7+s_{3})s_{4}+2{s_{4}}^{2}+s_{2}(3+\\
 & &+4s_{3}+s_{4}))
+h(-3+s_{3}+2s_{1}(-1+s_{2}-s_{4})+(-7+s_{3})s_{4}-2{s_{4}}^{2}+\\
 &  & -s_{2}(3+2s_{3}+3s_{4})))x_{1}+(2{s_{1}}^{2}s_{2}
+s_{3}+s_{2}(-3+{s_{2}}^{2}+(-2+s_{2})s_{3})+\\
 &  & +h^{2}(-1+s_{2}-2s_{4})-3s_{4}+(s_{3}-s_{2}(7+5s_{3}))s_{4}
-(7+3s_{2}+2s_{3}){s_{4}}^{2}+\\
 &  & -2{s_{4}}^{3}+h(3+s_{1}+s_{2}-3s_{1}s_{2}
-2{s_{2}}^{2}+s_{3}-s_{2}s_{3}
+2(4+s_{1}+s_{2}+s_{3})s_{4}+\\
& &+4{s_{4}}^{2})+s_{1}(3{s_{2}}^{2}+s_{2}(2s_{3}+s_{4})
-(3+2s_{3}+s_{4})(1+2s_{4}))){x_{1}}^{2}){x_{2}}^{2}+\\
& &+((s_{1}+s_{4})(-h+s_{1}+s_{2}+s_{3}+s_{4}))
-(h^{2}s_{2}+2s_{1}s_{2}+{s_{1}}^{2}s_{2}+2{s_{2}}^{2}+\\
& & +{s_{2}}^{3}-2s_{3}-s_{1}s_{3}+s_{2}s_{3}
+4s_{1}s_{2}s_{3}+2{s_{2}}^{2}s_{3}-{s_{3}}^{2}+s_{2}{s_{3}}^{2}+\\
 &  & +(5{s_{2}}^{2}-s_{3}(5+s_{1}+s_{3})+s_{2}(2+5s_{1}
+6s_{3}))s_{4}+(4s_{2}-s_{3}){s_{4}}^{2}+\\
 &  & +h(s_{3}-2s_{2}(1+s_{1}+s_{2}+s_{3})+(-5s_{2}+s_{3})s_{4}))x_{1}+\\
& &+(h^{2}s_{2}+2{s_{2}}^{2}+{s_{2}}^{3}-2s_{3}+2s_{2}s_{3}+\\
 &  & +3{s_{2}}^{2}s_{3}-{s_{3}}^{2}+2s_{2}{s_{3}}^{2}+(2s_{2}(1+2s_{2})
+5(-1+s_{2})s_{3}-2{s_{3}}^{2})s_{4}+\\ 
& & +(3s_{2}-2s_{3}){s_{4}}^{2}+s_{1}s_{2}(2+s_{2}+2s_{3}+3s_{4})+\\
& &-h(-s_{3}+s_{2}(2+s_{1}+2s_{2}+3s_{3})
+4s_{2}s_{4}-2s_{3}s_{4})){x_{1}}^{2}){x_{2}}^{3})\,,
\end{array}
\end{equation}
\begin{equation}
\begin{array}{rcl}
f_{0,0} & = & \frac{1}{(x_{2}-1)^{2}}(s_{2}s_{4}(h^{2}-h+s_{1}-4hs_{1}
+3{s_{1}}^{2}+s_{2}-2hs_{2}+4s_{1}s_{2}+{s_{2}}^{2}+s_{3}+\\
 &  & -{s_{3}}^{2}+s_{4}-2hs_{4}+4s_{1}s_{4}+2s_{2}s_{4}+{s_{4}}^{2}
+2(hs_{1}-{s_{1}}^{2}-s_{1}s_{2}
-hs_{3}+\\
& & +2s_{1}s_{3}
+s_{2}s_{3}+{s_{3}}^{2}-s_{1}s_{4}+s_{3}s_{4}+(1-h+2s_{1}+s_{2}+s_{4})\\
& &  (-h+s_{1}+s_{2}+s_{3}+s_{4})x_{1})x_{2}
+\times (-{s_{1}}^{2}-s_{1}(1+2s_{2}+4s_{3}+2s_{4})+\\
& &+2s_{1}(1+s_{2}+2s_{3}+s_{4})x_{1}+h^{2}(2x_{1}-1)+\\
& &+(s_{2}+s_{3}+s_{4})(-1-s_{2}-s_{3}-s_{4}+2(1+s_{2}
+2s_{3}+s_{4})x_{1})+\\ 
& &+h(1+2s_{1}+2s_{2}+2s_{3}+2s_{4}-2(1+s_{1}
+2s_{2}+3s_{3}+2s_{4})x_{1})){x_{2}}^{2}))-q_{4}\,,
\end{array}
\end{equation}
\begin{equation}
\begin{array}{rcl}
f_{0,1} & = & -\frac{1}{x_{2}-1}(s_{4}({s_{1}}^{2}(1+2x_{1}x_{2}
-{x_{2}}^{2})+s_{1}(1-h+2s_{2}-s_{3}+s_{4}+(-2-h+\\
 &  & +4s_{2}+3s_{3}+s_{4}+2(1-h+2s_{2}+s_{4})x_{1})x_{2}
-(1+5s_{2}+s_{3}+2s_{4}+\\
 &  & +2h(x_{1}-1)-2(-2+4s_{2}
+2s_{3}+s_{4})x_{1}){x_{2}}^{2}-(-2+s_{2}+s_{3}){x_{2}}^{3})+\\
 &  & +{s_{2}}^{2}(1+(2x_{1}-1)x_{2})(1+x_{2}(3+x_{2}))
+(1+h-s_{3}-s_{4})x_{2}\\
& & \times (h+h(2x_{1}-1)x_{2}
-(2+s_{4}-2x_{2})(1-x_{2}+2x_{1}x_{2})+\\
& & +s_{3}(1+x_{2}(-2+x_{2}-2x_{1}x_{2})))+\\
 &  & +s_{2}((1+(-1+2x_{1})x_{2})(1
+s_{4}+x_{2}+4s_{4}x_{2}+(-3+s_{4}){x_{2}}^{2})+\\
 &  & -h(1+(2x_{1}-1)x_{2}(1+x_{2}(4+x_{2}))+\\
& &+s_{3}(-1+x_{2}(3-2{x_{2}}^{2}
+4x_{1}x_{2}(2+x_{2})))))))\,,
\end{array}
\end{equation}
\begin{equation}
\begin{array}{rcl}
f_{2,0} & = & \frac{1}{(x_{2}-1)^{2}x_{2}}(x_{1}((h-s_{1}-s_{2}
+s_{3}-s_{4})(2+s_{1}-s_{2}+s_{4}+\\
 &  & +(-4+s_{1}+s_{2}-s_{4})x_{1})
+(-4h+4s_{1}-2hs_{1}+2{s_{1}}^{2}+4s_{2}+s_{1}s_{2}-{s_{2}}^{2}+\\
 &  & -6s_{3}-3s_{1}s_{3}+3s_{2}s_{3}+4s_{4}
-2hs_{4}+4s_{1}s_{4}+s_{2}s_{4}-3s_{3}s_{4}+2{s_{4}}^{2}+\\
&&-(h^{2}-s_{1}s_{2}-{s_{2}}^{2}+3s_{1}s_{3}+3s_{2}s_{3}-{s_{3}}^{2}+
5s_{1}s_{4}+4s_{2}s_{4}-3s_{3}s_{4}+5{s_{4}}^{2}+
\\ &  & +11(s_{1}+s_{2}
-s_{3}+s_{4})-h(11+s_{1}+6s_{4}))x_{1}+\\
 &  & -(-h+s_{1}+s_{2}+s_{4})(s_{1}+s_{2}-2(2+s_{4})){x_{1}}^{2})x_{2}
-({s_{1}}^{2}+2s_{1}s_{2}+{s_{2}}^{2}+\\
 &  & -3s_{1}s_{3}
+3s_{2}s_{3}+2s_{1}s_{4}+2s_{2}s_{4}-3s_{3}s_{4}+{s_{4}}^{2}-h(2
+s_{1}+s_{2}+s_{4})+\\
& &+2(s_{1}+s_{2}-3s_{3}+s_{4})+h(7+2s_{1}+3s_{2}-s_{3}+5s_{4})x_{1}+\\
 &  & -({s_{1}}^{2}+2{s_{2}}^{2}
-s_{3}(13+2s_{3})+7s_{4}-6s_{3}s_{4}+4{s_{4}}^{2}+\\
 &  & +s_{1}(7+3s_{2}+s_{3}+5s_{4})+s_{2}(7+4s_{2}+6s_{4}))x_{1}
+h^{2}(x_{1}-1)x_{1}+\\
& & -h(5+s_{1}+2s_{2}+s_{3}+4s_{4}){x_{1}}^{2}
+({s_{2}}^{2}-3s_{3}+5s_{4}-s_{3}s_{4}+3{s_{4}}^{2}+\\
& &+s_{1}(5+s_{2}
+2s_{3}+3s_{4})+s_{2}(5+3s_{3}+4s_{4})){x_{1}}^{2}){x_{2}}^{2}+\\
 &  & -(x_{1}-1)(-hs_{2}+s_{1}s_{2}+{s_{2}}^{2}-2s_{3}-s_{1}s_{3}+s_{2}s_{3}
+s_{2}s_{4}-s_{3}s_{4}+\\
&& -(s_{1}s_{2}+{s_{2}}^{2}-4s_{3}+s_{2}s_{3}
 -{s_{3}}^{2}+h(s_{3}-s_{2})+s_{2}s_{4}-3s_{3}s_{4})x_{1}){x_{2}}^{3}))\,,
\end{array}
\end{equation}
\begin{equation}
\begin{array}{rcl}
f_{0,2} & = & s_{4}x_{2}(2-h+2s_{1}+2s_{2}-s_{3}+s_{4}+(-6-s_{1}+3s_{3}+\\
& &+2(-h+2(1+s_{1}+s_{2})+s_{4})x_{1})x_{2}+\\
 &  & -(s_{1}+h(2x_{1}-1)-(2(-2+s_{2}+s_{3})+s_{4})(-1+2x_{1})){x_{2}}^{2})\,,
\end{array}
\end{equation}
\begin{equation}
\begin{array}{rcl}
f_{1,1} & = & \frac{1}{x_{2}-1}(hs_{1}-{s_{1}}^{2}-2s_{1}s_{2}+s_{1}s_{3}
+hs_{4}-2s_{1}s_{4}-2s_{2}s_{4}+s_{3}s_{4}-{s_{4}}^{2}-2hx_{1}+\\
 &  & +2s_{1}x_{1}+hs_{1}x_{1}-{s_{1}}^{2}x_{1}+2s_{2}x_{1}+hs_{2}x_{1}
-2s_{1}s_{2}x_{1}-{s_{2}}^{2}x_{1}-2s_{3}x_{1}+\\
 &  & +s_{1}s_{3}x_{1}+2s_{4}x_{1}-2hs_{4}x_{1}
+2s_{1}s_{4}x_{1}+2s_{2}s_{4}x_{1}
-2s_{3}s_{4}x_{1}+2{s_{4}}^{2}x_{1}+\\
&&-(-(s_{1}+s_{4})(2-2h+2s_{1}+3s_{2}
+3s_{3}+2s_{4})+(h^{2}+s_{1}+3s_{2}+\\
& &+4s_{1}s_{2}+2{s_{2}}^{2}-3s_{3}+3s_{1}s_{3}+3s_{2}s_{3}
 -{s_{3}}^{2}+(5+6s_{1}+10s_{2}-6s_{3})s_{4}+\\
& &+5{s_{4}}^{2}-h(3+s_{1}+3s_{2}+6s_{4}))x_{1}+((-2+s_{1}+s_{2})
(-h+s_{1}+s_{2})+\\ 
&&+(-2+4h-5s_{1}-5s_{2})s_{4}
-4{s_{4}}^{2}){x_{1}}^{2})x_{2}+(-(s_{1}+s_{4})(4-h+s_{1}+\\
 &  & -3s_{3}+s_{4})+(h^{2}+s_{1}+{s_{1}}^{2}+3s_{2}+5s_{1}s_{2}
+2{s_{2}}^{2}+s_{3}+s_{1}s_{3}-2{s_{3}}^{2}+\\
 &  & +h(-3-2s_{1}-3s_{2}+s_{3}-4s_{4})+11s_{4}+4(s_{1}+s_{2}-2s_{3})s_{4}
+3{s_{4}}^{2})x_{1}+ \\
&&+(-h^{2}+s_{1}-s_{2}-4s_{1}s_{2}-3{s_{2}}^{2}+s_{3}-2s_{1}s_{3}
-4s_{2}s_{3}-5s_{4}+\\
& &-3s_{1}s_{4}-4s_{2}s_{4}+5s_{3}s_{4}-{s_{4}}^{2}
+h(1+s_{1}+4s_{2}+s_{3}+2s_{4})){x_{1}}^{2}){x_{2}}^{2}+\\
& &+(-2+s_{2}
+s_{3})(x_{1}-1)(s_{1}+s_{4}-(-h+s_{2}+s_{3}+3s_{4})x_{1}){x_{2}}^{3})\,,
\end{array}
\end{equation}

\begin{equation}
\begin{array}{rcl}
f_{3,0} & = & \frac{1}{x_{2}(1-x_{2})}{x_{1}}^{2}(x_{1}-1)(-h+s_{1}
+s_{2}-s_{3}+s_{4}-(s_{1}+s_{2}-2s_{3}+s_{4}+\\
 &  & +h(x_{1}-1)-(s_{1}+s_{2}+s_{4})x_{1})x_{2}+s_{3}(x_{1}-1){x_{2}}^{2})\,,
\end{array}
\end{equation}
\begin{equation}
\begin{array}{rcl}
t_{0,3} & = & -s_{4}(x_{2}-1){x_{2}}^{2}(1+(2x_{1}-1)x_{2})\,,
\end{array}
\end{equation}
\begin{equation}
\begin{array}{rcl}
f_{2,1} & = & \frac{1}{x_{2}-1}(x_{1}(1-2s_{2}+s_{3}-2x_{1}+(3s_{1}
+3s_{2}-2s_{3}+s_{4})x_{1}+\\
 &  & +(s_{3}(6x_{1}-3)+x_{1}(3-4s_{1}
+s_{4}-2x_{1}+3s_{1}x_{1})+\\
 &  & +s_{2}(3+x_{1}(-5+3x_{1})))x_{2}+(x_{1}-1)(3+3s_{3}(x_{1}-1)
-s_{1}x_{1}+2s_{4}x_{1}){x_{2}}^{2}+\\
 &  & -(-2+s_{2}+s_{3})
(x_{1}-1)^{2}{x_{2}}^{3}-h(x_{2}+2x_{1}-1)(1+(x_{1}-1)x_{2})))\,,
\end{array}
\end{equation}
\begin{equation}
\begin{array}{rcl}
f_{1,2} & = & x_{2}(-(h-2s_{2}+s_{3})x_{1}+s_{1}(1+2x_{1}-x_{2})
(1+(x_{1}-1)x_{2})+\\
& & +x_{1}x_{2}(-3+3s_{3}-hx_{1}+2s_{2}x_{1}
  -(3+h-2s_{2}-2s_{3})(x_{1}-1)x_{2})+\\ 
& & + s_{4}(x_{1}-1)(x_{2}-1)
(1+(-1+3x_{1})x_{2}))\,,
\end{array}
\end{equation}
\begin{equation}
\begin{array}{rcl}
f_{3,1} & = & (x_{1}-1){x_{1}}^{2}(1+(x_{1}-1)x_{2})\,,
\end{array}
\end{equation}
\begin{equation}
\begin{array}{rcl}
f_{1,3} & = & -x_{1}(x_{2}-1){x_{2}}^{2}(1+(x_{1}-1)x_{2})\,,
\end{array}
\end{equation}
\begin{equation}
\begin{array}{rcl}
f_{2,2} & = & x_{1}(1+x_{1}(x_{2}-2)-x_{2})x_{2}(1
+(x_{1}-1)x_{2})\,,
\end{array}
\end{equation}
\begin{equation}
\begin{array}{rcl}
f_{4,0} & = & f_{0,4}=0\,.
\end{array}
\end{equation}

If we take homogeneous spins $s=s_{1}=s_{2}=\ldots s_{N}$ our equations
simplify and the coefficients for $\hat{q}_{3}$ look like
\begin{equation}
\begin{array}{rcl}
t_{0,0} & = & \frac{2i}{1-x_{2}}(h-1)(h-4s)s(x_{1}-1)x_{2}
\end{array}-q_{3}\,,
\end{equation}
\begin{equation}
\begin{array}{rcl}
t_{1,0} & = & \frac{1}{x_{2}(x_{2}-1)}(i(2(h-2s)(s+(s-1)x_{1})-2(-s(1-2h+2s)+\\
 &  & +(-1+h-4s)(-1+h-2s)x_{1}+(s-1)(1+2s){x_{1}}^{2})x_{2}+\\
 &  & -((h-2)(h-1)(x_{1}-1)x_{1}+4s^{2}x_{1}(3x_{1}-1)-2s(h-1+\\
 &  & -3(h-2)x_{1}+(4h-7){x_{1}}^{2})){x_{2}}^{2}))\,,
\end{array}
\end{equation}
\begin{equation}
\begin{array}{rcl}
t_{0,1} & = & -i((1+h)x_{2}(-2+h+2x_{2})+4s^{2}(1+x_{2})(1+x_{1}x_{2})
-2s(-1+h+\\
 &  & +2hx_{2}+(1+h+2x_{1}){x_{2}}^{2}))\,,
\end{array}
\end{equation}
\begin{equation}
\begin{array}{rcl}
t_{2,0} & = & \frac{i}{x_{2}(x_{2}-1)}(x_{1}-1)x_{1}
(-h+2s+2(-1+h-2s+2x_{1})x_{2}+\\
 &  & +(2-h+2s+2(h-2-4s)x_{1}){x_{2}}^{2})\,,
\end{array}
\end{equation}
\begin{equation}
t_{0,2}=-i(x_{2}-1)x_{2}(-2+h-4s+(h-2(s-2+sx_{1}))x_{2})\,,
\end{equation}
\begin{equation}
\begin{array}{rcl}
t_{1,1} & = & -2i(x_{1}(1+x_{2})(h+x_{2}-x_{1}x_{2})+s(-1+x_{2}+\\
 &  & +{x_{1}}^{2}(x_{2}-1)x_{2}-x_{1}(1+x_{2})(3+x_{2})))\,,
\end{array}
\end{equation}
\begin{equation}
\begin{array}{rcl}
t_{3,0} & = & -i(x_{1}-1)^{2}{x_{1}}^{2}\,,
\end{array}
\end{equation}

\begin{equation}
\begin{array}{rcl}
t_{0,3} & = & -i(x_{2}-1)^{2}{x_{2}}^{2}\,,
\end{array}
\end{equation}
\begin{equation}
\begin{array}{rcl}
t_{2,1} & = & i(x_{1}-1)x_{1}(2x_{1}x_{2}+x_{2}-1)\,,
\end{array}
\end{equation}
\begin{equation}
\begin{array}{rcl}
t_{1,2} & = & ix_{1}(x_{2}-1)x_{2}(x_{2}(x_{1}-1)-2)
\end{array}
\end{equation}
and for $\hat{q}_{4}$
\begin{equation}
\begin{array}{rcl}
f_{0,0} & = & \frac{1}{(x_{2}-1)^{2}}(s^{2}(h^{2}(1+x_{2})
(1+(2x_{1}-1)x_{2}) +\\
& &-h(1+8s)(1+x_{2})(1+(2x_{1}-1)x_{2})
+2s(2(1+x_{2})(1-x_{2}+2x_{1}x_{2})+\\ 
& &+s(7+x_{2}(2-9x_{2}+16x_{1}(x_{2}+1))))))-q_{4}\,,
\end{array}
\end{equation}
\begin{equation}
\begin{array}{rcl}
f_{1,0} & = & \frac{1}{(x_{2}-1)^{2}x_{2}}(h^{2}x_{1}x_{2}(-1
+(x_{1}-1)(-1+s(x_{2}-1))x_{2})+h(x_{1}(3x_{2}-2)\\
 &  & \times(1+(x_{1}-1)x_{2})-2s^{2}(x_{1}-1)(x_{2}-1)(1+2x_{1}x_{2}
+(4x_{1}-1){x_{2}}^{2})+\\
 &  & +s(2+x_{2}(2(x_{2}-2)+x_{1}(x_{2}-10)(x_{2}-1)-{x_{1}}^{2}(2
+(x_{2}-11)x_{2}))))+\\
 &  & +2s(2s^{2}(x_{1}-1)(x_{2}-1)(1+x_{2}+3x_{1}x_{2}
+(5x_{1}-2){x_{2}}^{2})+\\
 &  & +s(-2+x_{2}(3+10x_{1}(x_{2}-1)-{x_{2}}^{2}+{x_{1}}^{2}(3
+(x_{2}-12)x_{2})))+\\
 &  & +x_{1}(2+x_{2}(-5+x_{2}(2+x_{2})-x_{1}(-3+x_{2}(4+x_{2}))))))\,,
\end{array}
\end{equation}
\begin{equation}
\begin{array}{rcl}
f_{0,1} & = & \frac{1}{x_{2}-1}(s(-(1+h)x_{2}(h-2+2x_{2})(1+(2x_{1}-1)x_{2})
-4s^{2}(1+x_{2}(4+x_{2}\\
 &  & \times(3+2x_{2})+x_{1}(3+x_{2})(1+3x_{2})))
+2s(-1+h(1+(2x_{1}-1)x_{2})(1+\\
 &  & +x_{2}(4+x_{2}))+x_{2}(-1+(7-5x_{2})x_{2}
+2x_{1}(-1+x_{2}(4x_{2}-1))))))\,,
\end{array}
\end{equation}
\begin{equation}
\begin{array}{rcl}
f_{2,0} & = & \frac{1}{(x_{2}-1)^{2}x_{2}}(x_{1}(-4s-h^{2}x_{1}x_{2}(1
+(x_{1}-1)x_{2})+2s(3x_{2}-{x_{2}}^{3}+\\
 &  & -2{x_{1}}^{2}x_{2}(x_{2}(3+x_{2}))+x_{1}(x_{2}-1))(-4+x_{2}(7+3x_{2}))+\\
 &  & -s(-(x_{2}-1)^{3}+8{x_{1}}^{2}{x_{2}}^{2}+x_{1}(1+x_{2}(7
+(x_{2}-9)x_{2}))))+\\
 &  & +h((1+(x_{1}-1)x_{2})(2-2x_{2}+x_{1}(5x_{2}-4))+s(-(x_{2}-1)^{3}+\\
 &  & +8{x_{1}}^{2}{x_{2}}^{2}+x_{1}(1+x_{2}(7+(x_{2}-9)x_{2}))))))\,,
\end{array}
\end{equation}
\begin{equation}
\begin{array}{rcl}
f_{0,2} & = & sx_{2}(-h(1+x_{2})(1+(2x_{1}-1)x_{2})+2(1-3x_{2}
+2x_{2}(x_{1}+x_{2}-2x_{1}x_{2})+\\
 &  & +s(2+x_{2}(1-3x_{2}+5x_{1}(1+x_{2})))))\,,
\end{array}
\end{equation}
\begin{equation}
\begin{array}{rcl}
f_{1,1} & = & \frac{1}{(x_{2}-1)}(-h^{2}x_{1}x_{2}(1+(x_{1}-1)x_{2})
+2s(-3s+2x_{1}+sx_{1}+(2+3(x_{1}-1)x_{1}+\\
 &  & +s(4+x_{1}(5x_{1}-13)))x_{2}-(2(2+(x_{1}-4)x_{1})
+s(-1+x_{1}(8x_{1}-5))){x_{2}}^{2}+\\
 &  & -(s-1)(x_{1}-1)(5x_{1}-2){x_{2}}^{3})+h(-x_{1}(1+(x_{1}-1)x_{2})(2
+x_{2}(2x_{2}-1))+\\
 &  & +2s(1+x_{2}(-2+x_{2}+x_{1}(5-x_{1}+4(x_{1}-1)x_{2}
+(x_{1}-1){x_{2}}^{2})))))\,,
\end{array}
\end{equation}

\begin{equation}
\begin{array}{rcl}
f_{3,0} & = & \frac{1}{x_{2}(x_{2}-1)}(x_{1}-1){x_{1}}^{2}(h
+h(x_{1}-1)x_{2}+s(-2+x_{2}(1+x_{2}-x_{1}(3+x_{2}))))\,,
\end{array}
\end{equation}
\begin{equation}
\begin{array}{rcl}
f_{0,3} & = & -s(x_{2}-1){x_{2}}^{2}(1+(2x_{1}-1)x_{2})\,,
\end{array}
\end{equation}
\begin{equation}
\begin{array}{rcl}
f_{2,1} & = & \frac{1}{x_{2}-1}(x_{1}(1-s-2x_{1}+5sx_{1}
+x_{1}(3-2x_{1}+s(6x_{1}-2))x_{2}+\\
 &  & +(x_{1}-1)(3+s(4x_{1}-3)){x_{2}}^{2}-2(s-1)(x_{1}-1)^{2}{x_{2}}^{3}+\\
 &  & -h(x_{2}+2x_{1}-1)(1+(x_{1}-1)x_{2})))\,,
\end{array}
\end{equation}
\begin{equation}
\begin{array}{rcl}
f_{1,2} & = & x_{2}(-x_{1}(h+(3+h)x_{2})(1+(x_{1}-1)x_{2})+s(2+2x_{1}-4x_{2}+\\
 &  & x_{1}(7+x_{1})x_{2}+(x_{1}-1)(7x_{1}-2){x_{2}}^{2}))\,,
\end{array}
\end{equation}
\begin{equation}
\begin{array}{rcl}
f_{3,1} & = & (x_{1}-1){x_{1}}^{2}(1+(x_{1}-1)x_{2})\,,
\end{array}
\end{equation}
\begin{equation}
\begin{array}{rcl}
f_{1,3} & = & -x_{1}(x_{2}-1){x_{2}}^{2}(1+(x_{1}-1)x_{2})\,,
\end{array}
\end{equation}
\begin{equation}
\begin{array}{rcl}
f_{2,2} & = & x_{1}(1+x_{1}(x_{2}-2)-x_{2})x_{2}(1+(x_{1}-1)x_{2})\,,
\end{array}
\end{equation}
\begin{equation}
\begin{array}{rcl}
f_{4,0} & = & f_{0,4}=0\,.
\end{array}
\end{equation}




\newpage
\begin{thebibliography}{10}

\bibitem{gell}
{M. Gell-Mann, M.L. Goldberger, F.E. Low, E. Marx, F.Zachariasen}.
\newblock {Elementary particles of Conventional Field Theory as Regge poles.
  III}.
\newblock {\em Phys. Rev. B}, 133:145, 1964.

\bibitem{gell2}
{M. Gell-Mann, M.L. Goldberger, F.E. Low, E. Marx, F.Zachariasen}.
\newblock {Elementary particles of Conventional Field Theory as Regge poles.
  IV}.
\newblock {\em Phys. Rev. B}, 133:161, 1964.

\bibitem{Gribov:1968fc}
V.~N. Gribov.
\newblock {A Reggeon diagram technique}.
\newblock {\em Sov. Phys. JETP}, 26:414--422, 1968.

\bibitem{Fadin:1975cb}
V.~S. Fadin, E.~A. Kuraev, and L.~N. Lipatov.
\newblock {On the Pomeranchuk singularity in asymptotically free theories}.
\newblock {\em Phys. Lett.}, B60:50--52, 1975.

\bibitem{Bartels:1977hz}
J.~Bartels.
\newblock {High-Energy behavior in a nonabelian gauge field theory}.
\newblock {\em Phys. Lett.}, B68:258, 1977.

\bibitem{Cheng:1977gt}
H.~Cheng and C.~Y. Lo.
\newblock {High-energy amplitudes of Yang-Mills theory in arbitrary
  perturbative orders. 1}.
\newblock {\em Phys. Rev.}, D15:2959, 1977.

\bibitem{Bartels:1980pe}
J.~Bartels.
\newblock {High-energy behavior in a nonabelian guage theory. 2. First
  corrections to $T({n \rightarrow m})$ beyond the leading $\ln s$
  approximation}.
\newblock {\em Nucl. Phys.}, B175:365, 1980.

\bibitem{Kwiecinski:1980wb}
J.~Kwiecinski and M.~Praszalowicz.
\newblock {Three gluon integral equation and odd C singlet Regge singularities
  in QCD}.
\newblock {\em Phys. Lett.}, B94:413, 1980.

\bibitem{Jaroszewicz:1980mq}
T.~Jaroszewicz.
\newblock {Infrared divergences and Regge behavior in QCD}.
\newblock {\em Acta Phys. Polon.}, B11:965, 1980.

\bibitem{'tHooft:1973jz}
G.~'t~Hooft.
\newblock {A planar diagram theory for strong interactions}.
\newblock {\em Nucl. Phys.}, B72:461, 1974.

\bibitem{Lipatov:1990zb}
L.~N. Lipatov.
\newblock {Pomeron and odderon in QCD and a two-dimensional conformal field
  theory}.
\newblock {\em Phys. Lett.}, B251:284--287, 1990.

\bibitem{Lipatov:1993qn}
L.~N. Lipatov.
\newblock {High-energy asymptotics of multicolor QCD and two- dimensional
  conformal field theories}.
\newblock {\em Phys. Lett.}, B309:394--396, 1993.

\bibitem{Lipatov:1993yb}
L.~N. Lipatov.
\newblock High-energy asymptotics of multicolor qcd and exactly solvable
  lattice models.
\newblock {\em JETP Lett.}, 59:596--599, 1994.

\bibitem{Faddeev:1994zg}
L.~D. Faddeev and G.~P. Korchemsky.
\newblock {High-energy QCD as a completely integrable model}.
\newblock {\em Phys. Lett.}, B342:311--322, 1995.

\bibitem{Balitsky:1978ic}
L.~N.~Lipatov I.~I.~Balitsky.
\newblock {The Pomeranchuk singularity in Quantum Chromodynamics}.
\newblock {\em Sov. J. Nucl. Phys.}, 28:822--829, 1978.

\bibitem{Kuraev:1977fs}
E.~A. Kuraev, L.~N. Lipatov, and V.~S. Fadin.
\newblock {The Pomeranchuk singularity in nonabelian gauge theories}.
\newblock {\em Sov. Phys. JETP}, 45, 1977.

\bibitem{Jaroszewicz:1980rw}
T.~Jaroszewicz.
\newblock {High-energy multi - gluon exchange amplitudes}.
\newblock Triest preprint IC/80/175.

\bibitem{Janik:1998xj}
R.~A. Janik and J.~Wosiek.
\newblock {Solution of the odderon problem}.
\newblock {\em Phys. Rev. Lett.}, 82:1092--1095, 1999.

\bibitem{Gauron:1987jt}
P.~Gauron, B.~Nicolescu, and L.~Szymanowski.
\newblock {A possible field theoretical description of the odderon}.
\newblock IPNO/TH 87-53.

\bibitem{Lukaszuk:1973nt}
L.~Lukaszuk and B.~Nicolescu.
\newblock {A possible interpretation of p p rising total cross- sections}.
\newblock {\em Nuovo Cim. Lett.}, 8:405--413, 1973.

\bibitem{Derkachov:2002pb}
S.~E. Derkachov, G.~P. Korchemsky, and A.~N. Manashov.
\newblock {Noncompact Heisenberg spin magnets from high-energy QCD. III:
  Quasiclassical approach}.
\newblock {\em Nucl. Phys.}, B661:533--576, 2003.

\bibitem{Korchemsky:2001nx}
G.~P. Korchemsky, J.~Kotanski, and A.~N. Manashov.
\newblock {Compound states of reggeized gluons in multi-colour QCD as ground
  states of noncompact Heisenberg magnet}.
\newblock {\em Phys. Rev. Lett.}, 88:122002, 2002.

\bibitem{Derkachov:2002wz}
S.~E. Derkachov, G.~P. Korchemsky, J.~Kotanski, and A.~N. Manashov.
\newblock {Noncompact Heisenberg spin magnets from high-energy QCD. II:
  Quantization conditions and energy spectrum}.
\newblock {\em Nucl. Phys.}, B645:237--297, 2002.

\bibitem{Jaroszewicz:1982gr}
T.~Jaroszewicz.
\newblock {Gluonic Regge singularities and anomalous dimensions in QCD}.
\newblock {\em Phys. Lett.}, B116:291, 1982.

\bibitem{Lipatov:1985uk}
L.~N. Lipatov.
\newblock {The bare Pomeron in Quantum Chromodynamics}.
\newblock {\em Sov. Phys. JETP}, 63:904--912, 1986.

\bibitem{Korchemsky:2003rc}
G.~P. Korchemsky, J.~Kotanski, and A.~N. Manashov.
\newblock {Multi-Reggeon compound states and resummed anomalous dimensions in
  QCD}.
\newblock {\em Phys. Lett.}, B583:121--133, 2004.

\bibitem{DeVega:2001pu}
H.~J. De~Vega and L.~N. Lipatov.
\newblock {Interaction of reggeized gluons in the Baxter-Sklyanin
  representation}.
\newblock {\em Phys. Rev.}, D64:114019, 2001.

\bibitem{deVega:2002im}
H.~J. De~Vega and L.~N. Lipatov.
\newblock {Exact resolution of the Baxter equation for reggeized gluon
  interactions}.
\newblock {\em Phys. Rev.}, D66:074013, 2002.

\bibitem{Braun:1994ll}
{M. A. Braun}.
\newblock {\em {The interaction of reggeized gluons and Lipatov's hard pomeron.
  -- {\rm lectures}}}.
\newblock University of Santiago de Compostela, 17506 Santiago de Compostela,
  Spain, 1994.

\bibitem{Baxter}
{R. J. Baxter}.
\newblock {\em {Exactly Solved Models in Statistical Mechanics}}.
\newblock Academimc Press, London, 1982.

\bibitem{Kuraev:1976ge}
E.~A. Kuraev, L.~N. Lipatov, and V.~S. Fadin.
\newblock {Multi - Reggeon processes in the Yang-Mills theory}.
\newblock {\em Sov. Phys. JETP}, 44:443--450, 1976.

\bibitem{Bartels:1973pn}
J.~Bartels.
\newblock {A field - theoretic study of the Regge - eikonal model. 1}.
\newblock {\em Ann. Phys.}, 94:1, 1975.

\bibitem{Kaidalov:1982xg}
A.~B. Kaidalov.
\newblock {The quark - gluon structure of the Pomeron and the rise of inclusive
  spectra at high-energies}.
\newblock {\em Phys. Lett.}, B116:459, 1982.

\bibitem{Kovchegov:2002qu}
Yu.~V. Kovchegov.
\newblock {High energy QCD and the large $N_c$ limit}.
\newblock 2002.
\newblock hep-ph/0202238.

\bibitem{Bartels:1993ke}
J.~Bartels and M.~G. Ryskin.
\newblock {Absorptive corrections to structure functions at small $x$}.
\newblock {\em Z. Phys.}, C60:751--756, 1993.

\bibitem{Bartels:1994jj}
J.~Bartels and M.~Wusthoff.
\newblock {The triple Regge limit of diffractive dissociation in deep inelastic
  scattering}.
\newblock {\em Z. Phys.}, C66:157--180, 1995.

\bibitem{Mueller:1996hm}
A.~H. Mueller.
\newblock {Limitations on using the operator product expansion at small values
  of $x$}.
\newblock {\em Phys. Lett.}, B396:251--256, 1997.

\bibitem{Derkachov:2001yn}
S.~E. Derkachov, G.~P. Korchemsky, and A.~N. Manashov.
\newblock {Noncompact Heisenberg spin magnets from high-energy QCD. I: Baxter
  Q-operator and separation of variables}.
\newblock {\em Nucl. Phys.}, B617:375--440, 2001.

\bibitem{CFT}
{P. Di Francesco, P. Mathieu, D. S\'en\'echal}.
\newblock {\em {Conformal Field Theory}}.
\newblock Springer, New York, 1982.

\bibitem{Zuber:1995rj}
J.~B. Zuber.
\newblock {An introduction to Conformal Field Theory}.
\newblock {\em Acta Phys. Polon.}, B26:1785--1813, 1995.

\bibitem{Sklyanin:1991ss}
E.~K. Sklyanin.
\newblock {Quantum inverse scattering method. Selected topics}.
\newblock 1991.
\newblock hep-th/9211111.

\bibitem{Faddeev:1994nk}
L.~D. Faddeev.
\newblock {Algebraic aspects of Bethe Ansatz}.
\newblock {\em Int. J. Mod. Phys.}, A10:1845--1878, 1995.

\bibitem{Faddeev:1996iy}
L.~D. Faddeev.
\newblock {How Algebraic Bethe Ansatz works for integrable model}.
\newblock 1996.
\newblock hep-th/9605187.

\bibitem{Faddeev:1979gh}
L.~D. Faddeev, E.~K. Sklyanin, and L.~A. Takhtajan.
\newblock {The quantum inverse problem method. 1}.
\newblock {\em Theor. Math. Phys.}, 40:688--706, 1980.

\bibitem{Ewerz:2003xi}
C.~Ewerz.
\newblock {The odderon in Quantum Chromodynamics}.
\newblock 2003.
\newblock hep-ph/0306137.

\bibitem{Korchemsky:1995be}
G.~P. Korchemsky.
\newblock {Quasiclassical QCD Pomeron}.
\newblock {\em Nucl. Phys.}, B462:333--388, 1996.

\bibitem{Korchemsky:1997ve}
G.~P. Korchemsky.
\newblock {WKB quantization of Reggeon compound states in high-energy QCD}.
\newblock 1997.
\newblock hep-ph/9801377.

\bibitem{Wosiek:1996bf}
J.~Wosiek and R.~A. Janik.
\newblock {Solution of the odderon problem for arbitrary conformal weights}.
\newblock {\em Phys. Rev. Lett.}, 79:2935--2938, 1997.

\bibitem{Praszalowicz:1998pz}
M.~Praszalowicz and A.~Rostworowski.
\newblock {Spectrum of the odderon charge for arbitrary conformal weights}.
\newblock {\em Acta Phys. Polon.}, B30:349--357, 1999.

\bibitem{Kotanski:2001iq}
J.~Kotanski and M.~Praszalowicz.
\newblock {Solutions of the quantization conditions for the odderon charge
  $q_3$ and conformal weight h}.
\newblock {\em Acta Phys. Polon.}, B33:657--682, 2002.

\bibitem{Lipatov:1998as}
L.~N. Lipatov.
\newblock {Duality symmetry of Reggeon interactions in multicolour QCD}.
\newblock {\em Nucl. Phys.}, B548:328--362, 1999.

\bibitem{Navelet:1997xn}
H.~Navelet and R.~Peschanski.
\newblock {Conformal invariance and the exact solution of BFKL equations}.
\newblock {\em Nucl. Phys.}, B507:353--366, 1997.

\bibitem{Bartels:1999yt}
J.~Bartels, L.~N. Lipatov, and G.~P. Vacca.
\newblock {A new odderon solution in perturbative QCD}.
\newblock {\em Phys. Lett.}, B477:178--186, 2000.

\bibitem{Bartels:2001hw}
J.~Bartels, M.~A. Braun, D.~Colferai, and G.~P. Vacca.
\newblock {Diffractive $\eta_c$ photo- and electroproduction with the
  perturbative QCD odderon}.
\newblock {\em Eur. Phys. J.}, C20:323--331, 2001.

\bibitem{Vacca:2000bk}
G.~P. Vacca.
\newblock {Properties of a family of n reggeized gluon states in multicolour
  QCD}.
\newblock {\em Phys. Lett.}, B489:337--344, 2000.

\bibitem{Kovchegov:2003dm}
Yu.~V. Kovchegov, L.~Szymanowski, and S.~Wallon.
\newblock {Perturbative odderon in the dipole model}.
\newblock {\em Phys. Lett.}, B586:267--281, 2004.

\bibitem{Korchemsky:1994um}
G.~P. Korchemsky.
\newblock {Bethe ansatz for QCD pomeron}.
\newblock {\em Nucl. Phys.}, B443:255--304, 1995.

\bibitem{Korchemsky:1999is}
G.~P. Korchemsky and J.~Wosiek.
\newblock {New representation for the odderon wave function}.
\newblock {\em Phys. Lett.}, B464:101--110, 1999.

\bibitem{Engel:1997cg}
R.~Engel, D.~Yu. Ivanov, R.~Kirschner, and L.~Szymanowski.
\newblock {Diffractive meson production from virtual photons with odd
  charge-parity exchange}.
\newblock {\em Eur. Phys. J.}, C4:93--99, 1998.

\bibitem{Czyzewski:1996bv}
J.~Czyzewski, J.~Kwiecinski, L.~Motyka, and M.~Sadzikowski.
\newblock {Exclusive $\eta_c$ photo- and electroproduction at HERA as a
  possible probe of the odderon singularity in QCD}.
\newblock {\em Phys. Lett.}, B398:400--406, 1997.

\bibitem{Bartels:2003zu}
J.~Bartels, M.~A. Braun, and G.~P. Vacca.
\newblock {The process $\gamma^* + p \to \eta_c + X$: A test for the
  perturbative QCD odderon}.
\newblock {\em Eur. Phys. J.}, C33:511--521, 2004.

\bibitem{Korchemsky:1996kh}
G.~P. Korchemsky.
\newblock {Integrable structures and duality in high-energy QCD}.
\newblock {\em Nucl. Phys.}, B498:68--100, 1997.

\bibitem{Bender:1969si}
C.~M. Bender and T.~T. Wu.
\newblock {Anharmonic oscillator}.
\newblock {\em Phys. Rev.}, 184:1231--1260, 1969.

\bibitem{Turbiner:1987kt}
A.~V. Turbiner and A.~G. Ushveridze.
\newblock {Spectral singularities and quasiexactly solvable quantal problem}.
\newblock {\em Phys. Lett.}, A126:181, 1987.
\newblock ITEP-87-55.

\bibitem{Bender:1992bk}
C.~M. Bender and A.~V. Turbiner.
\newblock {Analytic continuation of eigenvalue problems}.
\newblock {\em Phys. Lett.}, A173:442, 1993.
\newblock WU-HEP-92-13.

\bibitem{Braun:1999te}
V.~M. Braun, S.~E. Derkachov, G.~P. Korchemsky, and A.~N. Manashov.
\newblock Baryon distribution amplitudes in {QCD}.
\newblock {\em Nucl. Phys.}, B553:355--426, 1999.

\bibitem{Belitsky:1999ru}
A.~V. Belitsky.
\newblock {Integrability and WKB solution of twist-three evolution equations}.
\newblock {\em Nucl. Phys.}, B558:259--284, 1999.

\bibitem{Takhtajan:1979iv}
L.~A. Takhtajan and L.~D. Faddeev.
\newblock The quantum method of the inverse problem and the heisenberg xyz
  model.
\newblock {\em Russ. Math. Surveys}, 34:11--68, 1979.

\bibitem{KBI}
N.~M. Bogolyubov, A.~G. Izergin, and V.~E. Korepin.
\newblock {\em {Quantum inverse scattering method and correlation functions}}.
\newblock Univ. Press, Cambridge, 1993.

\bibitem{Levin:2001cv}
E.~Levin and K.~Tuchin.
\newblock {Nonlinear evolution and saturation for heavy nuclei in DIS}.
\newblock {\em Nucl. Phys.}, A693:787--798, 2001.

\bibitem{Bartels:1992ym}
J.~Bartels.
\newblock {Unitarity corrections to the Lipatov pomeron and the small $x$
  region in deep inelastic scattering in QCD}.
\newblock {\em Phys. Lett.}, B298:204--210, 1993.

\bibitem{Bartels:1993ih}
J.~Bartels.
\newblock Unitarity corrections to the lipatov pomeron and the four gluon
  operator in deep inelastic scattering in qcd.
\newblock {\em Z. Phys.}, C60:471--488, 1993.

\bibitem{Schafer:1991na}
A.~Schafer, L.~Mankiewicz, and O.~Nachtmann.
\newblock {Double diffractive $J/\psi$ and phi production as a probe for the
  odderon}.
\newblock {\em Phys. Lett.}, B272:419--424, 1991.

\bibitem{Barakhovsky:1991ra}
V.~V. Barakhovsky, I.~R. Zhitnitsky, and A.~N. Shelkovenko.
\newblock {Odderon: A Sharp signal at HERA}.
\newblock {\em Phys. Lett.}, B267:532--534, 1991.

\bibitem{Ginzburg:1993gy}
I.~F. Ginzburg, D.~Yu. Ivanov, and V.~G. Serbo.
\newblock {Pomeron and odderon in photon initiated reactions and jet production
  in gamma gamma collisions}.
\newblock {\em Phys. Atom. Nucl.}, 56:1474--1480, 1993.

\bibitem{Ginzburg:1991hd}
I.~F. Ginzburg and D.~Yu. Ivanov.
\newblock {Semihard production of tensor mesons in gamma gamma collisions and
  the perturbative odderon}.
\newblock {\em Nucl. Phys. Proc. Suppl.}, 25B:224--233, 1992.

\bibitem{Motyka:1998kb}
L.~Motyka and J.~Kwiecinski.
\newblock {Possible probe of the {QCD} odderon singularity through the
  quasidiffractive eta/c production in gamma gamma collisions}.
\newblock {\em Phys. Rev.}, D58:117501, 1998.

\bibitem{Dosch:2002ai}
H.~G. Dosch, C.~Ewerz, and V.~Schatz.
\newblock {The odderon in high energy elastic p p scattering}.
\newblock {\em Eur. Phys. J.}, C24:561--571, 2002.

\bibitem{Bartels:2004ef}
J.~Bartels, L.~N. Lipatov, and G.~P. Vacca.
\newblock {Interactions of Reggeized gluons in the Moebius representation}.
\newblock {\em Nucl. Phys.}, B706:391--410, 2005.

\bibitem{Balitsky:1995ub}
I.~Balitsky.
\newblock {Operator expansion for high-energy scattering}.
\newblock {\em Nucl. Phys.}, B463:99--160, 1996.

\bibitem{Kovchegov:1999yj}
Yu.~V. Kovchegov.
\newblock {Small-$x$ $F_2$ structure function of a nucleus including multiple
  pomeron exchange}s.
\newblock {\em Phys. Rev.}, D60:034008, 1999.

\bibitem{Kovchegov:1999ua}
Yu.~V. Kovchegov.
\newblock {Unitarization of the BFKL pomeron on a nucleus}.
\newblock {\em Phys. Rev.}, D61:074018, 2000.

\bibitem{Levin:1999mw}
E.~Levin and K.~Tuchin.
\newblock {Solution to the evolution equation for high parton density QCD}.
\newblock {\em Nucl. Phys.}, B573:833--852, 2000.

\bibitem{Levin:2000mv}
E.~Levin and K.~Tuchin.
\newblock {New scaling at high energy DIS}.
\newblock {\em Nucl. Phys.}, A691:779--790, 2001.

\bibitem{Mueller:1993rr}
A.~H. Mueller.
\newblock {Soft gluons in the infinite momentum wave function and the BFKL
  pomeron}.
\newblock {\em Nucl. Phys.}, B415:373--385, 1994.

\bibitem{Mueller:1994jq}
A.~H. Mueller and B.~Patel.
\newblock {Single and double BFKL pomeron exchange and a dipole picture of
  high-energy hard processes}.
\newblock {\em Nucl. Phys.}, B425:471--488, 1994.

\bibitem{Mueller:1994gb}
A.~H. Mueller.
\newblock {Unitarity and the BFKL pomeron}.
\newblock {\em Nucl. Phys.}, B437:107--126, 1995.

\bibitem{Chen:1995pa}
Z.~Chen and A.~H. Mueller.
\newblock {The Dipole picture of high-energy scattering, the BFKL equation and
  many gluon compound states}.
\newblock {\em Nucl. Phys.}, B451:579--604, 1995.

\bibitem{McLerran:1993ka}
L.~D. McLerran and R.~Venugopalan.
\newblock {Gluon distribution functions for very large nuclei at small
  transverse momentum}.
\newblock {\em Phys. Rev.}, D49:3352--3355, 1994.

\bibitem{McLerran:1993ni}
L.~D. McLerran and R.~Venugopalan.
\newblock {Computing quark and gluon distribution functions for very large
  nuclei}.
\newblock {\em Phys. Rev.}, D49:2233--2241, 1994.

\bibitem{McLerran:1994vd}
L.~D. McLerran and R.~Venugopalan.
\newblock {Green's functions in the color field of a large nucleus}.
\newblock {\em Phys. Rev.}, D50:2225--2233, 1994.

\bibitem{Hatta:2005as}
Y.~Hatta, E.~Iancu, K.~Itakura, and L.~McLerran.
\newblock Odderon in the color glass condensate.
\newblock 2005.
\newblock hep-ph/0501171.

\bibitem{Balitsky:1998ya}
I.~Balitsky.
\newblock {Factorization and high-energy effective action}.
\newblock {\em Phys. Rev.}, D60:014020, 1999.

\bibitem{Balitsky:2001gj}
I.~Balitsky.
\newblock {High-energy QCD and Wilson lines}.
\newblock 2001.
\newblock hep-ph/0101042.

\bibitem{Fadin:1998py}
V.~S. Fadin and L.~N. Lipatov.
\newblock {BFKL pomeron in the next-to-leading approximation}.
\newblock {\em Phys. Lett.}, B429:127--134, 1998.

\bibitem{Levin:1992mu}
E.~M. Levin, M.~G. Ryskin, and A.~G. Shuvaev.
\newblock {Anomalous dimension of the twist four gluon operator and pomeron
  cuts in deep inelastic scattering}.
\newblock {\em Nucl. Phys.}, B387:589--616, 1992.

\bibitem{Gorsky:2002ju}
A.~Gorsky, I.~I. Kogan, and G.~Korchemsky.
\newblock {High energy QCD: Stringy picture from hidden integrability}.
\newblock {\em JHEP}, 05:053, 2002.

\bibitem{Staruszkiewicz93}
{A. Staruszkiewicz}.
\newblock {\em {Algebra i Geometria}}.
\newblock NKF, Krak\'ow, 1993.

\end{thebibliography}
\end{document}